\documentclass{JINST}

\usepackage{graphicx}
\usepackage{amsmath}
\usepackage{amssymb}

\title{\Large{Liquid noble gas detectors for low energy particle physics}}

\author{Vitaly Chepel$^a$\thanks{Corresponding author.} ~and Henrique
Ara\'{u}jo$^b$\\ 
\llap{$^a$}LIP--Coimbra \& Department of Physics,
University of Coimbra, \\
P-3004-516 Coimbra, Portugal\\ 
Email: \email{vitaly@coimbra.lip.pt}\\
\llap{$^b$}High Energy Physics Group, Imperial College London,\\
Blackett Laboratory, London SW7 2AZ, United Kingdom\\ 
}

\abstract{We review the current status of liquid noble gas radiation
detectors with energy threshold in the keV range, which are of interest
for direct dark matter searches, measurement of coherent neutrino
scattering and other low energy particle physics experiments.
Emphasis is given to the operation principles and the most important
instrumentation aspects of these detectors, principally of those
operated in the double-phase mode. Recent technological advances and
relevant developments in photon detection and charge readout are
discussed in the context of their applicability to those experiments.}


\begin{document}

\newpage
\section{Introduction}
\label{sec:Introduction}

Liquefied noble gases have attracted the attention of experimental
physicists since the middle of the 20th century
\cite{Hutchinson48,Marshall53,Northrop56,Northrop58}. The unique
combination of their scintillation properties with the fact that
electrons released in the ionization process can remain free to drift
across long distances favorably distinguishes these liquids from other
dense detector media. Another notable property of the `noble liquids'
is the possibility of extracting electrons to the gas phase, where the
ionization signal can be amplified through secondary scintillation or
avalanche mechanisms. These properties have been extensively studied
over the years and significant progress has been made in understanding
the underlying physics as well as on development of the associated
technologies, notably gas purification, material cleaning, cooling,
photon and charge detection, and low noise electronics. A wide
spectrum of applications has been considered involving both precise
particle tracking (owing to the low electron diffusivity) and
calorimetry/spectroscopy (good energy resolution): in high energy
physics, $\gamma$-ray astronomy, neutrinoless $\beta$$\beta$-decay,
medical imaging and, more recently, dark matter (DM) searches and
coherent neutrino scattering (CNS) detection.

The latter two applications have much in common from the
instrumentation point of view, as we shall discuss below: a low energy
threshold and low intrinsic background are important requisites in
both cases. In addition, both are low energy processes which benefit
from coherence of the scattering across all nucleons in the nucleus,
thus enhancing expected event rates very significantly. For that
reason, we shall consider them in close connection, despite the fact
that direct DM searches are more mature than CNS experiments: while
large scale DM detectors using liquid xenon and argon are already
running or in advanced stage of construction, the possibility of
detecting CNS with those media is still under consideration, and
several important questions remain open.

Although this review gives a reasonably comprehensive account of
efforts to measure low energy signals with liquefied noble gas
detectors, its main aim is to show how the noble liquid technology
works for this purpose and to discuss some recent advances and
improvement attempts. The authors apologize in advance for inevitable
incompleteness. More information can be found in other recently
published monographs and review papers
\cite{BarabashBolozdynya93,LopesChepel05,AprileBook06,AprileDoke09,BolozdynyaBook10,Akimov11}.

The article is organized as follows. After motivating the search for
particle dark matter and coherent neutrino-nucleus scattering in
Section~\ref{sec:ScatteringDMnu},
we examine in the following section the main
characteristics of the signals expected in liquid xenon and liquid
argon targets and how those interactions might be detected. We devote
Section~\ref{sec:RelevantProperties}
to a detailed description of the physics involved in the
response mechanisms. In  
Section~\ref{sec:StateOfTheArt}
we review the devices used to detect
scintillation light and ionization charge, both those which are well
established as well as others under development. In  
Section~\ref{sec:DirectDMExperiments}
we examine how the different DM search programs have implemented the
noble liquid technologies and indicate their status at the time of
writing. A brief overview of ongoing efforts towards a first detection
of CNS is presented in  
Section~\ref{sec:NeutrinoDetection}.
We conclude in 
Section~\ref{sec:Conclusion},
highlighting the progress in sensitivity achieved by direct DM
searches with noble liquids in the last two decades.

\section{Elastic scattering of dark matter particles and neutrinos off nuclei}
\label{sec:ScatteringDMnu}

There are several observational evidences for the existence of
non-luminous (i.e.~not observable electromagnetically), non-baryonic
matter in the Universe. The exact nature of this dark matter is not
known yet. It is recognized that it may exist in various hypothetical
forms; one possibility --- favored by current cosmological models --- is
a new, neutral particle, frozen-out as a thermal relic in the early
Universe, with mass in the GeV--TeV range and interacting with
ordinary matter through the weak force. These so-called Weakly
Interacting Massive Particles (WIMPs) might be the lightest
supersymmetric particles; in particular, the neutralino --- arising in
some flavors of Supersymmetry --- is the prime candidate for direct
detection.

A simple model for the distribution of this `cold' dark matter in our
galaxy suggests that WIMPs should gather in a spherical isothermal
halo --- much larger than the visible baryon disc --- with a mean
particle speed of 270~km/s. The solar system moves through this halo
with a speed of $\simeq$220~km/s. Therefore, in the Earth's frame one
should observe WIMPs coming preferentially from a certain direction in
the sky, with that velocity on average (plus a small seasonal
variation due to the Earth's rotation around the Sun with a speed of
about 30~km/s). Since the WIMP velocity dispersion is comparable to
our orbital velocity around the galactic center, such directional
effects are not large, and even a stationary observer in the DM halo
might be able to detect WIMPs. 

Naturally, this `canonical' dark halo offers too simplistic a view of
galaxy dynamics, one which has been questioned by n-body simulations
in particular. These propose a more `lumpy' distribution of galactic
dark matter and the possibility that the dark component may be
co-rotating with the baryon disc to some degree. However, we adopt the
simpler halo model for the purpose of this article. For a recent and
comprehensive review of dark matter the reader is referred
to~\cite{Bertone10}.

Elastic scattering of WIMPs off the atomic nuclei of a target material
should result in nuclear (atomic) recoils \cite{GoodmanWitten85}. With
an adequate detection technique, the minute energies transferred to
recoiling nuclei can be measured, allowing WIMPs to be observed. This
method is usually referred to as {\em direct dark matter detection}
(reviewed also, e.g.,
in~\cite{Akimov01,Sumner02,Gaitskell04,Ryabov08,FigueroaFeliciano11}). In
contrast, {\em indirect detection} relies on the observation of WIMP
annihilation products arising elsewhere in the Milky Way or beyond
\cite{Ryabov08,Cirelli12}. Discovery of the nature of dark matter
would not only solve an astrophysical puzzle dating back to the early
1930s --- and change radically how we view our own galaxy --- but it
would arguably be one of the most important steps in our understanding
of the Universe.

In turn, the interest in measuring coherent neutrino scattering off
atomic nuclei is twofold. From a fundamental physics viewpoint, it is
important to observe a process for which the Standard Model offers a
concrete prediction. This would allow several tests of the theory. For
example, observation of neutrino oscillations with CNS would provide
evidence for the existence of sterile neutrinos, since this is a
flavor-blind process. Measurement of the CNS cross section could shed
light on the value of the neutrino magnetic momentum, and so reveal
how neutrinos couple to the high magnetic fields which exist in some
astrophysical environments. It would contribute to our understanding
of neutrino dynamics in neutron stars and supernovae, where CNS adds
to the neutrino opacity, besides providing a high-rate mechanism for
supernova detection. Many other applications could be listed
(see~\cite{Barbeau07}, for example). In addition to these scientific
applications, detection of neutrinos through this scattering mechanism
could bring about new neutrino technologies, such as nuclear reactor
monitoring (nuclear safety, non-proliferation), geological prospecting
and perhaps even neutrino telecommunications.

The effect of coherence over all nucleons in the nucleus is the
enhancement of the scattering cross section by a factor of $N^2$ ($N$
is the neutron number) with respect to neutrino-nucleon scattering
\cite{Drukier84}. If the coherence condition $qR \lesssim 1$ is
fulfilled, with $R$ the nuclear radius and $q$ the transferred
momentum, then one can gain a factor of $N \sim 100$ in scattering
rate by using a neutron rich target. However, this enhancement works
well only for low energy neutrinos, roughly $\lesssim$50~MeV. The CNS
cross section (and the rate per kg) is also several orders of
magnitude higher than that of elastic neutrino-electron scattering ---
the process used in several neutrino detectors. The relatively large
cross section allows consideration of kg-scale, rather than
tonne-scale, detectors. This is what makes coherent scattering on
nuclei so attractive in spite of the challenge of measuring extremely
low recoil energies. The search for a spin-independent WIMP-nucleus
interaction benefits from the coherence enhancement in a similar way,
as outlined in the next section.

\section{Overview of detection principles}
\label{sec:OverviewDetectionPrinciples}

A nuclear recoil with energy in the keV range is the only signature of
the interaction of both WIMPs and neutrinos scattering elastically off
atomic nuclei. The recoil energy spectrum due to galactic WIMPs is
approximately exponential. Although it depends on the (unknown) WIMP
mass and atomic number of the target, most recoils have energies well
below $\sim$100~keV~\cite{LewinSmith96}. Therefore, expected WIMP
count rates depend critically on the detector energy threshold, which
must be as low as possible. The recoil spectrum expected from coherent
neutrino scattering has roughly a similar shape and can extend up to
tens of keV for higher energy neutrinos (from decay of relativistic
ions or decay of stopped pions, for example
\cite{Bueno06,Scholberg06}); on the other hand, the spectrum lies
mostly in the sub-keV region in the case of the reactor antineutrinos
\cite{Hagmann04}.

Liquid xenon (LXe) and liquid argon (LAr), as well as liquid neon
(LNe), have been proposed for direct dark matter search detectors. At
present, several liquid xenon detectors (with target masses of order
10--100~kg) have been built and operated for dark matter searches and
larger ones are in the process of construction or underground
deployment. Liquid argon projects have made slower progress after
initial demonstration of capability for WIMP searches, but systems are
now under construction with tonne-scale targets. Neon offers some
interesting complementarity in direct searches and is the subject of
R\&D work. We review briefly some of the existing projects in
Section~\ref{sec:DirectDMExperiments}.

The present predominance of liquid xenon detectors is explained by
several advantages of this medium, namely higher WIMP-nucleus
interaction cross section (the spin-independent cross section
$\sigma_{\chi,\mathrm{Xe}}^\mathrm{SI}$ includes a factor of $A^2$,
where $A$ is the atomic number), absence of long-lived radioisotopes
in natural xenon\footnote{The 2$\nu\beta\beta$-decay mode of
$^{136}$Xe has been measured recently \cite{Ackerman11,Gando12}, but
this produces an extremely low event rate at energies relevant for DM
and CNS searches. Also, $^{85}$Kr ($\beta$-decay, $T_{1/2}\sim
10.8$~years) is a common contaminant, but can be removed by cryogenic
distillation \cite{Abe09} or adsorption-based chromatography
\cite{Bolozdynya07}.}  and presence of odd-neutron isotopes with
non-zero nuclear spin, which makes the detector sensitive also to a
spin-dependent interaction component. High $Z$ and density combined
into a continuum (unsegmented) active medium allow for a compact
self-shielding detector geometry. In turn, liquid argon detectors
benefit from higher electron recoil discrimination capability due to a
large difference between the decay times of the two scintillation
components (expanded in 
Section~\ref{sec:EmissionMechanisms}
in more detail). The much lower
cost of argon and the relative ease of contaminant purification are
the other advantages of liquid argon. Unfortunately, natural
atmospheric argon contains the radioisotope $^{39}$Ar, with a specific
activity of 1~Bq/kg; sourcing from specific underground sites is
required to mitigate against this problem \cite{AcostaKane08}.

As for coherent scattering neutrino experiments, low $A$ liquids
(argon and especially neon) are favored for low neutrino energies, due
to the fact that the nuclear recoil energy is inversely proportional
to the mass of the target nucleus, $E_r^{max} \propto E_\nu^2/A$.  For
example, for neutrinos with energy $E_\nu=3$~MeV (close to the average
energy of reactor antineutrinos) the maximum recoil energy is
$\approx$1~keV for neon, $\approx$0.5~keV for argon and merely
$\approx$150~eV for xenon. This means that the signals one can expect
from ionization or scintillation detectors are of the order of a few
electrons or photons, respectively. Thus, the requirement of a low
detection energy threshold is even more demanding for CNS than for DM
search experiments, unless WIMPs are much lighter than presently
thought. On the other hand, the scattering cross section increases
with the square of the neutron number $N$ in the nucleus, $\sigma_\nu
\propto E_\nu^2N^2$, thus giving advantage to xenon as far as the
interaction probability is concerned. For a given target mass, the
scattering rate is proportional to $E_\nu^2N^2/A$. Xenon detectors can
be advantageous if the source provides neutrinos with energy 
above~$\sim$30~MeV, so that the maximum recoil energy becomes 
$>$15~keV~\cite{Bueno06}, \cite{Scholberg06}.

By any standard, scattering cross sections are small for both
processes. The scalar WIMP-nucleon cross section is unknown, but from
existing experimental constraints it is smaller than $\sigma_{\chi,n}
\sim2\cdot10^{-45}$~cm$^2$ ($2\cdot 10^{-9}$~pb). Due to coherence,
the cross section scales up with $A^2$ for the whole nucleus and gains
also a kinematic factor $(A(m_n+m_\chi)/(Am_n+m_\chi))^2$, where $m_n$
is the nucleon mass and $m_\chi$ is the WIMP mass. Assuming
$m_\chi$=100~GeV, present experimental limits imply
$\sigma_{\chi,\mathrm{Xe}} < 10^{-37}$~cm$^2$ and
$\sigma_{\chi,\mathrm{Ar}} < 2\cdot 10^{-39}$~cm$^2$ for scattering
off the two nuclei.\footnote{In reality, this is measured in the
opposite way: the scattering cross section on xenon
($\sigma_{\chi,\mathrm{Xe}}$) provides the present experimental
constraint, and the WIMP-nucleon cross section ($\sigma_{\chi,n}$) is
calculated via a hypothetical particle model (typically a neutralino);
scattering rates on other nuclei, such as argon, can then be
inferred.}

\begin{figure}[t]
\includegraphics[height=6.9cm]{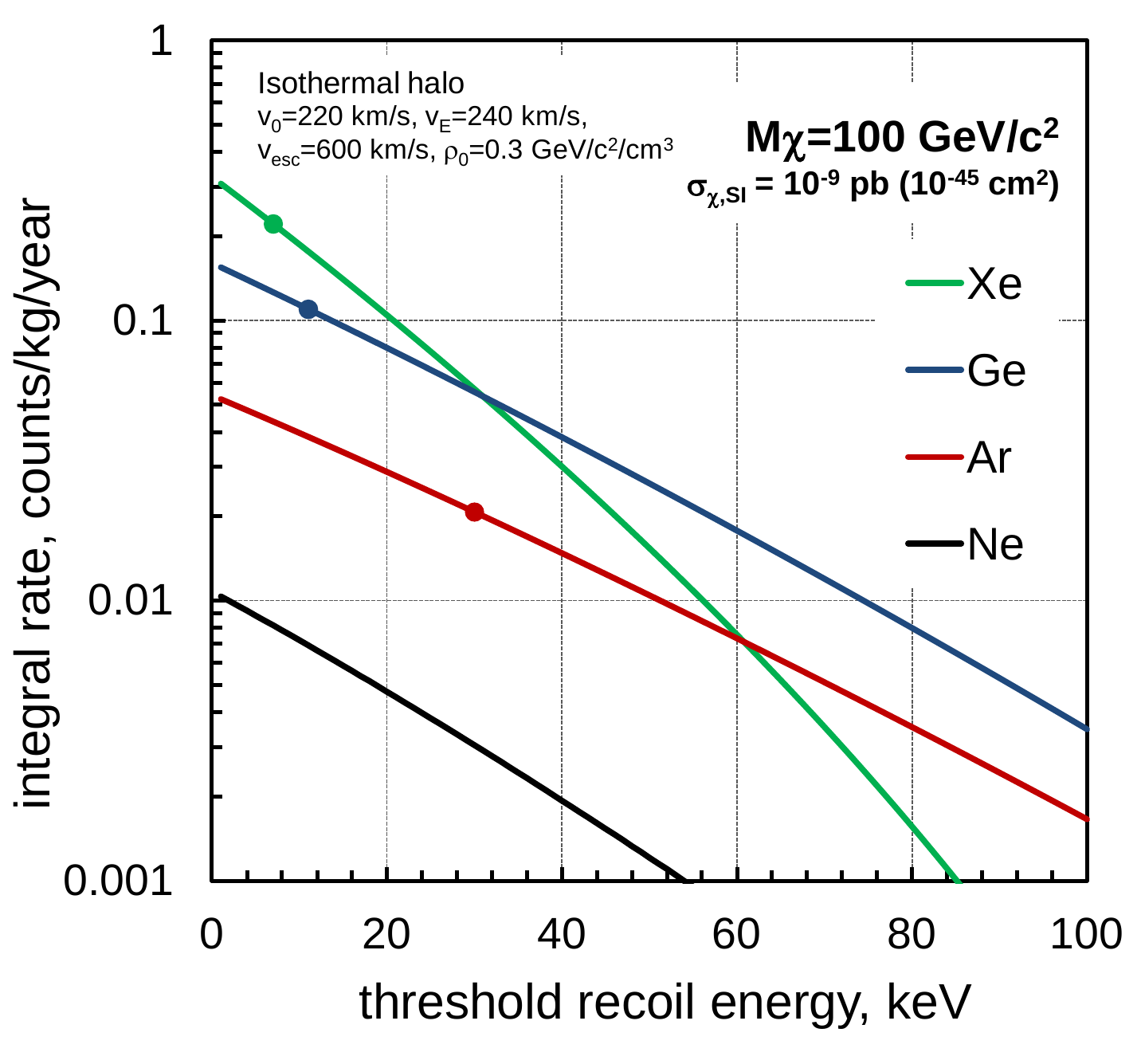}
\includegraphics[height=6.9cm]{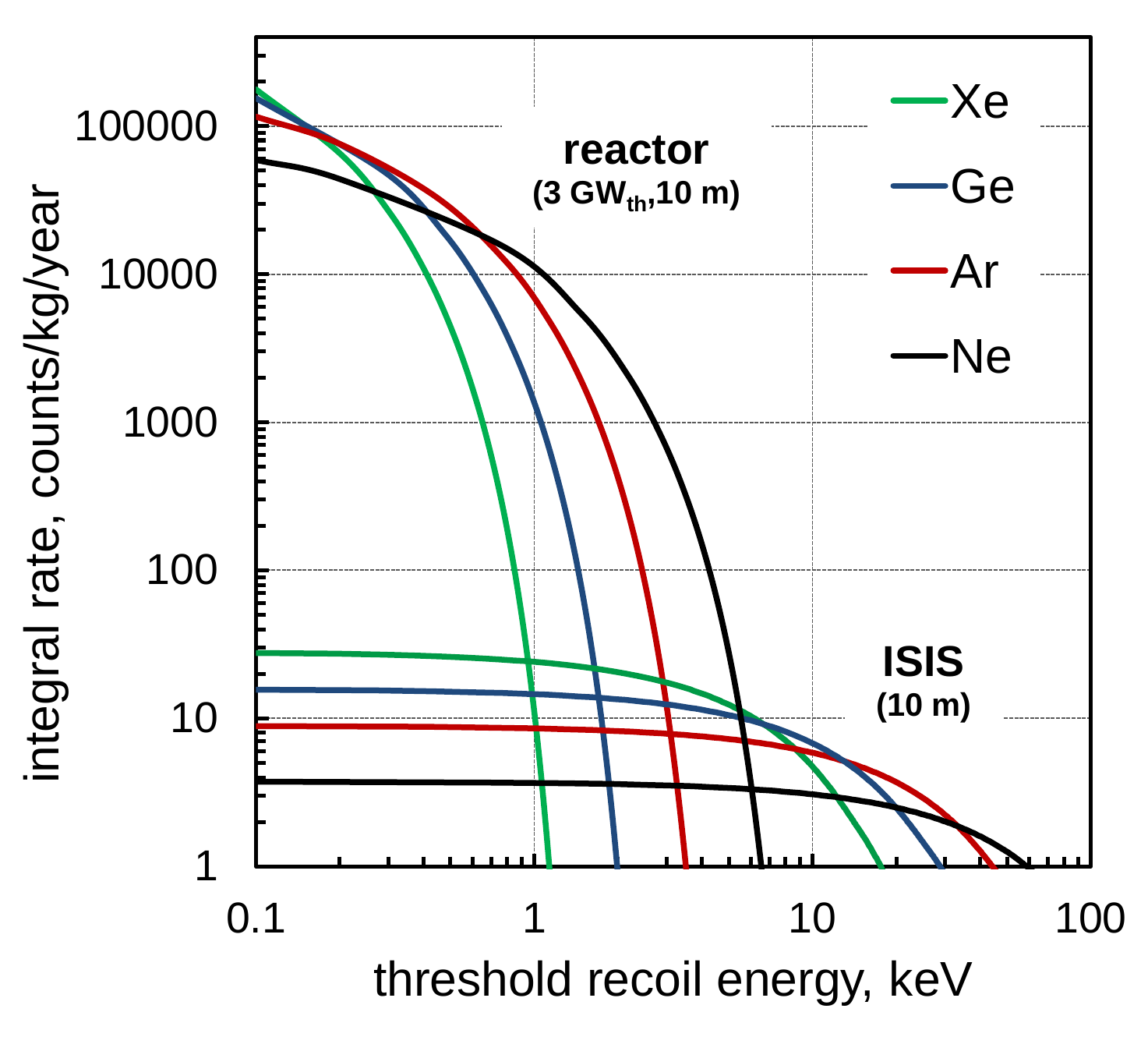}
\caption{Predicted integral spectra for WIMP elastic scattering (left)
and for coherent neutrino-nucleus elastic scattering (right) for Xe,
Ge, Ar and Ne (in order of decreasing rate at zero threshold). Both
plots assume perfect energy resolution. Dark matter rates are for a
100~GeV/c$^2$ WIMP with 10$^{-45}$cm$^2$~(10$^{-9}$~pb) interaction
cross section per nucleon, calculated as per \cite{LewinSmith96} with
the halo parameters shown; the markers indicate typical WIMP-search
thresholds for each technology. CNS rates are calculated at 10~m from
a 3~GW$_{\mathrm{th}}$ nuclear reactor ($4\cdot 10^{13}~\nu/$cm$^2$/s) and at
the same distance from the ISIS neutron spallation source (thanks to
E.~Santos), where 3 neutrino flavors result from pion and muon decay
at rest ($1\cdot 10^{7}~\nu/$cm$^2$/s for all flavors
\cite{Burman96}).}
\label{fig:WIMPrates}
\end{figure}

At the same WIMP mass of 100~GeV, a simple estimate of expected
scattering rates can be obtained from the particle flux in our
galactic neighbourhood, where the WIMP density is $\rho_0\sim
0.3$~GeV/cm$^3$~\cite{Cerdeno10} and the mean particle speed is
$\bar{v}_\chi \sim 270$~km/s (assuming a stationary Earth in the
galactic frame for simplicity). The WIMP flux is therefore
$\Phi_\chi\sim\rho_0\bar{v}_\chi/m_\chi \sim
10^5$~s$^{-1}$cm$^{-2}$. From this one can calculate the expected
total scattering rate for a given target species $T$ as
$R_T=\sigma_{\chi,T}\Phi_\chi N_A / A_T$, which results in a maximum
of $\sim$0.1 event per year per 1~kg of xenon, and nearly an order of
magnitude lower for argon. The left panel in 
Figure~\ref{fig:WIMPrates} 
shows scattering
rates above detection threshold for several noble gas elements and for
germanium, the most competitive cryogenic bolometer material.

These calculations must take into account the weak nuclear form
factor, which decreases with transferred momentum much more rapidly
for xenon than for argon
\cite{LewinSmith96,Helm56,Engel91,Duda07}. The Xe form factor drops
practically to zero for recoil energies of 100~keV, while it is still
$\approx$0.5 for argon recoils of that energy --- this behavior is
apparent in 
Figure~\ref{fig:WIMPrates} (left), where the full calculations following
\cite{LewinSmith96} are presented. This is fortunate since WIMP-search
thresholds are anyway higher in LAr as discussed later. Such low
scattering rates justify why dark matter search experiments require
large detector masses, with tonne-scale experiments being built at
present. It is also important to keep in mind that these estimates are
made under the assumption that the WIMP mass is about 100~GeV. One
must also point out that the rates mentioned above are {\em
scattering} rates. Actual {\em detection} rates will also depend on
detector efficiency (and energy resolution) as a function of recoil
energy.

For neutrinos, the scattering cross section can be calculated in the
Standard Model framework \cite{Drukier84,Freedman74}. It is different
for neutrons and protons due to different coupling to up and down
quarks: for neutrons it is $\sigma_{\nu,n} \approx 0.42\cdot
10^{-44}(E_\nu/\mathrm{MeV})^2$~cm$^2$, whereas for protons it is a
factor of $\sim$200 smaller. Therefore, the effect of coherence over
the whole nucleus is an enhancement factor of $N^2$. For example, for
10~MeV neutrinos, the cross section for scattering on a Xe nucleus 
is~\mbox{$\sigma_{\nu,\mathrm{Xe}}\sim 2\cdot 10^{-39}$~cm${^2}$}; 
for Ar it is 
an order of magnitude smaller, $\sigma_{\nu,\mathrm{Ar}}\sim 2\cdot
10^{-40}$~cm${^2}$. Although these values are even smaller than those
expected for WIMPs, significantly higher fluxes can be obtained with
neutrinos from artificial sources ($\sim$10$^{13}$~cm$^{-2}$s$^{-1}$
at a distance of $\sim$10~m from a nuclear reactor, to give one
example). Calculated rates as a function of threshold for two neutrino
sources are shown in 
Figure~\ref{fig:WIMPrates} (right). In addition, `on/off'
experiments are also possible in this instance, which is a significant
advantage for controlling systematic uncertainties. Therefore,
detectors with a mass of the order of kilograms can, in principle,
provide a reasonable rate. However, one must not neglect the fact
that, contrary to WIMP searches, where only a few events with correct
signature could constitute a discovery in a nearly background-free
experiment conducted underground, a neutrino experiment in a surface
laboratory must accumulate enough recoil signals to produce a
statistically significant distribution in energy (or in the number of
ionization electrons, as only few-electron signals can be expected for
MeV neutrinos~\cite{Hagmann04,Akimov09,Santos11}).

The low scattering rate makes the background issue of extreme
importance. Background reduction (passive shielding, low radioactivity
environment and radio-clean construction) and its active
discrimination in the experimental setup are essential. In the case of
direct dark matter searches in underground laboratories, two kinds of
background can be distinguished: one resulting in electron recoils and
the other leading to production of nuclear (atomic) recoils in the
sensitive volume of the detector. Most electron recoils are produced
by $\gamma$-ray interactions, both from the environment and from
radioactivity of detector components, and from decays within the
detector or contamination of internal surfaces. The intrinsic
radioactivity is mostly due to the U/Th decay chains, but other
radioisotopes can be important (e.g.~$^{40}$K, $^{60}$Co,
$^{137}$Cs). Ultimately, elastic scattering of solar p-p neutrinos off
electrons will provide a challenging background for next-generation
detectors, with hundreds of events/year expected in a tonne-scale
target~(see, e.g.~\cite{McKinseyDoyle99,Suzuki01}).

Nuclear recoil background is the most dangerous one since some of the
resulting signals are indistinguishable from those that WIMPs or
neutrinos would produce. The dominant source is the elastic
scattering of neutrons arising from either ($\alpha$,n) reactions
(induced by $\alpha$-particles from the U and Th decay chains) or from
spontaneous fission of $^{238}$U. Trace amounts of these nuclides are
naturally present in the surrounding environment and in detector
components, despite careful screening and purification of
materials. Another source of neutrons is the interaction of
atmospheric muons, some of which penetrate deep underground, in the
rock and $\gamma$-ray shielding around the experiment. Low-A shielding
is essential for moderation of external neutrons; an active neutron
veto surrounding the detector (e.g.~Gd-loaded scintillation counters
\cite{Akimov10b,Ghag11}) can also be very effective at tagging
radioactivity as well as muon-induced neutrons.

Fiducialization of the sensitive volume by excluding from the data
interactions in the outer regions of the target is crucial for
background reduction, not only those effects mentioned above but also
nuclear recoils from $\alpha$-decays on detector surfaces. This relies
on the ability to reconstruct particle interaction sites in three
dimensions.  Owing to their high spatial resolution, double-phase
noble element detectors are capable of very effective fiducialization
even in modest-sized targets.

While elastic neutron single-scatter events cannot be discriminated
from WIMP or CNS signals in the detector, electron recoils have a
different signature and therefore can be identified and rejected to a
significant extent. The early dark matter detectors relied on the fact
that the relative contributions of slow and fast components to the
observed scintillation signal are different for nuclear recoils and
for electrons. This was the case in the DAMA/LXe and ZEPLIN-I liquid
xenon scintillation detectors \cite{Bernabei02,Luscher04,Alner05}. The
difference between the two effective decay time constants in liquid
xenon is not large: $\sim$32~ns and $\sim$18~ns were measured for
electrons and for nuclear recoils, respectively, in the energy range
of 10~keV to 20~keV \cite{Alner05,Spooner01}. Considering, in
addition, the small number of scintillation photons and the relatively
small VUV light detection efficiency, it is understandable that a
discrimination efficiency of only $\sim$50\% was achieved. The current
generation of liquid xenon detectors does not rely on pulse shape
discrimination for the primary scintillation signal. However, for
liquid argon targets this method is much more effective and can
increase the discrimination power very significantly. Indeed, the two
scintillation decay time constants differ very much in liquid argon:
$\sim$1,600~ns versus $\sim$7~ns (see 
Section~\ref{sec:EmissionMechanisms}).

\begin{figure}[tp]
\centerline{\includegraphics[width=.8\textwidth]{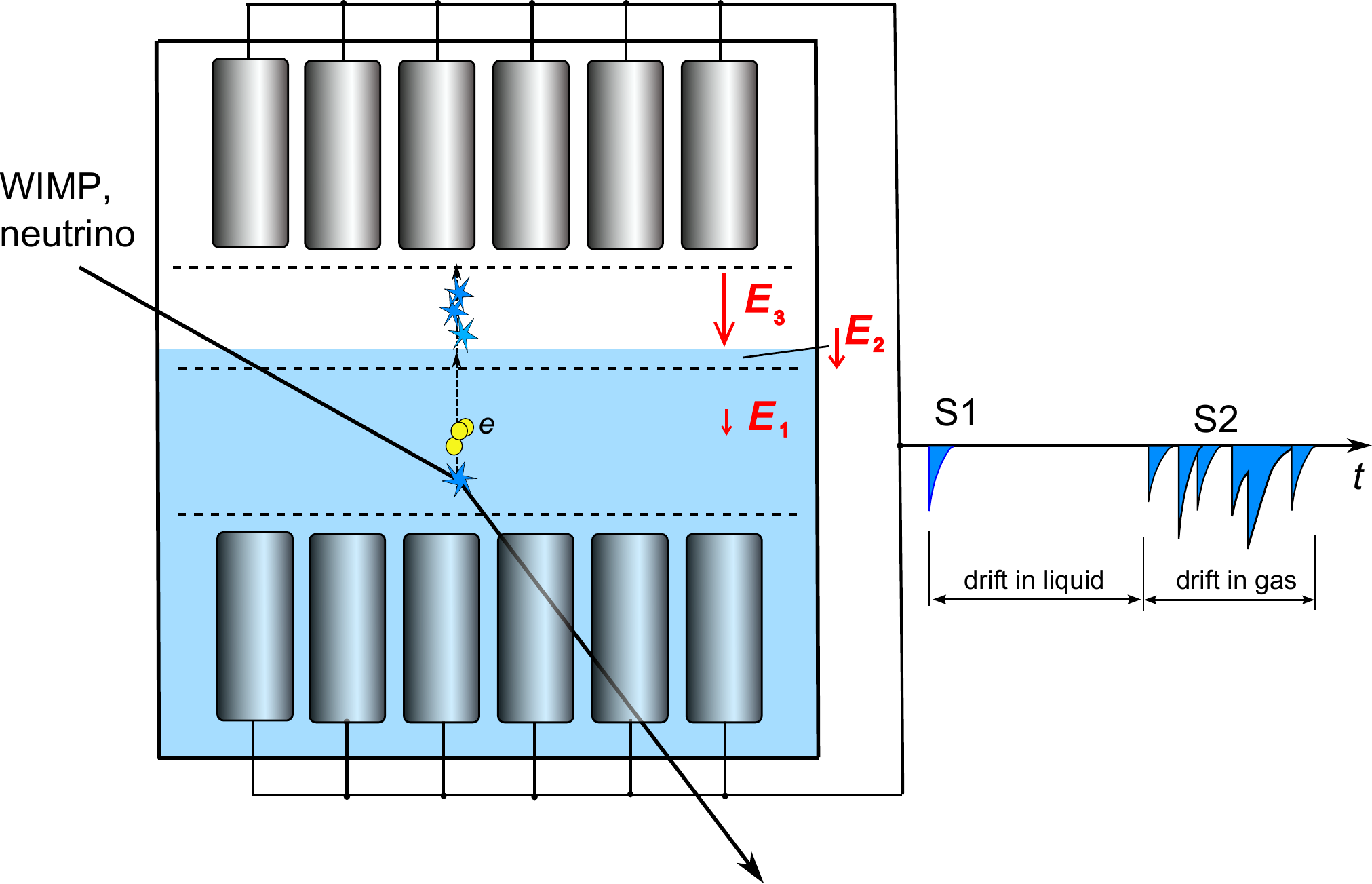}}
\caption{Illustration of the operation principle of a double-phase
electroluminescence detector. The number of electrodes may vary: some
detectors do not feature a grid under the liquid surface; additional
grids are often placed in front of the photomultipliers in order to
shield them from the external electric field. Only one array of
photomultipliers is used in some chambers: either immersed into the
liquid and looking upwards, or placed in the gas phase viewing
downward.}
\label{fig:TwoPhaseXenon}
\end{figure}

Much higher discrimination efficiency can be achieved if, in addition
to primary scintillation, the ionization signal is also measured. Most
noble liquid detectors dedicated to dark matter searches share a
common operation principle. It is based on electron emission from the
liquid, followed by secondary scintillation (also known as
electroluminescence) in the uniform electric field in the gas phase,
as illustrated in 
Figure~\ref{fig:TwoPhaseXenon}. 
The foundations for this technology were
laid by Dolgoshein, Lebedenko and Rodinov in 1970 \cite{Dolgoshein70},
who demonstrated the possibility of particle detection in a
double-phase argon chamber; by Barabash and Bolozdynya some two
decades later \cite{Barabash89}, who first proposed the use of
double-phase detectors for WIMP searches; and in work of the ICARUS
group at CERN, who proposed to measure both scintillation and
ionization signals for background discrimination of WIMP signals from
electron and $\gamma$-ray background in the early 1990s
\cite{Benetti93b}. Following preliminary work with several small
prototype chambers built by different groups, these three crucial
components were implemented in ZEPLIN-II --- the first double-phase
dark matter search detector operated underground \cite{Alner07}.

\begin{figure}[ht]
\centerline{\includegraphics[width=.6\textwidth]{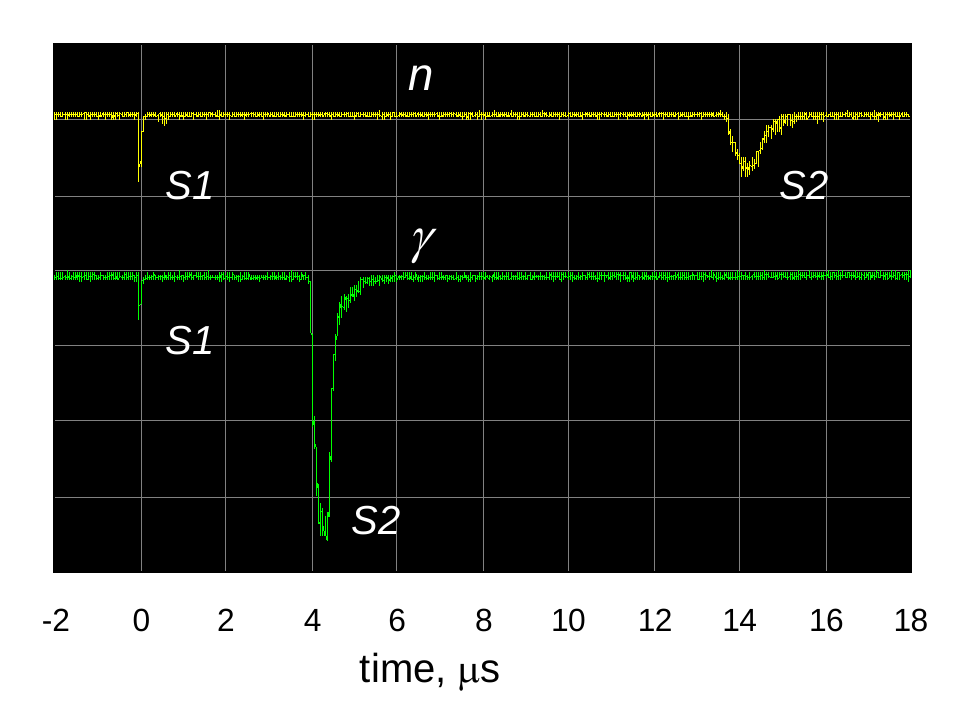}}
\caption{Signals from a xenon nuclear recoil after elastic scattering
of a neutron (upper trace) and from a $\gamma$-ray (lower trace)
measured in ZEPLIN-III \cite{Akimov07}. The label `S1' designates prompt
scintillation in the LXe, while `S2' denotes secondary scintillation
produced in the gas phase by the electrons extracted from the
liquid. (Courtesy ZEPLIN-III Collaboration.) }
\label{fig:Z3Signals}
\end{figure}
 
The interaction of a dark matter particle with the detector takes
place in the bulk of the liquefied xenon or argon. The particle
scatters elastically off an atomic nucleus transferring to it part of
its kinetic energy. In the energy range considered ($\sim$1--100~keV)
the initial velocity of the nuclear recoil is comparable to that of
the atomic electrons, so that the recoiling atom conserves most of its
electrons and moves through the liquid as a positive ion with low
effective charge (or even as a neutral atom). The moving ion
continuously exchanges electrons with other atoms along its way and
its average charge in xenon is $\sim$0.1$e$ at 1~keV and $\sim$1$e$
for 100~keV (for argon and neon these values are approximately twice
as large). As the medium consists of atoms of the same species, the
primary recoil can transfer a significant fraction of its kinetic
energy in each collision, thus losing rapidly its `projectile'
identity and producing a cascade of secondary recoils of comparable
energy which interact with the medium in the same way.  

The interaction mechanism of nuclear recoils with matter differs from
that of electrons, as besides the energy loss to atomic electrons it
involves energy transfer to atomic nuclei (essentially generating
heat, which does not contribute to the observed signal in these
detectors). Consequently, the track structure is different for
electrons and nuclear recoils and one can expect that the charge
recombination along the particle track also behaves differently. This
is indeed verified~\mbox{experimentally}. For a given electric field
strength, the fraction of electrons escaping recombination with
positive ions is higher for electrons than for nuclear recoils. This
affects directly the size of the ionization signals but also
influences, in the opposite way, the amplitude of the scintillation
signal. Therefore, one can roughly say that in the case of nuclear
recoils more energy is drawn into the luminous signal than into the
(observable) ionization signal, while for electrons the situation is
reversed, as shown in 
Figure~\ref{fig:Z3Signals}. 
The plot shows two signals measured by
an array of photomultipliers (PMTs) with an Am-Be neutron source. The
lower trace corresponds to a $\gamma$-ray interaction and the upper
one is due to a neutron elastically scattered on a nucleus. The first
short pulse in each signal (usually known as `S1') is due to primary
scintillation of the liquid xenon, while the second one (`S2')
corresponds to the secondary scintillation produced in the gas by the
electrons extracted from the liquid. Although both signals are
detected in the form of light, S2 is frequently called the
`ionization' signal to stress that its origin is in the ionization
process and its size (pulse area) is proportional to the number of
electrons extracted from the particle track. Note that the time delay
between these two signatures provides a very accurate measurement of
the depth of the interaction site in the liquid. 

\begin{figure}[ht]
\centerline{\includegraphics[width=.5\textwidth]{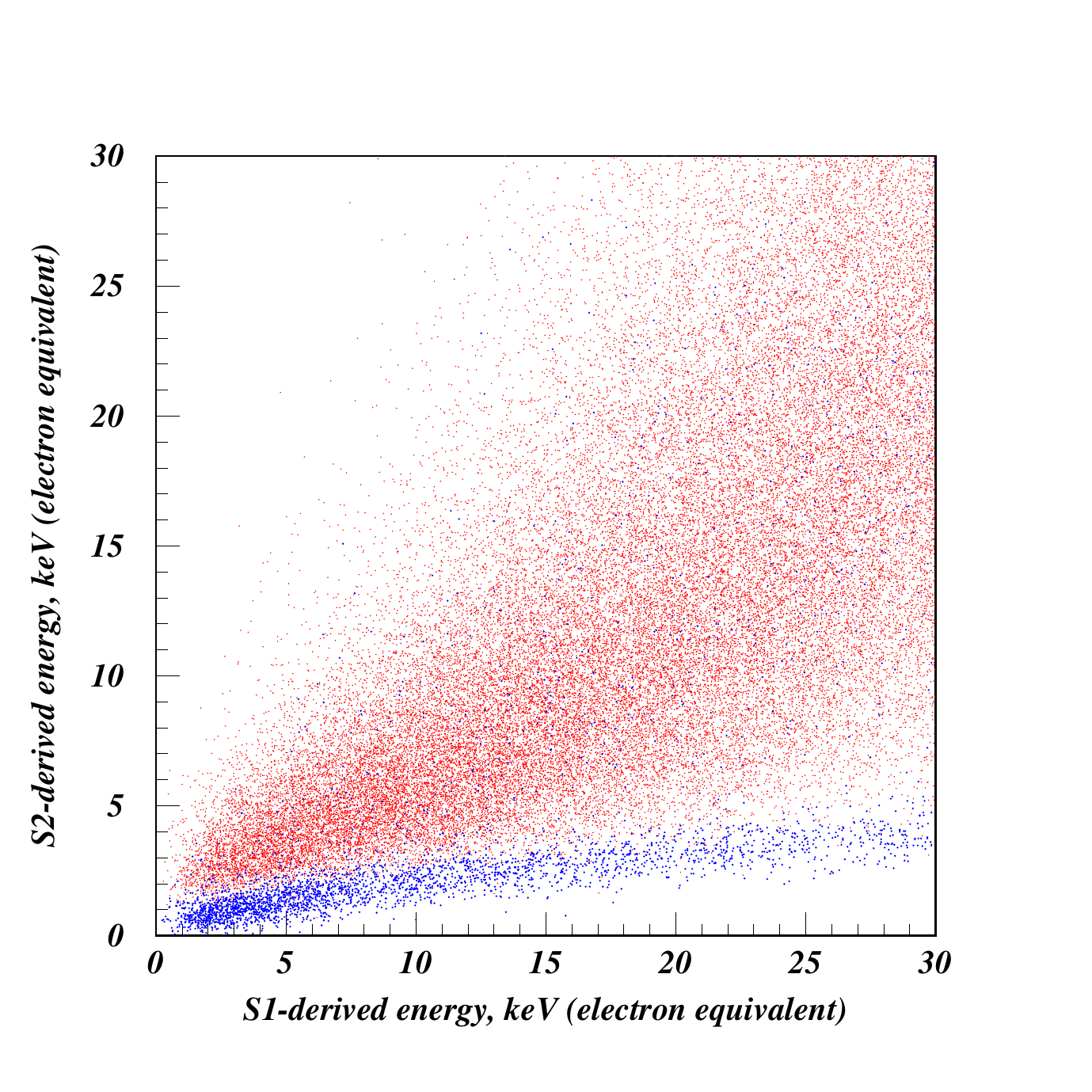}}
\vspace{0.3cm}
\caption{Scatter plot of the ionization signal measured from
proportional scintillation (S2) versus the prompt scintillation signal
(S1), both expressed in electron equivalent energy, for
Compton-scattered $^{137}$Cs $\gamma$-rays (upper population, in red)
and for xenon recoils induced by elastic scattering of Am-Be neutrons
(blue), acquired during the ZEPLIN-III calibration runs at 3.9~kV/cm
drift field in the liquid. (Courtesy ZEPLIN-III Collaboration.)}
\vspace{1cm}
\label{fig:LXeDiscrimination}
\end{figure}

A complete picture of the energy transfer mechanisms as well as energy
sharing between different channels for low energy particles is still
missing, both for nuclear recoils and for electrons. However, it is
well established experimentally that the distributions of the
collected charge versus the number of detected primary scintillation
photons (both depending on the deposited energy) appear quite
different for the two particles (see 
Figure~\ref{fig:LXeDiscrimination}). This constitutes the
basis for discrimination of the electron/$\gamma$-ray background in
the current double-phase dark matter detectors. This method was
initially tested with $\alpha$-particles and $\gamma$-rays in liquid
xenon \cite{Benetti93b}, with the ionization signal observed through
secondary scintillation in the liquid; it was further investigated
with double-phase xenon in \cite{Suzuki98} and \cite{Wang98}, until it
was conclusively demonstrated in \cite{Aprile06}.

\begin{figure}
\centerline{\includegraphics[width=\textwidth]{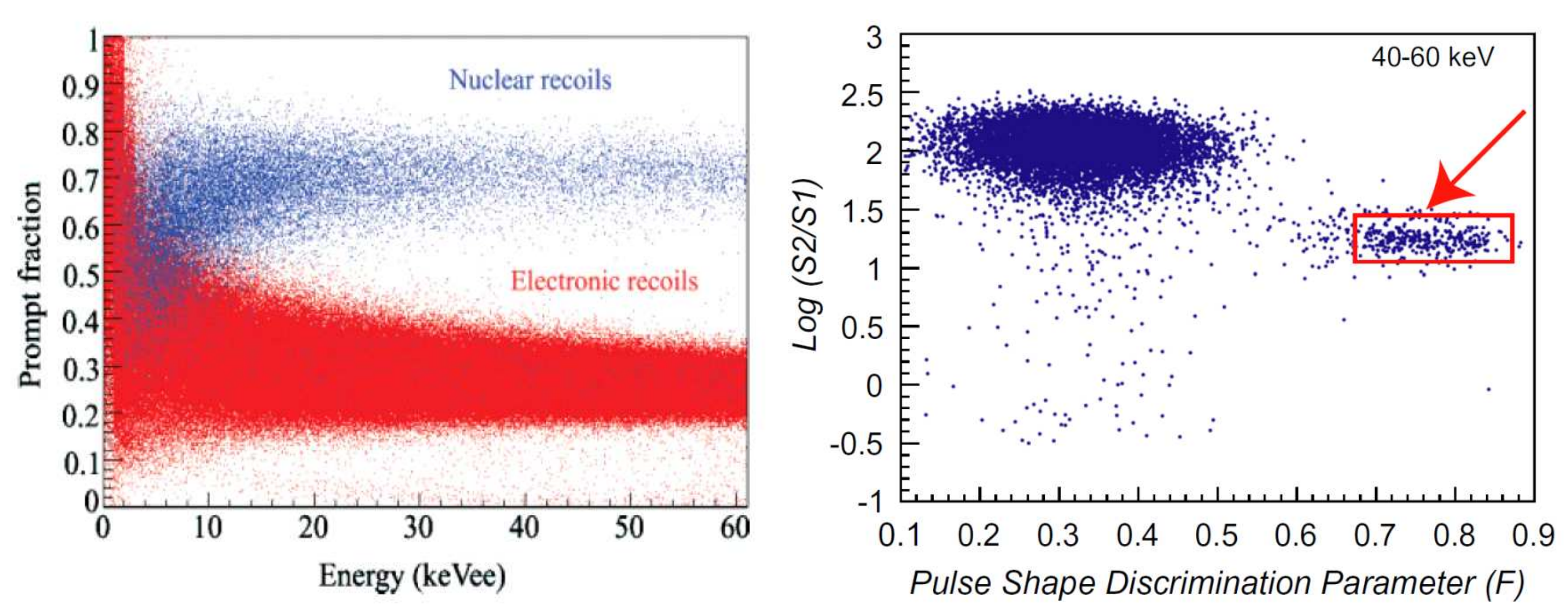}}
\vspace{0.3cm}
\caption{Discrimination in liquid argon. Left: pulse shape
discrimination in a single-phase chamber yielding 4.85~photoelectrons
per keV (electron equivalent); the nuclear recoils are produced with a
D-D neutron generator \cite{Lippincott08}. (Courtesy D.~McKinsey;
copyright (2008) by The American Physical Society.) Right: dual
parameter discrimination in the 2.3-liter WARP detector; nuclear
recoils with energy 40--60~keV from an Am-Be neutron source are
indicated by the red region (from \cite{Benetti08}; with permission
from Elsevier). In both panels the scintillation pulse shape is
described by a `prompt fraction', i.e.~the ratio of signal integrated
within a narrow prompt window (90~ns and 200~ns, respectively) to the
total pulse area (several $\mu$s).}
\label{fig:LArDiscrimination}
\end{figure}

\begin{figure}[ht]
\centerline{\includegraphics[width=0.6\textwidth]{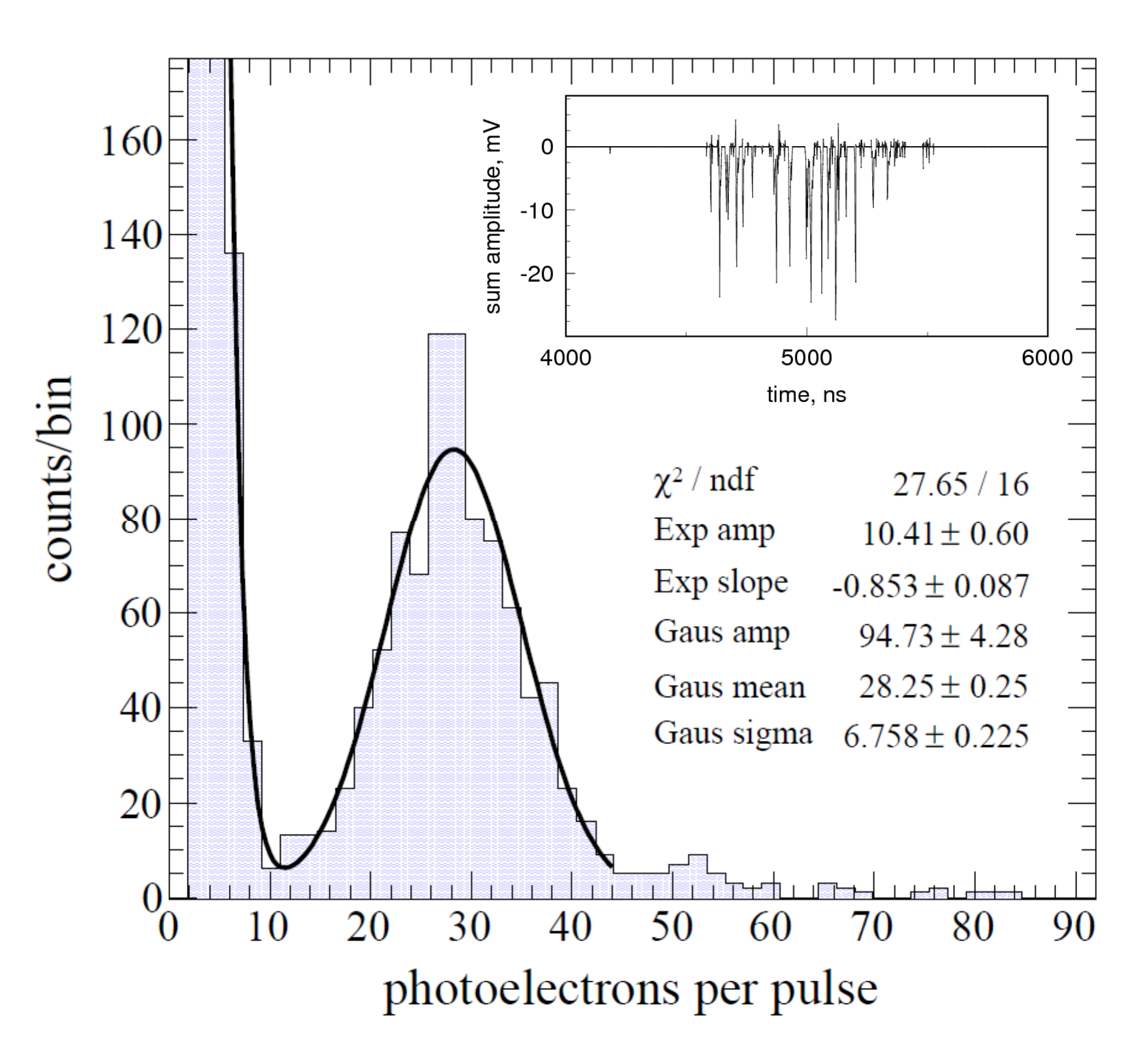}}
\caption{Pulse size distribution of the secondary scintillation signal
due to cross-phase emission of a single electron in ZEPLIN-III
(adapted from \cite{Santos11}); the mean signal contains
28~photoelectrons. The inset shows the summed waveform from
31~photomultipliers; each pulse of $\sim$10~ns width represents a
single photoelectron measured at the output of one PMT. (Courtesy
ZEPLIN-III Collaboration; with permission from Springer Science and
Business Media.) }
\label{fig:SEinZ3}
\end{figure}

In liquid argon, pulse shape discrimination on the primary
scintillation signal is extremely effective at rejecting electronic
interactions above $\sim$10~keV; this is confirmed by data from a
single phase detector shown in 
Figure~\ref{fig:LArDiscrimination} (left). In addition, the
charge/light ratio measured in double-phase configuration can be used
to enhance the electron recoil rejection, as illustrated in 
Figure~\ref{fig:LArDiscrimination}
(right). Very high discrimination efficiency is essential in LAr to
mitigate against the $^{39}$Ar background.

The sensitivity of the scintillation channel is lower than that of the
ionization by as much as an order of magnitude. Therefore, the minimum
detectable energy is limited by the S1 signal in experiments where the
presence of both signals is required. Strictly, the minimum detectable
energy is not well defined: due to significant dispersion at low
energies, the probability to observe energy depositions below an
established energy threshold is not zero. The nuclear recoil energy
threshold for the purpose of WIMP searches is chosen by each
experiment as a compromise between the count rate expected in the
signal region and acceptable background discrimination efficiency
(which decreases for low energies, as illustrated in 
Figures~\ref{fig:LXeDiscrimination} and~\ref{fig:LArDiscrimination}.
For example, the ZEPLIN-III and XENON100 experiments used similar
thresholds of $\approx$10~keV for nuclear recoils
\cite{Akimov12a,Aprile11b}. Recoil energy can be reconstructed both
from S1 and S2 signals, none of the methods being
straightforward. Since direct, {\it in situ} calibration of the
nuclear recoil response with MeV neutrons has so far proved
impractical, the energy scale is usually obtained indirectly by
irradiating the detector with 122~keV $\gamma$-rays from $^{57}$Co. If
primary scintillation is used as a measure of recoil energy, the
relative scintillation efficiency for electrons and nuclear recoils
must be known, from ancillary measurements, as well as its dependence
on the recoil energy and field strength (see 
Section~\ref{sec:PrimaryScintillation}).
In the case of the S2 signal, one has to correct the energy scale for the
differences in charge yield from electron and nuclear recoil tracks,
also dependent on the field and energy.\footnote{In the large
detectors now under construction, calibration with $^{57}$Co becomes
problematic due to thicker vessels and self-shielding of the inner
detector volume. It may become necessary to refer the nuclear recoil
response to short-lived, internally-dispersed sources, e.g.~$^{83m}$Kr
($T_{1/2}$=1.8~hr), which emits 32.1~keV and 9.4~keV conversion
electrons~\cite{Kastens09,Manalaysay10}. However, an anomalously high 
yield was recently measured for the 9.4~keV electrons at zero field, 
explained by the presence of positive ions left over in LXe after the 
preceding 32.1~keV transition~\cite{Aprile12d};
further study of the  $^{83m}$Kr source is therefore desirable.}

\begin{figure}[t]
\centerline{\includegraphics[width=.5\textwidth]{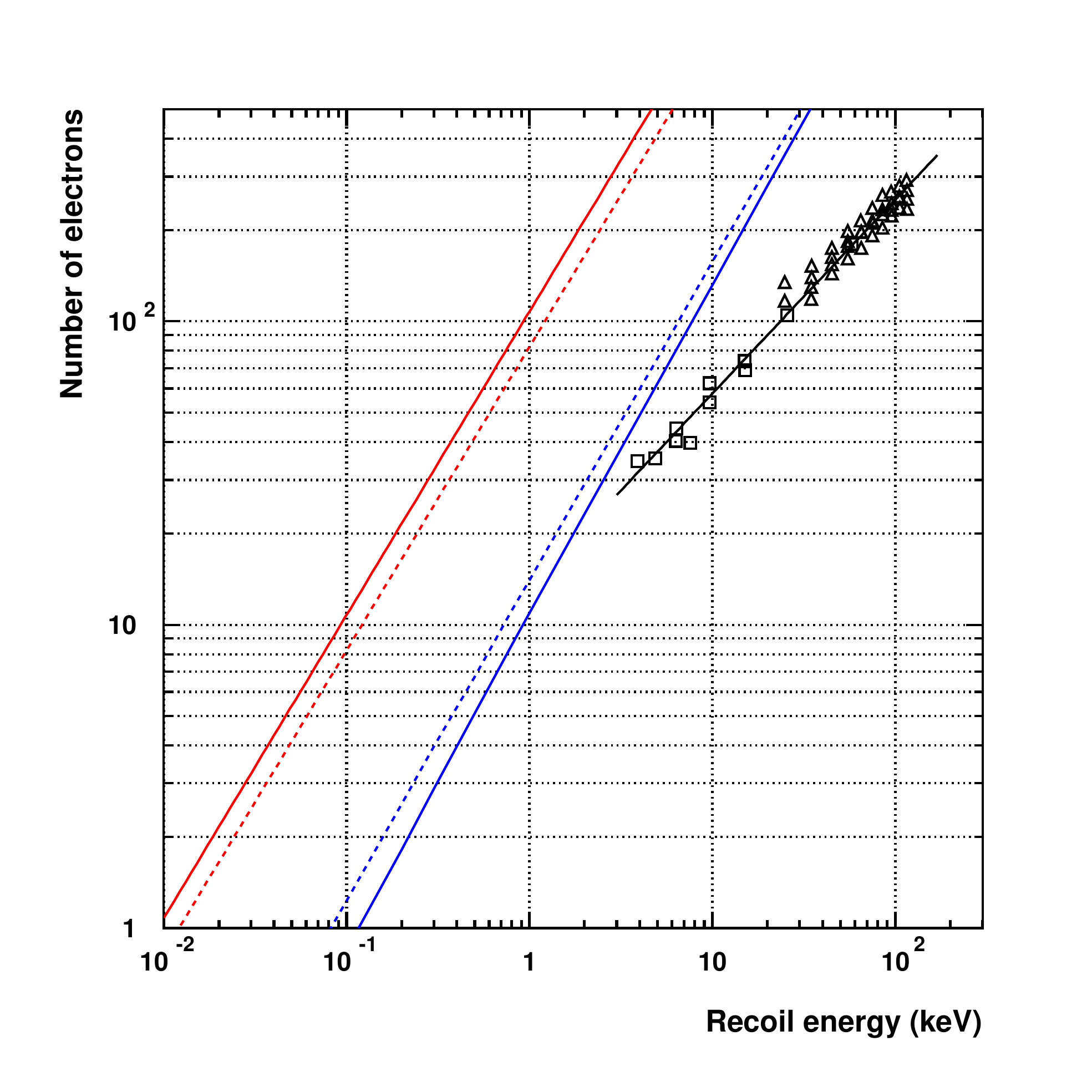}}
\vspace{-0.25cm} 
\caption{Number of electrons extracted from tracks of Xe recoils in
liquid xenon. Experimental data from \cite{Aprile06} (triangles) and
\cite{Manzur10} (squares); the black line is an empirical
fit. Electric field strength varies from 0.1~kV/cm to 4~kV/cm. Color
lines: red -- ionization potential (band gap) limit; blue -- SRIM
\cite{SRIM} prediction for the energy transferred to atomic electrons.
Solid lines correspond to the solid state model (band gap
$E_g$=9.62~eV); dashed lines are for a gas model (ionization potential
$I$=12.13~eV). The ionization potential and band gap values are from
\cite{Doke05}.  
\vspace{-0.2cm}   }
\label{fig:ChargeYield}
\end{figure}

Remarkably, the ionization signal measured through secondary
scintillation in the gas is sensitive to single electrons extracted
from the liquid \cite{Edwards08,Sorensen08,Santos11}. An example of
such a signal in ZEPLIN-III is shown in the inset of 
Figure~\ref{fig:SEinZ3}. 
Each pulse on the waveform corresponds to a single photoelectron emitted
from a PMT photocathode (outputs of all PMTs are summed in this
plot). On average, about 30~photoelectrons are detected per extracted
electron \cite{Santos11}. The main histogram in the same figure shows
the size distribution of those signals, very clearly separated from
the exponential-like noise. 

In order to express the sensitivity to single electrons in terms of
recoil energy, the electron extraction efficiency from the recoil
tracks must be known; this includes the probability to escape
recombination and the transfer efficiency from liquid to gas. Both are
functions of electric field strength; recombination depends also on
the recoil energy. Any losses during the charge drift to the liquid
surface must also be considered. Using $\sim$7~e/keV as a reference
for LXe \cite{Aprile06,Sorensen09,Manzur10,Horn11}, one can estimate
that the nuclear recoil threshold in the ionization channel could be
as low as a few hundred eV 
(Figure~\ref{fig:ChargeYield}).


\section{Relevant properties of the liquefied rare gases}
\label{sec:RelevantProperties}

\subsection{Particle energy transfer to the liquid}
\label{sec:EnergyTransfer}

The energy transferred by a particle to the medium, $E_0$, is split
between three channels --- ionization, excitation and heat. For
electrons, the following equation can be written \cite{Platzman61}:
\begin{equation}
  E_0=N_i\,E_i + N_{ex}\,E_{ex} + N_i\,\epsilon \,,
  \label{eqn:Platzman}
\end{equation}
where $E_i$ and $E_{ex}$ are the mean energies spent to ionize or to
excite an atom; $N_i$ and $N_{ex}$ are the mean numbers of ionized and
excited atoms, respectively; and $\epsilon$ is the mean energy of
sub-excitation electrons immediately after the last collision to
result in either excitation or ionization. Below such energy free
electrons will participate only in elastic collisions with atoms, thus
increasing the temperature of the medium. The value of $\epsilon$ has
been estimated to lie between 6.3~eV and 7.7~eV for LAr and between
4.65~eV and 5.25~eV for LXe, depending on the theoretical model
\cite{Doke76}. In more recent work, the lowest of the two values for
each liquid has been adopted (e.g.~\cite{Doke05}).

In the case of nuclear recoils, a significant fraction of the particle
energy is spent in nuclear collisions which do not result in
excitations or ionizations. Therefore, an additional term should be
considered in equation~(\ref{eqn:Platzman}) to account for recoil
elastic scattering. We shall discuss various aspects related to this
topic in Sections~\ref{sec:PrimaryScintillation},
\ref{sec:IonizationCharge} and \ref{sec:CombinedSignal}.

Returning to equation~(\ref{eqn:Platzman}) for electrons, one notices
that, in the gas phase, the energy quantities can be conveniently
referred to the atomic ionization potential, $I=15.75$~eV for argon
and $I=12.13$~eV for xenon:
\begin{equation}
  \frac{E_0}{I}=N_i\frac{E_i}{I} + N_{ex}\frac{E_{ex}}{I} 
  + N_i\frac{\epsilon}{I} \,.
  \label{eqn2}
\end{equation}
It has been found that the ratio $E_i/I$ exceeds unity by a few
percent due to a small probability to produce multiple ionizations or
to form an excited ion.

The liquefied noble gases (Ar, Kr and Xe, at least) exhibit a band
structure of electronic states, so that the atomic ionization
potential should be replaced by the band gap, $E_g$, and
equation~(\ref{eqn2}) can be rewritten for the liquid phase as:
\begin{equation}
  \frac{E_0}{E_g}=N_i\frac{E_i}{E_g} + N_{ex}\frac{E_{ex}}{E_g} 
  + N_i\frac{\epsilon}{E_g} \,.
  \label{eqn3}
\end{equation}
The band gap was found to be $E_g = 14.2$~eV for solid Ar and 9.28~eV
for solid Xe \cite{Baldini62} (see also Table~2 in 
Section~\ref{sec:IonizationCharge},
where ionization processes are discussed in more detail). For liquid
xenon a band gap of 9.22~eV was measured \cite{Asaf74}.

Using the definition of the $W$-value, $W=E_0/N_i$, equation~(\ref{eqn3})
can now be written as:
\begin{equation}
  \frac{W}{E_g}=\frac{E_i}{E_g} + \frac{E_{ex}}{E_g}\cdot\frac{N_{ex}}{N_i} 
  + \frac{\epsilon}{E_g} \,.
  \label{eqn4}
\end{equation}
These ratios for liquid argon, krypton and xenon were evaluated by
several authors. Although a general agreement seems to have been
achieved, some values are not concordant. In particular, a ratio
$N_{ex}/N_i = 0.06$ has been calculated for LXe, although 0.2 is more
consistent with experimental data. Their average, 0.13, is often used
for some estimates. For LAr better agreement has been obtained around
the ratio $N_{ex}/N_i \approx 0.2$. We refer the reader to
Refs.~\cite{Doke05,Takahashi75,Doke02} for a comprehensive set of
parameters, some of which are reproduced in Table~2 in this work.

For nuclear recoils, the ratio of the number of primary excitations to
that of ionizations in liquid xenon was found to be much higher than
for electrons: $N_{ex}/N_i \sim 1$~\cite{Sorensen11,Dahl09}.

In the condensed medium, charge recombination along the particle track
plays a very important role in determining the response to
radiation. Therefore, the observable ionization and scintillation
signals depend on the applied electric field and, in most cases, are
not equal to those that one might expect on the basis of the $N_i$ and
$N_{ex}$ values. Due to recombination, the number of electrons
collected at an anode is $<N_i$ and depends both on the ionizing
particle (species and energy) and on the applied electric field. For
some particles, such as 1~MeV electrons, the number of collected
electrons approaches $N_i$ already at fields of $\sim$1~kV/cm, while
for heavily ionizing particles \mbox{($\alpha$-particles, for
example)} it can be significantly smaller even at some tens of kV/cm.
For the very low particle energies relevant to DM and CNS searches ---
both regarding electron and nuclear recoils --- it is found that the
dynamics of the recombination process becomes qualitatively different
from that described by the models applicable to MeV particles and
above. 

In the absence of electric field and for medium ionization density,
which on a first approach can be characterized by the linear energy
transfer (LET), practically all electron-hole pairs created initially
recombine and give rise to recombination luminescence; in this
instance, the scintillation yield is at its maximum, and the number of
emitted photons is $N_{ph}\approx N_i+N_{ex}$.  For low density
tracks, some electrons escape recombination even at zero field, and so
$N_{ph}<N_i+N_{ex}$. On the other extreme are the tracks surrounded by
a high density of ionized and excited species. In this region, the
luminescence signal can be suppressed because of a high probability of
collision between excited species, leading to so-called `bi-excitonic
quenching'~\cite{Hitachi83,Hitachi92}. We shall discuss these effects
in more detail in
Sections~\ref{sec:EmissionMechanisms}~and~\ref{sec:Recombination}.

Regarding the energy transfer from nuclear recoils, two substantial
differences with respect to high energy particles must be stressed.
Firstly, although some of the energy of the impinging ion goes
ultimately into electronic excitations and ionization, a large
fraction is spent in collisions in which the target atom recoils as a
whole. Therefore, a significant fraction of the particle energy
appears as kinetic energy of the target atoms (i.e.~heat) and is not
visible either as light or as charge. This mechanism of energy
transfer is referred to as nuclear energy loss.  In the formalism of
stopping power it is described by a separate term: $dE\!/\!dx =
(dE\!/\!dx)_e + (dE\!/\!dx)_n$, where the index $n$ denotes energy
losses through nuclear collisions and $e$ refers to the electronic
component.  The second difference lays in a distinct
ionization/excitation pattern that exists in the medium at the instant
when the energy of all secondary particles drops below the excitation
threshold --- frequently called the `track structure', except that in the
case of very low energies this pattern does not resemble a track, but
rather a cloud of electrons, positive ions and excited species (see
Section~\ref{sec:Recombination}).

A comprehensive theory of the energy loss of ions with velocity
comparable to that of atomic electrons is not sufficiently developed
to provide a solid theoretical background for exact
calculations~\cite{Mangiarotti07}. This certainly constitutes a
handicap in understanding the physics associated with WIMP detection.
The theoretical basis was laid out by
Lindhard~$et\,al.$~\cite{Lindhard63b,Lindhard68}, who calculated the
nuclear stopping power using the Thomas-Fermi screened potential to
describe the interaction between the ion and a target atom. The
electronic stopping was obtained by considering an ion moving through
a constant density electron gas. The dependence on the ion and target
atom charge was included again using the Thomas-Fermi
description. Lindhard himself considered the treatment of electronic
stopping the most approximate of the two.  In fact, the charge of an
ion moving through a medium is not constant due to electron exchange
with other atoms and accurate determination of the effective charge is
a difficult task.  This formulation is usually referred to as the
\mbox{Lindhard--Scharff--Schi{\o}tt} (LSS) theory.

As in WIMP detectors the medium consists of atoms of the same species
as the nuclear recoil, the latter very rapidly loses its `projectile'
identity producing a cascade of secondary recoils which undergo the
same interactions as the primary particle.  In order to evaluate the
total energy partition between the electronic and nuclear components
for an impinging recoil of a given energy (Lindhard's partition
function~\cite{Lindhard63}), the whole cascade must be taken into
account. In spite of all difficulties, the LSS theory provides a
useful framework and describes reasonably well the energy losses by
nuclear recoils, at least in what the scintillation efficiency of LXe
is concerned (see~\cite{Hitachi05,Mangiarotti07}).  We
recommend~\cite{Mangiarotti07} for a concise analysis of Lindhard's
model and more recent works related to it, in view of WIMP detection
with liquefied noble gases.

The SRIM/TRIM code~\cite{SRIM} 
provides information on ranges and stopping powers,
$(dE\!/\!dx)_e$ and $(dE\!/\!dx)_n$,
using parameterizations adjusted to the 
existing experimental data, mostly for light ions; 
it can also simulate the whole cascade. For xenon, it results in somewhat
low values for the electronic-to-nuclear loss ratio, which predicts 
a lower-than-observed scintillation yield for nuclear recoils
even if no quenching is taken into account. 
A comparative analysis of SRIM/TRIM and LSS model predictions 
can be found in~\cite{Mangiarotti07}.~\!\footnote{
The 2003 version of SRIM/TRIM code was used 
in~\cite{Mangiarotti07}; newer versions are now available. }

\subsection{Primary scintillation}
\label{sec:PrimaryScintillation}

\subsubsection{Emission mechanism and yields}
\label{sec:EmissionMechanisms}

The emission mechanisms and scintillation yields
have been studied primarily (and better understood)
for particles of relatively high energies: $\sim$1~MeV electrons 
and $\gamma$-rays as well as $\alpha$-particles and relativistic ions.
Information is rather scarce in the range $\lesssim$100~keV, 
although some progress has been made in recent years. 
Therefore, we begin by discussing the results obtained with
fast electrons and $\alpha$-particles and then move to
low-energy electrons and nuclear recoils. 

In general terms, the mechanism of primary scintillation is similar
for liquid argon and xenon. Imagining the liquid as a compressed gas
(which is not strictly correct given the acknowledged existence of a
band structure), one can say that most of the observed light is
emitted by diatomic excited molecules which are formed in two distinct
processes. One is excitation of rare gas atoms by electron impact with
subsequent formation of strongly bound diatomic molecules in the
excited state (excimers), similarly to what happens in the gas phase
(using the terminology of solid state physics, one would refer to
these states as `excitons'):
\begin{align*}
& e^- + R \to R^* + e^-       && \mathit{impact\,\,excitation} \\
& R^* + R \to R_2^{*,v}       && \mathit{excimer\,\,formation} \\
& R_2^{*,v} + R \to R_2^* + R && \mathit{relaxation} \\
& R_2^* \to R + R + h\nu      && \mathit{VUV\,\,emission}
\end{align*}
Here, $R$ = Ar or Xe and the superscript $v$ is used to distinguish
excited states with vibrational excitation ($R_2^{*,v}$) from purely
electronic excitation with $v=0$ ($R_2^*$). The vibrational relaxation
is mostly non-radiative, as shown in the above equations, but emission
of infrared photons is also possible. The scintillation photons, with
wavelength in the vacuum ultraviolet (VUV) region for all rare gases,
are emitted in a transition from one of the two lowest electronic
excited states $^3\Sigma_u^+$, or $^1\Sigma_u^+$ with $v=0$, to the
ground state $^1\Sigma_g^+$. The transition occurs at short
interatomic distances, where the ground state potential is repulsive,
resulting in dissociation of the molecule. Although the two
transitions are spectroscopically indistinguishable, their decay times
are quite different, especially in liquid argon. Direct transition
from the triplet state $^3\Sigma_u^+$ is forbidden, but the decay
becomes possible owing to the mixing between $^3\Sigma_u^+$ and
$^1\Pi_u$ states through spin-orbital coupling~\cite{Kubota78}. This
can result in rather long life-times as is the case in LAr
($\sim$1~$\mu$s). As the coupling becomes stronger for molecules with
higher atomic number, the decay time is significantly shorter for LXe
($\sim$27~ns). Conversely, the triplet lifetime is even longer in LNe
(15~$\mu$s~\cite{Nikkel08}) and reaches $\sim$13~seconds in
LHe~\cite{McKinsey99}.

Emission in the visible and near infrared (NIR) wavelength region (up
to 2000~nm) has also been observed in LAr and LXe
\cite{Bressi00,Bressi01,Heindl10,Buzulutskov11}.\footnote{The origin
of the emission in the near UV, visible and infrared regions in these
liquids is not fully understood. Observed spectra bear some similarity
to those measured in the gas phase, where some lines have been
identified as due to atomic transitions from the higher excited levels
to the first excited state $3p^54s$~\cite{Lindblom88,Bressi01a}, but
with significant suppression of the intensity for
$\lambda>1300$~nm~\cite{Bressi01}. A weaker intensity (by a factor of
$\sim$100) in LAr and LXe compared to that in the gas may be
understood in view of the much higher collision rate in the condensed
phase, leading to efficient non-radiative quenching of the highly
excited states.} Its intensity, however, is much lower than in the VUV
region in those liquids (and falls also far short of the NIR emission
in the respective gas). For liquid argon, emission in the range
$400-1000$~nm represents only $\sim$1\% of the VUV
yield~\cite{Heindl10}. In \cite{Buzulutskov11}, a value of
0.51$\pm$0.09 infrared photons per keV is reported for LAr for
$\lambda=690-1000$~nm, a result generally consistent with that
in~\cite{Heindl10}. For LXe, an estimate of $\gtrsim$0.2~photons/keV
can be obtained combining results from \cite{Bressi01} and
\cite{Belogurov00}.

\begin{figure}[ht]
\centerline{\includegraphics[width=.85\textwidth]{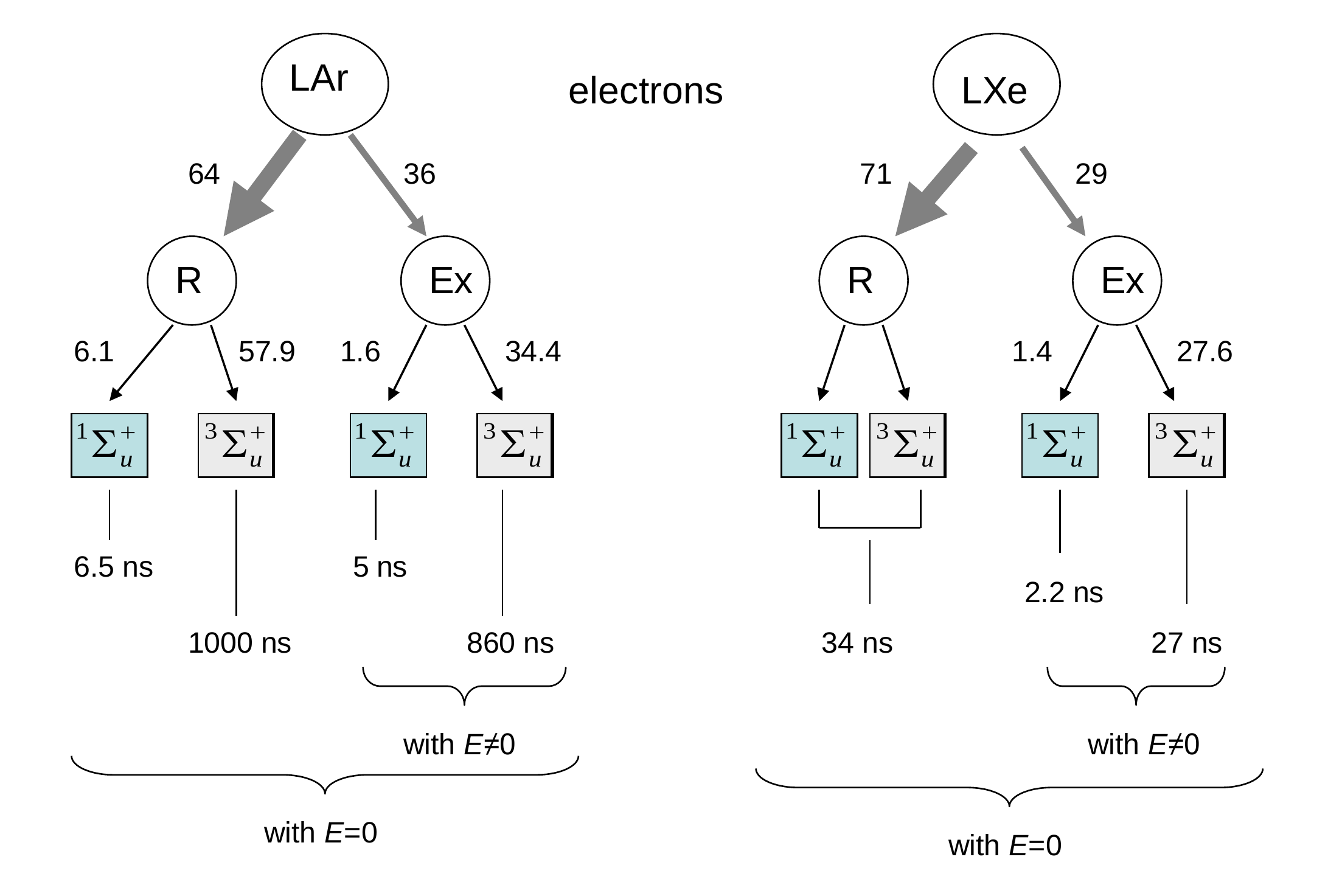}}
\caption{Observable distribution (percent) of the luminous energy for fast
electrons. $R$ and $Ex$ stand for recombination and direct excitation
channels, respectively. Estimates are from data in~\cite{Kubota78}
and~\cite{Kubota82}. The share between the direct excitation and the
recombination channel is from~\cite{Kubota79}.\newline
*)  Other values can be found in the literature (see also~\cite{Morikawa89} for earlier measurements). 
On time constants, 7.0~ns has been reported for the fast component 
in LAr~\cite{Hitachi83}, whereas the slow component has been also 
measured as 1600~ns~\cite{Hitachi83}, 1463~ns~\cite{Lippincott08} and 1260~ns~\cite{Acciarri10a} (all at zero electric field);  
for LXe, singlet and triplet decay times of 4.3~ns and 22.0~ns were reported
in~\cite{Hitachi83} along with an apparent scintillation decay time of 45~ns
also at zero field.
Other singlet/triplet intensity ratios have been published:
0.31/0.69~\cite{Kubota79} and 0.23/0.77~\cite{Hitachi83} in LAr, and
0.36/0.64 in LXe~\cite{Kubota79}. 
}
\label{fig:LuminousElectrons}
\end{figure}

The alternative VUV luminescence process to direct excitation involves
recombination of the ionization electrons with positive ions. The
recombination occurs mostly with molecular ions formed shortly
($\sim$ps) after ionization of atoms of the liquid by a
particle~\cite{Doke02,Doke80}:
\begin{align*}
&e^- + R \to R^+ + 2e^-        && \mathit{ionization} \\
&R^+ + R + R \to R_2^+ + R     && \\
&e^- + R_2^+ \to R^{**} + R    && \mathit{recombination} \\
&R^{**} + R \to R^{*}+ R+\text{heat}       && \\
&R^{*} + R + R \to R_2^{*}+ R+\text{heat}  && \\
&R_2^* \to R + R + h\nu        && \mathit{VUV\,\,emission}
\end{align*}
The final stage of this sequence is similar to the direct excitation
channel and consequently wavelengths and decay times are
similar. However, the relative population of the singlet and triplet
excited states is not necessarily the same. Moreover, the
recombination process preceding the excimer formation can introduce a
considerable time delay. This is the case in liquid xenon, where
recombination is rather slow compared with the excimer de-excitation
times, which leads to the appearance of a non-exponential third
component in the decay curve. For liquid xenon, apparent time
constants of 34~ns to 45~ns have been
reported~\cite{Kubota78,Hitachi83} (see 
Figure~\ref{fig:LuminousElectrons}).
However, a longer tail of up to
$\sim$2~$\mu$s, at the level of $\sim$1\% of the initial amplitude of
the scintillation signal, could also be observed in the study reported
in~\cite{Kubota79}. This component disappeared completely when an
electric field of a few kV/cm was applied, which confirms its
recombination origin.

In liquid argon no significant changes in the shape of the decay curve
under electric field have been observed --- a fact explained by the
relatively fast ($\sim$1~ns) recombination, so that the timing of the
photon emission is mostly determined by the decay time of excitons
formed shortly after ionization~\cite{Kubota79}. Measurements of the
electron thermalization time (0.9~ns for LAr and 6.5~ns for
LXe~\cite{Sowada82}) support this conclusion. We point out that the
recombination dynamics is especially sensitive to the purity of the
liquid, as contaminant atoms and molecules can provide more effective
thermalization of ionization electrons than elastic scattering off
xenon atoms --- thus hindering their escape from the influence of the
positive ions.

It has long been observed that, for electrons under a field of the
order of $\sim$10~kV/cm, the scintillation yield decreases by a factor
of up to $\approx$3 in both liquids, which means that only
$\approx$1/3 of the scintillation light emitted in the absence of an
applied field is due to direct excitation, the remaining $\approx$2/3
resulting from recombination. 
The relative contribution of the recombination component to the total light 
yield was measured to be (67$\pm$2)\% for
liquid argon and (74$\pm$2)\% for liquid xenon~\cite{Kubota78a}. 
Similar results, 64\% and 71\% for argon and
xenon, respectively, were obtained in~\cite{Kubota79}. The observed
light yield decreases with field more rapidly in xenon than in
argon. For example, in xenon it drops by a factor of 2 already at
$\sim$250~V/cm, while in argon $\sim$1~kV/cm is necessary to produce
the same effect.
 
The diagram shown in 
Figure~\ref{fig:LuminousElectrons}
 represents the distribution of the total
number of emitted scintillation photons between different excitation
channels when the liquid is excited by fast electrons with typical
energies from 0.5~MeV to 1~MeV. The diagram is based on data published
in~\cite{Kubota78,Kubota79,Kubota82}; other results are referred 
in the figure caption. Note the small
contribution of the transitions from the singlet state to the direct
excitation channel (only about 1.5\% of the total number of emitted
photons). In argon, the fast component is enhanced through the
recombination channel resulting in about 8\% of the total number of
photons emitted at zero field. In LXe, however, the fast component is
practically unobservable for electrons unless an electric field is
applied to suppress recombination luminescence. 
It is important to point out again that impurities can
affect (sometimes significantly) the luminescence properties of the
medium, namely the light yield, emission spectrum and observed decay
times. This may explain some discrepancies (especially noticeable 
for the slow component in liquid argon) between the early
measurements and more recent ones, when more advanced purification
techniques and better control of the impurity content were employed. 
This was observed in~\cite{Acciarri10a,Acciarri10b} where 
the addition of small concentrations of N$_2$ and O$_2$ 
($\gtrsim$0.1~ppm) effectively quenched Ar$^{*}_{2}$ 
triplet states  and significantly shortened 
the observed decay time for the slow component, 
reducing the overall light yield (especially significant in the case of nitrogen). 

\begin{figure}[ht]
\centerline{\includegraphics[width=.6\textwidth]{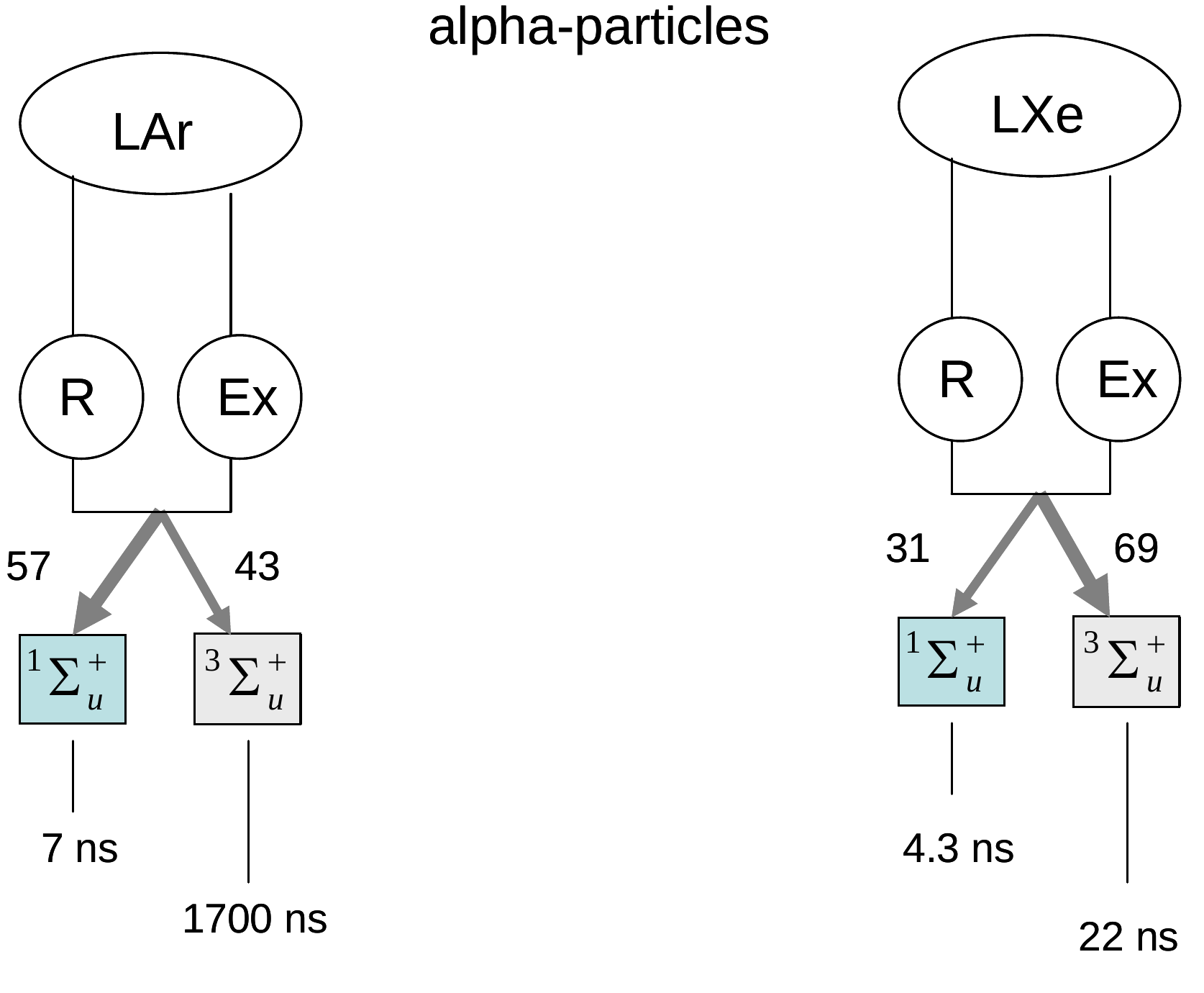}}
\caption{Observable distribution (percent) of the luminous energy for
$\alpha$-particles, based on results in~\cite{Hitachi83}. For liquid
xenon, different estimates for the fast/slow intensity ratio have also
been obtained: 0.53/0.47 from~\cite{Kubota80} and 0.77/0.23
from~\cite{Kubota82}.}
\label{fig:LuminousAlphas}
\end{figure}

For particles with higher LET, such as
$\alpha$-particles or fission fragments, no difference in the decay
time constants has been observed compared to those for
electrons~\cite{Kubota80,Kubota82,Hitachi83}. This means that the
species emitting the scintillation photons are the same for all
exciting particles. It also indicates that there is no substantial
non-radiative quenching of the $^3\Sigma_u^+$ and $^1\Sigma_u^+$
states (which, being present, would shorten their effective
lifetimes). However, the relative contribution of each component to
the observed scintillation signal does depend on the particle type and
energy. A strong enhancement of the fast component with respect to the
slow one has been observed with 5~MeV $\alpha$-particles in liquid
xenon~\cite{Kubota80,Kubota82,Hitachi83} (see 
Figure~\ref{fig:LuminousAlphas}).

The linear energy transfer for $\alpha$-particles is higher by a
factor of $\sim$100 than for fast electrons. Therefore, the density of
the ionized and excited species along the particle track is also much
higher for $\alpha$-particles, which leads to stronger and faster
recombination. No long ($\sim\mu$s) tail nor the $\sim$40~ns
recombination component have been observed in LXe with
$\alpha$-particles, which means that, in this instance, the
recombination time is shorter than the excimer decay time constants
even in liquid xenon, contrary to what happens for
electrons~\cite{Kubota80}. Moreover, it is an experimental fact that
only a few percent of the charge initially created along
$\alpha$-particle tracks can be extracted even with fields of
$\sim$10~kV/cm (for electrons, this would collect nearly 100\% of the
ionization). Therefore, the two scintillation mechanisms (direct
excitation and recombination) cannot be distinguished in
practice. This is represented in the diagram in 
Figure~\ref{fig:LuminousAlphas}
by the joining of the two channels together.

Strong recombination does not explain {\it per se} why the relative
contribution of the fast component is enhanced for
$\alpha$-particles. To our knowledge, a comprehensive explanation for
this observation is still missing, although a number of hypotheses
have been suggested~\cite{Hitachi83}. One of the plausible 
mechanisms is quenching of singlet states in superelastic collisions
with thermal electrons resulting in transitions from the singlet  to triplet state.
These transitions are more probable in the case of electrons than for 
$\alpha$-particles due to slower recombination along the electron tracks. 

An even stronger contribution of the fast component to the total light
yield was found for fission fragments, which have LET values
$\sim$10$^5$ times higher than those for electrons. According
to~\cite{Hitachi83}, in liquid argon the number of photons emitted in
transitions from the singlet state for electrons, $\alpha$-particles
and fission fragments is about 23\%, 57\% and 75\% of the total,
respectively. In liquid xenon, the singlet component contributes with
about 60\% in the case of $\alpha$-particles and fission fragments and
only $\sim$5\% in the case of relativistic electrons.

With regard to the scintillation decay time, a similar situation is
found in the case of low energy nuclear recoils: the effective decay
time for recoils is also shorter than that for electrons and
$\gamma$-rays. This was the basis of $\gamma$-ray background
discrimination in early dark matter detectors using liquid xenon as a
target~\cite{Alner05,Bernabei02}. For example, decay times of 33~ns
and 19~ns were observed for $\sim$20~keV electrons and xenon recoils
(after elastic scattering of neutrons),
respectively~\cite{Alner05}. In liquid argon, the enhancement of the
singlet component for nuclear recoils is of the same order of
magnitude as in xenon (or even less significant) but the very large
difference between the singlet and triplet time constants allows much
more efficient pulse shape discrimination to be
used~\cite{Lippincott08,Boulay09}. (Liquid neon and helium are even
more promising from this point of view.)


\begin{figure}[t]
\centerline{\includegraphics[width=.77\textwidth]{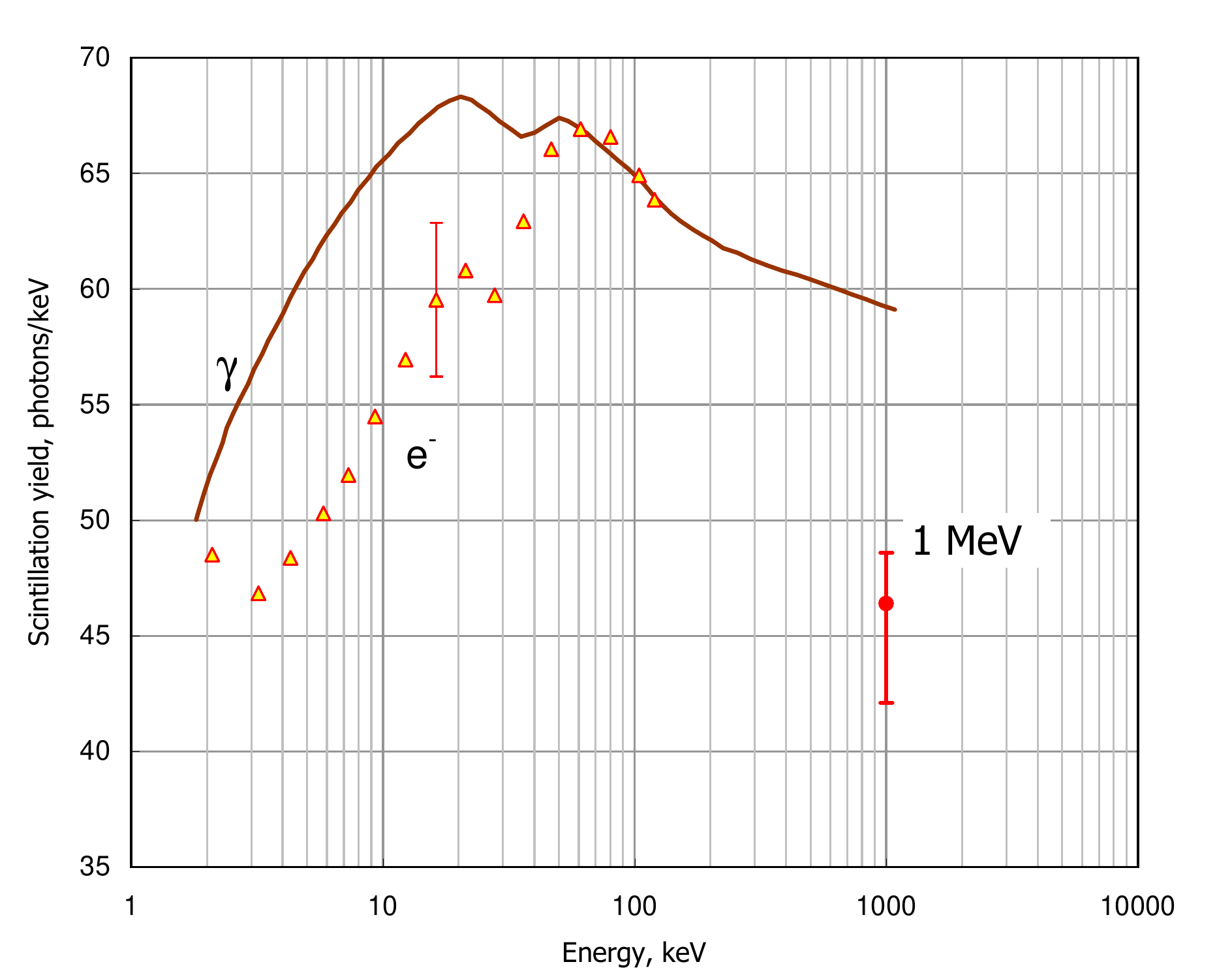}}
\caption{Scintillation yield of LXe as a function of energy for
electrons and $\gamma$-rays.  Solid line -- evaluated yield (absolute)
for $\gamma$-rays based on a comprehensive analysis of the existing
experimental data in~\cite{Szydagis11}. Triangles -- relative
measurements with Compton electrons in~\cite{Aprile12d}; these data
were re-normalized by us to the evaluated curve at $\approx$120~keV
(see main text).  
The circle is for 1~MeV electrons from~\cite{Doke02}.
\vspace{0.2cm}
}
\label{fig:ScintillationYieldLowEnergy}
\end{figure}

\vspace{5mm}
In general, the liquefied noble gases are very good scintillators. For 1 MeV electrons 
and \mbox{$\gamma$-rays}, yields of 40~to~50~photons/keV, i.e.~comparable 
to that of NaI(Tl), have been reported for LXe (see~\cite{Doke02,Chepel05} 
and references therein). At lower energies, the light yield increases 
rather significantly, especially in xenon: in the energy region 
between 10 and 100 keV LXe can produce nearly 70~photons/keV. 
A comprehensive analysis of the existing data for $\gamma$-ray scintillation
in liquid xenon has been presented in~\cite{Szydagis11}. 
Those authors also developed a simulation
model consistently describing most of the available datasets. Their results 
are shown by the brown curve in Figure~\ref{fig:ScintillationYieldLowEnergy}. 
Recently, the relative scintillation efficiency for low energy electrons has also been reported~\cite{Aprile12d}. The electrons were produced in 
Compton scattering of 662~keV $\gamma$-rays in a liquid xenon target. 
These data, re-normalized by us to the absolute light yield 
from~\cite{Szydagis11} at $\approx$120~keV, 
are also shown in Figure~\ref{fig:ScintillationYieldLowEnergy} 
(red yellow-filed triangles). When considering these two datasets 
one must bear in mind that absorption of $\gamma$-rays in the liquid 
results in not one but several electrons of different energies
(photoelectron, those resulting from absorption of xenon fluorescence, 
Auger electrons). 
Therefore, direct comparison of the light yields 
for electrons and $\gamma$-photons of the same energy is not generally correct 
at low energies\footnote{Although it actually works out correctly at $\approx$120~keV for xenon.}.
However, the similarity of the trends with energy between 
the two datasets gives an important insight into the physics of the scintillation
process, as we shall discuss in the context of recombination in
Section~\ref{sec:Recombination}. There are also important practical 
implications that follow from the comparison shown in
Figure~\ref{fig:ScintillationYieldLowEnergy}: the difference in light yields 
must be taken into account when performing detector calibration with low 
energy electrons or $\gamma$-rays. 

It should be remarked that absolute measurements of the scintillation
yield in liquefied rare gases are difficult, in large part due to the
short wavelength of scintillation photons (178.1~nm for liquid xenon
and 129.6~nm for argon --- excited with
$\alpha$-particles~\cite{Jortner65}).  For xenon, quartz-windowed PMTs
can be used, but for argon wavelength shifters are usually required
($p$-terphenyl, tetraphenyl butadiene (TPB) or other). Poor knowledge
of the angular reflectivity of different materials in this wavelength
region and absolute calibration of the VUV photon detectors pose
additional difficulties. (See Section~\ref{sec:LightPropagation} for
more information on material reflectivities.)

The scintillation efficiency, or the number of photons per unit
deposited energy, is frequently expressed via its reciprocal quantity,
the $W_s$-value, similarly to the $W$-value for charge. $W_s$ is
defined by dividing the energy deposited by a particle in the liquid,
$E_0$, by the number of emitted photons, i.e.~$W_s=E_0/N_{ph}$. The
number of scintillation photons emitted from a particle track depends
strongly on the electric field, being maximum at $E = 0$. Therefore,
$W_s$ is defined for zero field (contrary to the $W$-value, which is
defined for an infinite field).

Assuming that all the excited atoms give one VUV photon (after
undergoing the excimer formation\footnote{Formation of self-trapped
excitons, using the terminology of solid state physics.} phase
described above) and that all initially created electron-ion pairs
recombine and also give a photon, one can write for the number of
emitted photons $N_{ph}=N_{ex}+N_i$. From this equation, and recalling
that $W=E_0/N_i$, one can establish a simple relationship between the
two $W$-values:
\begin{equation}
   W_s^{min} = \frac{W}{1+N_{ex}/N_i} \,.
  \label{eqn5}
\end{equation}

This $W_s^{min}$ corresponds to the minimum possible energy needed to
produce a scintillation photon (maximum $N_{ph}$) when there are no
quenching processes, which depends very much on the track structure
and ionization/excitation density on its different parts. We also note
that $W_s$ reflects the efficiency of energy transfer to the medium
through both direct excitation and ionization (in contrast with the
$W$-value for ionization, which determines $N_i$ only). This has
important consequences for improving energy resolution and linearity
of detector response to electrons, as discussed in 
Section~\ref{sec:CombinedSignal}.

The available data on $W_s$ for different particles are shown in
Table~1. Compilations of the existing experimental and theoretically
estimated $W_s$ values for liquid xenon can be found
in~\cite{Doke99,Chepel05,Szydagis11,Horn11}. One can see that, in
spite of the existence of a few discordant points, there is a general
agreement, within $\approx$10\%, between the various datasets, except
probably for low energy nuclear recoils, where some significant
statistical and systematic uncertainties still remain (although the
situation has improved in recent times).

\begin{table}[ht]
  \label{tab:WS}
  \centering
  \caption{Energy expended per scintillation photon for different particles.}
  \vspace{2mm}
  \small
  \begin{tabular}{|l|c|c|c|c|}
    \hline
    \rule{0pt}{3ex} 
    \ \ \ \ \ \ \ \ \ \ \ Particle  & Energy    & LET, MeV/(g$\cdot $cm$^{2}$)  &  $W_s$, eV (LAr)  & $W_s$, eV (LXe)         \\ 
    \hline 
    \rule{0pt}{4ex} 
    No quenching; $W_s^{min}$ 	& --            & --                   	& $19.5\pm1.0^{\ a)}$       & $13.8\pm0.9^{\ a)}$     \\
    ~                           & ~             & ~                     & $19.8^{\ b)}$             & $13.0^{\ b)}$           \\ 
    ~                           & ~             & ~                     & $18.4^{\ b)}$             & $14.7\pm1.5^{\ c)}$     \\ 
    ~                           & ~             & ~                     & ~                         & $13.45\pm0.29^{\ d)}$   \\ 
    ~                           & ~             & ~                     & ~                         & $13.7\pm0.2^{\ e)}$     \\ 
    \rule{0pt}{3ex}
    Relativistic electrons      & 1 MeV         & $\approx 1$           & $25.1\pm2.5^{\ c)}$       & $23.7\pm2.4^{\ c)}$     \\ 
    ~                           & ~             & ~                     & $24.4^{\ a)}$             & $21.6^{\ a)}$           \\ 
    ~                           & ~             & ~                     & ~                         & $22.5\pm2.5^{\ f)}$     \\ 
    ~                           & ~             & ~                     & ~                         & $<35^{\ g)}$            \\ 
    ~                           & ~             & ~                     & ~                         & $42\pm6^{\ h)}$         \\ 
    ~                           & ~             & ~                     & ~                         & $67\pm22^{\ i)}$        \\ 
    \rule{0pt}{3ex}
    Low energy electrons        & 20 -- 100 keV & $\sim$7 to 2          & --                        & $18.3\pm1.5^{\ f)}$     \\ 
    ~                           & ~             & ~                     & ~                         & $14.2^{\ j)}$           \\ 
    ~                           & ~             & ~                     & ~                         & $12.7\pm1.3^{\ k)}$     \\ 
    ~                           & ~             & ~                     & ~                         & $29.6\pm1.8^{\ l)}$     \\ 
    \rule{0pt}{3ex}
    $\alpha$-particles    & $\approx 5$~MeV     & $\sim4\times10^{2}$   & $27.1^{\ a)}$             & $17.9 ^{\ a)}$          \\
    ~                           & ~             & ~                     & $27.5 \pm2.8  ^{\ c)}$    & $19.6 \pm2.0  ^{\ c)}$  \\
    ~                           & ~             & ~                     & ~                         & $16.3 \pm0.3  ^{\ m)}$  \\
    ~                           & ~             & ~                     & ~                         & $17.1 \pm1.4  ^{\ f)}$  \\
    ~                           & ~             & ~                     & ~                         & $39.2  ^{\ n)}$         \\
    \rule{0pt}{3ex}
    Relativistic heavy ions     & $\sim 1$ GeV/amu & $\sim10^2$ to $10^3$ & $19.4 \pm2.05  ^{\ c)}$ & $14.7 \pm1.5  ^{\ c)}$  \\
    \rule{0pt}{3ex}
    Nuclear recoils$^*$         & 60 keV        & $2.9/4.0\times10^3$   & $\sim100^{\ p)}$ (exp)    & $95\pm20^{\ r)}$ (exp)  \\
    ~                           & ~             & ~  & $\sim90^{\ q)}$ (theor)   & $\sim77^{\ s)}$ (theor) \\
    \rule{0pt}{2ex} 
    ~                           & 20 keV        & $2.6/2.7\times10^3$   & $\sim100^{\ p)}$ (exp)    & $110\pm20^{\ r)}$ (exp) \\
    ~                           & ~             & ~  & $\sim105^{\ q)}$ (theor)  & $\sim86^{\ s)}$ (theor) \\
    \rule{0pt}{2ex} 
    ~ 	                        & 5 keV         & $1.9/1.5\times10^3$   & $\sim100^{\ p)}$ (exp)    & $160\pm40^{\ r)}$ (exp) \\
    ~                           & ~             & ~  & $\sim140^{\ q)}$ (theor)  & --                      \\
    \rule{0pt}{3ex}
    Fission fragments           & $\sim 1$ MeV/amu & $\sim10^4$		& $\sim110^{\ t)}$          & $60^{\ u)}$             \\
    \hline
  \end{tabular}
  \footnotesize 
  \parbox{14cm}
	 { \rule{0pt}{4ex}
	   $^{*}$ For nuclear recoils the total LET values, estimated from range tables
 in~\cite{SRIM}, are presented; the electronic part can be roughly obtained by multiplying 
 by $\sim 0.2$. Values in column 3 are for argon/xenon.\\
	   $^{a)}$ From \cite{Doke02}: evaluated from several estimates and considered by those authors to be the most probable values.	\\
	   $^{b)}$ From \cite{Doke02}: other estimates.	\\
	   $^{c)}$ \cite{Doke99,Doke90}; 
	   $^{d)}$ \cite{Shutt07};
	   $^{e)}$ \cite{Dahl09};	\\
	   $^{f)}$ \cite{Chepel05};
	   $^{g)}$ \cite{Lavoie76}; 
	   $^{h)}$ \cite{Sonsbeek95}; 
	   $^{i)}$ \cite{Braem92}; 
	   $^{j)}$ \cite{Seguinot92}; 
	   $^{k)}$ \cite{Seguinot95};
	   $^{l)}$ \cite{Belli93};
	   $^{m)}$ \cite{Miyajima92};
	   $^{n)}$ \cite{Aprile90}.	\\
	   $^{p)}$ Using measured scintillation efficiency referred to
	   122~keV $\gamma$-rays \cite{Gastler11} and $W_s^{min}$ from
	   \cite{Doke02}.  \\
	   $^{q)}$ Using theoretical quenching factor from
	   \cite{Suzuki11}, \cite{Hitachi04} and $W_s^{min}$ from
	   \cite{Doke02}.  \\
	   $^{r)}$ Estimated by the authors using the relative
	   scintillation efficiency with respect to 122~keV
	   $\gamma$-rays from data compiled in \cite{Horn11} (see our
	   Figure~\ref{fig:LXeLeff}), with $W_s^{min}$ from \cite{Doke02}. \\
	   $^{s)}$ Using a theoretical estimate for the quenching
	   factor from \cite{Hitachi05} and $W_s^{min}$ from
	   \cite{Doke02}.  \\
	   $^{t)}$ \cite{Doke88} with $W_s^{min}$ from \cite{Doke02}. \\
	   $^{u)}$ From \cite{Doke99}.
	 }
\end{table}

\begin{figure}[t]
\centerline{\includegraphics[width=.95\textwidth]{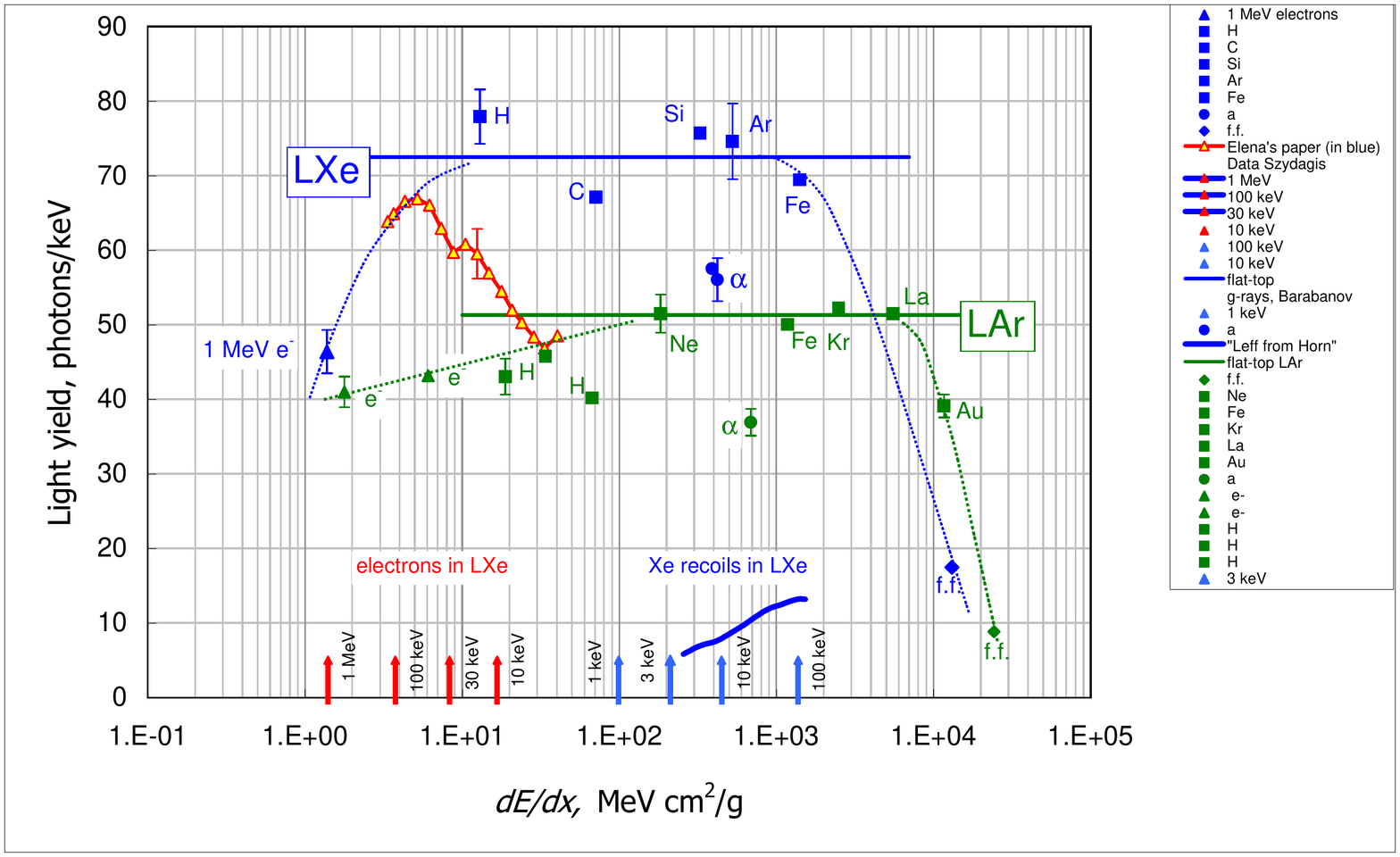} }
\caption{
Scintillation yield of LXe and LAr as a function of linear
energy transfer for various particles.  Data points (in blue for LXe,
and in green for LAr) are after~\cite{Doke02} (Figures 2,~4 and Table~III); 
not all data points are shown for clarity.  For the same reason,
only a typical error bar is shown for each dataset.  Dashed lines are
for guiding the eye only, no underlying model is assumed; the top
plateau corresponds to a minimum energy to produce a scintillation
photon $W_s^{min}$=13.8~eV for LXe and 19.5~eV for LAr according
to~\cite{Doke02,Doke05}.  Red arrows indicate average $dE\!/\!dx$
values for electrons in LXe calculated as the initial particle energy
(indicated next to arrow) divided by the range from
ESTAR~\cite{ESTAR}; blue arrows indicate $(dE\!/\!dx)_{e}$ for Xe
recoils in LXe calculated in a similar way using stopping power and
range tables from SRIM~\cite{SRIM} and the Lindhard partition function
from~\cite{Lindhard63}.  
The red yellow-filed triangles represent the relative measurements 
with Compton electrons reported in~\cite{Aprile12d}; 
we re-scaled their energy to $dE\!/\!dx$ using ESTAR as above, 
and \mbox{re-normalized} their response to that of $\gamma$-rays at 
$\approx$120~keV using the evaluated curve from~\cite{Szydagis11} 
as in Figure~\protect\ref{fig:ScintillationYieldLowEnergy}. 
}
\label{fig:ScintillationYield}
\end{figure}

\begin{figure}
\centerline{\includegraphics[width=.7\textwidth]{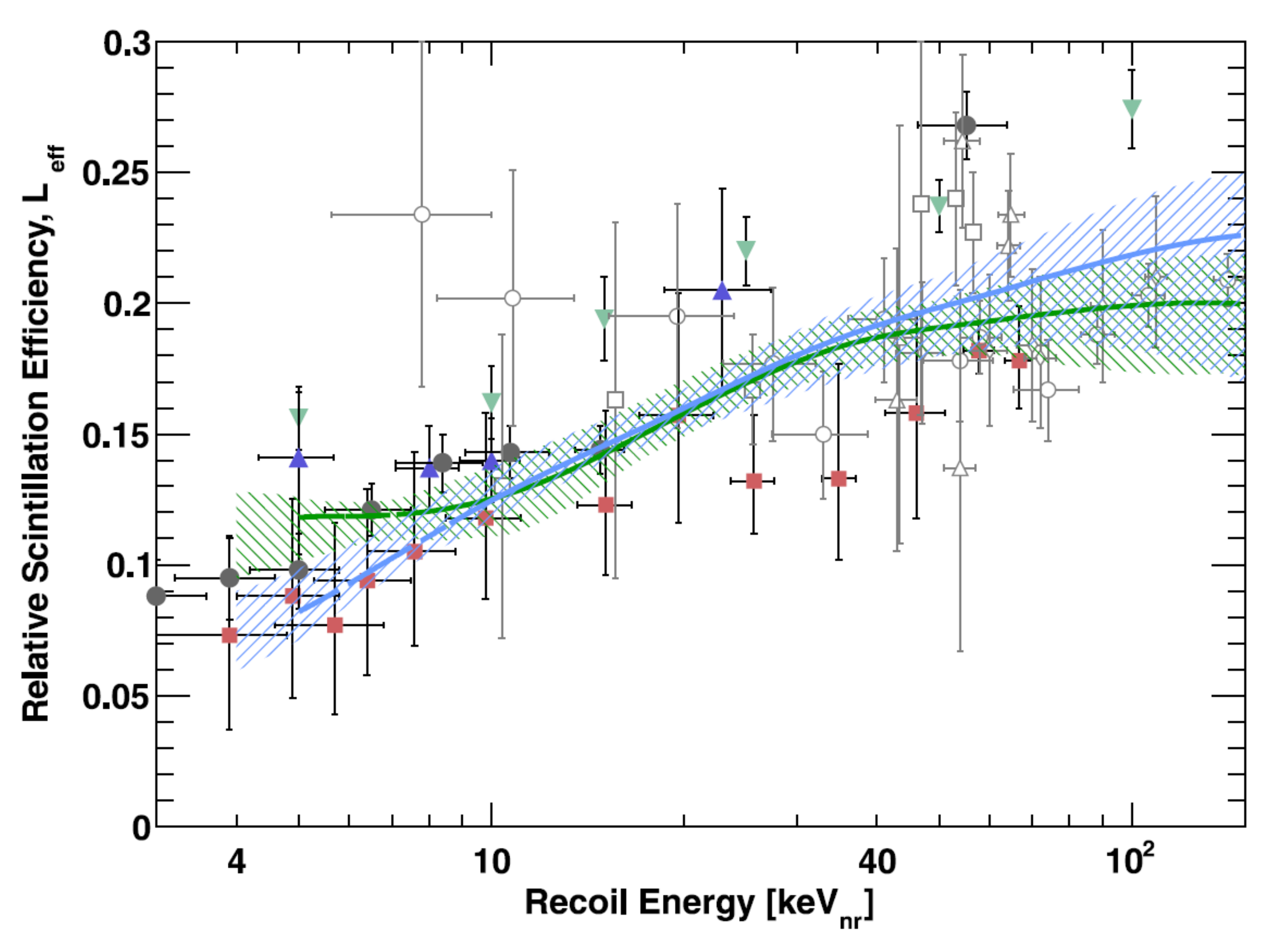}}
\caption{Energy-dependent relative scintillation yield for nuclear
recoils in liquid xenon; data from mono-energetic neutron beam
measurements are labelled by the following markers:
($\blacksquare$)~\cite{Manzur10}; ($\bullet$)~\cite{Plante11};
($\blacktriangle$)~\cite{Aprile09}; ($\diamond$)~\cite{Arneodo00};
($\triangle$)~\cite{Akimov02}; ($\circ$)~\cite{Chepel06};
($\square$)~\cite{Aprile05}. Indirect measurements from Monte Carlo:
($\blacktriangledown$)~\cite{Sorensen09}; the two lines are the
indirect measurements obtained from the first and second runs of
ZEPLIN-III, with 68\% CL bands shown in green
(\textbackslash\textbackslash\textbackslash) and blue (///),
respectively~\cite{Horn11}. (From~\cite{Horn11}; with permission from
Elsevier.) }
\label{fig:LXeLeff}
\end{figure}

Figure~\ref{fig:ScintillationYield} shows a compilation of
scintillation yields of LXe and LAr for various particles of different
energies as a function of LET. 
Although no unified picture on the LET dependence of the scintillation
efficiency emerges from the figure, some observations can be made.  
It is apparent that a $dE\!/\!dx$-based model is at odds with the
experimental data for low energy electrons. The behavior of low
energy nuclear recoils is similarly discrepant; an indicative curve
for the scintillation yield of Xe recoils with energy
$\lesssim$100~keV is also shown in Figure~\ref{fig:ScintillationYield}
(more precise data are shown in Figure~\ref{fig:LXeLeff} and discussed
later in this section). A significant reduction of scintillation yield 
is obvious in this case. 

The LET dependence of the scintillation efficiency of liquid argon and
xenon for high energy particles is discussed
in~\cite{Doke02,Doke99,Doke88,Doke90,Tanaka01}.  The scintillation
yield is approximately constant for intermediate LET values, in the
range between $\sim$10$^2$ and $\sim$10$^3$~MeV/(g$\cdot$cm$^2$), but
decreases for lower and higher ionization densities. This trend can be
observed both in LXe and LAr.  According to those studies, the
`flat-top response' in the region of intermediate LET corresponds to
the situation when each of the excited and ionized species created by
a particle gives a photon, and therefore $W_s=W_s^{min}$ in this
region. The $W_s^{min}$ value for LAr was calculated to be
19.5$\pm$1.0~eV, in good agreement with experiment (see Table~1). In
LXe $W_s^{min}$ is lower and, furthermore, some disagreement remains
between the values obtained from theoretical estimates and
experimental data.  Assuming $N_{ex}/N_i=0.06$, a value of
$W_s^{min}=14.7\!\pm\! 1.5$~eV was calculated in~\cite{Doke02} while
experimental data point to about 13.0~eV, which translates to a ratio
of $N_{ex}/N_i \approx 0.20$ instead.  Therefore, an evaluated value
of $W_s^{min}=13.8\!\pm\! 0.9$~eV was presented by those authors.
More recent measurements resulted in 13.45$\pm$0.29~eV~\cite{Shutt07}
and 13.7$\pm$0.2~eV~\cite{Dahl09} for $W_s^{min}$ in liquid xenon.  We
refer to Table~1 for other references and also to~\cite{Szydagis11}
for a recent compilation.

Reduction of the scintillation yield at low LET values is attributed
to a higher probability for an electron to escape recombination
even at zero field. This is the case for $\sim$1~MeV electrons and
$\gamma$-rays. In the high LET region above 
$\sim$10$^3$~MeV/(g$\cdot$cm$^2$)
a decrease is observed too, but for a different reason. 
The density of excited species along the
particle track becomes high enough to render non-negligible the
probability of collisions between them. The mechanism of bi-excitonic
quenching has been proposed in~\cite{Hitachi83}. This mechanism
implies autoionization of one of the two colliding free excitons
($R^*\!+\!R^* \to R \!+\! R^+\!+\!e^-$) at the beginning of the exciton
formation process. Although the formed $R^+$ ion has a good chance of
recombining with an electron to produce a new excited state, this
would result, at best, in one emitted photon instead of two --- one
from each of the two excitons initially created if these underwent a
normal process\footnote{
When mentioned in the context of nuclear recoils,
this kind of quenching mechanism is usually referred to as 
`electronic quenching' to distinguish from `nuclear quenching' 
which is the fraction of the particle energy transferred to atoms
of the medium in elastic collisions, i.e.~lost to heat.} 
  $R^*\!+\!R \to R_2^* \to R\! +\! R\! +\! h\nu$. 

\vspace{-0.25mm}
Low energy electrons and nuclear recoils
($\lesssim$100~keV) fall out of the trend observed in
Figure~\ref{fig:ScintillationYield} for higher energies.  This fact
can be explained by a different track structure and therefore a
different recombination mechanism. We shall postpone this discussion
until Section~\ref{sec:Recombination}, but we do consider here the
experimental data (and some interpretations) on the scintillation
yield for nuclear recoils --- a case of special importance for DM and
CNS search experiments.

\vspace{-0.25mm}
A significantly reduced light yield has been found for low energy
xenon ions traveling in LXe and argon ions in LAr.  Xenon recoils
below 140~keV are found to expend at least $W_s \sim 100$~eV per
scintillation photon produced, a much higher energy than, for
instance, for electrons.  In the dark matter community, the recoil
scintillation efficiency is usually referenced to the scintillation
yield for 122~keV $\gamma$-rays from a $^{57}$Co source, a convenient
energy for detector calibration (one should, however, keep in mind
that the light yield for $\gamma$-rays of this energy is $\approx$12\%
lower than the maximum possible yield, while quenching calculations
usually use the maximum as reference).  According to this common
definition, the scintillation efficiency is
$L_{eff}(E)=W_{s,e}(\mathrm{122~keV})/W_{s,\mathrm{Xe}}(E)$ at zero
electric field (the indices `$e$' and `Xe' correspond to electron and
xenon recoils, respectively).

\vspace{-0.25mm}
Most experimental measurements of the scintillation efficiency for
nuclear recoils to date were carried out with different setups but
using the same method --- elastic scattering of mono-energetic $\sim$MeV
neutrons off a LXe
target~\cite{Manzur10,Arneodo00,Akimov02,Aprile05,Aprile09,Plante11,Chepel06}.
The scattered neutrons are detected at a fixed angle, thus allowing
the energy transferred to Xe atoms to be determined kinematically.

\vspace{-0.25mm}
An alternative, indirect method is afforded by modern Monte Carlo
codes, notably GEANT4 \cite{agostinelli03}, which can model the
elastic scattering of neutrons as well as detector response effects
very accurately. The energy-dependent scintillation efficiency can be
extracted by fitting the scintillation response obtained with a
broadband source (e.g.~Am-Be) with simulated spectra of deposited
energy folded with all efficiencies and energy resolution (including
the sought quenching effect), as demonstrated in~\cite{Horn11}. The
beam measurements do not rely directly on Monte Carlo, but they
require small chambers with high light yield, usually operated at zero
electric field to maximize light yield and light collection. The
indirect method can be applied to calibration data from real dark
matter experiments and provides, therefore, a useful validation of
nuclear recoil detection efficiencies. The most recent datasets,
namely~\cite{Manzur10}, \cite{Plante11}, and~\cite{Horn11} report a
nuclear recoil scintillation efficiency which decreases gently with
energy, as shown in 
Figure~\ref{fig:LXeLeff}.

As far as liquid argon is concerned, fewer experimental data exist on
the relative scintillation efficiency for nuclear recoils. A value of
0.28 has been reported for 65~keV argon recoils referred to
$\gamma$-rays of 20~keV~\cite{Brunetti05}. However, the measurements
were carried out with an applied electric field, so that the
recombination component of primary scintillation was partly
suppressed, the effect being more significant for $\gamma$-rays than
for nuclear recoils. In more recent measurements, a constant
scintillation efficiency of 0.25$\pm$0.02 (relative to 122~keV
$\gamma$-rays) has been obtained for argon recoils in the range from
20 to 250~keV~\cite{Gastler11}. Even newer results were reported
in~\cite{Regenfus12}; these are compatible with a flat interpretation
of $L_{eff}$ with a mean value of 0.29$\pm$0.03 above 20~keV. In both
cases a slight increase in efficiency was observed below 20~keV
(similarly to some measurements in xenon~\cite{Chepel06}).

Some recent data exist also for liquid neon in the range
30--370~keV~\cite{Lippincott11}; an average efficiency of 0.24 is
reported above 50~keV.  Similarly to liquid argon, an upturn at low
energies is also observed in this work (although the authors point out
that the uncertainties on the lowest energy points are also the most
significant).

The quenching effect for xenon recoils has been assessed theoretically
in~\cite{Hitachi05}. The theoretical prediction, based on the concept
of bi-excitonic quenching for the electronic component and using
Lindhard's theory~\cite{Lindhard63,Lindhard63b} to calculate the
contribution from nuclear collisions,\footnote{ The observed reduction
of the light yield with respect to its maximum value due to the
increasing contribution of the nuclear component is sometimes referred
to as `nuclear quenching' (misleadingly, in fact, because strictly
speaking there is no quenching of the excited species, only a lower
probability for their formation).  } agrees quite well with
experimental data. The model also predicts a gradual decrease of the
scintillation efficiency with decreasing recoil energy. It was
concluded that for 60~keV xenon recoils electronic quenching is
responsible for the reduction of the scintillation efficiency by a
factor of $\approx$1.5, while nuclear collisions contribute with
another factor of $\approx$3.2. Applied to liquid argon, a similar
behavior of the quenching factor with recoil energy is
predicted~\cite{Suzuki11,Hitachi04}: $\approx$0.26 for 100~keV recoils
with gradual decrease to $\approx$0.19 for 20~keV and, further, to
$\approx$0.15 for 5~keV. We recall that the quenching factor is
defined with respect to the maximum scintillation yield, which is
observed with relativistic light ions, i.e.~at intermediate LET
values; the light yield for $\sim$100~keV electrons is somewhat lower
(Figure~\ref{fig:ScintillationYield})
thus resulting in an efficiency referred to 122~keV
$\gamma$-rays of $L_{eff} \approx 0.32$, $\approx$0.23 and
$\approx$0.18 for 100~keV, 20~keV and 5~keV recoil energies,
respectively.

Another phenomenological approach to the problem was proposed
in~\cite{Mei08}. This study combined Lindhard's theory with Birks'
saturation law, which was rather successful in describing the LET
dependence for organic scintillators. This combination results in a
less pronounced dependence of $L_{eff}$ on recoil energy at higher
values. For xenon, the predicted relative scintillation efficiency is
constant down to approximately 20~keV recoil energy, at which point it
begins to decrease slowly. This approach seems to be better suited to
describe experimental data for argon and neon than for xenon.

The authors of~\cite{Manzur10} have found that their dataset on
$L_{eff}$ for LXe at zero field is better described if, in addition to
nuclear and bi-excitonic quenching as proposed in~\cite{Hitachi05}, a
non-negligible electron escape probability is assumed for nuclear
recoils tracks. This assumption was supported by a very weak
dependence of the extracted charge on applied electric field (see
Figure~\ref{fig:ChargeYieldVsField}
and 
Section~\ref{sec:IonizationYield}).
Considering a constant ratio
$N_{ex}/N_i=0.06$, both for electrons and nuclear recoils, they
arrived at rather high values for the fraction of escaping electrons,
$N_{esc}/N_i$, which increases with decreasing recoil energy from
$\sim$0.15 for 70~keV up to 0.7$\pm$0.2 for 4~keV.

None of the above models foresees an increase of the scintillation
efficiency at low recoil energies as observed in LAr.

\subsubsection{The role of recombination}
\label{sec:Recombination}

In view of the previous discussion, it is manifest that the
recombination light contributes significantly to the total
scintillation yield in the liquefied noble gases.  Recombination in
these liquids is rather strong and it is a complex process which
depends on a large number of factors: the initial distribution of ions
and electrons immediately after the particle passage, including all
secondary ionizations and in some cases also excitations; the speed
with which subexcitation electrons lose their energy to phonons and
the distance traveled to reach thermal equilibrium with the medium;
electron mobility and diffusion rate; collision frequency and the
probability of recombination to occur when an electron encounters a
positive ion.  In this section the interested reader will find a more
detailed discussion of recombination models which describe the
scintillation and ionization responses.

The scintillation efficiency as a function of LET for different
particles and energies shown in Figure~\ref{fig:ScintillationYield}
reflects well the complexity associated with recombination.  For high
energy particles a general trend was recognized early on (see
~\cite{Doke02,Doke99} and references therein): a flat-top response at
intermediate values of LET with some reduction in the low LET region
and a significant drop for high LET fission fragments, similar for LXe
and LAr. Clearly, low energy electrons --- and $\gamma$-rays --- 
as well as nuclear recoils fall out of that trend, and this has been 
recognized only recently.  In this section we search for an explanation 
for such irregular behavior starting, as before, from higher energies since
recombination along the tracks of these particles is better
understood.

A common feature of the interaction of high energy particles with
matter is a relatively well defined track along which the ionizations
and excitations are distributed.  The distribution of ionization
electrons, ions and excited species at the instant when all electrons
slow down to sub excitation energy is frequently called the `track
structure' and this determines to a large extent the subsequent
behavior of the charge carriers.  The track structure varies
significantly for different particles and LET values, thus hindering
the development of a universal theory of recombination in liquid
radiation detectors. Nevertheless, one can always distinguish a
cylindrically symmetric principal trace and secondary branches due to
$\delta$-rays; these can be partially cylinder-shaped or form more
complex structures, especially around the end point.

Thus, according to~\cite{Kubota79}, the track of a 1~MeV electron (low
LET) can be regarded as a column of widely spaced positive ions with
average distance between them close to the Onsager radius, $r_c =
e^2/(4\pi \epsilon_0 \epsilon k T)$ ($r_c=125$~nm for liquid argon and
49~nm for xenon), where $\epsilon $ is the dielectric constant and
other symbols have their usual meaning. Comparing these values with
the mean distance between atoms, $\sim$0.4~nm, it is clear that ions
are separated from each other by hundreds of neutral atoms.  At the
surface of a sphere with the Onsager radius, the potential energy of
the Coulomb attraction of the electron to its parent ion is equal, by
definition, to the kinetic energy of the thermal motion. Thus, the
electrostatic attraction to the ion dominates within the sphere, while
outside it the thermal motion is likely to draw the electron away from
the ion. The thermalization length for electrons is estimated as
$\sim$4.5~$\mu$m in LXe~\cite{Mozumder95b} and $\sim$1.7~$\mu$m in
LAr~\cite{Mozumder95a}, meaning that a good fraction of the electrons
can be found out of the reach of positive ions when thermalization is
achieved. In the absence of an external electric field, these will
partly diffuse to the chamber walls thus escaping recombination (these
escaping electrons are considered to be the cause of the decrease in
scintillation efficiency at low LET values discussed earlier --- see
Figure~\ref{fig:ScintillationYield}). The remaining fraction of
electrons governed by diffusion will sooner or later meet an ion and
recombine.

The mean volume density of free electrons along the track of 1~MeV
electrons can be estimated from~\cite{Kubota79} as
$\sim$2$\cdot$10$^{-7}$~nm$^{-3}$ for liquid argon and
$\sim$3$\cdot$10$^{-6}$~nm$^{-3}$ for liquid xenon (or even much lower
taking into account thermalization distances
from~\cite{Mozumder95a,Mozumder95b}).  These values are to be compared
with densities of up to $\sim$10$^{-2}$~nm$^{-3}$ to
$\sim$10~$^{-1}$~nm$^{-3}$ in the track core of an $\alpha$-particle
or a fast heavy ion (estimated
from~\cite{Hitachi92},~\cite{Miyajima74}
and~\cite{Northcliffe70}). For these particles, the positive ions form
a continuous line of positive charges with a mean separation,
projected to the particle trajectory, of about 0.22~nm and 0.13~nm for
LAr and LXe, respectively, so that an electron ejected from an atom
finds itself in a strong cylindrical field from which it will have
difficulty escaping. According to~\cite{Hitachi92}, an electron with
sub-excitation energy in liquid argon can stray from the track core up
to a distance of only some 5~nm, whereupon it turns back, being
attracted by the column of positive ions, crosses it and continues to
oscillate around the track axis until it recombines with one of the
ions. It has been shown in~\cite{Hitachi92} that about 10 passes are
sufficient for an electron to recombine with large probability, and
this takes in fact a very short time of $\sim$0.4~ps. This suggests
that, on $\alpha$-particle tracks, electrons recombine long before
they reach thermal equilibrium with the liquid, in contrast with the
low LET particles.

\enlargethispage{\baselineskip} 
The track structure of $\alpha$-particles and fast ions with similar
LET values in LAr was analyzed in~\cite{Hitachi92} leading to the
conclusion that more energy is deposited in the core track and less in
the penumbra in the case of $\alpha$-particles than for
ions. Moreover, the core radius was estimated as~$\sim$0.4~nm for
$\alpha$-particles versus $\sim$6~nm for ions meaning higher
ionization/excitation density along $\alpha$-particle tracks.  This
should lead to complete recombination which does not result, however,
in maximum scintillation yield in either LAr or LXe --- a fact
explained by bi-excitonic quenching, similarly to what happens for
particles with very high LET values such as fission fragments
(note the discordant points for $\alpha$-particles in 
Figure~\ref{fig:ScintillationYield}).

This shows how much the ionization/excitation pattern at the particle
interaction site can influence the observed scintillation signal (as
well as the ionization one, since the recombination strength naturally
affects the amount of charge that can be extracted from the particle
track --- we shall discuss this issue in
Section~\ref{sec:IonizationYield}).

Concerning the scintillation for low-energy electrons and
$\gamma$-rays (the former shown as yellow-filled red triangles in
Figure~\ref{fig:ScintillationYield}) we note that the
data agree reasonable well with the above trend for energies
$\gtrsim$80~keV, but this is not so for lower energies.  The higher
energy part can be suitably fitted with a Birks-type
law~\cite{Szydagis11} in terms of LET, as proposed in~\cite{Doke88}
for light ions and $\sim$1~MeV electrons:

\begin{equation}
  \frac{dY(E)}{dE}=\frac{A\,\frac{dE(E)}{dx} }{1+B\,\frac{dE(E)}{dx} } +
  \eta_{0} \,.
  \label{eqn:Birks}
\end{equation}
In this model, the scintillation yield due to recombination was
considered to be a sum of two components: recombination of geminate
and non-geminate types. Geminate recombination and the direct
excitation component do not depend on LET and are both included in the
constant term $\eta_0$. An electron is considered to undergo geminate
recombination if it thermalizes within the Onsager
sphere. Recombination of all remaining electrons is included instead
into the first term, which was obtained under the assumption that
recombination can be treated in terms of concentrations of electrons
($n_{-}$) and ions ($n_{+}$): $dn_{\mp}/dt \!\!=\!\! - \alpha n_{-}
n_{+}$, where $\alpha$ is the recombination coefficient, thus
presuming that this process is volumetric in nature. Further
simplification was made by setting $n_{-}\!\!=\!\!n_{+}\!\!= n$.  A
link to $dE\!/\!dx$ is obtained by assuming proportionality, $n
\propto dE\!/\!dx$, which is, strictly speaking, valid only for a
cylindrical geometry (i.e.~long tracks). Indeed, a function in the
form of equation~(\ref{eqn:Birks}) was successfully applied for
$E_{\gamma} \gtrsim$~80~keV but failed for lower energies where the
ionization pattern resembles more a spherical blob than a
cylinder~\cite{Szydagis11}. For these energies those authors used the
Thomas-Imel box recombination model which was rather successful in
parameterizing specifically the charge yield as a function of electric
field~$\mathcal{E}$~\cite{Thomas87}:\footnote{In fact, this model was
originally developed to describe the field-dependent charge yield for
low-LET particles such as MeV electrons. A better fit of the
experimental data was obtained, however, by adding a second term to
equation~(\ref{eqn:Thomas}), similar to the existing one, but
depending on an additional recombination parameter
$\xi'$~\cite{Thomas88}; this recognizes that the principal trace can
appear surrounded by low-energy $\delta$-electrons creating ionization
`blobs' with a different recombination strength.}

\begin{equation}
  \frac{Q(\mathcal{E})}{Q_0}=\frac{1}{\xi} ln(1+ \frac{1}{\xi}) \, ,
  \label{eqn:Thomas}
\end{equation}
with $\xi = N_0 \alpha /4a^2 \mu \mathcal{E} = \kappa / \mathcal{E}$;
here, $N_0$ is the total number of electrons and $\mu$~is the electron
mobility, assumed to be independent of field.  In this model,
recombination is also described through the term $\alpha n^{2}$ and,
to provide a connection between the electron/ion concentrations and
the total charge created by a particle, a box of arbitrary size $a$
containing that charge (assumed uniformly distributed) is defined.
Then, $\kappa = N_0 \alpha /4a^2 \mu$ is used as a single adjustable
parameter.

The brown curve in Figure~\ref{fig:ScintillationYieldLowEnergy} 
corresponds to a combination
of these two models --- Birks' for `long' tracks and Thomas-Imel for
`short' ones --- with which the authors of~\cite{Szydagis11} succeeded
in describing consistently most of the existing data on scintillation
yield of liquid xenon for \mbox{$\gamma$-rays}. To adapt the Thomas-Imel
model to zero-field conditions, $\mu \mathcal{E}$ was replaced by some
constant velocity so that the adjustable parameter becomes $\xi$ and
not $\kappa$.  In order to account for energy partition between
various channels (photo- and Compton electrons, X-rays, Auger
electrons) a detailed GEANT4 Monte Carlo simulation was used and the
most adequate model was chosen for each electron track having the
thermalization length scale as a cross-over distance.

Qualitatively, the decrease in scintillation efficiency on the low
energy (high LET) side can be understood as the increasing probability
for electrons to escape recombination. For sufficiently low energies,
the dimensions of the ionization region are determined by the
thermalization length $r_{th}$ rather than by the electron range and
so the charge concentration becomes a linear function of deposited
energy $n_{-} \approx (E/W) ( 3/4 \pi r_{th}^3)$ (here $W$ is the
$W$-value for charge --- see Section~\ref{sec:IonizationYield}).
However, this simple picture is in tension with the basic presumption
of the Thomas-Imel model which considers no electron diffusion.
Without diffusion, no escape from recombination is possible unless an
electric field is applied.

The combined model was also able to describe the field dependence of the 
scintillation yield. The geminate (Onsager) recombination component
included in the term $\eta_0$ in equation~(\ref{eqn:Birks}) was found to be
significant at zero field, but more easily suppressed by an external electric 
field than the non-geminate part~\cite{Dahl09,Szydagis11}.


To reproduce the field dependence in the modified Thomas-Imel
parameterization, some field effect on $\xi$ had to be empirically
re-introduced, although it was found to be much weaker than in the
original model: $\xi \propto \mathcal{E}^{-0.1}$,
approximately~\cite{Dahl09,Szydagis11} (originally, $\xi \propto
\mathcal{E}^{-1}$~\cite{Thomas87}).

In spite of physical inconsistencies in the existing recombination
models, the approach proposed in~\cite{Szydagis11} provides a very
useful framework for modeling detector response to low energy
backgrounds and to $\gamma$-rays used for calibration of DM detectors.
These inconsistencies reflect in fact our poor understanding of the
recombination process in noble liquids. Critical reviews of these and
other models can be found
in~\cite{Holroyd89,Hatano94,Schmidt97a,Bartczak05,Dahl09}.  Original
papers such as~\cite{Tachiya87,Shinsaka88,Mozumder95a,Mozumder95b} may
also constitute useful reading.

\vspace{5mm}
Even less is known about recombination along nuclear recoil tracks
with energy relevant for DM and CNS searches.  It is clear that the
contribution from nuclear collisions cannot be neglected, especially
in liquid xenon, but the details of the track structure are not
sufficiently clear. The Bohr impulse principle is not applicable in
the case of ions moving through a medium with velocity comparable to,
or lower than, that of atomic electrons (see~\cite{Mangiarotti07}).
Besides, as the medium consists of atoms of the same species, the
primary recoil can transfer a significant fraction of its kinetic
energy in each collision, thus losing rapidly its `projectile'
identity and producing a cascade of secondary recoils of comparable
energy which interact with the medium in the same way. Consequently,
the spacial distribution of ionizations and excitations will be quite
different from what is expected along the tracks of other particles.

\begin{figure}
\centerline{\includegraphics[width=.73\textwidth]{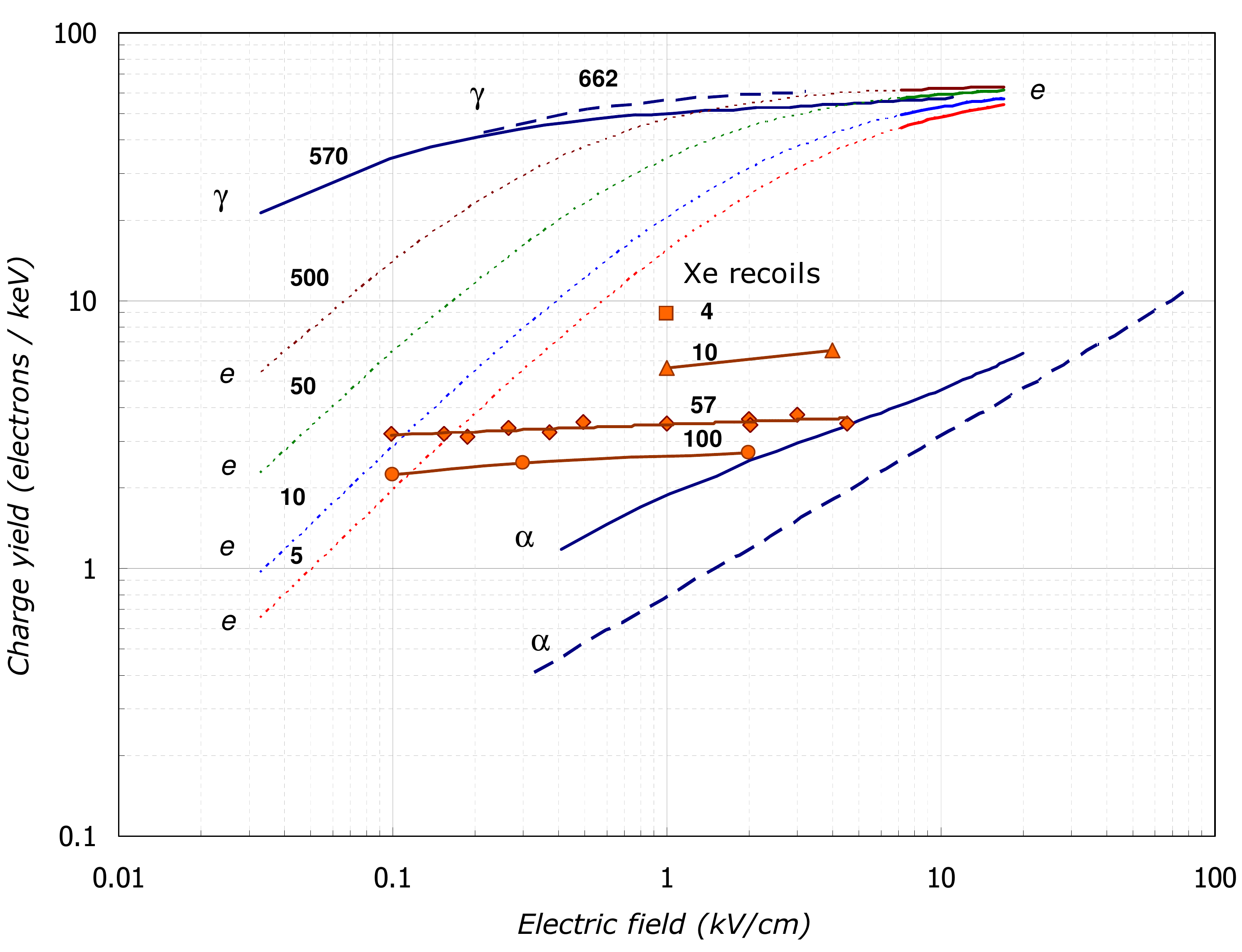}}
\caption{Free electron yield from the tracks of different particles as
a function of electric field in liquid xenon. Numbers next to the data
represent particle energy in keV. For $\alpha$-particles the data are:
solid line from~\cite{Aprile91b}, dashed line
from~\cite{Pruett67}. Red symbols indicate charge yield from xenon
recoil tracks from~\cite{Aprile06} and~\cite{Manzur10} (not all data
are shown). For $\gamma$-rays: solid line~\cite{Aprile91}, dashed
line~\cite{Barabash85}. For electrons: calculated using recombination
coefficients $k$ from ~\cite{Voronova89}. The coefficients were
determined from the charge yield measurements with $\mathcal{E}$ in
the range 7--15~kV/cm (in this field region, the data are represented
by solid lines), assuming $Q(\mathcal{E})=Q_0/(1+k/\mathcal{E})$.
Dotted lines correspond to extrapolation to the lower field region
using the above equation. We point out the difference between these
calculations for 500~keV and the experimental data for 570~keV and
662~keV, which may indicate a systematic shift. However, the important
information here is the dependence of the charge yield on electron
energy, which shows the opposite tendency to that observed for nuclear
recoils.}
\vspace{5mm}
\label{fig:ChargeYieldVsField} 
\end{figure}

\enlargethispage{\baselineskip}
Recombination affects both scintillation and charge yields --- the two
signals that constitute the basis of background rejection in most dark
matter detectors.  These signals are complementary and depend on
electric field in opposite ways.  It is therefore important to study
them simultaneously.  The charge yield from xenon recoils tracks has
been measured directly in dedicated
experiments~\cite{Aprile06,Manzur10} and also indirectly assessed from
calibration data of real dark matter
detectors~\cite{Sorensen09,Sorensen10b,Horn11}.  A relatively high
charge yield and weak electric field dependence has been found,
against expectations, as shown in Figure~\ref{fig:ChargeYieldVsField}.
Although data with simultaneous measurement of scintillation and
ionization are still scarce and a detailed discussion of this topic is
premature, a few interpretations of the existing data can be
mentioned.

\begin{figure}
\centerline{\includegraphics[width=.7\textwidth]{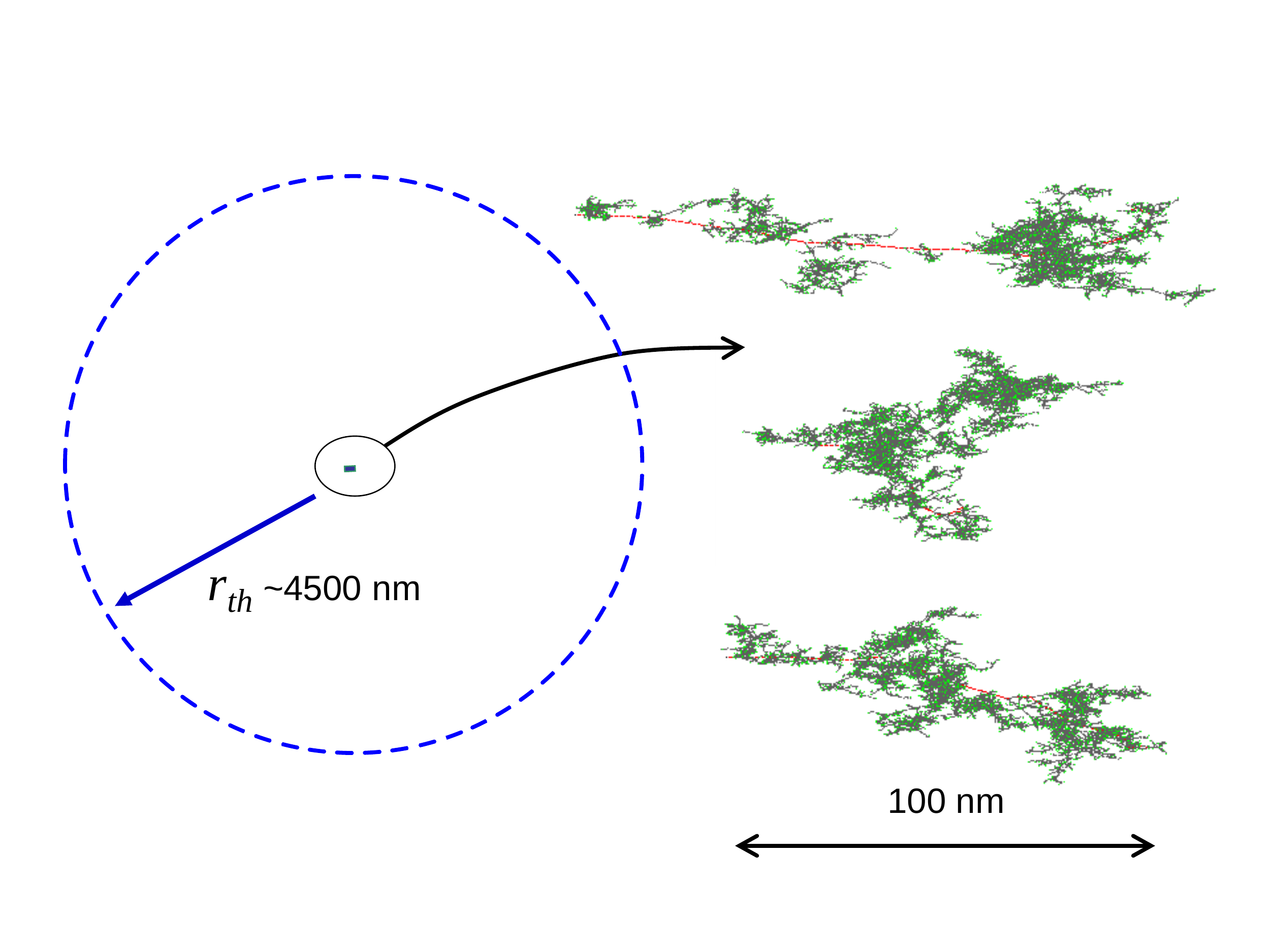}}
\caption{Xe recoil tracks in liquid xenon simulated with 
TRIM~\cite{SRIM} compared with the thermalization distance
(calculated in~\cite{Mozumder95b}). The projectile trajectory 
is shown in red; 
trajectories of secondary atoms/ions are in gray with end points in green (this
shows the number of secondary atomic recoils in the cascade). 
No electron tracks are shown.}
\label{fig:NRTrack} 
\end{figure}

Most of the energy of a Xe ion or atom moving in liquid xenon (this is
what is meant by a `nuclear recoil') is lost in elastic collisions
with other atoms.  The topology of the cascade represents a complex
and very ramified structure with tens or even hundreds of branches
formed by secondary recoils (see Figure~\ref{fig:NRTrack} for some
examples).  The transverse dimensions of the cascade are, in most cases,
comparable to that along the initial direction of primary recoil
(although long tracks resembling, to some extent, a high energy
particle also can appear with some probability).  Therefore,
recombination models based on cylindrical symmetry are hardly
applicable to nuclear recoils. The extent of these cascades is in the
$\lesssim$100~nm scale, which is much smaller than the electron
thermalization distance ($r_{th} \sim$4~$\mu$m in liquid
xenon~\cite{Mozumder95b}).  Therefore, the distribution of thermalized
electrons looks more like a sphere surrounding a tree-like core of
positive ions at its center with size $\lesssim$1/40 of that of the
sphere. In this sense, the recombination should be similar to that for
low energy electrons with the difference that, for the same particle
energy, only a fraction $\sim$0.2 goes into ionizations/excitations in
the case of nuclear recoils.

This picture is qualitatively consistent with the trend observed for
recoils of different energies in Figure~\ref{fig:ChargeYieldVsField}.
More electrons per keV can be extracted for lower recoil energies
because less charge is distributed over approximately a constant
volume with radius $\sim r_{th}$.  Thus, the electron density should
be roughly proportional to the recoil energy, something which is not
fully confirmed by the data.  The energy dependence of Lindhard's
energy partition function may partially explain the absence of strict
proportionality since the electronic component of the energy losses
decreases with decreasing energy~\cite{Lindhard63}).

A loose distribution of the thermalized ionization electrons may
explain the relatively large charge yield from nuclear recoil tracks
and its weak variation with field: practically all electrons escape
recombination even at very low field ($\sim$0.1~kV/cm), as argued
in~\cite{Manzur10}.  Considering a constant ratio $N_{ex}/N_{i} =
0.06$, both for electrons and nuclear recoils, those authors concluded
that the fraction of escaping electrons increases with decreasing
recoil energy from $\sim$0.15 for 70~keV to 0.7$\pm$0.2 for 4~keV.

The measured charge yield, although higher than one might expect on
the basis of the $dE\!/\!dx$ argument (cf.~$\alpha$-particles), is
still lower than for electrons even taking into account the energy
partition between electronic and nuclear parts. This may be explained
by assuming a much higher $N_{ex}/N_i$ ratio for nuclear recoils as
proposed in~\cite{Dahl09} and~\cite{Sorensen11} where it was estimated
to be $N_{ex}/N_i \sim 1$ \ (cf.~$\sim$0.1~for other particles --- see
Table~2). This would also explain lower S2/S1 ratios observed for
nuclear recoils than in the case of electrons.  As remarked
in~\cite{Dahl09}, a possible mechanism for more energy to be channeled
into excitation than ionization might be the lowering of some atomic
levels during the interpenetration of atomic shells of colliding xenon
atoms~\cite{Fano65}. Hypotheses of possible field or energy dependence
of the ratio $N_{ex}/N_i$ \,have also been proposed in~\cite{Dahl09,
Bezrukov11,Sorensen11}.

Explanation of the trend for the recoil scintillation efficiency
observed in Figure~\ref{fig:LXeLeff} in terms of escaping electrons
also seems plausible: the lower the recoil energy the more electrons
escape recombination and less light is emitted per unit energy.  In
this respect, the importance of bi-excitonic quenching, proposed to
approximate Lindhard's prediction for the electronic component of
energy transfer to the observed scintillation yield~\cite{Hitachi05},
might be questioned. On the other hand, one should bear in mind that
excitons do not migrate as fast as electrons. Their distribution
should accompany the cascade topology and therefore the existence of
high excitonic densities locally cannot be excluded.

The possibility of a more complex recombination picture cannot be
ruled out either, for example, the existence of regions with very high
ionization density from which extraction of electrons with field is
difficult. This may be the reason for a weak dependence of the charge
yield with field: when all escaped electrons have already been
collected and those in highly ionized regions require much higher
field to be extracted. A weak dependence of the scintillation signal
on the applied field for nuclear recoils~\cite{Aprile06} can also be
qualitatively explained on this grounds.  In this case, there may be
no need to consider $N_{ex}/N_i$ to be higher for nuclear recoils than
for other particles.

\subsubsection{Light propagation}
\label{sec:LightPropagation}

\begin{figure}
\centerline{\includegraphics[width=.7\textwidth]{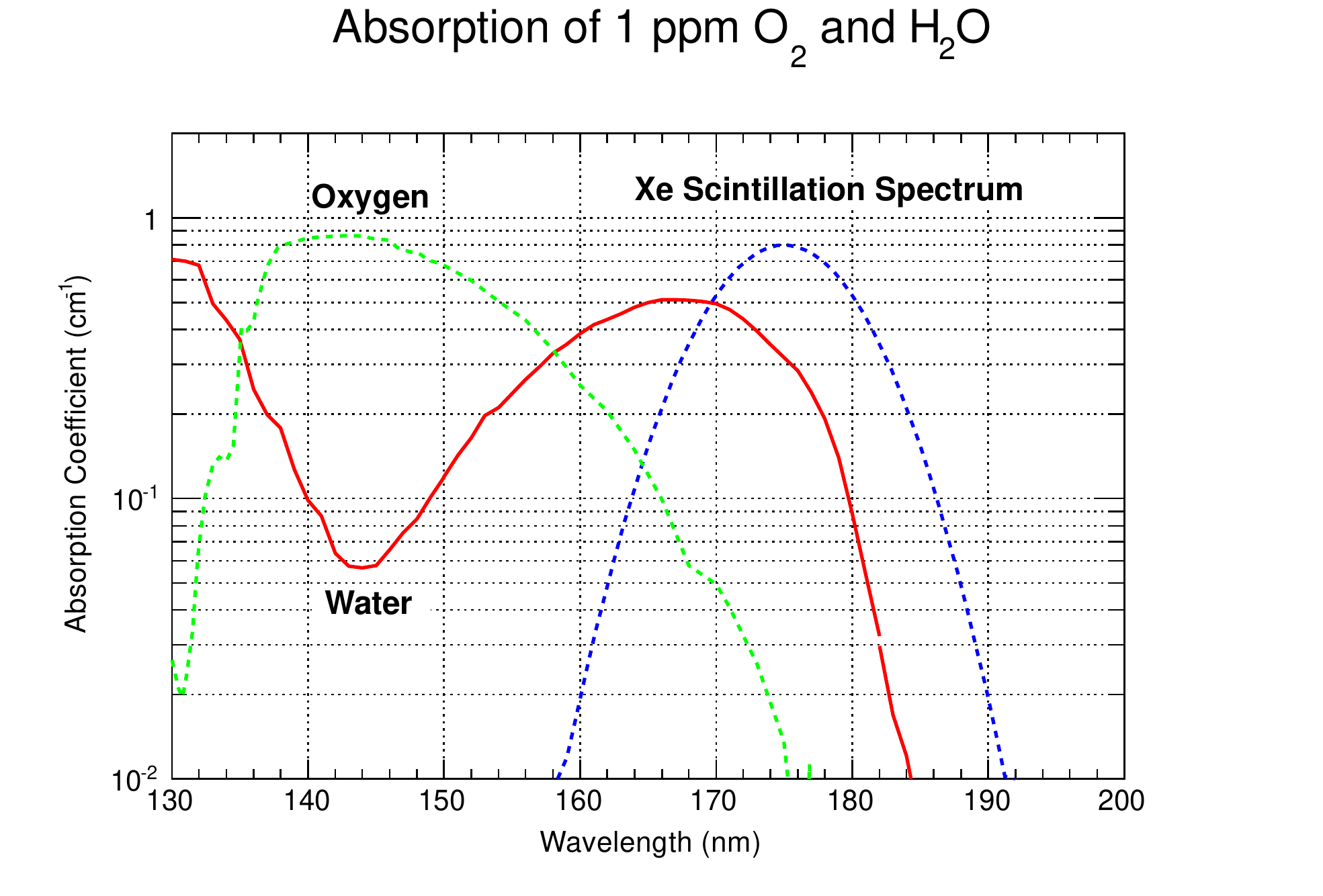}}
\caption{VUV absorption coefficients for 1~ppm water vapor and 1~ppm
oxygen in liquid xenon calculated in~\cite{Baldini05} (we note that in
the original publication the $y$-axis is wrongly labelled in
`m$^{-1}$' and this is corrected here to `cm$^{-1}$'; we thank
G.~Signorelli for pointing this out). (With permission from
Elsevier.)}
\label{fig:AbsO2H2O}
\end{figure}

Beyond the generation of scintillation photons one must consider which
conditions may affect their propagation in the liquid up to their
eventual detection. Two processes contribute to a finite photon
attenuation length in noble liquids: absorption, which leads to loss
of scintillation photons, and Rayleigh scattering, which is an elastic
process. Absorption is mostly due to impurities --- the
excimer-mediated scintillation mechanism described in
Section~\ref{sec:EmissionMechanisms} ensures that self-absorption of
VUV photons is very unlikely. Water vapor is one of the main culprits;
for example, 1~ppm H$_2$O in liquid xenon will absorb most
scintillation light in under 10~cm~\cite{Baldini05} (see
Figure~\ref{fig:AbsO2H2O}).

Separate measurement of the two light attenuation components is
difficult. Usually, a combined effect is observed, which is
characterized by a total attenuation length, $L$. The attenuation
length is related to the absorption and scattering lengths ($L_a$ and
$L_s$, respectively) via the equation $L^{-1} = L_a^{-1} +
L_s^{-1}$. The value of $L_a$ can be assessed if reflection on all
passive (non-sensitive) surfaces is close to 100\%. For example, $L_a
> 100$~cm has been reported for liquid xenon~\cite{Baldini05}. On the
other hand, in experiments where no special provision has been made to
enhance reflections, or where these were intentionally suppressed,
attenuation lengths between 30 and 50~cm have been obtained by several
authors~\cite{Braem92,Chepel94,Ishida97,Solovov04b}. Assuming that
$L_a > L_s$, this gives an estimate for the scattering length, $L_s
\approx L$. It has also been noticed that the measured $L$ increases
if the photon wavelength is shifted to longer values (as observed
in~\cite{Ishida97} when a few percent of Xe was added to LAr or LKr),
also indicating a significant contribution from Rayleigh scattering in
the pure liquid. For liquid xenon, a Rayleigh scattering length of
$\sim$30--50~cm is generally accepted, resulting both from the
measurements described above and from a calculation of scattering on
density fluctuations (30~cm~\cite{Seidel02}). These values are also
consistent with those estimated from the measured refractive
index~\cite{Baldini06,Solovov04b}. For liquid argon, an experimental
value of $L_s=66$~cm has been reported~\cite{Ishida97}, while
theoretical calculations predict 90~cm~\cite{Seidel02}.

As discussed in Section~\ref{sec:LiquidPurity}, the purity
requirements demanded by ionization readout are even more stringent,
but loss of scintillation performance is increasingly important in
large noble liquid detectors. Elastic scattering of scintillation
photons is not as problematic in small chambers (the total light yield
is preserved), but in very large, scintillation-only systems
reconstruction of the interaction position becomes more difficult.

\begin{figure}
\includegraphics[width=.5\textwidth]{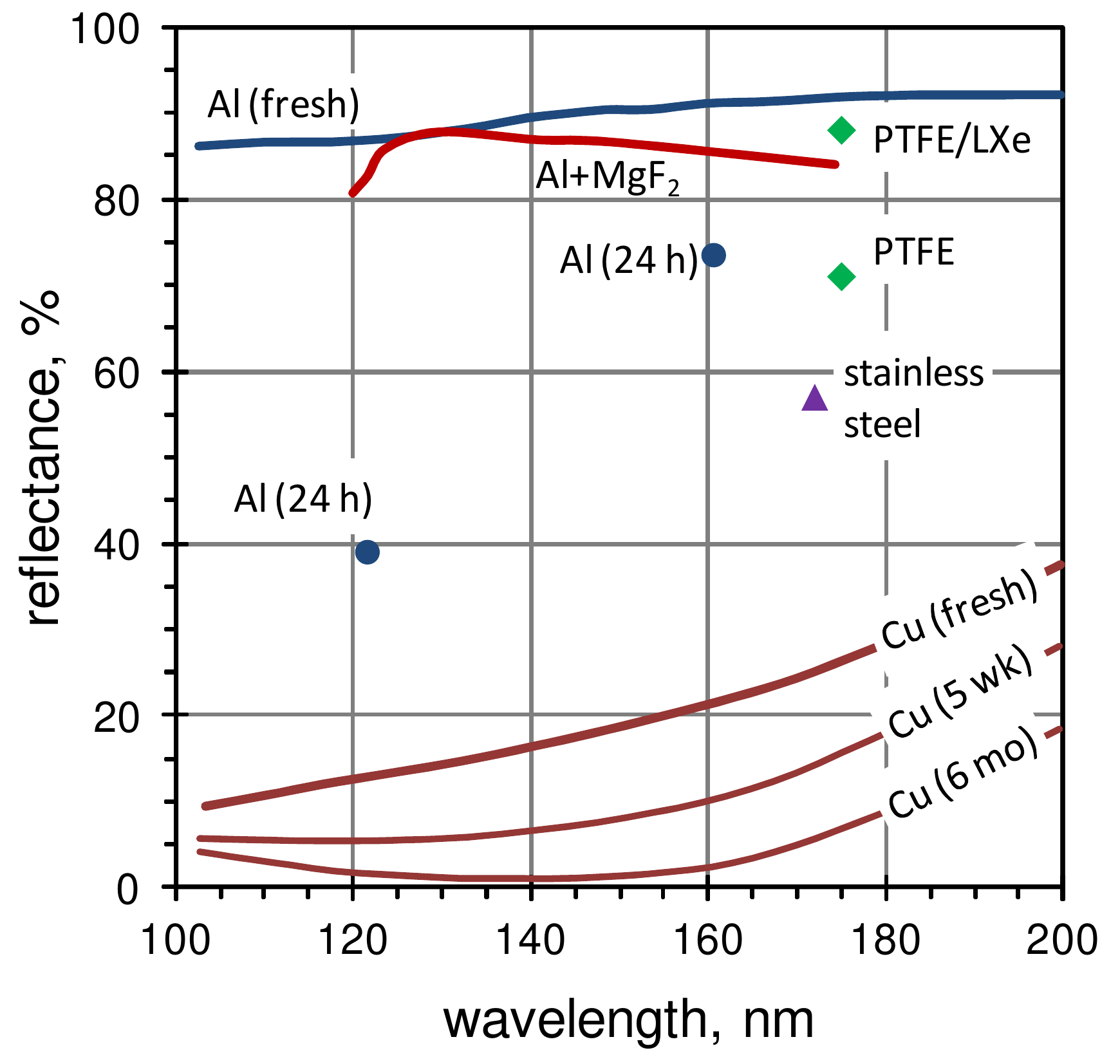}
\includegraphics[width=.5\textwidth]{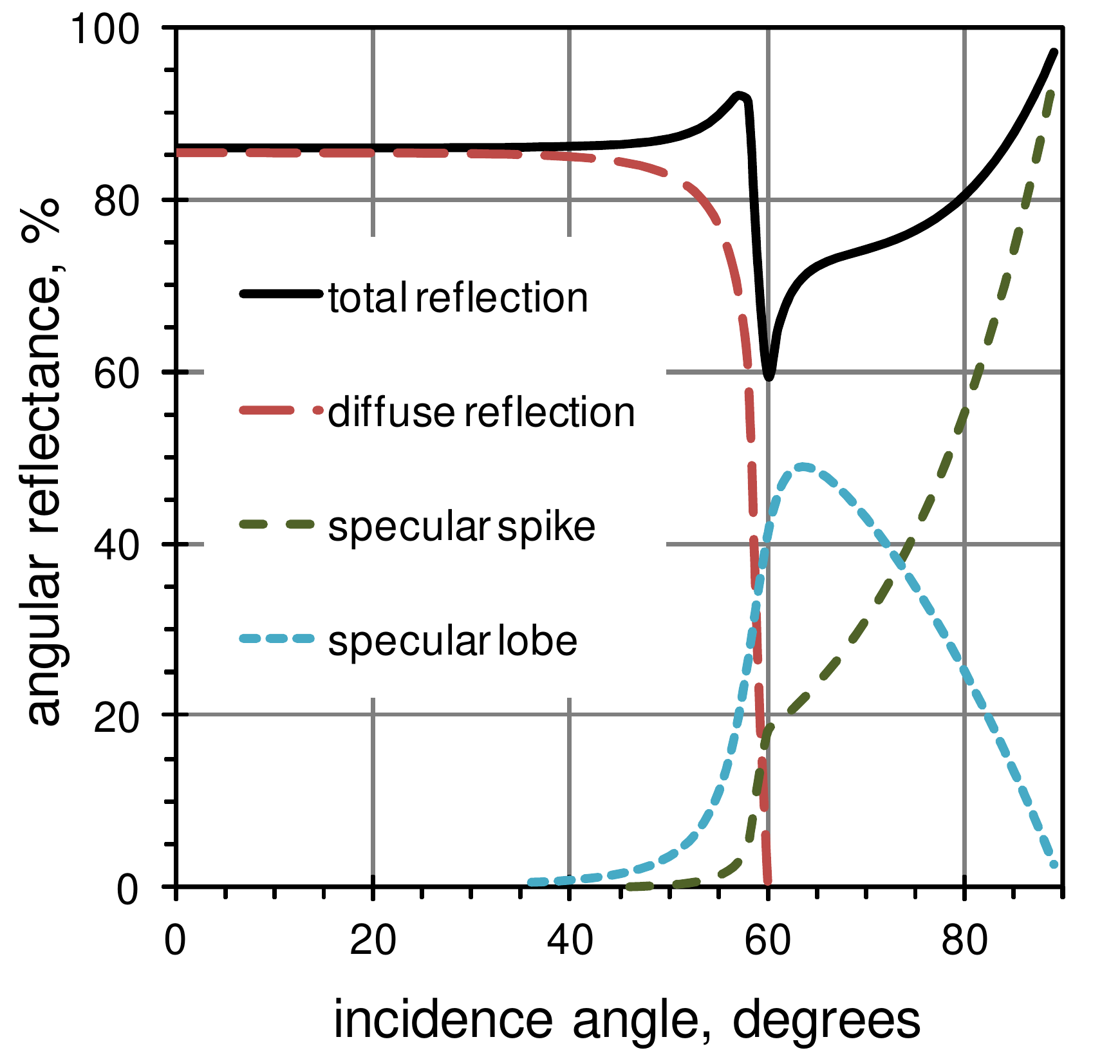}
\caption{The VUV reflectance of some materials employed internally in
noble liquid chambers is shown left. Data for Al~\cite{Madden63} and
Cu~\cite{Canfield65} are for freshly evaporated films and the ensuing
degradation from build-up of the native oxides; a logarithmic decrease
of reflectivity up to 3~years after Al evaporation was reported
in~\cite{Hass56}. Data for Al passivated with MgF$_2$ (both deposited
by evaporation) is from~\cite{Perea07}; a similar coating on a
chemically-etched surface was used in the 80~kg LXe detector
in~\cite{Minamino12}. Data for stainless steel is
from~\cite{Bricola07}. Measurements with pressed, polished PFTE
illuminated with Xe scintillation light in vacuum were reported
in~\cite{Silva10}; an angle-averaged value is indicated here. A model
extrapolation for the angular reflectance of a PTFE/LXe interface is
shown on the right (\cite{Silva09}, courtesy C.~Silva). The increase
in specular reflection relative to the interface with vacuum increases
the angle-averaged reflectance to approximately 90\%. Some
experimental measurements suggest even higher reflectivity for the
PTFE/LXe interface~\cite{Yamashita04,Neves05,Akerib12c}.}
\label{fig:Reflectances}
\end{figure}

High reflectivity of detector components at the scintillation
wavelength is also very important for efficient detection of the
scintillation light. Data on reflectivity of some relevant materials
for $\lambda <200$~nm are shown in Figure~\ref{fig:Reflectances}.
Among the metals, freshly deposited Al presents the highest
reflectance although this rapidly degrades when exposed to air. A
MgF$_2$ coating can protect the Al surface from oxidation and preserve
high reflectance in the VUV for long periods of time. As for
dielectric materials, much attention has been paid to PTFE because it
is known to be compatible with the high purity and low radioactivity
requirements, and is therefore a material of choice for electrical
insulation and light enhancement in DM detectors.  The exact values of
PTFE reflectivity for xenon scintillation light are still being
discussed; in fact, they depend on the manufacturing technology and
surface treatment. Importantly, it is found that the reflectivity of a
PTFE surface in contact with liquid xenon is enhanced quite
significantly relative to that in vacuum (see
Figure~\ref{fig:Reflectances},~left). Values in excess of 90--95\%
have been reported for the PTFE/LXe
interface~\cite{Yamashita04,Neves05,Akerib12c}.

A detailed knowledge of the angular reflectivity profiles is also
important for correct modeling of the detector response. Recently,
there have been new efforts to determine the reflectivity profiles of
some materials of interest for detector construction at the xenon
emission wavelength~\cite{Silva07,Silva10,Silva10b}; some results are
shown on the right panel of Figure~\ref{fig:Reflectances}.

\subsection{Ionization charge}
\label{sec:IonizationCharge}

\subsubsection{Ionization yields and transport properties in the liquid}
\label{sec:IonizationYield}

The number of primary ionizations, $N_i$, created in the liquid by a
particle is significantly larger than the number of excitations,
$N_{ex}$, for electrons and comparable in the case of nuclear recoils
(see Table~2). For the ionization signal to be detected, the following
conditions must be met: i)~charge carriers must escape recombination;
ii)~their mobility in the electric field must be high; iii)~the
probability to form low mobility states along the drift path must be
low, and iv)~a high gain amplification mechanism is required in order
to form a measurable signal.

The ionization process results initially in the creation of free
electrons and positive ions $R^+$; the treatment of dielectrics in
solid state physics provides the relevant framework here, and in this
context we would refer to these positive ions as `holes'. Holes are
rapidly localized through formation of molecular ions $R_2^+$ within
picoseconds (see~\cite{Kubota79} and references therein). The
electrons lose their kinetic energy in inelastic collisions, producing
more ionizations and excitations, and in elastic collisions with
atoms, producing heat (or `phonons', to carry the solid state analogy
further), until thermal equilibrium with the medium is reached.

\begin{table}[ht]
  \label{tab:Ionization}    
  \centering
  \vspace{10mm}
  \caption{Some properties of LAr and LXe relevant for particle
  detection. $W$ is the $W$-value and $F$ is the Fano factor, both for
  electrons; $N_{ex}/N_i$ is the ratio of the number of primary
  excitations to the number of ionizations in the particle track;
  $\upsilon_e$ is the electron drift velocity, $\mu$ is the mobility given
  for several ions (TMSi stands for trimethylsilane). The ion drift
  velocity is a linear function of electric field, $\upsilon\! =\! \mu \mathcal{E}$. 
  This   is also true for electrons at low field, until their energy begins
  to deviate from thermal.}
  \vspace{5mm}
  \begin{tabular}{|l|c|c|}
    \hline
    \rule{0pt}{3ex} 
    ~                              	& LAr                      	& LXe                             \\
    \hline
    \rule{0pt}{3ex}
    $E_g$, eV                           &                               &                                 \\
    $\quad$solid			& $14.2^{\ a)}$          	& $9.28^{\ a)}$                   \\ 
    $\quad$liquid		        & --				& $9.22\pm0.01^{\ b)}$            \\
    \rule{0pt}{3ex}
    $W$, eV                      	& $23.6\pm0.3^{\ c)}$ 		& $15.6\pm0.3^{\ d)}$             \\ 
    \rule{0pt}{3ex}
    $F$                            	& $0.116^{\ e)}$		& $0.059^{\ e)}$                  \\ 
    \rule{0pt}{3ex}
    $N_{ex}/N_i$                 	&                               &                                 \\ 
    $\quad$electrons             	& $0.21^{\ f)}$              	& $0.13^{\ f)}$                   \\ 
    $\quad$nuclear recoils$\qquad$     	& --                         	& $\sim1^{\ g)}$                  \\ 
    \rule{0pt}{3ex}
    $\upsilon_e$, cm/s      	                &                               &                                 \\ 
    $\quad$$\mathcal{E}\!=$1~kV/cm           	& $2.6 \times 10^{5}$ $^{ h)}$  & $2.25 \times 10^{5}$ $^{ h)}$   \\ 
    $\quad$$\mathcal{E}\!=$10~kV/cm    	        & $4.9 \times 10^{5}$ $^{  h)}$ & $2.8 \times 10^{5}$ $^{  h)}$   \\ 
    \rule{0pt}{3ex}
    $\mu$, cm$^2$/(V$\cdot$ s)          &                               &                                 \\ 
    $\quad$Ar$^+$ in LAr or Xe$^+$ in LXe                      & $0.2 \times 10^{-3}$ $^{ i)}$ & $3.5 \times 10^{-3}$ $^{  j)}$  \\ 
    $\quad$TMSi$^+$                     & --                     	& $0.2 \times 10^{-3}$ $^{  k)}$  \\ 
    $\quad$O$_2^-$                      & $0.2\times 10^{-3}$ $^{  i)}$ & $0.7 \times 10^{-3}$ $^{  k)}$  \\
    \hline
  \end{tabular}
  \footnotesize
  \parbox{10cm} {
  \vspace{1mm}
    $^{a)}$\cite{Baldini62}; see~\cite{Rossler76,Schmidt97a,Steinberger05} for other measurements
    and references. The agreement between data for the band gap for the solid state is better than 0.5\%; \\
    $^{b)}$\cite{Asaf74}; $^{c)}$\cite{Miyajima74}; $^{d)}$\cite{Takahashi75}; $^{e)}$\cite{Doke76};\\
    $^{f)}$\cite{Doke02}; for LXe, an average value between that
		expected from theory (0.06) and that suggested by
		experiment (0.2) is given; for LAr, these values
		agree; \\ 
		$^{g)}$ \cite{Sorensen11,Dahl09}; $^{h)}$
		\cite{Miller68}; $^{i)}$ \cite{Dey68}; $^{j)}$
		\cite{Hilt94}; $^{k)}$ \cite{Hilt94b}. }
\vspace{5mm}
\end{table}

Recombination prevents some of the charge created initially from being
collected by an electric field; this fraction depends very much on the
type of particle and its energy, as discussed in
Section~\ref{sec:Recombination}.  For high energy particles,
$\gtrsim$100~keV, which leave a well-defined `long' track in the
liquid, the $dE\!/\!dx$ approach works relatively well: the higher the
$dE\!/\!dx$ value the stronger the recombination. Thus, for the
low-LET $\sim$1~MeV electrons ($dE\!/\!dx \sim$1~MeV\,cm$^2$/g) about
90\% of the charge created initially can be extracted by applying a
field of the order of a few kV/cm~\cite{Kubota79,Barabash85,Aprile91},
while for $\alpha$-particles ($dE\!/\!dx \sim$500~MeV\,cm$^2$/g) a
field of $\sim$20~kV/cm in liquid argon, or $\sim$80~kV/cm in liquid
xenon, is required to collect about 20\% of the
charge~\cite{Pruett67,Aprile91b}.
Figure~\ref{fig:ChargeYieldVsField} on page \pageref{fig:ChargeYieldVsField}
shows some of the available data on the charge yield as a function 
of electric field for different particles.

For low energy particles which are of interest for DM and CNS
searches, the ionization pattern presents no cylindrical symmetry and
therefore $dE\!/\!dx$ is no longer a good parameter to describe the
charge yield.  For example, higher yield is observed for nuclear
recoils than for $\alpha$-particles in liquid xenon in spite of
comparable $dE\!/\!dx$ values for both projectiles. The yield
dependence on the applied field is also quite different as
Figure~\ref{fig:ChargeYieldVsField} attests. Although some absolute
figures for the charge yield for xenon recoils are known, it
is not clear yet how much ionization is originally created in the
liquid by these particles.

The usual way to characterize a medium with respect to its ionization
`capacity' is through the energy expended per electron-ion or
electron-hole pair $W=E_0/N_i$, where $E_0$ is the energy transferred
to the medium by a particle and $N_i$ is the number of initially
created electron-hole pairs. The Platzman
equation~(\ref{eqn:Platzman}) allows one to express it through other
quantities characterizing the detection medium (see
Section~\ref{sec:EnergyTransfer}).  The $W$-value is a convenient
practical parameter but its measurement is not easy. Experimentally,
absolute charge measurements are inherently challenging for low energy
interactions, although the single electron signal measured in
double-phase detectors provides an ideal calibration
standard. However, the main problem rests with the fact that the
charge observed at an electrode is not necessarily equal to that
initially created by the passing particle. This can often be due to
attachment of electrons to impurities, but the main reason is
recombination, which in principle exists at any finite field
strength. Therefore, in order to determine the number of electrons
created promptly by a particle, $N_i=Q_0/e$, one needs to extrapolate
the extracted charge, measured as a function of electric field,
$\mathcal{E}$, to an infinitely large field. In turn, that requires
the assumption of a model describing this dependence. In the absence
of a complete model, it is generally accepted that the
parameterization $Q(\mathcal{E})=Q_0/(1+k/\mathcal{E})$ works well for
the purpose, although the physical assumptions made to derive this
expression are not strictly valid for liquefied rare gases. This
equation was first obtained by Jaff\'{e} to describe recombination in
a column of positive and negative ions with equal
mobilities~\cite{Jaffe13}, which is far from being the case for the
noble liquids (see Table~2).  Besides, recombination is treated as a
perturbation in this approach, which might be acceptable for gases but
certainly not for liquids.~\footnote{The constant $k$ in the above
parameterization characterizes the recombination strength.
Representative values are: for a few hundred keV electrons and
$\gamma$-rays $k \approx 0.56$~kV/cm in LAr and $\approx$0.42~kV/cm in
LXe~\cite{Shibamura75}; $k\approx 2.4$~kV/cm for 15.3~keV
X-rays~\cite{Voronova89}; $k \sim 470$~kV/cm for $\alpha$-particles
(obtained with a different parameterization, but one in which $k$
still combines with the field strength in the same way, i.e.~also
through the ratio $k/\mathcal{E}$~\cite{Thomas87}). A more complete
compilation can be found in~\cite[p.340]{LopesChepel05}. }

Assuming the above dependence for $Q(\mathcal{E})$, linear fits of
$1/Q$ against $1/\mathcal{E}$ provide $1/Q_0$ as the vertical axis
intercept, thus allowing $N_i$ to be determined. The $W$-values for
liquid argon and xenon have been measured
in~\cite{Miyajima74,Takahashi75,Obodovski79}.  Table~2 presents
commonly adopted values for $W$, although other values have been also
reported (for example $W$=13.6$\pm$0.2~eV for liquid
xenon~\cite{Obodovski79}).

Electrons which escape recombination drift relatively quickly in heavy
noble liquids~\cite{Miller68,Gushchin82a}: at a field of 1~kV/cm their
drift velocities in LAr and LXe are very similar (2.6~mm/$\mu$s and
2.25~mm/$\mu$s, respectively). At higher fields, the drift velocity
increases more rapidly in liquid argon, reaching 4.9~mm/$\mu$s
compared with 2.8~mm/$\mu$s in xenon at $\mathcal{E}$ = 10~kV/cm (see
Table~2). In argon, it continues to increase at least up to 100~kV/cm,
while in xenon it saturates already at 10~kV/cm~\cite{Miller68}. (In
liquid neon and helium a gas bubble is formed around the drifting
electron, thus reducing its mobility dramatically ---
see~\cite{Khrapak05}.)

The drift velocity of positive carriers at ordinary fields is a factor
of $\sim$10$^5$ lower than that of electrons and, therefore, their
motion is not detectable in most experiments. It has been noticed that
the mobility of positive carriers resulting from the ionization of
xenon atoms is significantly higher than that of other positive ions
--- a fact attributed to hole-type conductivity~\cite{Hilt94,Hilt94b}.
The same does not happen in liquid argon, in which the same mobility
has been measured for positive carriers and O$_2^-$
ions~\cite{Dey68}. This indicates that positive ions actually drift
physically in liquid argon, similarly to the transport in gases, while
in liquid xenon it is the vacancy that moves, very much like in solids
(for comparison, in solid xenon and argon the hole mobility is about
2$\cdot$10$^{-2}$~cm$^2$/(V$\cdot$s) for both
elements~\cite{LeComber75}).

The high mobility of negative charge carriers can decrease
significantly in the presence of even a very small amount of
electronegative impurities, such as O$_2$, water and some other
molecules. These molecules can capture free electrons and form
negative ions with extremely low mobility (for example,
$\mu(\mathrm{O}_2^-) \approx 0.2 \cdot 10^{-3}$~cm$^2$/(V$\cdot$s) in
LAr and 0.7$\cdot$10$^{-3}$~cm$^2$/(V$\cdot$s) in
LXe~\cite{Hilt94,Hilt94b} --- see Table 2 --- resulting in drift
velocities of a few mm/s at practical electric fields). The
probability for an electron to be captured by an impurity species
depends on its concentration, the reaction rate constant and the path
length of the electron on its way to the anode. Therefore,
purification of liquefied rare gases to better than ppb level is a
critical issue. (Impurities can also affect the scintillation light
yield and time constants as mentioned previously.)

Another aspect to consider is electron diffusion during their drift in
the electric field. Diffusion in noble liquids is much lower than in
the respective gas and this is why measurements of the diffusion
coefficients are difficult. For electrons in thermal equilibrium with
the medium and at zero field, diffusion is isotropic and is
characterized by a diffusion coefficient $D$, which is related to the
zero-filed electron mobility $\mu_0$ through the Einstein equation
$eD\! / \!\mu_0 \!=\! kT$.  With increasing fields, the electron
energy rises and their mobility begins to deviate from $\mu_0$.  The
equation can be modified to $eD/\mu \!=\! F \langle \epsilon \rangle$,
where $\langle \epsilon \rangle$ is the mean electron energy and $F$
is a constant depending on the electron energy distribution (for
example, for Maxwell's distribution function \mbox{$F\!=\! 2/3$}).  In
the presence of an electric field, however, the diffusion process is
no longer isotropic and occurs predominantly in the plane
perpendicular to the electric field. Two diffusion coefficient are
therefore required: $D_T$ for transverse diffusion, and $D_L$ for
longitudinal diffusion. For liquid xenon, a ratio $D_L/D_T \sim 0.1$
is rather well verified for fields $\gtrsim$1~kV/cm~\cite{Doke80}; for
a field of 730~V/cm, $D_L/D_T \sim 0.15$ has been
obtained~\cite{Sorensen11b}.  In the zero-field limit this ratio
should approach unity.

Absolute values for diffusion coefficients in liquid xenon are not
well measured but some indicative figures can be presented: $D_T
\sim$100~cm$^2$/s has been reported for 1~kV/cm, decreasing to
$\sim$50~cm$^2$/s for 10~kV/cm; for $D_L$ the values are an order of
magnitude lower~\cite{Doke80,AprileDoke09}.  Theoretical estimates
in~\cite{Atrazhev05} result in $D_T\approx 85$ to 100~cm$^2$/s which
vary only very weakly with field in the range from 100~V/cm to
10~kV/cm.  Recent measurements indicate $D_T\approx$30~cm$^2$/s with
variation of about $\pm$5~cm$^2$/s in the field range of 0.5 to
1.2~kV/cm~\cite{Chen11}.  For longitudinal diffusion, a value $D_L=(12
\pm 1)$~cm$^2$/s has been reported~\cite{Sorensen11b}.

Less information is available for liquid argon.  Transverse diffusion
is studied experimentally in~\cite{Shibamura79}.  The authors present
values for the characteristic energy of electrons $eD\!/\!\mu$ between
0.1~eV and 0.4~eV for fields 2 to 10~kV/cm, which are consistent with
some earlier measurements and also with predictions from Lekner's
theory. In terms of diffusion coefficient this translates to $\sim
3$~cm$^2$/s for 1~kV/cm and $\sim 16$~cm$^2$/s for 10~kV/cm.  Somewhat
higher values were reported in~\cite{Gushchin82a} for low fields.
Those authors report 13~cm$^2$/s for 1~kV/cm and 9~cm$^2$/s for
300~V/cm.  The effect of the liquid temperature was also studied in
this work.  The ICARUS group reported on the longitudinal diffusion
coefficient in LAr at fields of $\sim$100~V/cm :
$D_L=$(4.8$\pm$0.2)~cm$^2$/s~\cite{Cennini94}; this value was also
referred in a more recent publication from this group~\cite{Amerio04}.

The impact of electron diffusion on the performance of DM and CNS
search detectors in not expected to be very significant (for current
designs, at least).  For a 1~m drift in LXe in a field of 1~kV/cm one
can estimate that an initially point-like electron cloud will spread
over a distance $\sigma_x=(2DL/\upsilon_e)^{1/2}\approx4$~mm. This
smearing can even be turned into an advantage when considering very
low detection thresholds for nuclear recoils, as argued
in~\cite{Sorensen11b}: for very low recoil energies with no observable
S1 signal, the width of the S2 signal can provide information on the
drift time of the electrons and hence help fiducialize the sensitive
volume.

\subsubsection{Liquid purity}
\label{sec:LiquidPurity}
\vspace{0.4cm}

From a practical point of view, the purity problem has several
facets. Firstly, one must consider the purity of the gas to be
liquefied. The purity supplied by the manufacturer is an important
starting point but never sufficient --- the contamination is usually
of order parts per million (ppm). Therefore, additional on-site
purification is required. Much effort has been put into this over the
last decades and numerous purifiers have been developed; nowadays
commercial getters (operating hot or cold) can deliver sufficient
purity for most applications: free electron lifetimes of $\sim$1~ms
can now be achieved, providing $>$1~m drift distances. There are also
a number of distillation and vapor evacuation tricks that can help
remove inert impurities present in elevated quantities and which are
not removed efficiently by most getters. A common problem, which still
has no simple and reliable solution, is to know exactly which
impurities actually contaminate the system. The impurity monitoring is
best done with a residual gas analyzer using mass spectrometry. Its
sensitivity is limited by a high partial pressure of the bulk gas;
however, using a cold trap between the gas inlet and the mass
spectrometer can improve the sensitivity to ppb
levels~\cite{Dobi12}. Oxygen and water are the most common
electronegative contaminants, but it is not unusual to measure
electron lifetimes which decrease with electric field, just opposite
to what would be expected for these species~\cite{Bakale76}; N$_2$O is
often blamed for this behavior but, to our knowledge, the evidence for
its presence in most systems is not conclusive. The presence of
nitrogen has been detected in some setups where a significant
reduction of secondary light in the gas phase was observed; in
double-phase systems, a significant nitrogen concentration can build
up in the thin gas layer above a large LXe target to quench the
secondary scintillation completely.

No less important is the cleanliness of the experimental setup,
including detector components and the gas handling system. Standard
ultra-high vacuum (UHV) techniques should be adhered to whenever
possible, to minimize contamination of the gas due to outgassing of
materials, or from inappropriate component handling, cleaning or
storage. A rigorous selection of `wetted' materials is essential, and
special surface treatment of metallic surfaces to reduce porosity and
diffusion is very desirable. The choice of electrical insulators
should be subject to special attention: the large surface and bulk
porosity of many such components can lead to a high outgassing
rate. Implementation of rigorous storage, handling and cleaning
protocols is essential: the system must be assembled in a clean
environment, after ultrasonic cleaning of all components using the
right solvents in the correct sequence.

After ensuring that the system is leak-tight, baking under high vacuum
and at maximum allowed temperature for long periods (days to weeks,
depending on dimensions and system complexity) is the next important
step. This is not always feasible for detectors operated underground
(for safety reasons) or when delicate internal components are involved
(e.g.~some photo-sensors and vacuum seal techniques). Systems built
from low-outgassing materials and following strict gas handling
procedures have sustained adequately large electron lifetimes for
longer than a year without re-purification~\cite{Majewski12}. On the
other hand, some of the largest noble liquid detectors today contain
large amounts of fluoropolymers such as PTFE to enhance VUV
reflectivity, and this leads to very high outgassing rates. In this
instance, continuous pump-driven purification of the gas is often
used: some gas is taken from the detector, recirculated through the
purification system and condensed into the detector again;
purification directly in the liquid phase is also possible but
technically more challenging.

Finally, the issue of radio-purity is intimately connected with this
discussion. For example, most commercial getters are not suitable for
application in rare event searches due to very significant radon
emanation rates. $^{222}$Rn atoms wash into the detector where they
decay, and (metallic) radioactive progeny will subsequently plate-out
internal surfaces; some are $\alpha$ emitters, which can cause nuclear
recoils into the active volume. Radon mitigation is also related to
component cleaning and storage (typical Rn concentrations in air give
1--100~Bq/m$^3$), especially in systems which cannot fiducialize the
active target very accurately (such as those relying only on
scintillation).

For further information on purification methods, material selection
and techniques for obtaining acceptable purity of the liquid the
reader is referred, for example,
to~\cite{BarabashBolozdynya93,AprileBook06,Schinzel05} and references
therein. We can also
refer~\cite{Benetti93a,Chepel94,Akimov07,Babussinov10,Aprile12c} for
some gas purification systems.

\subsection{Combined signal}
\label{sec:CombinedSignal}

\begin{sloppypar}
When detecting either scintillation or ionization due to a particle,
what is actually observed is an electrical signal which is
proportional either to the number of emitted VUV photons,
\mbox{$\mathrm{S1} \propto N_{ph}= N_{ex} + rN_i$}, or the number of
free electrons, $\mathrm{S2} \propto N_{e}= (1-r)N_i$, where $r$ is
the charge recombination fraction for a given particle species, energy
and electric field (note that S2 may be measured by means other than
proportional scintillation). In the limit of infinite electric field,
all initially created electrons escape recombination and therefore
direct excitation is the sole mechanism leading to photon emission; in
general, $r>0$, i.e.~the charge signal is smaller and the
scintillation signal is larger due to recombination.
\end{sloppypar}

Both $N_{ph}$ and $N_e$ are subject to statistical fluctuation ---
however, not independently. They are constrained by the total energy
transferred to the medium by the particle, as indicated by
equation~(\ref{eqn:Platzman}) in the case of electrons. Therefore,
fluctuations of a linear combination of S1 and S2 with suitably chosen
coefficients are expected to be much smaller than fluctuations of each
of the two signals separately. The idea of using the combined signals
to form an improved energy estimator was originally proposed
in~\cite{Seguinot92} but it was not sufficiently appreciated at the
time and was implemented only much later~\cite{Conti03}. Presently,
LXe dark matter experiments use a combined signal mostly to
characterize detector performance and electron backgrounds, exploiting
the improved linearity and energy resolution thus made possible.

It has been proposed in~\cite{Shutt07} that the best estimator for the
energy of an electron recoil event can be obtained by recognizing that
$\ N_i + N_{ex} = N_{ph} + N_e \ $ for any value of $r$. Making this
replacement in the definition of $W_s^{min} = E_0/(N_i+N_{ex})$ leads
to the recombination-independent sum 
\mbox{$E_0 = (N_{ph} + N_e) \cdot W_s^{min}$}.  
Note that $W_s^{min}$ corresponds to the `flat top'
response in 
Figure~\ref{fig:ScintillationYield} on page~\pageref{fig:ScintillationYield},
i.e~when complete recombination occurs at zero
field (see also~\cite{Doke02}); it can be measured from fixed energy
interactions (e.g.~$^{57}$Co $\gamma$-rays) by varying the electric
field applied to the liquid, which effectively scans through $N_{ph}$
and $N_e$ whilst keeping their sum constant (see Table~1 for
$W_s^{min}$ values).

As for the energy resolution, we recall that the observables S1 and S2
are proportional --- but not identical --- to $N_{ph}$ and $N_e$,
respectively; the number of quanta actually detected can be
substantially lower than the number released at the particle track ---
especially in the case of scintillation --- and this implies that
minimizing the variance of the combined signal may require different
coefficients to those suggested by the equation for $E_0$ above. By
exploiting a combined energy estimator, $\sigma/E \approx 3.4$\% for
122~keV $\gamma$-rays~\cite{Solovov12}, $\sigma/E \approx 3$\% for
570~keV~\cite{Conti03} and $\sigma/E \approx 1.7$\% for 662~keV
$\gamma$-rays~\cite{Aprile07} have been achieved in liquid xenon. The
resolution improvement is more pronounced at high energies but still
significant below
100~keV~\cite{Solovov12,Aprile11a,Araujo12,Aprile12c}. For example,
for 40~keV $\gamma$-rays XENON100 reported $\sigma/E \approx 16.2$\%
when measured with the primary scintillation signal only, and
$\approx$9\% for the combined signal~\cite{Aprile12c}. In
$\beta$$\beta$-decay experiments any improvement in energy resolution
is extremely important; EXO-200 uses the combined
scintillation/ionization signal to achieve $\sigma/E = 1.67$\% at the
$Q$-value in $^{136}$Xe (2,458~keV)~\cite{Auger12}.

In WIMP and CNS searches, precise reconstruction of nuclear recoil
energies is even more important, but unfortunately harder to achieve
at present. Currently, most double-phase experiments utilize the S1
response together with independent measurements of the relative
scintillation efficiency for nuclear recoils in this reconstruction,
as discussed in 
Section~\ref{sec:PrimaryScintillation}.
Therefore, this ignores information
encoded in the ionization response. In~\cite{Sorensen11} it is argued
that a combined energy scale should also be applied in this instance,
but specifically to the fraction $f_n$ of nuclear recoil energy
transferred to electronic excitations (i.e.~including ionization and
atomic excitation, but excluding the energy lost in elastic collisions
with atoms): $E_0=(N_{ph}+N_e) \cdot W_s^{min}/f_n$. In that study the
authors find that $f_n$ as predicted by Lindhard's theory is in
general agreement with experimental data for LXe. More measurements
with simultaneous acquisition of scintillation and ionization signals,
especially reaching below 4~keV (see 
Figure~\ref{fig:LXeLeff}),
are required to
consolidate this approach.

\subsection{Electron emission from liquid to gas}
\label{sec:ElectronEmission}

With double-phase detectors, the combination of large active detector
masses with exquisite sensitivity to two response channels --- at the
level of single carriers in ionization and a few photons in
scintillation --- is bringing about a steep improvement in sensitivity
for dark matter searches and may soon allow a first detection of
coherent neutrino scattering. Ultimately, this is due to the
possibility to extract electrons across the liquid surface. In
addition to providing amplification of the ionization signal in the
gas, this also allows the ionization response to be transduced into an
optical signal, so that the same photon detectors can be used for both
signatures. The fact that free electrons can cross, under moderate
electric field, the liquid/gas boundary has been known for more than
60~years~\cite{Hutchinson48}. Owing to its application to particle
detection, electron emission has been extensively studied for several
decades (for example,~\cite{Dolgoshein70},~\cite{Bolozdynya95} and
many others --- see~\cite{BolozdynyaBook10} for more references). The
decisive impetus to this technique came in the last decade from the
growing interest in direct detection of WIMP dark matter, which led to
several programs worldwide.

\begin{figure}
\centerline{\includegraphics[width=.52\textwidth]{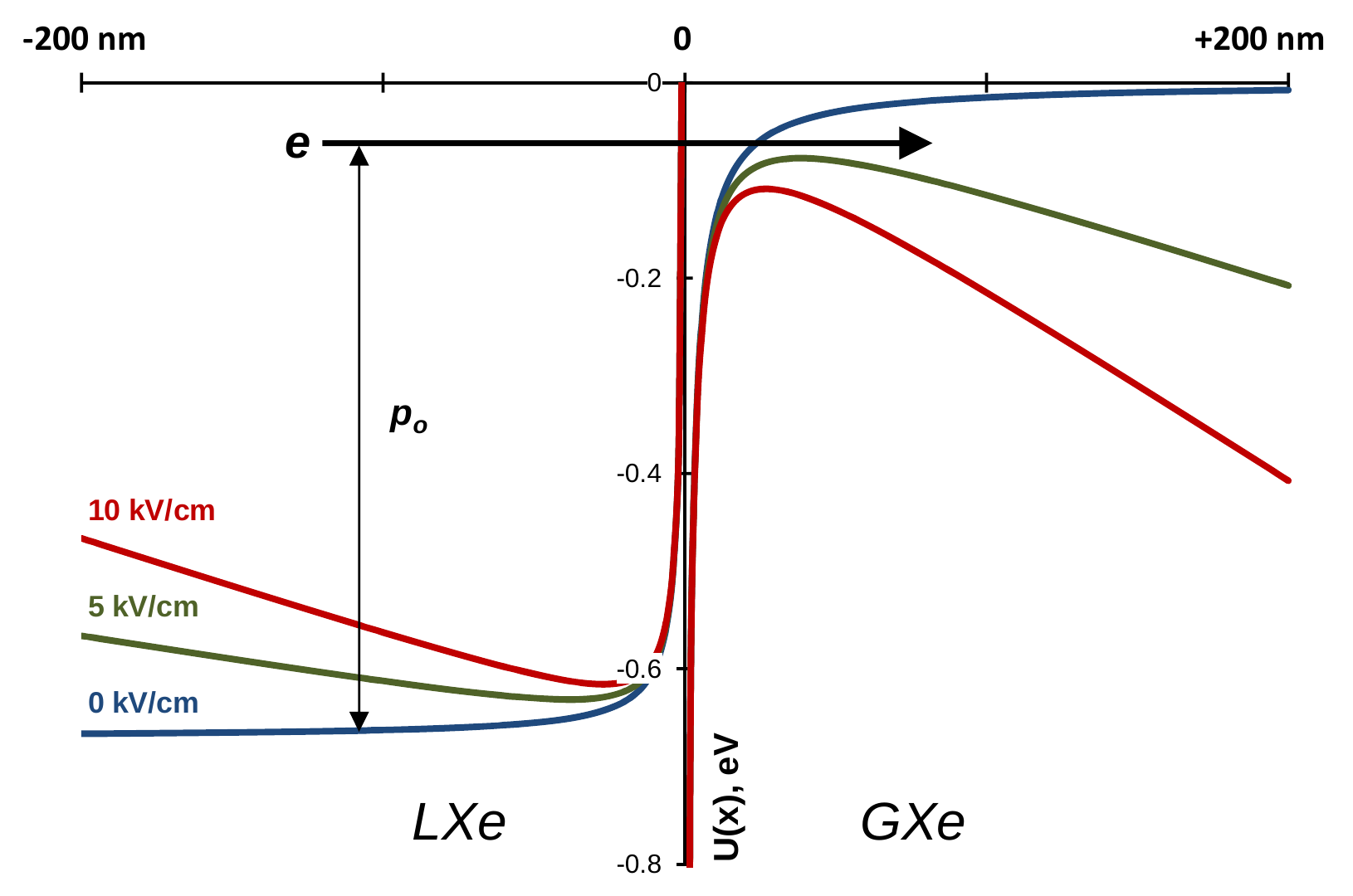}}
\vspace{4mm}
\caption{Illustration of the electron emission process in double-phase
xenon; the figure shows the potential energy of excess electrons near
the liquid-gas interface calculated from the model
in~\cite{Bolozdynya99}, with different electric field strengths
indicated for the liquid. }
\vspace{1mm}
\label{fig:EmissionProcess}
\end{figure}

The potential energy of a free electron in the liquefied heavy rare
gases is somewhat lower than in the gas or vacuum (by 0.2~eV in liquid
argon and 0.67~eV in xenon --- see $V_0$ value in Table~3). Therefore,
it is energetically advantageous for electrons to remain in the liquid
phase. However, some electrons in the upper tail of the Maxwellian
velocity distribution can overcome the barrier if there is even a weak
electric field forcing them to approach the surface. With increasing
drift field strength, the thermal equilibrium between electrons and
the liquid is broken and the mean electron energy starts to increase
--- already at a few tens of V/cm~\cite{Sakai05}. Another effect of
the external field is some reduction of the height of the potential
barrier at the liquid-gas interface. The barrier is the sum of two
components: the potential step of height $V_0$ and the image potential
of a charge placed above the liquid surface. The distance at which the
resulting potential curve in the gas phase reaches its maximum also
depends on the field strength, as illustrated in
Figure~\ref{fig:EmissionProcess}. 
The maximum approaches the liquid surface with increasing field, thus
reducing the probability of the electron back-scattering into the
liquid (for example, at 1~kV/cm the maximum is $\approx$60~nm from the
surface, while at 5~kV/cm it is only $\approx$20~nm away). Finally,
under an applied field the electron velocity distribution in the
liquid gains some anisotropy towards the surface, also resulting in a
higher emission probability.

It is usually considered that electron emission occurs through two
mechanisms: emission of `hot' electrons and `thermal' emission. The
first mechanism refers to the case when the mean electron energy is
higher than the mean thermal energy and also greater than $V_0$, so
that a significant fraction of the electrons approaching the liquid
surface has sufficient kinetic energy to overcame the potential
barrier and thus pass immediately into the gas phase. Many electrons,
however, do not cross the surface barrier at the first attempt. They
are reflected back into the bulk of the liquid and, after a number of
scatterings, return to the surface guided by the field. Again, those
which do have energy above the barrier cross it, but the remainder
return to the bulk and so forth. This process is similar to thermal
evaporation from the tail of a Maxwell-Boltzmann distribution, hence
the name `thermal emission' for the second mechanism. We note,
however, that the mean energy of these electrons is not necessarily
equal to the thermal value.

\begin{table}[hb]
  \label{tab:EmissionParameters}    
  \centering
  \caption{Electron emission coefficients for LAr and LXe. $V_0$ is
    the electron energy in the liquid relative to the vacuum level;
    $E_{th}$ is the threshold field for electron emission; $E_c$ is
    minimum field strength for which the mean electron energy begins
    to deviate from thermal; $\langle\epsilon_{th}\rangle$ is the mean
    electron energy at $E = E_{th}$; $\langle\epsilon_{sat}\rangle$ is
    the mean electron energy at the field corresponding to saturation
    of the emission curve (Figure~\protect\ref{fig:EmissionProbability}). 
    Data from~\cite{Schmidt97a,Bolozdynya95,Sakai05,Gushchin82b}.
}
  \vspace{3mm}
  \begin{tabular}{|l|c|c|}
    \hline
    \rule{0pt}{3ex}
    ~                       & LAr           & LXe             \\ \hline
    \rule{0pt}{3ex}
    $V_0$, eV               & $-0.2\pm0.03$ (83 K) & $-0.67\pm0.05$ (161 K) \\ 
    \rule{0pt}{3ex}
    $E_{th}$, V/cm          & 250           & 1750            \\ 
    \rule{0pt}{3ex}
    $E_{c}$, V/cm           & $\sim$60      & $\sim$20        \\ 
    \rule{0pt}{3ex}
    $\langle\epsilon_{th}\rangle$, eV  & $\sim$0.02    & $\sim$0.3  \\ 
    \rule{0pt}{3ex}
    $\langle\epsilon_{sat}\rangle$, eV & $\sim$0.12    & $\sim$0.4  \\
    \hline
  \end{tabular}
\end{table}

\begin{figure}
\centerline{\includegraphics[width=.55\textwidth]{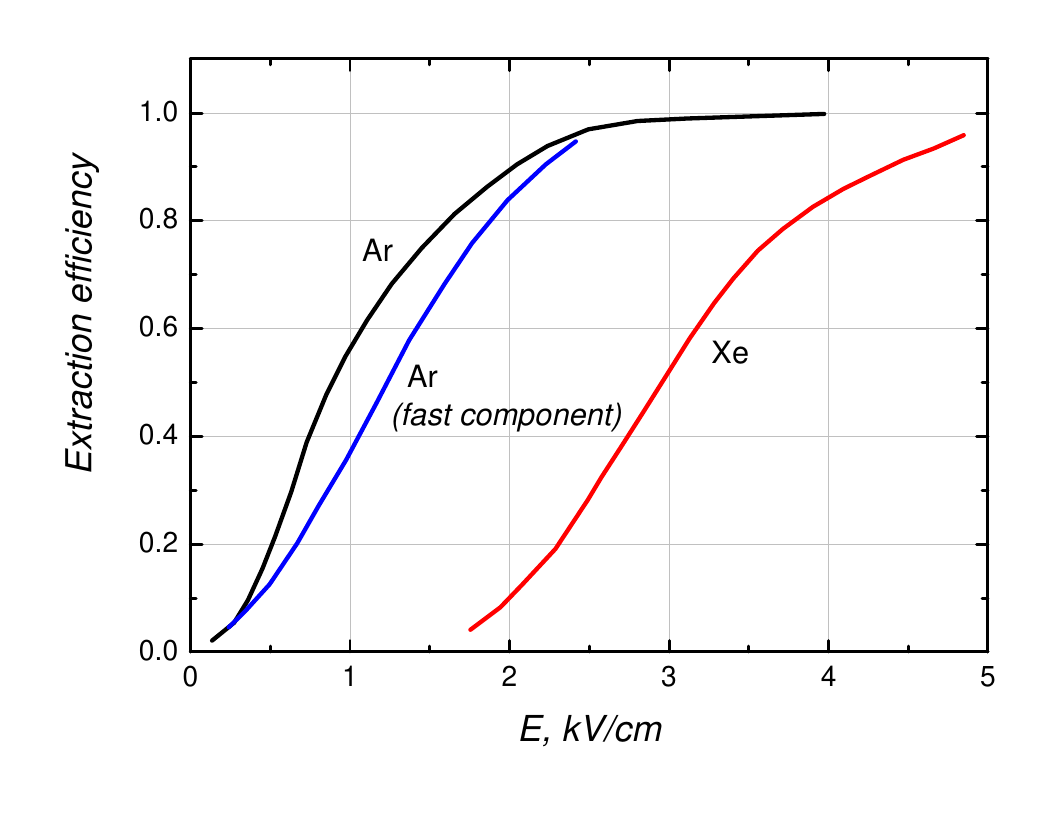}}
\caption{Probability of electron emission from liquid to gas as a
function of electric field. Re-drawn from data in \cite{Gushchin82b}.}
\label{fig:EmissionProbability}
\end{figure}

The relative contribution from these two processes depends on the
$V_0$ value (Table~3) as well as on the field strength. In liquid
argon, two very distinct time constants have been observed for the
emission time \cite{Gushchin82b,Borghesani90,Bondar09b}: a fast
component of the order of 1~ns or less and a very slow emission up to
$\sim$1~ms at $E \sim 100$~V/cm, the latter explained by electron
trapping under the surface~\cite{Borghesani90}. The time constant of
the slow component depends on the field as $1/E$ as suggested by the
thermal emission model. A similar phenomenon, although less
pronounced, has also been reported in liquid xenon
detectors~\cite{Santos11,Angle11}. In liquid xenon, the surface
barrier is higher than in argon and, therefore, a lesser contribution
from thermal emission is expected, and the longer emission time is
limited, in practice, by electron attachment to impurities. A much
higher emission threshold for xenon is also consistent with the above
considerations (see 
Figure~\ref{fig:EmissionProbability},
where the emission efficiency is
presented as a function of electric field).

We recommend \cite{Bolozdynya95,Bolozdynya99,BolozdynyaBook10} and
references therein for additional information.

\subsection{Electrons in the gas phase}
\label{sec:ElectronsInGas}

Once in the gas phase, electrons can easily be accelerated by the
applied electric field to sufficient energies to excite the gas atoms
(and thus produce secondary scintillation, also known as
electroluminescence) or even to ionize them resulting in a cascade of
secondary electrons (avalanche). The latter process is widely used in
gaseous detectors and has been very well studied in various gas
mixtures. It allows very high amplification gains to be achieved and
signals due to a single initial electron to be detected (however, not
with 100\% efficiency). Charge multiplication has also been tried in
the gas phase of double-phase systems, but unsuccessfully. Discharging
occurred already at gains of a few hundred~\cite{Zaklad74} (see
also~\cite{Bolozdynya85}). As the excitation energy is lower than the
ionization potential and Ar and Xe are both very efficient
scintillators, plenty of photons are produced in the avalanche. Such
VUV photons cause photoelectric feedback effects at cathode electrodes
with high probability, resulting in secondary avalanches. Strong
purity concerns and low temperature do not allow much scope for
traditional solutions --- such as adding quenching
molecules. Nevertheless, signal amplification through electron
multiplication in the gas layer of a double-phase system continues to
attract interest. As has been shown recently, structures such as GEM,
THGEM, LEM or similar devices may offer a promising solution for this
problem. This topic is discussed in 
Section~\ref{sec:AlternativeTechniques}.

Contrary to electron multiplication, secondary scintillation does not
seem to have any significant drawbacks except the need to detect VUV
photons. It does not require any admixtures and is in fact most
efficient in the pure rare gases. It was proven to work in
medium-sized detectors in the early 1980s~\cite{Egorov82,Egorov83}
(see~\cite{BolozdynyaBook10} for more references). The
electroluminescence response has now been shown to provide stable
operation in underground experiments over periods of
$\sim$1~year~\cite{Aprile12a,Akimov12a}.

A single electron extracted from the liquid can produce hundreds of
photons along its drift path. In the ZEPLIN-III configuration, for
example, one electron extracted into the gas produced some 300~VUV
photons, resulting in a total signal of about 30~photoelectrons at the
PMT photocathodes~\cite{Santos11} (see also 
Figure~\ref{fig:SEinZ3}).

The mechanism of secondary scintillation is well understood and is
similar in argon and xenon~\cite{Leite80}. Atoms are initially excited
by electron impact to one of the lowest excited states, $^3\!P_2$,
$^3\!P_1$, $^3\!P_0$ and $^1\!P_1$ (all of these corresponding to the
electronic configuration $3p^54s^1$). Excitation to higher levels,
namely those corresponding to the configuration $3p^54p^1$, is also
possible but less probable. Relaxation of these states occurs through
the transition $4p\to 4s$ with emission of an infrared photon, or
non-radiatively via collisions with other atoms. At low gas pressure
(less than a few mbar), collisions between atoms are rare so that the
excited atoms have enough time to decay to the ground state $^1S_0$
(if the transition is not forbidden by the selection rules, as is the
case of $^3\!P_2$ and $^3\!P_0$, which form metastable states). The
allowed transitions $^3\!P_1 \to ^1\!S_0$ and $^1\!P_1 \to ^1\!S_0$
occur with emission of VUV photons with wavelengths of about 107~nm
and 105~nm in argon, and 147~nm and 130~nm in xenon, with a narrow
spectrum. As the gas density increases, the collision frequency also
increases and formation of diatomic excimers Ar$_2^*$ or Xe$_2^*$ in
$^3\Sigma_u^+(0_u^+)$ and $^1\Sigma_u^+(1_u)$ becomes more
probable. Therefore, a wider molecular continuum at longer wavelengths
starts to appear in the emission spectra~\cite{Policarpo81}. At gas
pressures of 1~bar and above the atomic lines are very much suppressed
so that the emission spectrum shows only the second continuum, which
corresponds to transitions from the lowest vibrational levels of the
$^3\Sigma_u^+(0_u^+)$ and $^1\Sigma_u^+(1_u)$ states to the ground
level. The observed spectrum is rather similar to that of primary
scintillation in the liquid except for a small difference in the peak
position and its width, especially in xenon (in argon, $\lambda_{liq}
= 129.6$~nm, $\Delta\lambda_{liq} \approx 10$~nm, $\lambda_{gas} =
128$~nm, $\Delta\lambda_{gas} \approx 10$~nm; in xenon, $\lambda_{liq}
= 178.1$~nm, $\Delta\lambda_{liq} \approx 14$~nm, $\lambda_{gas} =
171$~nm, $\Delta\lambda_{gas} \approx 12$~nm --- data
from~\cite{Jortner65} for liquids, and~\cite{Takahashi83} for
gases). The difference in wavelength is due to the fact that the
exciton energy levels in the liquid are slightly shifted down with
respect to the excimer levels in the gas.

The number of photons emitted by an electron in a uniform electric
field is proportional to the drift path length. The light yield per cm
is well described by a linear function of $E/n$~\cite{Conde77}:
\begin{equation}
   \frac{1}{n}\frac{dN_{ph}}{dx} = a\frac{E}{n}-b
   \quad \text{[photons$\cdot$cm$^2$/e]} \,,
  \label{eqn:SecLight}
\end{equation}
where $E$ is the field strength (in V/cm) and $n$ is the number of
rare gas atoms per cm$^3$ (related to the gas density $\rho$ through
$n=N_A\rho/A$, being $N_A$ Avogadro's number and $A$ the atomic
number); \mbox{$a$ and $b$ are} gas-specific empirical
coefficients. Secondary scintillation is a threshold process,
requiring a minimum field $E/P \approx 1.0\!\pm\!0.3$~kV/cm/bar at
room temperature (if the temperature is different, $P$ should be
treated as the equivalent pressure at room temperature for the same
gas density). In terms of number density, the threshold is $E/n
\approx (4 \!\pm\!1) \cdot 10^{-17}$~V$\cdot$cm$^2$/atom. The stated
uncertainty reflects the variability of data published by different
authors. A recent compilation of the coefficients $a$ and $b$ for
xenon can be found in~\cite{Monteiro07}. Despite more than 30 years of
such measurements, the light yields reported by different authors
differ by a factor of up to 2 even for recent data, with most lying
below the values predicted theoretically~\cite{Fraga90,Santos94}. This
can be explained by the difficulty in achieving precise absolute
calibration of the photon detectors used in the measurements,
accounting correctly for light reflections and non-uniformity of
both light collection and photoelectric conversion efficiency across
the PMT photocathode or photodiode.

The secondary scintillation light yield of xenon was found to increase
with decreasing temperature. In saturated xenon gas at
$-$90$\mathrm{^o}$C it was measured to be a factor of 1.5 higher than
at room temperature~\cite{Fonseca04}. This might be attributed to the
increasing probability of direct excitation of pre-formed Xe$_2$
molecules (and probably more complex aggregates Xe$_n$ with $n>2$),
the concentration of which increases when approaching the saturation
point~\cite{Egorov82}.

Concerning argon, scarce data are available on secondary scintillation
in this gas even at room temperature and no data exist for saturated
vapor, to our knowledge. Recent measurements in argon gas at room
temperature~\cite{Monteiro08} show that the light yield can also be
described by a linear function of $E/n$, as above, with coefficients
$a$ and $b$ given in Table~4. On average, the light yield in argon is
a factor of about 2 lower than in xenon.

\begin{figure}[ht]
  \centerline{\includegraphics[width=.8\textwidth]{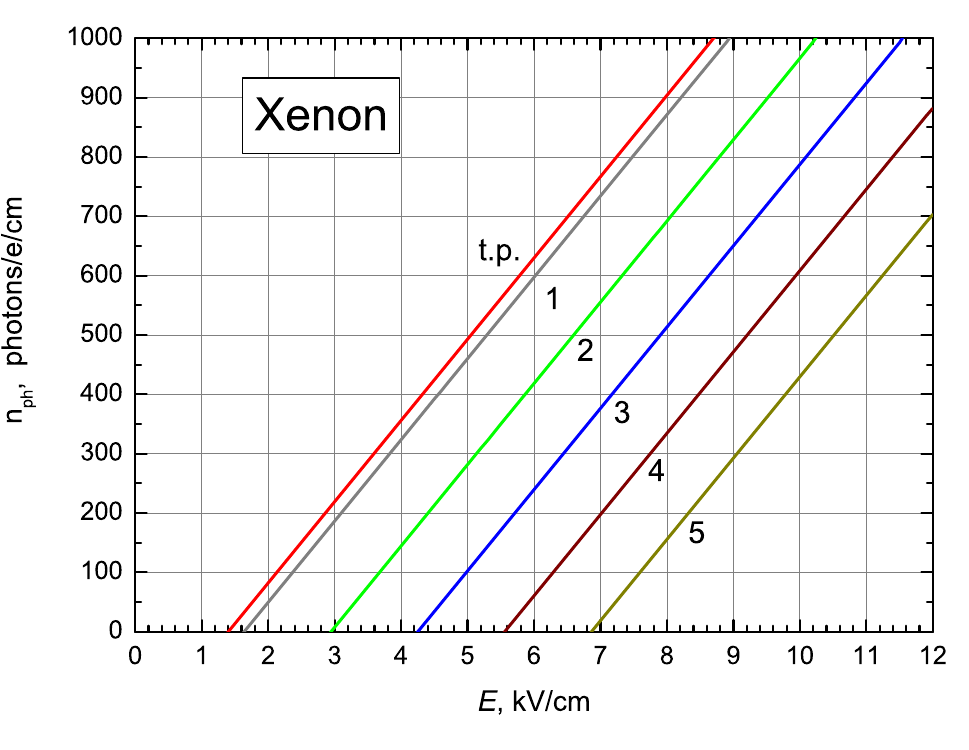}}
  \centerline{\includegraphics[width=.8\textwidth]{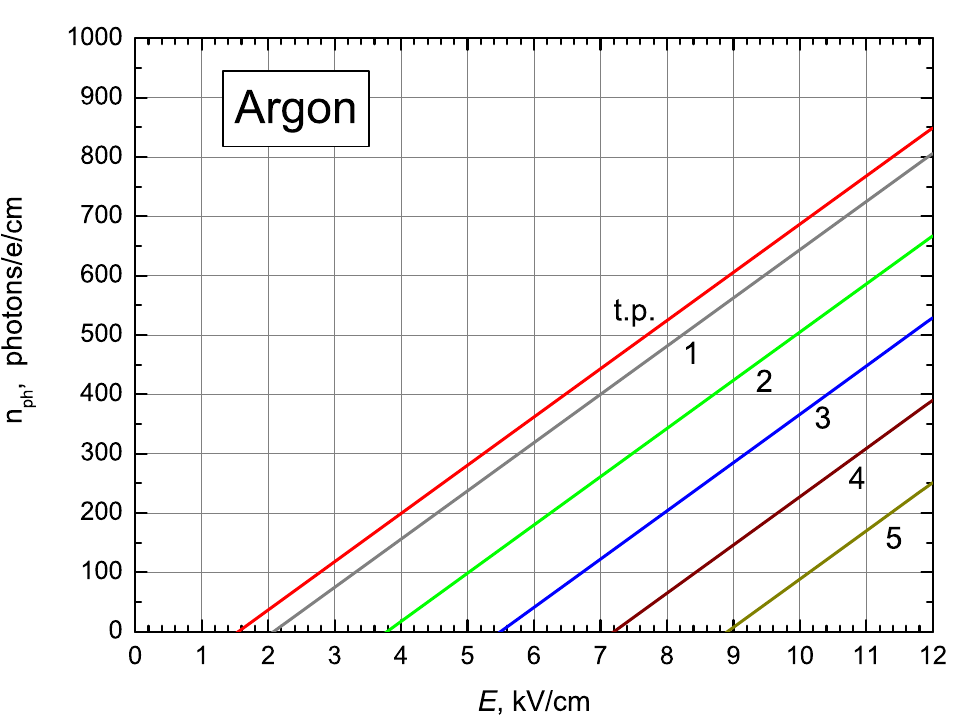}}
  \caption{Number of secondary scintillation (electroluminescence)
    photons generated by an electron traveling 1~cm in saturated gas
    at different gas pressure (indicated next to each curve, in bar)
    as a function of electric field; `t.p.' stands for triple point
    and corresponds to a pressure of 0.6889~bar for Ar and 0.8175~bar
    for Xe). For xenon, the data are from~\cite{Fonseca04}, taken in
    double-phase at $T=183$~K; for argon, data are
    from~\cite{Monteiro08} measured at room temperature. Thermodynamic
    data are from~\cite{NIST}.}
  \label{fig:SecScint}
\end{figure}

For practical purposes, a more convenient parameterization of the
light yield as a function of field strength and gas pressure is:
\begin{equation}
   \frac{dN_{ph}}{dx} = \alpha E - \beta P - \gamma
   \quad \text{[photons/(e$\cdot$cm)]} \,,
  \label{eqn:SecLight1}
\end{equation}
with $E$ in V/cm and $P$ in bar. It takes into account the fact that
the density of the saturated vapor is described by a linear function
of pressure, $\rho(P) = a_0 + a_1 P$, which is a good approximation up
to at least 10~bar~\cite{NIST}. The coefficients are shown in Table~4,
and the number of photons generated by one electron over a distance of
1~cm is plotted as a function of $E$ in 
Figure~\ref{fig:SecScint}.

\begin{table}[ht]
  \label{tab:ab}
  \centering
  \caption{Secondary scintillation coefficients in
  equations~(\protect \ref{eqn:SecLight}) and (\protect \ref{eqn:SecLight1}); 
  $a$ and $b$ are according to
  \cite{Monteiro08} for argon (measured at room temperature) and
  \cite{Fonseca04} for xenon (measured in saturated xenon vapor in a
  double-phase chamber). The coefficients $\alpha$, $\beta$ and
  $\gamma$ are calculated by the authors with densities and pressure
  from~\cite{NIST}.}
  \vspace{5mm}
  \begin{tabular}{|l|c|c|}
    \hline
    \rule{0pt}{3ex}
    ~                                  & Ar                   & Xe                   \\ 
    \hline
    \rule{0pt}{3ex}
    $a$, V$^{-1}$                      & 0.0813               & 0.137                \\ 
    \rule{0pt}{3ex}
    $b$, cm$^{2}$                     & $1.90\times10^{-18}$ & $4.70\times10^{-18}$ \\ 
    \rule{0pt}{3ex}
    $\alpha$, V$^{-1}$                 & 0.0813               & 0.137                \\ 
    \rule{0pt}{3ex}
    $\beta$, bar$^{-1}\cdot$cm$^{-1}$  & 139                  & 177                  \\ 
    \rule{0pt}{3ex}
    $\gamma$, cm$^{-1}$                & 30.6                 & 45.7                 \\
    \hline
  \end{tabular}
  \vspace{5mm}
\end{table}

Secondary light emission in pure noble gases in the visible and near
infrared regions has recently received some interest in view of
advances in solid state photon detectors sensitive at these
wavelengths. Published data on the light yield are scarce and not
consistent, indicating, however, that it is significantly lower than
in the VUV region.

For Ar at atmospheric pressure, a yield of
$\sim$3~photons/(e$\cdot$cm) was measured at $E=6.3$~kV/cm
($E/n=25\cdot 10^{-17}$~V$\cdot$cm$^2$) in the range 690--1000~nm, in
good agreement with calculations~\cite{Fraga00}. It should be noted
that at this field there is already some electron multiplication with
a gain of $\sim$9. For unity gain ($E/n < 7\cdot
10^{-17}$~V$\cdot$cm$^2$, i.e.~$E<1.7$~kV/cm), the maximum light yield
achieved is $\sim$0.2~photons/(e$\cdot$cm). In a more recent study
\cite{Buzulutskov11}, values of $\sim$60 and
$\sim$2~photons/(e$\cdot$cm) have been reported for
$E/n=25\cdot10^{-17}$~V$\cdot$cm$^2$ and
$7\cdot10^{-17}$~V$\cdot$cm$^2$, respectively. The measurements were
done at $T$=163~K and $P$=0.6~bar, i.e.~in argon gas of approximately
the same density as in the studies reported in~\cite{Fraga00}. For
comparison, the VUV yield is $\sim$90~photons/(e$\cdot$cm) for
\mbox{$E/n=7\cdot 10^{-17}$~V$\cdot$cm$^2$~\cite{Monteiro08}. }

It was also found that the yield in the NIR region, $1/n\cdot
dN_{ph}/dx$, as a function of $E/n$ follows the same linear law which
has been established for the VUV light through
equation~(\ref{eqn:SecLight}), but with smaller slope and about three
times higher threshold of $E/n=6.5\cdot 10^{-17}$~V$\cdot$cm$^2$,
\mbox{i.e.~$E \approx 1.7$~kV/cm} (about 570~V/cm for VUV; $P=$1~bar). Note the proximity of the threshold
to the minimum field required for electron multiplication. These facts
indicate that the excitation to higher atomic states, resulting in the
emission of NIR photons, becomes appreciable only near the ionization
threshold.

In Xe gas, secondary scintillation in the wavelength regions
3--14~$\mu$m~\cite{Carugno98} and
\mbox{0.7--1.6~$\mu$m}~\cite{Belogurov00} has been observed, but
unfortunately not quantified.

\newpage

\section{State-of-the-art technologies and methods}
\label{sec:StateOfTheArt}

\subsection{Detection of the VUV light}
\label{sec:DetectionVUV}

\subsubsection{Photomultiplier tubes}
\label{sec:PMTs}

All liquefied rare gases scintillate predominantly in the vacuum
ultraviolet region. Liquid xenon has the longest wavelength of 178~nm,
while argon emits at 130~nm~\cite{Jortner65} and neon at about
80~nm~\cite{Schneider74}. Xenon and argon light can be directly
detected with photomultiplier tubes (PMTs) if a special entrance
window transparent to the short wavelengths is used. For xenon light,
a quartz window will be sufficient, while a magnesium fluoride or
lithium fluoride window is required to detect the light from liquid
argon. Quartz window PMTs are more expensive than ordinary tubes with
borosilicate glass window, but not prohibitively so, and can be
manufactured with relatively large diameter. Photomultipliers with
MgF$_2$ window are much more expensive and the window diameter is
usually smaller, two inches at most. In addition, no ultra-low
background PMTs have been commercialized with these windows.

Therefore, for detection of liquid argon scintillation the most common
solution is to use wavelength shifting materials, which absorb VUV and
re-emit the light at longer wavelength. For example, $p$-terphenyl
emits mostly between 280~nm and 350~nm, diphenyloxazolyl-benzene
(POPOP) shifts the light to the region 340--420~nm, diphenyloxazole
(PPO) emits between 305~nm and 365~nm~\cite{Berlman65}. Tetraphenyl
butadiene (TPB), which emits in the blue wavelength region from 400 to
470~nm~\cite{Lally96,McKinsey97,Gehman11}, is a popular choice for
large liquid argon detectors~\cite{Benetti03,Boccone09}. A more
complete list of wavelength shifting materials can be found
in~\cite[p.76]{BolozdynyaBook10}. The use of a wavelength shifter in
liquid xenon detectors can enhance the photon detection efficiency by
$\sim$20\%, but it seems that the risk of contamination of the liquid,
which can lead to a substantial reduction of the electron lifetime,
should not be neglected~\cite{Bolozdynya08} (this work indeed reports
that $p$-terphenyl can dissolve in liquid xenon).

When using photomultipliers for these applications, one should take
into account that the electrical resistivity of the photocathode
materials increases when the temperature decreases. This can lead to
saturation of the photocathode under illumination and to partial or
even total loss of sensitivity of the PMT due to photocathode
charging. This effect is more accentuated for bialkali materials,
which have the highest photocathode sheet resistance, while
multialkali photocathodes are affected much less by the low
temperature~\cite{Ichige93}. 
\begin{figure}[ht]
\centerline{\includegraphics[width=.92\textwidth]{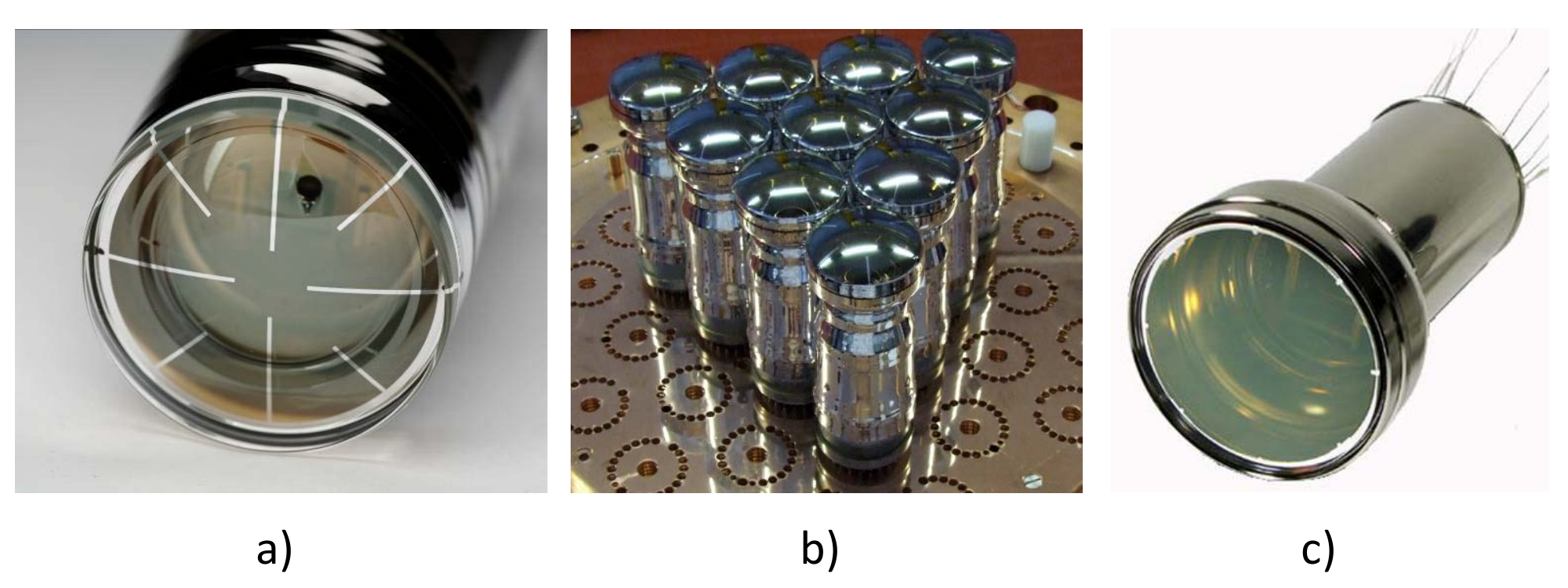}}
\caption{(a) Hamamatsu R7724Q-MOD tailored for LXe operation,
  featuring metal strips evaporated under the quartz window to
  compensate for the increase in resistivity of the bialkali
  photocathode at low temperature; (b) ETEL D766Q low background
  photomultipliers mounted on a stack of voltage distribution plates
  (courtesy ZEPLIN-III Collaboration); (c) Extremely low radioactivity
  3-inch Hamamatsu R11410 (courtesy Hamamatsu Photonics).}
\vspace{2mm}
\label{fig:PMTs}
\end{figure}
The critical temperature, at which a sharp drop in sensitivity of
bialkali photocathodes is observed, is related to the illumination
intensity and the photocathode diameter~\cite{Araujo97,Araujo98}. For
example, it was possible to operate a PMT with 25~mm photocathode
diameter at photocurrents of up to 60~pA down to $-125^\circ$C, while
for a \mbox{2-inch} PMT at 6~pA photocurrent the sensitivity started
to decrease sharply already at $-100^\circ$C. In spite of that,
bialkali photomultipliers were found to be suitable for most
applications using liquid xenon as a scintillator. Most
photomultiplier tubes intended for low temperature applications were
provided until recently with radial metal strips deposited under the
photocathode and connecting the central photocathode region with the
peripheral conducting ring, to which the power supply is connected
(Figure~\ref{fig:PMTs}\,a). This provides a faster compensation of the
charge emitted from the photocathode under illumination, so that
higher irradiances can be withstood at lower temperatures without
degradation of performance. For example, 2-inch PMTs R2154 from
Hamamatsu Photonics with metal fingers under the bialkali photocathode
were successfully operated in liquid xenon~\cite{Neves05}; it was
shown in~\cite{Araujo04} that 2-inch photomultipliers with metal
fingers from ETEL (model D730Q/9829QA) can withstand a photocurrent of
up to $\sim$100~pA at LXe temperatures.

The quantum efficiency of the photocathode can also be affected by
temperature, the effect being different at different wavelengths. An
increase in quantum efficiency by $\sim$10--20\% at xenon wavelength
has been observed for a batch of 35 bialkali PMTs (ETEL model
D730Q/9829QA) upon cooling to $-100^\circ$C~\cite{Araujo04}. A similar
increase by 20\% had been previously noticed with a R1668 Hamamatsu
photomultiplier at 170~nm but not at 185~nm, for which almost no
change was observed~\cite{Araujo98}. In the same work, a much more
significant variation of the radiant sensitivity at 170~nm, by up to a
factor of 2 (also not present at 185~nm) was observed with a Philips
XP2020Q tube. A comparable (but wavelength dependent) decrease in the
overall PMT response has also been observed for other PMTs (Hamamatsu
R8778, ETL D742~\cite{Hollingworth04}). A recent study of the
Hamamatsu R8520 1-inch PMT ~\cite{Aprile12b} revealed an improvement
in QE of $\sim$5--11\% upon cooling. These results highlight the
importance of calibration of the photomultiplier tubes under the exact
conditions in which they will be used in real experiments.

In spite of the risk of saturation (notably with the higher photon
rates required during calibration), bialkali photomultipliers are the
most frequent choice in liquid xenon DM search experiments because of
their high quantum efficiency and low dark noise. In most cases 1- or
2-inch tubes have been used, usually with conductive metal fingers,
for operation while immersed in the liquid phase
(ZEPLIN-III~\cite{Araujo04}, XENON10~\cite{Yamashita05},
XENON100~\cite{Aprile12c}, XMASS~\cite{Suzuki08},
LUX350~\cite{Akerib12e}). Larger tubes (130-mm diameter ETEL D742Q)
were successfully used in ZEPLIN-II~\cite{Alner07}. These had a thin
platinum underlay to reduce the effect of photocathode charging at low
temperature. Unfortunately, this came at the expense of quantum
efficiency (only 17\% for xenon scintillation).

In liquid argon detectors, PMTs have to operate at even lower
temperature (87~K). An 8-inch model (ETEL 9357FLA) with platinum
underlay as above was successfully used to detect liquid argon
scintillation~\cite{Benetti03}. With a TPB wavelength shifter on the
PMT glass window, a quantum efficiency of $\sim$20\% for liquid argon
scintillation light has been reported; a batch of 54 such tubes has
been tested with visible light at 77~K in~\cite{Ankowski06}. No
dramatic changes were detected for most PMTs. Compared to room
temperature, a gain drop was observed at 77~K, varying from sample to
sample in the range from 15\% up to a factor of 5. It was possible to
compensate the gain loss by increasing the anode voltages by
$\approx$100~V. A 2- to 3-fold increase of the dark count rate was
also observed, which led to some deterioration of the peak-to-valley
ratio in the single electron response. The transit time jitter was
measured to be twice shorter at 77~K than at room temperature; the
sensitivity decrease was 20\% at 470~nm; good linearity was found up
to 300~photoelectrons per 50~ns pulse at 10~kHz repetition rate
(corresponding to 4.8~pA average and 1~nA peak photocurrent) both at
room and low temperatures.

A similar 8-inch tube with bialkali photocathode and platinum underlay
from Hamamatsu Photonics, model R5912-MOD02, was used in contact with
liquid argon~\cite{Lippincott08}. This tube was also tested down to
29~K to investigate its applicability to detect liquid neon
scintillation~\cite{Nikkel07}. A sharp loss of gain from $\sim$10$^9$
to $\sim$10$^7$ was observed near 75~K followed by some recovery at
lower temperatures. Mechanical deformations in the dynode system were
suggested as the most likely reason for the gain variations. In the
same test, the relative photon detection efficiency (photocathode
quantum efficiency times photoelectron collection efficiency to the
first dynode) decreased by only 25\% and at a higher temperature ---
about 100~K. The dark count rate initially decreased down to 250~K,
then rose gradually with further cooling and eventually reached a
3-fold increase relative to the value measured at room temperature. 

A 3-inch diameter ETL~D747/9361~FLB photomultiplier tube was used
successfully to detect scintillation from liquid neon although the
tube was kept at about 155~K in this experiment~\cite{Nikkel08}.

Recently, a special type of bialkali photocathode has been developed
for low temperature applications~\cite{Nakamura10}. It is
characterized by a lower sheet resistance, obviating the need for
metal strips or platinum backing of the sensitive layer. A more recent
version of the 2-inch Hamamatsu R7724Q-MOD is an example designed for
operation at LXe temperatures with quantum efficiency of $\approx$25\%
at 178 nm. The 3-inch models R11065 and R11410 (shown in 
Figure~\ref{fig:PMTs}\,c)
have been developed specifically for low background experiments with
LXe and LAr~\cite{Hamamatsu2011,Akerib12b,Lung12}.

Another important achievement is the development of new
`superbialkali' photocathodes with significantly enhanced quantum
efficiency by Photonis and Hamamatsu~\cite{Nakamura10}. This progress
was achieved by using extremely pure photocathode materials and
process tuning. Hamamatsu quotes a typical value of 43\% at 350~nm for
Ultrabialkali (UBA) and 35\% for Superbialkali (SBA)
phototubes~\cite{Hamamatsu2011}. A peak value of 55\% is advertised by
Photonis~\cite{Renker09}. At the time of writing, there was no
information available on the quantum efficiency of these photocathodes
below 200~nm. Operation at low temperature is also not guaranteed due
to high sheet resistance. Deposition of a metal pattern, as for normal
bialkali photocathodes, might help extend the range to LXe
temperature.

\begin{figure}[ht]
\centerline{\includegraphics[width=.8\textwidth]{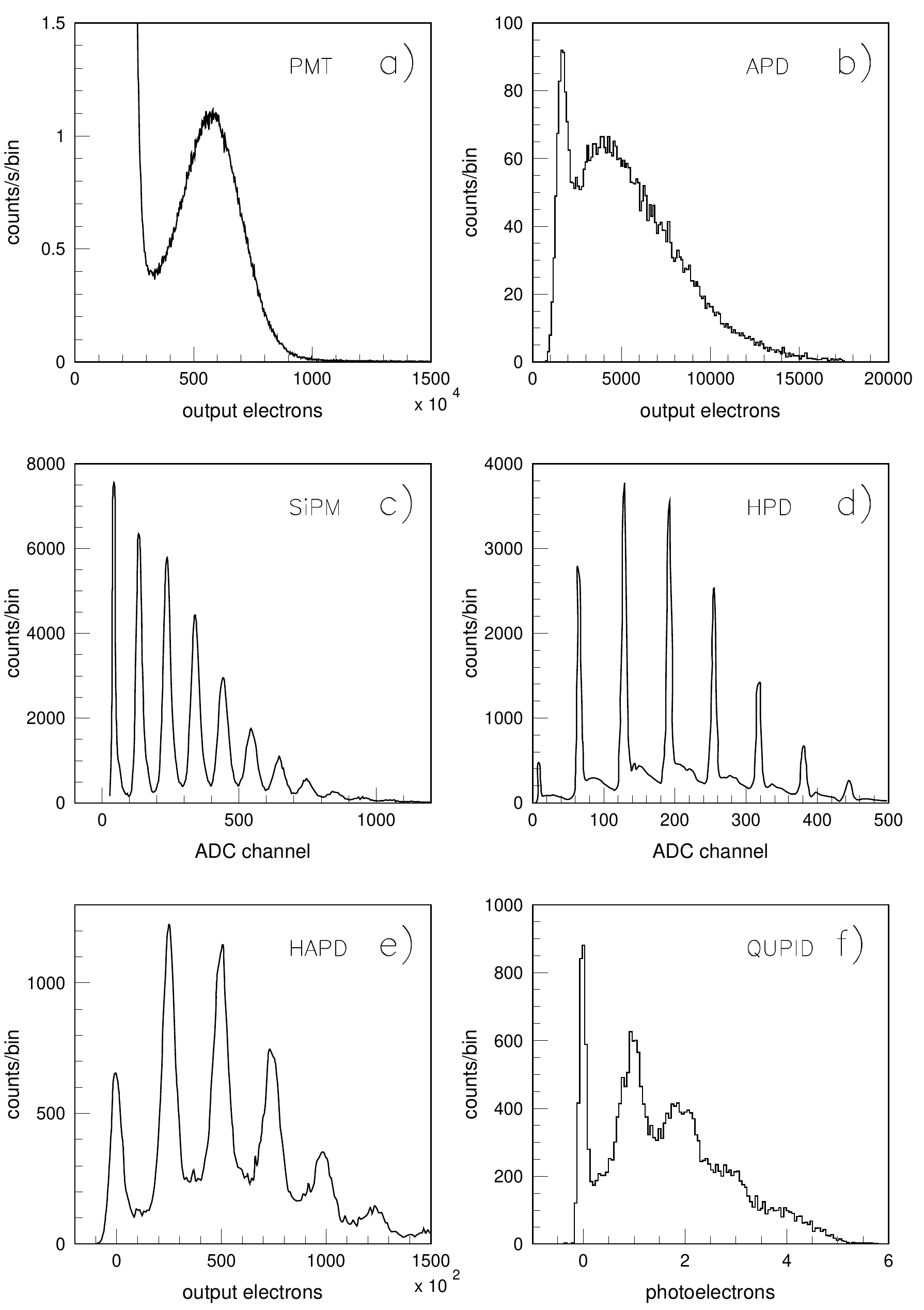}}
\caption{Low signal response of several photon detectors discussed in
  the text: (a) Single photoelectron response (dark counts) measured
  with a 2-inch Hamamatsu R7724Q-MOD photomultiplier with quartz
  window and Al fingers under a bialkali photocathode (courtesy
  V.~Solovov and L.~de~Viveiros); (b) Response of a 5-mm diameter
  LAAPD from Advanced Photonix to faint light pulses generating a mean
  of 4.3 electron-hole pairs after pedestal subtraction (adapted
  from~\cite{Solovov03}; with permission from Elsevier); 
  (c)  1$\times$1~mm$^2$ SiPM, Hamamatsu S10362-11-100U (courtesy
        F.~Neves) 
  (d) HPD from Delft Electronic Products, PP0270K with Amptek 250 
       preamplifier (adapted from~\cite{Datema97};
  with permission from Elsevier); 
 (e) 13-inch HAPD (adapted  from~\cite{Nakayama06}; with permission 
      from Elsevier); 
 (f) QUPID   (adapted from~\cite{Teymourian11}; with permission 
      from Elsevier).}
\label{fig:PhotonDetectorSERs}
\end{figure}

Good definition of the single photoelectron signal is also an
important requirement for reliable detection of faint luminous
signals. For a good photomultiplier, the spectrum of anode pulses due
to single photoelectrons emitted from the photocathode should present
a well defined peak, separated from the exponential noise with a
peak-to-valley ratio of about 2 or higher. As an example we present
the single electron spectrum measured with a 2-inch R7724Q-MOD tube
(Figure~\ref{fig:PhotonDetectorSERs}\,a).
A technique for {\it in-situ} calibration of single
photoelectron responses in photomultiplier arrays under exact data
conditions is described in~\cite{Neves10a}.

Radioactivity of the photomultiplier components is a key concern in
low background experiments. In fact, PMTs can dominate the total
radioactivity of a detector. For a model with glass envelope, the
emission rate of $^{40}$K $\gamma$-rays can be quite significant (up
to $\sim$10~Bq per device for normal glass~\cite{McAlpine}). Trace
amounts of radioisotopes from the U/Th decay chains contribute to both
$\gamma$-ray and neutron background (the latter through ($\alpha$,$n$)
reactions and spontaneous fission of $^{238}$U). Quartz has
$\sim$10$^4$ lower content of $^{40}$K and $\sim$10$^2$ less
$^{232}$Th and $^{238}$U than normal glass, and so PMTs with quartz
window are preferable from this point of view, too. An all-quartz
envelope would be ideal, but practically unattainable due to
difficulties in welding the contact pins. A significant effort is
being made by manufacturers and research teams to reduce the
background from photomultiplier tubes by rigorous choice of the raw
materials used for all components. For example,
\mbox{$\sim$10$^3$~$\gamma$-rays/day} and 
\mbox{$\sim$10$^{-3}$~neutrons/day} have
been achieved in the 2-inch
Hamamatsu~R8778~\cite{Fiorucci09,Viveiros10}, and a factor of
$\sim$3~higher for the ETEL~D766Q~\cite{Araujo12} (shown in
Figure~\ref{fig:PMTs}\,b).
A significant further reduction of radioactive
backgrounds has been reported recently: a prototype of a 3-inch
R11410-MOD, developed by Hamamatsu, exhibits 24 times lower content of
$^{238}$U, 9~times less $^{232}$Th and 8 times less $^{40}$K per PMT
when compared with the R8778; this would result in a reduction of
\mbox{PMT-induced} nuclear recoil background by a factor of 36 in the LUX350
experiment, adding to a further reduction of electron recoil
background~\cite{Akerib12b}.

Traces of radioactive nuclei are encountered also in the metal and
ceramic parts of the electron multiplication system. A typical 2-inch
photomultiplier contains about 70~g of glass, 20~g of metal and 10~g
of ceramic~\cite{McAlpine}. Hybrid photomultiplier vacuum tubes, which
do not use dynode structures for amplification and in which the
variety (and total mass) of the required materials can be
significantly reduced, are an attractive solution. In this type of
device, photoelectrons emerging from the photocathode are accelerated
in a strong electric field up to energies of 10~to~20~keV and focused
onto a silicon detector. As silicon is virtually free of radioactivity
and the mass of the photodiode can be very small, the background from
the inner part of the tube can be significantly reduced (by a factor
of $\sim$100 according to~\cite{Arisaka09}). In addition, all
non-metal parts can be made of quartz. We shall discuss some hybrid
devices in 
Section~\ref{sec:HybridDevices}.

The exciting and fast-paced history of recent developments in photon
detectors, including vacuum-based, solid-state and their combinations,
can be found in the review articles
\cite{Arisaka00,Renker07,Renker09,Renker09a}.

\subsubsection{Large area avalanche photodiodes}
\label{sec:LAAPDs}

Silicon photodiodes with intrinsic amplification are an attractive
alternative to photomultiplier tubes. Their great advantage is much
lower intrinsic radioactive background and smaller mass. Moreover,
higher quantum efficiency than can be achieved with PMTs is possible,
especially in the VUV wavelength region. Among the disadvantages one
should mention lower amplification gain, higher noise, and high cost
per unit area. At present, two types of silicon photodiode with
amplification are being considered as possible candidates to replace
photomultiplier tubes: Large Area Avalanche Photodiodes (LAAPD)
operating in proportional mode, and multipixel devices operating in
Geiger mode; the latter are frequently called Silicon Photomultipiers
(SiPM) or Multi-Pixel Photon Counters (MPPC).

A comprehensive review on advances in solid state photon detectors has
been published recently~\cite{Renker09a}. We therefore refer the
reader to this work for detailed information on the subject and will
consider here only some of the aspects that seem to us to be the most
relevant for applications in liquefied rare gas detectors.

In a silicon avalanche photodiode, photons are absorbed in an
intrinsic (undoped) absorption region which is sandwiched between a
metal contact and a $p-n$ junction. A reverse-bias voltage of
$\sim$0.2--2~kV is applied to the photodiode thus creating a field of
$>$10$^5$~V/cm across the depletion region near the
junction. Electrons drifting through this region acquire sufficient
energy to produce secondary electron--hole pairs by impact ionization
so that an avalanche is developed and thus the initial photocurrent is
amplified. In silicon, both electrons and holes can produce secondary
ionization.

The multiplication gain is a very steep function of the applied
voltage, and dependent also on temperature. Typically, gains
$\sim$100--1000 are achieved at room temperature, being limited by
breakdown. At low temperature, higher gains can be achieved: values of
4$\cdot$10$^3$ at 193~K~\cite{Solovov03} and even $\sim$2$\cdot$10$^4$
at 40~K~\cite{Yang03} have been reported in the literature (the APD
models were different in these works). The practical gain is, however,
limited by noise, which also increases with the applied voltage.

In avalanche devices noise has two components of distinct nature. One
is due to dark current fluctuations, while the other results from the
statistics of the multiplication process. The dark current increases
with voltage (i.e.~with gain) but, on the other hand, it depends
strongly on temperature. For example, a reduction by almost a factor
of 10$^5$ was observed with a 5~mm diameter LAAPD from Advanced
Photonix (API)~\cite{API} when cooled from room temperature down to
$-100^{\circ}$C~\cite{Solovov00}. At this temperature, corresponding
to the typical operation point of liquid xenon detectors, the dark
current was less than 1~pA, while in normal conditions it was measured
to be of the order of tens of nA. The electronic noise was found to be
practically independent of gain for $T<-40^{\circ}$C and gains
$>$5~\cite{Solovov03}, thus indicating that fluctuations of the dark
current do not contribute significantly in these regimes.

The random nature of the multiplication process results in additional
fluctuations of the output signal of avalanche devices, which is
characterized by the excess noise factor $F=\langle m^2 \rangle/M^2$,
where $m$ is the stochastic multiplication gain and $M=\langle m
\rangle$ is its mean value. For gains $M \gg 1$, the excess noise
factor is approximately linear with gain and therefore at high gains
and low temperature the noise is determined by gain fluctuations.

Several models of avalanche photodiode have been investigated as
candidates for detection of scintillation signals in liquefied rare
gases. LAAPDs from API with 5~mm and 16~mm diameter active areas were
found to perform very well whilst immersed into liquid
xenon~\cite{Solovov02,Ni05}. Scintillation signals from 5~MeV
$\alpha$-particles and conversion electrons were successfully detected
with energy resolution comparable to that measured with
photomultiplier tubes. The quantum efficiency for xenon scintillation
light was found to be $>$100\%~\cite{Solovov02}, suggesting a
non-negligible contribution from multiple electron-hole pair
production by a single VUV photon (each photon carries $\approx$7~eV,
while the Si band gap is 1.12~eV). The authors are not aware of any
published results on quantum efficiency for argon scintillation light;
however, efforts are ongoing to extend the sensitivity to shorter
wavelengths. In liquid xenon, a maximum gain of $\sim$1000 was
obtained with API devices~\cite{Solovov02}, above which degradation of
the energy resolution due to excess noise became noticeable. A gain of
$\sim$10$^4$ has been reported~\cite{Oberlack07,Shagin09} for
13$\times$13~mm$^2$ LAAPDs from Radiation Monitoring
Devices~\cite{RMD}. In all instances the photodiodes were directly
immersed into the LXe.

The same model from RMD was tested from 77~K down to liquid helium
temperature~\cite{Yang03}. A similar tendency of gain-voltage
dependence was observed --- the lower the temperature, the steeper the
dependence becomes, and the lower the breakdown voltage. Between 77~K
down to 40~K multiplication gains up to $\sim$2$\cdot$10$^4$ were
obtained. However, with further cooling breakdown appears at a bias
voltage of only 500~V, even before appreciable amplification sets
in. Moreover, an abrupt degradation of quantum efficiency to about
15\% of that at room temperature was observed between 50~K and
35~K. These measurements were carried out with a red~LED.

In contrast to photomultiplier tubes, avalanche photodiodes do not
present a well defined single photoelectron peak --- signal amplitudes
due to a single electron-hole pair have nearly exponential
distribution 
Figure~\ref{fig:PhotonDetectorSERs}\,b).
The reason resides in the multiplication
process, different for a photodiode and a PMT (quasi continuous with
gain of 2 at each step for the APD, versus a discrete process with
gain of 5--10 at each dynode in the case of a PMT). This makes
detection of single photons, a strong requirement for DM searches and
CNS experiments, more problematic as it becomes more difficult to
discriminate electronic and other exponential like noise without
losing a significant fraction of the single photon signals. Yet, good
discrimination seems possible given the fact that the quantum
efficiency of the photodiode can be as high as $\sim$100\% (for xenon
light at least). This can allow setting rather high a discrimination
threshold so that even if 2/3 of the single photon signals are below
threshold, the remaining 1/3 would provide the same single photon
detection efficiency as with a PMT with good single electron spectrum
and $\sim$30\% photocathode quantum efficiency~\cite{Chepel02} (equal
sensitive areas are assumed).

The first large-scale application of these devices was in the EXO-200
$\beta$$\beta$-decay experiment, which features Advanced Photonix
LAAPDs (model SD630-70-75-500) used as bare dies, without the standard
ceramic encapsulation; these experiments require also extremely low
background levels and excellent energy resolution around the $Q$-value
for $^{136}$Xe (2,458~keV), but not as low a scintillation threshold;
468 such devices operate submerged in the LXe to readout the
scintillation light from a chamber with 110~kg active
mass~\cite{Neilson09}.

\subsubsection{Silicon photomultipiers}
\label{sec:SiPMs}

An excellent single photoelectron response can be obtained with
silicon photodiodes operated in the Geiger mode 
Figure~\ref{fig:PhotonDetectorSERs}\,c).
The voltage applied across the depleted region in these devices is set
above the breakdown limit so that a single electron or hole entering
the multiplication region can initiate a breakdown, similarly to what
happens in a conventional gaseous Geiger counter. Thus, a large pulse
of fixed amplitude determined by the stored charge is produced. A
built-in resistor limits the restoration current so that the voltage
drops rapidly below the multiplication threshold and the discharge
ceases. In this mode, two or more electrons at the beginning of the
multiplication region would produce signals of the same amplitude as a
single electron, typically yielding 10$^4$ to 10$^6$ electrons. The
solution for obtaining proportionality between the output signal and
the number of photons is to pixellate the device and make each pixel
work as an independent counter~\cite{Bondarenko00}. Then, the number
of detected photons is simply proportional to the number of pixels
which present signals or, more conveniently, to the amplitude of the
sum signal over all pixels.

\begin{figure}[ht]
\centerline{\includegraphics[width=.72\textwidth]{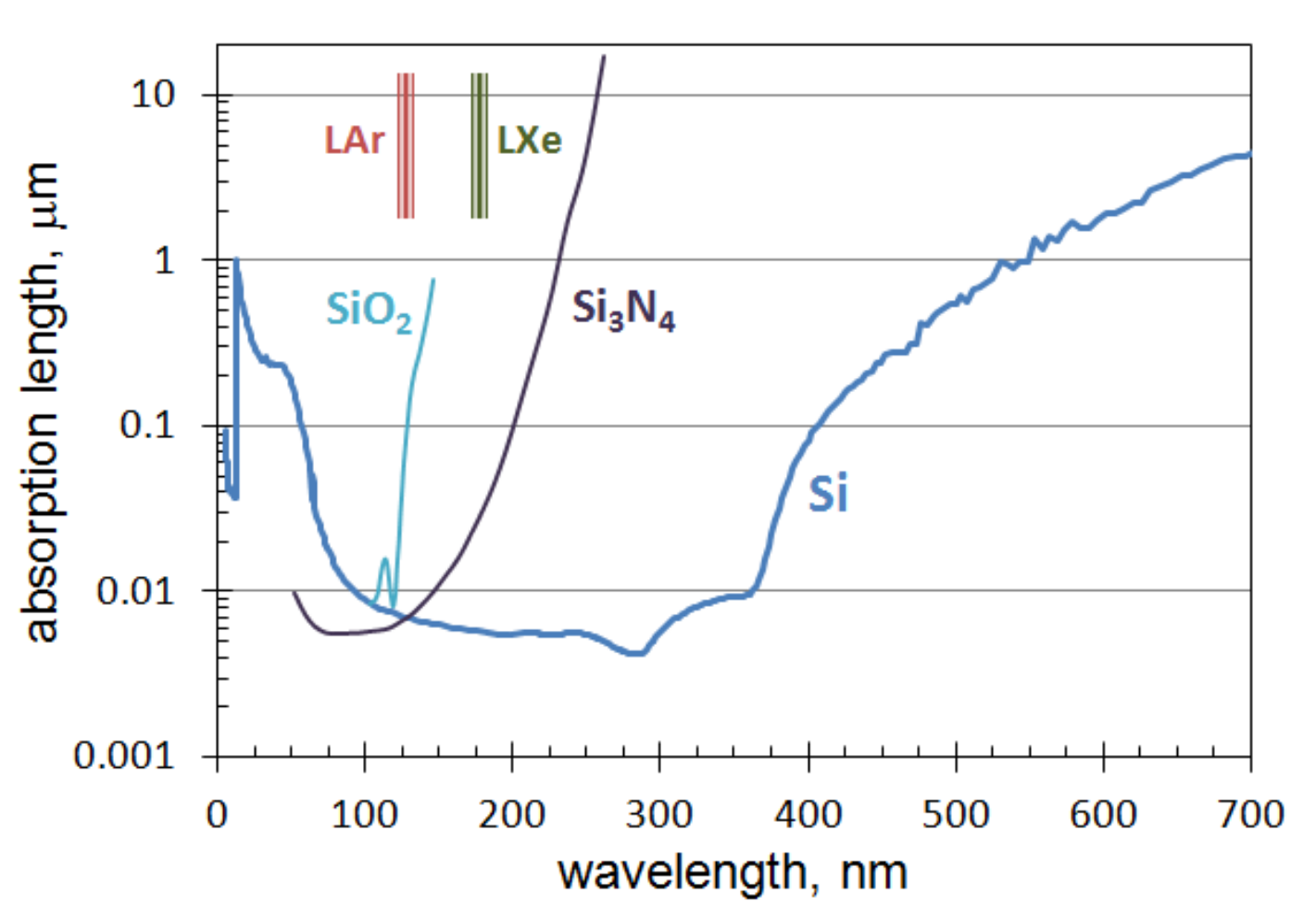}}
\caption{Absorption length in bulk silicon and its thermal oxide and
  nitride, which are typical device passivation materials (note that
  other native silicon oxides, with different absorption spectra, can
  also form). Calculated with extinction coefficients
  from~\cite{Photonics} and~\cite{Philipp82}.}
\label{fig:SiAbsLength}
\end{figure}

In silicon, the photon attenuation length is $>$100~nm in the visible
region for $\lambda \gtrsim 400$~nm, but it drops abruptly for shorter
wavelengths, being only $\sim$6~nm in the region of $\lambda$ between
300~nm to 150~nm 
(Figure~\ref{fig:SiAbsLength}).
For a conventional $n^+/p/p^+$ structure
(in the sequence encountered by incident photons), the short
wavelength photons are absorbed at the very front of the undepleted
$n^+$ region, which has typically a thickness of $\sim$100~nm and
where recombination is quite strong~\cite{Piemonte06}. This is one of
the main reasons for the low quantum efficiency of these devices for
short wavelengths. Photon absorption in the native SiO$_2$ and
protective Si$_3$N$_4$ layers frequently used in these devices is also
critical, especially for argon light.

The SiPM is a pixel device and, therefore, some dead space between
pixels is necessary to obtain electrical and optical isolation (the
latter is to avoid optical cross-talk induced by photons generated
during the multiplication process). Although the quantum efficiency of
each individual pixel, i.e. the probability of conversion of a photon
into an electron-hole pair, can be as high as 50\% to 80\% in the
visible wavelength region, from a practical point of view a more
relevant quantity is the photon detection efficiency (PDE), defined as
$\eta_\mathrm{SiPM}=Q \!\cdot\! F\!\cdot\! \epsilon_d$ , 
where $Q$ is the quantum
efficiency, $F$ is the so called `fill factor' (the ratio of the
sensitive area to total area) and $\epsilon_d$ is the probability for
a charge carrier to trigger an avalanche. For example, Hamamatsu
Photonics quotes fill factors of 0.3--0.8 for different
models~\cite{HamamatsuMPPC} depending on the pixel size (generally
higher for larger pixels).

The probability to initiate an avalanche depends on the wavelength
(and on the applied voltage, naturally) and is different for electrons
and holes. For holes, $\epsilon_d$ is about half of that for
electrons~\cite{Piemonte06, Oldham72} and here is another factor
affecting the sensitivity in the short wavelength region. Visible
photons are absorbed predominantly in the wide absorption region of
intrinsic silicon or lowly doped $p$-type material, which sits beyond
the multiplication region, and the avalanche is triggered by an
electron upon reaching the depletion region (traveling backwards
relative to the photon); in this instance $\epsilon_d$ approaches
unity at sufficiently high voltage. On the other hand, short
wavelength photons are absorbed at short distances from the
surface. The electron drifts away from the junction and does not reach
the multiplication region at all. The avalanche can still develop
triggered by the hole (if it does not recombine instead of entering
the multiplication region) but with a lower probability.

For visible light, a photon detection efficiency of up to
$\eta_\mathrm{SiPM}\sim 0.6$ can be achieved. This should be compared
with the corresponding value for a photomultiplier tube:
$\eta_\mathrm{PMT}=Q\cdot \epsilon_1$ , where $\epsilon_1$ is the
collection efficiency of photoelectrons to first dynode, typically
$\epsilon_1 \approx 0.7-0.8$. Hence, for a good PMT ($Q$=30\%) the
photon detection efficiency is $\eta_\mathrm{PMT}\sim$0.20--0.25. This
comparison shows that silicon photomultipliers have a great advantage
over vacuum tubes (equal areas being assumed). At present, the area of
commercially available single-crystal SiPMs is at best
6$\times$6~mm$^2$. Arrays of 16~SiPMs and a matrix of 16~such arrays
(i.e.~256 SiPMs) became available recently~\cite{SensL}. The
61$\times$61~mm$^2$ total area is already comparable to that of a
conventional PMT. Hamamatsu offers arrays of 4$\times$4 and 8$\times$8
arrays of SiPMs 3$\times$3~mm$^2$ each (models S11064 and S11834).

The low temperature performance of a 1$\times$1~mm$^2$ SiPM from
SensL~\cite{SensL} has been studied in~\cite{Lightfoot08}. It was
found to perform well down to liquid nitrogen temperature at gains up
to 2$\cdot$10$^6$. Similarly to an APD operated in proportional mode,
the gain at fixed voltage also depends on temperature and, therefore,
a correction of the applied voltage on temperature is necessary in
order to keep the gain stable. However, in contrast with an APD, where
the gain is a steeper-than-exponential function of the voltage, in the
Geiger mode the gain depends on the voltage linearly (because the
output signal is limited by the charge stored in the cell capacitance)
and, therefore, its variation with temperature is not so dramatic as
for APDs. The gain was found to vary at the rate of about
0.5\%/$^{\circ}$C for gains of $M\sim 2\cdot10^6$. As liquefied gas
detectors require temperature stabilization with precision much better
than $\pm$1$^{\circ}$C anyway, the gain stability should not be a
serious issue during operation of the detector, although a gain
monitoring system and, probably, a temperature compensation circuit
might be desirable. To compare, a regular APD operating at $M=100$
exhibited a gain temperature coefficient of $\approx$4\% per
$^{\circ}$C, while for $M=1000$ this rose to as much as 15\% per
$^{\circ}$C~\cite{Chepel02}. Therefore, APDs require much more tight
temperature control and a gain stabilization mechanism becomes
indispensable.

The dark noise count rate of a SiPM at room temperature is very high
and very sensitive to the bias voltage. For example, values of
10$^6$~Hz and 10$^5$~Hz were measured with 1$\times$1~mm$^2$ devices
from SensL and Hamamatsu~\cite{Lightfoot08,Neves12}. Cooling results
in a significant reduction of the dark noise: by reducing the
temperature down to $-100^{\circ}$C the noise counts were suppressed
by a factor of $\sim$1000 in~\cite{Lightfoot08} and, for the other
model, to a level of $\sim$1~Hz~\cite{Neves12}. The effect of further
cooling was found to be much less significant~\cite{Lightfoot08}:
cooling down from $-100^{\circ}$C to $-200^{\circ}$C gives another
factor of $\sim$10 only.

Silicon photomultipliers from other manufacturers have also been
tested at low temperature. A 2$\times$2~mm$^2$ G-APD (Geiger APD,
according to the manufacturer's terminology) CPTA model \mbox{149-35}
(Russia) was successfully operated in liquid argon~\cite{Bondar11}; a
1$\times$1~mm$^2$ SiPM from \mbox{FBK-IRST} (Italy) was found to perform well
down to approximately $-150^{\circ}$C~\cite{Collazuol11}.

The sensitivity of silicon photomultipliers in the VUV wavelength
region has been questioned, with existing results being
contradictory. According to the manufacturers, there should be no
significant sensitivity below $\approx$400 nm or $\approx$300 nm
depending on the structure ($n^{+}/p/\pi/p^{++}$ or $p^{+}/n/\nu
/n^{++}$, respectively); this is largely related to the photon
absorption depth argument laid out above. Nevertheless, a number of
attempts to detect xenon light have been made.

One of the earliest versions of silicon photomultiplier from Pulsar
(Russia) has been tested in liquid xenon to detect primary
scintillation light from $\alpha$-particles~\cite{Aprile06a}. A
quantum efficiency of 22\% and a photon detection efficiency of
$\sim$5.5\% due to a poor fill factor has been estimated for the xenon
wavelength.

Three different models from Russian companies MEPhI/Pulsar and CPTA
have been tested with light emitted by gaseous and liquid xenon under
$\alpha$-particles. A photon detection efficiency of $<$1\% has been
reported~\cite{Akimov09b}. A windowless version of a 3$\times$3~mm$^2$
MPPC from Hamamatsu was tested with a VUV light source and a narrow
175~nm optical filter in~\cite{Neves12}. A total PDE of $\approx$2.0\%
has been measured against a calibrated photomultiplier tube, which
gives a quantum efficiency of $\approx$2.6\% taking into account the
fill factor of 78.5\%. The PDE obtained with the VUV light source
agreed well with that measured in liquid xenon excited by
$\alpha$-particles. The MPPC was immersed into the liquid in these
measurements~\cite{Neves12}.

A serious issue with SiPMs is afterpulsing due to release of the
charges trapped in the high field region during avalanche
development. The probability for an afterpulse to occur can be as high
as 20\% and can depend on temperature. For example, for FBK-IRST SiPMs
the afterpulsing probability was observed to be constant from room
temperature down to 120~K but increased by a factor of 8 with further
cooling down to 70~K~\cite{Collazuol11}. An active quenching circuit,
which reduces temporarily the bias voltage shortly after the main
pulse, can help mitigate this problem. At the same time, the
manufacturing process also seems to be evolving to reduce this
drawback.

\subsubsection{Hybrid devices}
\label{sec:HybridDevices}

A combination of a vacuum phototube with a silicon device acting as a
photoelectron detector can bring significant advantages from the point
of view of radioactive background from the photon detector, as already
mentioned in 
Section~\ref{sec:PMTs}.
The light sensitive part of this type of
device is a semitransparent photocathode deposited onto a quartz
window, just like in a conventional photomultiplier tube (and hence
the same variety of photocathode and window dimensions can be
used). Photoelectrons emitted from the photocathode are accelerated
(in vacuum) in a strong electric field, which focuses them onto the
silicon detector. Under 10~to~20~kV accelerating voltage each
photoelectron creates several thousand electron-hole pairs in Si, thus
providing a gain of $\sim$10$^3$~\cite{DeSalvo92}. These devices are
commonly known as Hybrid Photon Detectors (HPD).

The gain is not high and requires additional amplification, but with a
good low noise external preamplifier a very good spectrum for low
intensity light pulses can be obtained and individual photoelectrons
can be clearly distinguished~\cite{DeSalvo97,Datema97}. It should be
pointed out that the amplification mechanism in HPDs is based on
energy dissipation by an accelerated photoelectron in the anode rather
than on a sequential multiplication of electrons with a gain of 4--6
at each step as in photomultiplier tubes, thus reducing significantly
the statistical fluctuations. An example of such spectrum is shown in
Figure~\ref{fig:PhotonDetectorSERs}\,d.

Obviating the need for a resistive chain required for operation of
conventional PMTs is also an advantage, but the high voltage required
to bias an HPD is a clear drawback (although no significant current is
consumed). Several HPD models with windows up to 72~mm diameter have
been developed commercially, e.g.~by Photonis~\cite{Photonis},
including the multi-pixel model developed for the LHCb
experiment~\cite{Gys06}.

Clearly, a higher gain would bring a significant advantage to this
hybrid solution. This became possible with the development of
avalanche diodes. Using one of these devices instead of a silicon
diode adds another factor of $\sim$300 to the total gain, thus raising
it to $\sim$10$^5$. A 13-inch HAPD (Hybrid Avalanche Photo-Detector)
based on that principle has been developed by several Japanese
institutes together with Hamamatsu for use in large Cherenkov
detectors for neutrino experiments~\cite{Nakayama06}. It features good
timing resolution, photocathode uniformity and excellent single
electron response (Figure~\ref{fig:PhotonDetectorSERs}\,e).

\begin{figure}[ht]
\vspace{5mm}
\centerline{\includegraphics[width=.65\textwidth]{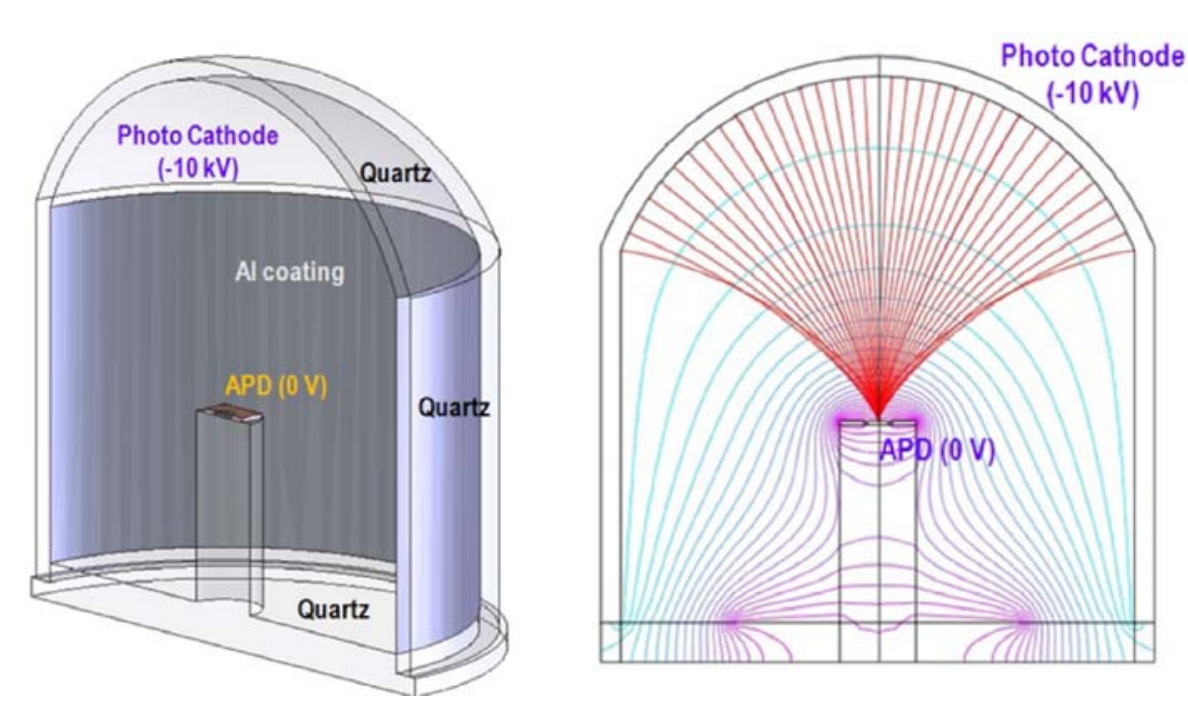}}
\caption{The Quartz Photon Intensifying Detector (QUPID) developed by
  a collaboration of the University of California, Los Angeles and
  Hamamatsu. (Courtesy K.~Arisaka.)}
\vspace{5mm}
\label{fig:QUPID}
\end{figure}

\vspace{-3mm}
A collaborative effort between UCLA and Hamamatsu has been developing
a dedicated hybrid photosensor for dark matter experiments with
special requirements of low radioactivity content, capability to
operate at liquid xenon temperature, high quantum efficiency and good
single electron response. This device is known as the `QUPID' ---
QUartz Photon Intensifying
Detector~\cite{Arisaka09,Fukusawa10,Teymourian11}. The QUPID is made
of a quartz tube of 71~mm diameter with a hemispherical photocathode
window 
(Figure~\ref{fig:QUPID}).
The photocathode is kept at $-6$~kV while the APD
is at ground potential. The sealing between the different parts is
made by indium bonding. The version described in~\cite{Teymourian11}
uses a special low temperature bialkali photocathode, which does not
require metal strips or an underlay to reduce the resistance, although
previous versions were provided with metal fingers under normal
bialkali photocathodes. A quantum efficiency of (34$\pm$2)\% has been
measured for xenon light at room temperature for several samples. The
QUPID has also been tested in liquid xenon at $-100^{\circ}$C and
1.6~bar being fully immersed into the liquid. The liquid xenon
scintillation signals from 122~keV $\gamma$-rays and
$\alpha$-particles were recorded~\cite{Teymourian11}. Tests at even
lower temperature described in~\cite{Fukusawa10} have confirmed that
the QUPID can operate down to $-175^{\circ}$C. An example of an
amplitude spectrum measured with weak light pulses, showing that single
photoelectrons are clearly resolved, is shown in 
Figure~\ref{fig:PhotonDetectorSERs}\,f.

\subsection{Alternative techniques for light and charge signal detection}
\label{sec:AlternativeTechniques}

\subsubsection{Electron multiplication with micro-pattern structures}
\label{sec:ElectronMultiplication}

The double-phase technology, with measurement of both primary
scintillation in the liquid and secondary scintillation in the gas
phase with photomultiplier tubes, has proven to be reliable and
capable of providing high nuclear/electron recoil discrimination
capability down to keV energies. Nevertheless, significant effort is
being put into development of alternative read-out configurations
aiming ultimately at reducing the cost of instrumenting very large
underground detectors, keeping low radioactive background associated
with the readout system as well as providing spatial resolution of
$\sim$1~mm, which is important for target fiducialization and
multi-vertex resolution (in dark matter searches) or for full track
reconstruction (for high energy neutrino experiments). This effort is
taking the following directions: a)~attempting to measure the
ionization signal in the gas phase through electron multiplication;
b)~developing new, low mass photon detectors for VUV light, other than
photodiodes, capable of substituting expensive photomultiplier tubes
(silicon photodiodes, discussed in the previous section, cannot be
considered as a cost effective alternative, at the time of writing at
least); c)~combining a micro-pattern device with an array of SiPMs (or
other photon detectors) to enhance the secondary scintillation signal
(in double- or single-phase configuration); and d)~search for single
phase solutions whilst retaining the concept of measuring two signals
for each interaction.

Most existing and planned experiments aiming at WIMP search with
liquid noble gas detectors rely on detection of the two signals ---
primary scintillation and ionization charge. The ionization signal
can, in principle, be measured in several ways, with secondary
scintillation in the gas phase being one but not the only option. An
alternative can be amplification of the ionization charge extracted
from the liquid through the avalanche process. In this respect, much
of the attention is turned to micro-pattern avalanche detectors of
various kinds, such as GEM and GEM-like detectors and Micromegas (GEM
stands for Gaseous Electron Multiplier and Micromegas for MICROMEsh
GAseous Structure). One should note that efficient detection of a
single electron extracted from a nuclear recoil track is very
desirable for sensitivity to a light WIMP scenario and is probably
indispensable for detection of coherent neutrino scattering. Existing
double-phase xenon DM detectors have already proven the capability to
measure single electrons emitted from the liquid
~\cite{Edwards08,Santos11}, with very good signal-to-noise ratio (see
Figure~\ref{fig:SEinZ3}),
and this sets the bar to judge any alternatives.

Two main problems are associated with operation of an avalanche device
in a pure double-phase medium: strong VUV photon emission results in
positive feedback and thus leads to early onset of discharging; and
high gas density, which affects the maximum achievable multiplication
gain. In the gaseous avalanche detectors the first problem is usually
mitigated by adding a small concentration of quenching molecules (at
percent level), which suppress VUV emission through excitation
relaxation (also through the Penning effect in some cases). However,
this does not appear viable in the double-phase detectors (at least
from our present understanding) due to the strong purity concern and
constraints on the quencher choice imposed by low temperature. As for
the second concern, the lowest saturated gas density achievable in a
liquid/gas system (at the triple point) is higher than that at NTP
conditions by a factor of 1.5 for xenon and 2.5 for argon. At more
convenient vapor pressures, say 1.5~bar, the equivalent gas densities
correspond to 2.7~bar and 5.1~bar for Xe and Ar at room temperature,
respectively~\cite{NIST}. Solid/gas detectors would allow operation at
much lower equilibrium pressures, but more significant challenges
arise in growing crystals with the required quality in very large
detectors.

The GEM structure and details on its design and operation are
thoroughly described in the literature, e.g.~\cite{Sauli97}.
Essentially, it is a two-electrode system formed by two copper layers
on either side of a thin ($\sim$50~$\mu$m) dielectric foil with
thousands of fine through-holes across the whole structure
($\sim$50~$\mu$m~diameter, at $\sim$150~$\mu$m~pitch). A significant
advantage of this approach is the possibility of stacking several
devices to achieve higher gains.

\begin{figure}[ht]
\centerline{\includegraphics[width=.6\textwidth]{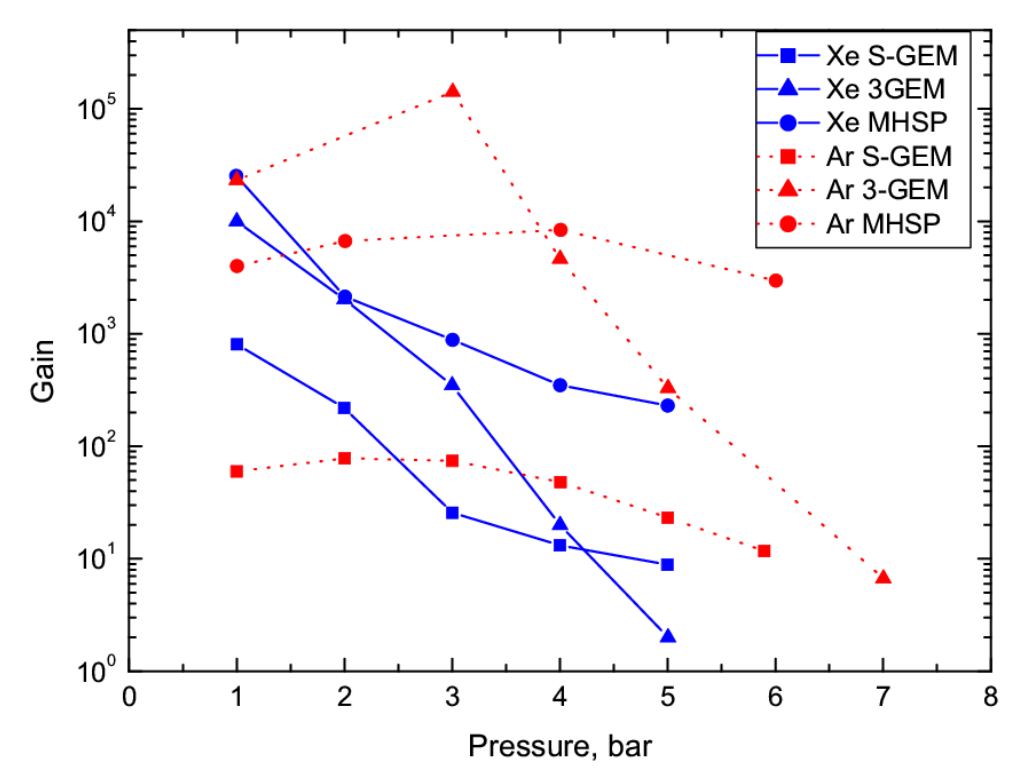}}
\caption{Maximum effective gain (i.e.~measured at an external
  collector) as a function of gas pressure for a single GEM
  (S-GEM)~\cite{Amaro07}, a cascade of 3 GEMs (3-GEM)~\cite{Bondar02}
  and a micro-hole strip plate (MHSP)~\cite{Amaro06}. (Adapted
  from~\cite{Amaro07}; with permission from Elsevier.)}
\vspace{0.2 cm}
\label{fig:GEMGains}
\end{figure}

The maximum attainable gain in pure noble gases is, in general, lower
than in gases with quenchers~\cite{Buzulutskov00} and decreases
rapidly with the gas pressure ~\cite{Bondar02,Amaro07} (see
Figure~\ref{fig:GEMGains}).
At several bar the advantage of using a triple-GEM
structure against a single GEM becomes less important as the
difference between the maximum attainable gains becomes smaller. For
xenon above 4~bar the single GEM configuration is even better than the
cascade of three GEMs (although gains are very low). The effect was
explained by the authors in terms of a more significant contribution
from the positive ion backflow to the secondary electron emission in a
cascaded structure~\cite{Buzulutskov02}. Better performance at high
pressures was found by combining a GEM with microstrip electrodes
etched on its back surface, so that the electron multiplication occurs
in two steps --- first in the GEM holes and then near the anode
strips~\cite{Amaro06}.

\begin{figure}[ht]
\centerline{\includegraphics[width=\textwidth]{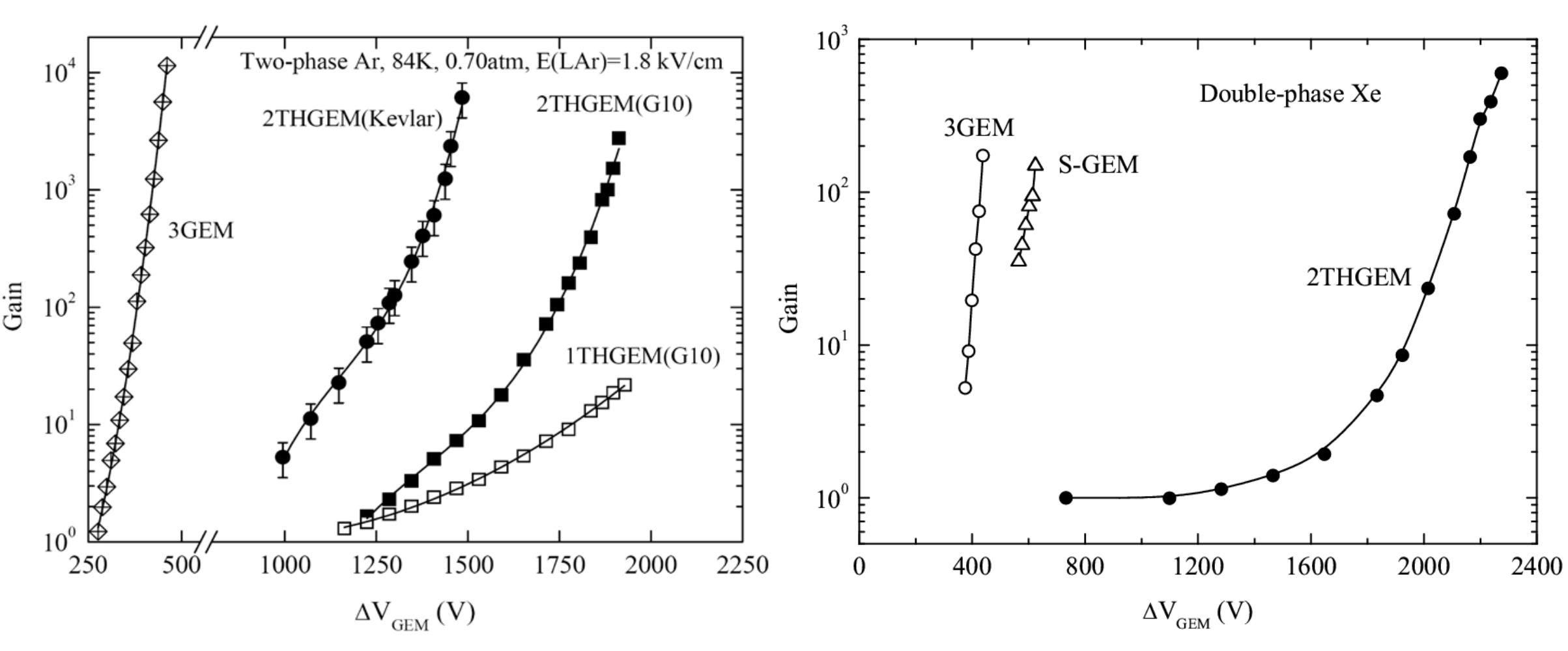}}
\caption{Gain as a function of voltage for triple-GEM and several
  configurations of THGEM. Left: in double-phase Ar
  (from~\cite{Bondar08}; with permission from IOP Publishing). Right:
  in double-phase Xe ($T=164$~K, $P=1$~bar); data for single GEM
  (S-GEM) from~\cite{Balau09}; for triple-GEM (3GEM)
  from~\cite{Bondar06} and for a double-THGEM system (2THGEM)
  from~\cite{Bondar11} (with permission from IOP Publishing).}
\vspace{0.2 cm}  
\label{fig:GEMsInTwoPhase}
\end{figure}

Operation of a triple-GEM combination in the gas phase of a
double-phase detector was studied
in~\cite{Bondar06,Bondar07,Bondar09a} (see also references therein and
a recent review article~\cite{Buzulutskov12}). It was found that
significantly higher maximum gain can be achieved in argon than in
xenon: $G_{max}\approx$3000 as opposed to $G_{max}\approx$200,
respectively~\cite{Bondar06}. Subsequently, the same authors achieved
gains of up to 10$^4$ in argon with an improved
setup~\cite{Bondar09a}. 
Figure~\ref{fig:GEMsInTwoPhase}
shows data for both media. 
A similar maximum gain, $G_{max}\approx$150, was measured with a
single GEM in high purity xenon ~\cite{Balau09} at the same gas
density as in~\cite{Bondar06} (at room temperature, the same density
would occur at $P_{eq}\approx$1.87~bar). It was also observed that the
liquid temperature had a significant effect on gain --- higher gains
could be achieved at lower temperature~\cite{Solovov07,Balau09}, which
is consistent with the vapor density considerations. An increase from
165~K to 171~K (from $P_{eq}\approx 1.87$~bar to $P_{eq}\approx
2.6$~bar) resulted in the decrease of $G_{max}$ from 150 to about
25. This highlights the importance of good temperature stabilization
both for avoiding discharges and for gain stability. It is worth
noting that the temperature stability requirement for an avalanche
device is more severe than in the case of secondary scintillation
given that secondary photon production is linear with $E/n$ ($n$ is
the number density) while the avalanche process depends on it
exponentially.

\begin{figure}[ht]
\vspace{3mm}
\centerline{\includegraphics[width=.6\textwidth]{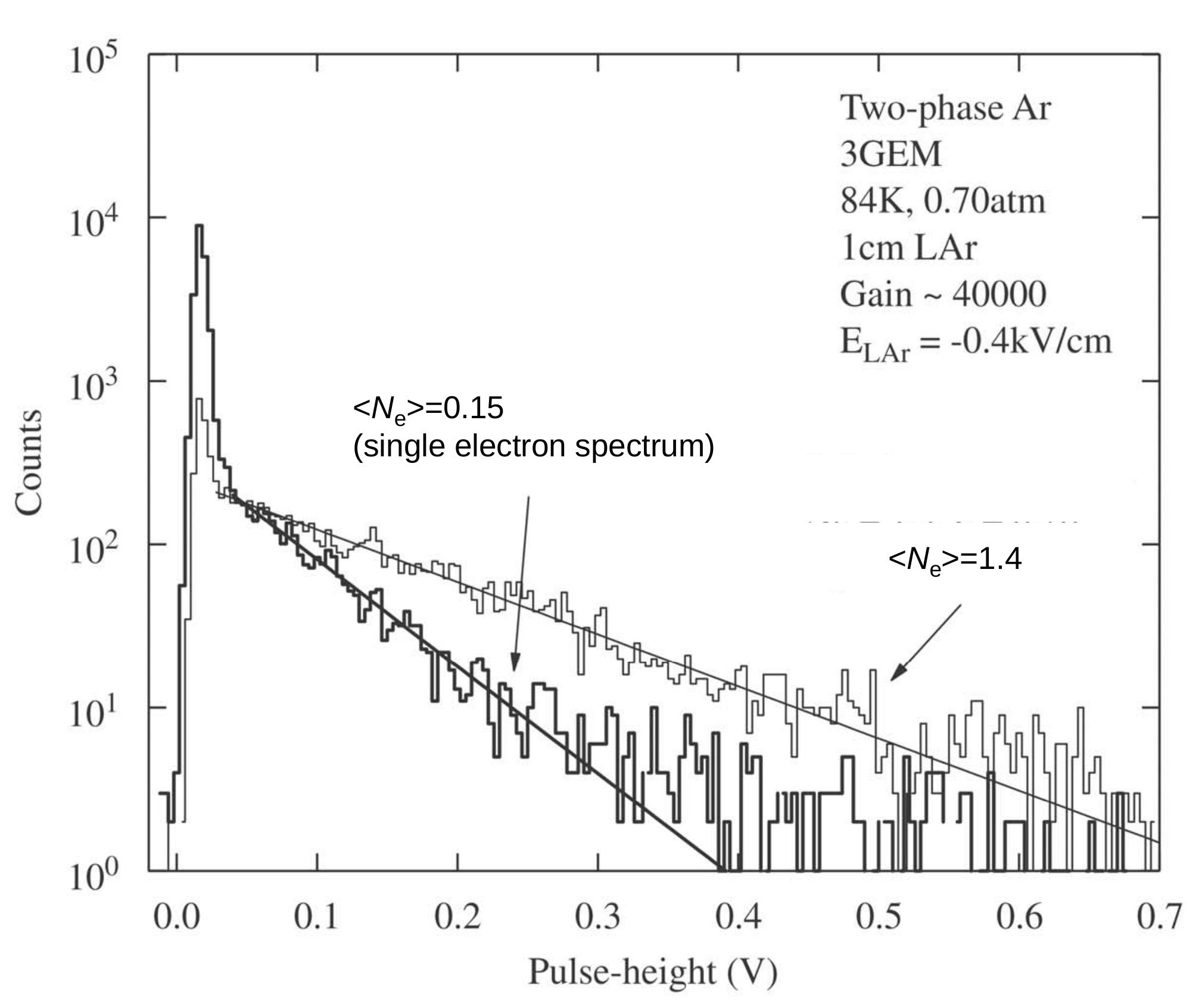}}
\caption{Pulse height spectra measured with triple-GEM in double-phase
  Ar in the few electron regime. The liquid under a reversed field
  ($E_\mathrm{LAr} = 0.4$~kV/cm) was irradiated with low intensity
  X-rays so that electrons at the input of the multiplication system
  are produced by photoelectric effect on the copper electrode of the
  lower GEM by the LAr scintillation. The authors estimate that, for
  each X-ray absorbed in LAr, $\sim$0.1 photoelectrons are produced on
  average. By adjusting the intensity of the X-ray tube they were able
  to vary the mean number of primary electrons at the entrance of the
  triple-GEM system ($\langle N_e \rangle$ in the plot). The gain is
  4$\cdot$10$^4$ in these measurements. (From~\cite{Bondar07}; with
  permission from Elsevier.)}
\vspace{3mm}
\label{fig:3GEMsInTwoPhaseAr}
\end{figure}

The maximum gain achieved to date with these structures operating in
double-phase xenon is far from sufficient for detection of single
electrons extracted from the liquid. On the other hand, one would
expect this to be possible in argon, where gains of $\sim$10$^4$ have
been realized. Indeed, single electron signals have been observed with
a triple-GEM structure~\cite{Bondar07}, although with an exponential
spectrum as expected from the statistics of the avalanche process, as
illustrated in 
Figure~\ref{fig:3GEMsInTwoPhaseAr}.
The authors estimate the probability to
detect a single electron as $\sim$50\%.

More robust `macroscopic' versions of GEM, proposed more
recently~\cite{Kim00,Periale02,Chechik04}, are similar to GEM with all
dimensions scaled up by a factor of 5 to 20. This device, known as LEM
(Large Electron Multiplier) or THGEM (THick GEM), can be produced both
by etching techniques and by purely mechanical means using precise CNC
machining. Other technologies are also being explored. These devices
operate at higher voltages than GEMs but are more resistant to
discharges. The LEM/THGEM structures can benefit from a better
confinement of the electron avalanche if channels with a smaller
aspect ratio are used. This reduces photon feedback and, therefore,
the probability of photon-induced discharges to occur. The extent of
the multiplication region is larger than in GEMs, which is also an
advantage because the same gain can be obtained at a lower field. We
refer to review articles~\cite{Shalem06a,Shalem06b,Breskin09} and
references therein for more detailed information on the operation and
present status of these devices.

LEM/THGEM structures have been extensively tested in various gas
mixtures achieving gains of up to 10$^5$ and even 10$^6$ at low
pressure~\cite{Chechik04}. They were also proven to operate in pure
noble gases at room temperature and pressures of a few
bar~\cite{Alon08}. In normal conditions, maximum gains of
$\sim$4$\cdot$10$^4$ were achieved with a double-THGEM system in
purified Ar, Xe and Ar-Xe (95\%+5\%) mixtures. In xenon, the
amplification gain was also measured as a function of pressure: as the
pressure increased the maximum gain dropped gradually down to a few
hundred for $P=2.9$~bar. For a single THGEM, the respective gain was
generally a factor of 2 to 4 lower than for a chain of two
devices. Much weaker dependence on pressure was found for the Ar+Xe
Penning mixture: similar maximum gains of $\sim$2$\cdot$10$^4$ are
reported both for 1 bar and for 2 bar.

Several types of THGEM were tested in double-phase argon at 84~K and
$P=0.7$~bar and compared with a triple-GEM system~\cite{Bondar08}:
THGEM with G10 as insulator, Kevlar-based THGEM and RETGEM (see next
paragraph). It was observed that only the G10-based THGEM showed
stable operation with a maximum gain of 200 for a single device and
$\sim$3$\cdot$10$^3$ for a double-THGEM system (compare with 10$^4$
for triple-GEM). About 18\% {\it r.m.s.}~energy resolution was
measured for 60~keV $\gamma$-rays ($\sim$1000 primary electrons in
LAr) with the triple-GEM and the double G10-based THGEM
assemblies. The two systems were shown to operate at low detection
threshold of 4 and 20 primary electrons for the triple-GEM and
double-THGEM, respectively, the spectra being exponential. A strong
charging effect was observed for the Kevlar-based THGEM both at room
temperature and in double-phase argon. The noise rates of GEM and
THGEM multipliers have also been measured in double-phase
argon~\cite{Bondar08}: for a threshold corresponding to 4 primary
electrons the noise rate was found to be $\sim$0.2~counts/s per cm$^2$
of detector active area; a value of 0.007~counts/(s$\cdot$cm$^2$) was
reported for a 20-electron threshold.

In a larger scale system, operation of a 10$\times$10~cm$^2$
double-LEM structure in pure Ar gas and in the double-phase regime has
been reported~\cite{Badertscher08}. The LEMs, placed 3~mm apart, were
1.6~mm thick with 0.5~mm diameter holes at 0.8~mm pitch. They were
operated with an electric field inside the holes of $\sim$25~kV/cm. A
gain of $\sim$10$^3$ in the gas at normal temperature and pressure was
reached, while in the double-phase system the LEMs were operated at a
gain of 10~\cite{Badertscher10}. It was found that segmentation of the
LEM upper electrode to obtain avalanche position significantly
distorted the electric field thus leading to premature
discharges. More recently, the same group reported on operation of a
single unsegmented LEM in double-phase argon at an effective gain of
30 (measured at a separate electrode placed behind the LEM to which
electrons are collected)~\cite{Badertscher11}. The collector (termed a
`projective readout anode') had two sets of orthogonal strips, which
allowed determination of the avalanche position. The LEM and the 2D
readout system were placed above a 30~cm thick LAr layer. At the time
of writing, no record of higher gains in a medium- or large-scale
double-phase argon system, in which electron drift across long
distances is required, could be found in the literature.

In order to increase the resilience of micro-hole structures to
discharges and thus increase the maximum achievable gain, the use of
resistive electrodes instead of metallic ones has been
proposed~\cite{Peskov07}. Thanks to the low conductivity of the
resistive electrodes, a spark leads to a rapid drop of local
voltage. The discharge current is thus limited and the ability to
produce damage is significantly reduced. Thick GEMs with resistive
electrodes received the abbreviated name RETGEM (Resistive Electrode
Thick GEM). Several versions have been developed and shown to operate
in pure argon and neon at gains of $\sim$6$\cdot$10$^3$ (single plate)
or $\sim$10$^5$ in a cascade of two in Ar, and up to
$\sim$7$\cdot$10$^5$ with a double RETGEM in Ne~\cite{Peskov07}. The
performance of 1~mm thick RETGEM with very thin copper electrodes
coated with a CuO resistive layer has also been studied in argon at
low temperature. The gain was observed to decrease gradually from
$\sim$6$\cdot$10$^3$ at room temperature to $\sim$10$^3$ at 100~K
($P=1$~bar in both cases) and to $\sim$400 in saturated gas at 89~K
($\sim$2$\cdot$10$^3$ in a cascade of two). On the other hand, the
authors of~\cite{Bondar08} could not observe any multiplication with a
\mbox{RETGEM} in a double-phase argon chamber, and suggested argon
condensation in the holes and increased resistance of the resistive
layers at low temperature as possible reasons. In neon, a gain of
$\sim$3$\cdot$10$^4$ was achieved at room temperature with a single
RETGEM with metal strips for position readout; the gain decreased only
slightly when the temperature was lowered to 78 K~\cite{Martinengo09}.

Charge multiplication has been observed also using Micromegas with
50~$\mu$m gap in the gas phase of a double-phase xenon chamber filled
with a Xe+2\%CH$_4$ mixture~\cite{Lightfoot05}. Micromegas stands for
MICROMEsh GAseous Structure~\cite{Giomataris96} and consists of two
parallel electrodes --- a micromesh cathode and an anode plate placed
at a short distance from each other --- so that the electron
multiplication takes place in a very strong quasi-uniform field
between the electrodes. A gain of about 500 was reported for the
double-phase Xe+2\%CH$_4$ system~\cite{Lightfoot05} albeit only for
short periods of time --- the amplification disappeared after
10~to~30~min of operation, a fact attributed by the authors to
condensation of xenon in the amplification gap. Xenon condensation in
the region of strong electric field (some 140~kV/cm in that
experiment) is favored by the high polarizability of xenon atoms.

Gas condensation in the holes of avalanche devices has been reported
as a likely reason for temporary failure of GEMs~\cite{Solovov07},
LEMs~\cite{Badertscher08} and RETGEMs~\cite{Bondar08}, for
double-phase argon as well as xenon. In some instances this could be
reversed with thermal cycling. Contrary to what was observed with
Micromegas, this seemed to occur mostly at the initial stage of the
chamber filling and could disappear after a couple of hours of
temperature stabilization. A carefully chosen temperature gradient in
the chamber may help, but a reliable recipe to avert this effect has
not been developed yet.

Generally, one can conclude that the avalanche micro-pattern devices
considered here work better in argon than in xenon and better still in
neon than in argon. This is not surprising given that the Townsend
coefficient in xenon is the lowest among the three gases due to the
higher ratio of the number of excitations to the number of ionizations
per collision~\cite{Kruithof40}. Another aspect to mention is the much
larger field required for electron extraction from liquid xenon than
from liquid argon~\cite{Gushchin79,Gushchin82b}. High electric field
in front of a GEM or similar structure placed above the liquid surface
reduces the efficiency of electron collection into the holes and
therefore the observable amplification gain is also lower.

More information can be found in a recently published comprehensive
review on operation of micro-pattern avalanche detectors at cryogenic
temperatures~\cite{Buzulutskov12}.

\subsubsection{Photon detection with micro-pattern structures}
\label{sec:PhotonDetectionWithMicropattern}

To find a feasible alternative to existing WIMP detector designs,
which use photomultiplier tubes to detect primary scintillation along
with the ionization signal via secondary scintillation in gas, one
must find a reliable way to detect the prompt VUV light emitted from
the liquid. Considerable effort has been put into trying to couple a
photocathode to an electron multiplication system, which can be
operated in gas (ideally, in the liquid too). This concept is usually
referred to in the literature as the `gaseous photomultiplier' (see,
for example, review papers~\cite{ChechikBreskin08,Buzulutskov08}).

The idea of such coupling is intrinsically a contradictory one since a
high photoelectric conversion coefficient, required for efficient
detection of incoming photons, would also mean high probability of
positive feedback due to photons emitted in the avalanche. One can
imagine several solutions to this problem. One is to suppress photon
emission during the avalanche development by a suitable choice of gas
composition and/or by shifting the wavelength of the unwanted photons
to the wavelength region where the sensitivity of the photocathode is
low (CsI, for example, has a sensitivity cutoff for $\lambda >
200$~nm). In the case of double-phase detectors this, however, would
mean separation of the avalanche region from the principal volume of
the detector with a sealed VUV-transparent
window~\cite{Periale07,Duval11a}. Although not impossible technically,
it remains to be established that this approach can outperform
traditional photomultiplier tubes. We shall discuss recent efforts in
this direction at the end of this section.

On the other hand, windowless solutions with both the avalanche device
and the photocathode operated in pure noble gas would suffer from
strong photon emission during the avalanche process, occurring at the
same wavelength to which the photocathode is supposed to be most
sensitive. In this situation, feedback might be reduced by preventing
illumination of the photocathode by the avalanche photons by way of a
careful geometrical arrangement. In other words, one requires the
photocathode to be `visible' for photons coming from the detector
volume but not for those generated in the avalanche. Some micro-hole
structures seem well suited for this purpose: the avalanches are
mostly confined within the holes while the photocathode can be
deposited on the surface of the structure turned down to the detector
volume, thus being geometrically shielded from the direct avalanche
photons. However, there is still the matter of reflected photons,
which becomes more problematic as one tries to enhance the VUV
reflectivity of all inner surfaces of the detector to improve the
light collection efficiency in the chamber.

Concerning the choice of photocathode material, work has focused
mostly on cesium iodide due to its sensitivity in the VUV wavelength
region but also for technological reasons: the relative easiness of
deposition of CsI on metallic and glass surfaces and its capability to
recover after exposure to air. Much work has been done since the first
successful experiments~\cite{Seguinot90,Anderson92} and the
literature on CsI is quite extensive~\cite{Breskin96}. The most
immediate application of these devices is in Cherenkov imaging
detectors~\cite{Piuz03}. Cesium iodide has been proven to operate in
several gas mixtures with quenchers as well as in pure noble gases and
liquids~\cite{Seguinot90,Breskin96,Buzulutskov00,
  Aprile94}. Practically all of the above mentioned microstructures
have been experimented in combination with a CsI photocathode (see,
for example,~\cite{ChechikBreskin08} and references therein).

\begin{figure}[ht]
\vspace{5mm}
\centerline{\includegraphics[width=.6\textwidth]{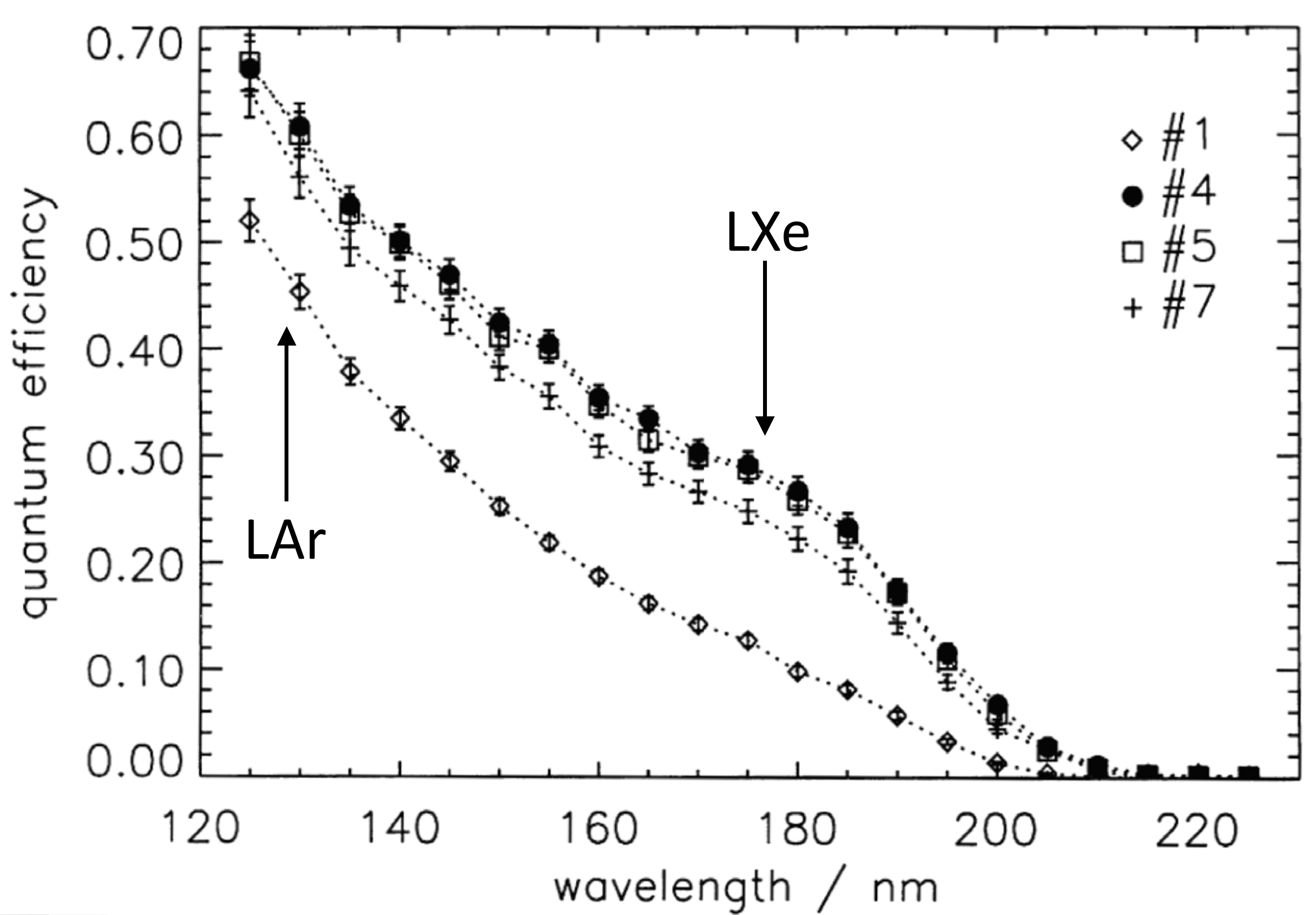}}
\caption{Quantum efficiency of reflective CsI photocathodes in vacuum
  as a function of the photon wavelength for four different
  samples. The arrows indicate the emission wavelengths for liquid
  argon and xenon. (Adapted from~\cite{Rabus99}; with permission from
  Elsevier.)}
\label{fig:CsIPhotocathode}
\end{figure}

The quantum efficiency of reflective CsI photocathodes in vacuum has
been measured to be up to 30\% for xenon light and up to 60\% for VUV
from argon, depending on the sample preparation~\cite{Rabus99} (see
Figure~\ref{fig:CsIPhotocathode}).
The photocurrent in a uniform electric field, measured
with an electrometer, was found to saturate at a field of about
1.7~kV/cm. A lower quantum efficiency of 15\% has also been reported
for 175~nm~\cite{Azmoun09}.

When the photocathode is operated in gas, photoelectrons can suffer
backscattering off the gas atoms and return to the
photocathode~\cite{Breskin95,Coelho07,Covita11}. Frequent collisions
with gas atoms can also result in insufficient energy gained from the
electric field between collisions to overcome the image potential of
the photocathode. Both factors lead to reduction of the photoelectron
extraction efficiency. Backscattering is more significant in noble
gases than in molecular gases. Its probability increases with the gas
density and the energy of incident photons. The latter is because the
elastic scattering cross section increases with electron energy in the
eV region~\cite{Theobald53,Peskov77,DiMauro96}. At atmospheric
pressure, the photoelectron collection efficiency in xenon was found
to be about half of that in argon and neon at fields of a few
kV/cm. For example, for $E=1$~kV/cm it is about 0.47 in Ar, 0.4 in Ne
and Kr, 0.25 in Xe and 0.22 in He~\cite{Coelho07} (measured as
gas/vacuum photocurrent ratio for a given illumination intensity and,
therefore, includes possible difference in quantum
efficiency). Measurements of quantum efficiency of semi-transparent
CsI photocathodes in vacuum have also shown some dependence on the
temperature. For 165~nm VUV light it changed from 33\% at room
temperature to 29\% at 130~K and to 26\% at 77~K~\cite{Periale04}.

In~\cite{Periale02}, several micro-hole systems, including LEMs, were
experimented within pure argon, krypton and xenon with CsI
photocathodes. A gain of 10$^4$ was obtained with LEM without a
photocathode (apparently in pure Xe, but the authors are not explicit
on this point), but with a CsI coating the maximum gain decreased by
an order of magnitude, which can be explained by ion feedback. VUV
photons from xenon scintillation were detected with $\sim$1\%
efficiency (including quantum efficiency of the photocathode, the
probability for a photoelectron to escape and to produce an avalanche
in the LEM hole). A better efficiency between 2.5\% and 8\% (for 130~K
and 293~K, respectively) was measured for 165~nm with CsI deposited on
top of a GEM and operated in pure Ar~\cite{Periale04}. An even higher
value of 15\% was reported by the same authors for a photocathode
deposited on a GEM operated in Ar+10\%CH$_4$ gas mixture at 1~bar and
at room temperature~\cite{Periale05}. At 170~K the efficiency
decreased to 11\%, the effect being attributed to back-diffusion of
the photoelectrons. The maximum achievable gain was found to be
practically independent on temperature, being about 500 for a double
GEM with CsI (when CsI was deposited on the surface of a capillary
plate, a gain of $\sim$10$^3$ was measured).

A triple-GEM structure with a CsI photocathode deposited on the bottom
surface of the first GEM was tested in double-phase argon at 84~K and
$P=0.7$~bar~\cite{Bondar07b, Bondar09a}. Primary scintillation from
60~keV $\gamma$-rays was detected at a gain of 1.4$\cdot$10$^3$,
however with surprisingly low efficiency (about 2~photoelectrons at
the entrance of the first GEM).

RETGEMs also were tested with CsI~\cite{Peskov07}, including a version
with segmented electrodes \mbox{S-RETGEM} (standing for
strip-RETGEM)~\cite{Peskov08}. In argon at room temperature and
$P=1$~bar, the \mbox{S-RETGEM} was operated with gains up to
3$\cdot$10$^3$. By reversing the drift field, to avoid the ionization
charge to drift to the multiplication structure and observe only
scintillation signals, a higher gain of up to 10$^4$ could be reached
with a single plate. Much higher gains were obtained with neon:
$\sim$2$\cdot$10$^4$ with a single GEM, $\sim$10$^5$ with single GEM
and reversed drift field, $\sim$3$\cdot$10$^5$ with double GEM and
reversed field (obviously, the reversed field option cannot be used if
one intends to detect the ionization signal). A detection efficiency
for 185~nm photons of $\sim$13\% was reported for Ar, Ne and Ar+CO$_2$
mixture.

We refer to the review article~\cite{Buzulutskov12} for more
information on these devices.

\vspace{2mm}

The development of sealed gaseous photomultipliers has also
progressed, using reflective or semi-transparent CsI (for VUV) or
bialkali photocathodes (for VUV or visible light) ---
see~\cite{Chechik03} and references therein. Recently, a 500 day run
with stable operation of a sealed device with bialkali photocathode
and a Pirex glass GEM was reported~\cite{Sumiyoshi11} (the authors
also refer to this glass GEM as a `micro blasting glass plate' ---
MB-GP). A Kapton-based GEM was found incompatible with bialkali
photocathodes due to the high absorption of alkali vapors by Kapton
during the photocathode activation. Gains $\sim$10$^5$ have been
reached with a double MB GP multiplication structure in
Ne/isobutane~(10\%) and Ne/CF$_4$~(10\%) mixtures without significant
ion or photon feedback. A quantum efficiency of 14\% was reported for
Ne, versus 20\% in vacuum at 350~nm. On the other hand, for a
semi-transparent CsI photocathode deposited onto a quartz window, and
the same multiplication structure and gas mixtures, the same group
reported similar gains but only $\sim$0.5\% quantum efficiency at
170~nm light~\cite{Tokanai11}. All the above results were obtained at
room temperature.

Detection of liquid xenon scintillation due to $\alpha$-particles with
a gaseous photomultiplier was reported in~\cite{Duval11a}. The
photomultiplier was coupled to a liquid xenon chamber through a
MgF$_2$ window and operated at 173~K under continuous flow of He or Ne
based gas mixtures at 1.1~bar; CH$_4$ and CF$_4$ were used as
admixtures to avoid condensation at low temperature. A reflective CsI
photocathode was deposited on one side of a THGEM. Two multiplication
structures were experimented with: double THGEM and THGEM followed by
a Parallel Ionization Multiplier (PIM) and a Micromegas. Gains up to
$\sim$10$^4$ were attained. For both structures, instabilities were
observed at low temperature, rather significant in the case of the
second configuration. The authors do not provide enough information to
allow the number of photoelectrons extracted from the photocathode to
be estimated, but refer that it was significantly lower than expected
assuming 22\% extraction efficiency. As a possible reason,
photocathode degradation due to condensation of impurities is
mentioned.

\subsubsection{Optical readout from micro-pattern structures}
\label{sec:OpticalReadoutWithMicropattern}

Given the difficulties involved in operating micro-pattern detectors
in the avalanche mode, attention has been recently turned to their
operation at lower voltages, insufficient for developing significant
avalanche but high enough to produce secondary scintillation in the
detector holes. The general idea is to obtain a higher light gain
(defined usually as the total number of photons per primary electron
extracted from the liquid) than in a uniform field as presently done
in the current double-phase DM detectors and to use highly segmented
solid-state photon detectors to enable efficient multi-vertex
resolution (especially important for future CNS experiments).

Production of the secondary light within the THGEM holes in
double-phase argon system was studied in~\cite{Lightfoot09}. A
1~mm$^2$ SiPM was positioned directly above the center of a 65~mm
diameter THGEM, aligned with one of the holes, 5~mm above it. In order
to render it sensitive to the argon VUV light, a special gel
containing TPB wavelength shifter, re-emitting the absorbed VUV light
at 460~nm, was used. The light yield per primary electron increased
exponentially with the voltage across the THGEM up to charge gain of
$\sim$100, above which a steeper-than-exponential rise was observed
(although the charge gain continued on an exponential trend)
accompanied by degradation of the energy resolution. According to
authors' estimates, $\sim$600 photons per primary electron were
generated in the THGEM holes at a gain of $\sim$100. Given the low
$\gamma$-ray energy used to ionize the liquid (5.9~keV from
$^{55}$Fe), charge gain determination at low gain values was
difficult. We cannot, therefore, estimate the light yield in the
absence of charge multiplication.

The authors of~\cite{Bondar10} studied avalanche
scintillation\footnote{The term `avalanche scintillation' is employed
  by the authors in order to distinguish the light emitted in the
  avalanche from that produced at lower fields below the charge
  multiplication threshold, commonly called secondary or proportional
  scintillation.} in double-phase argon using a double THGEM
multiplication structure and a 4.4~mm$^2$ SiPM (termed by the
manufacturer G-APD --- `Geiger APD') placed behind a glass window at a
distance of 4~mm from the top THGEM. The G-APD was sensitive only in
the visible and near infrared regions up to 950~nm with an average
quantum efficiency of 15\%. The chamber was irradiated with 60~keV
$\gamma$-rays from $^{241}$Am. For a charge multiplication gain of 400
(total gain for the two THGEMs), about 0.7~photoelectrons per primary
electron was measured. For a gain of 60 this value was 0.14,
indicating that the charge gain increased faster with voltage than the
light gain. From these data, one can estimate that each electron
extracted from the liquid resulted in $\sim$1800 photons emitted in
4$\pi$ if the system is operated at gain of 400 and $\sim$350 photons
for a gain of 60. Unfortunately, the authors are not explicit on the
charge calibration procedure that may result in by a factor of about
3~lower yields per primary electron taking into account the known
$W$-value for liquid argon. Even so, the avalanche light yield is very
high, comparable to that observed in~\cite{Lightfoot09} with a
wavelength shifter.

Using the same setup, avalanche scintillation in xenon was also
studied~\cite{Bondar11b}. The THGEMs were operated in gaseous Xe at
$T=$200~K ($-73^\circ$C) and $P=$0.73~bar (i.e.~not in saturated
vapor) at charge gains of up~to~600. The light yield in the NIR region
was found to be significantly lower than in argon. Thus, for a charge
gain of 350, each primary electron resulted in 0.071~photoelectrons in
the~SiPM (compare with 0.7~for~Ar). Taking into account all
inefficiencies, a value of $\sim$250~NIR~photons/primary~electron
emitted in 4$\pi$ can be estimated.

Secondary light from a GEM in double-phase xenon has been observed
with a LAAPD sensitive to xenon light~\cite{Solovov10}. The GEM
foil was placed 3~mm above the liquid surface and operated at vapor
pressure of 1.4~bar. The voltage across the GEM (400~V) was well below
the multiplication threshold (see 
Figure~\ref{fig:GEMsInTwoPhase},
right panel~\cite{Balau09}). Normalized to 4$\pi$, the number of secondary
photons produced in the GEM channels was a factor of $\sim$2 higher
than in the uniform field between the liquid surface and the GEM. The
authors estimate that number to be about 200~VUV photons per primary
electron~\cite{SolovovPrivate}.

An array of 19~G-APDs, 2$\times$2~mm$^2$ each, was used to detect
secondary light from a single THGEM and also from a stack of 2 THGEMS
in double-phase Xe~\cite{Akimov11d}. As the G-APDs are not
sensitive in the VUV region a 140~nm layer of $p$-terphenyl wavelength
shifter was deposited on a sapphire window placed in front of them. To
avoid contamination of xenon, the $p$-terphenyl layer was either
enclosed in a vacuum tight cell between two sapphire windows and
filled with Ar or a 1~$\mu$m poly-para-xylylene protective layer was
deposited on top of it. An electron lifetime in liquid xenon of
$\sim$10~$\mu$s was reported. The measured PDE agreed rather well with
that expected from the device specification ($\sim$15\% for 370~nm)
assuming 100\% re-emission efficiency for \mbox{$p$-terphenyl}. At the
voltages used in this work, a comparable (within a factor of 1.5)
number of secondary photons were produced in the THGEM channels at
1.6~kV and in the uniform field between the liquid surface and the
first THGEM (5~mm drift distance; 4.9~kV/cm). For a more compact 
\mbox{G-APD} packing the authors anticipate a total response of $\sim$10
photoelectrons per primary electron. Instabilities on THGEMs were also
reported.

The authors of~\cite{Akimov12d} are exploring new techniques for
rendering wavelength shifters harmless to the purity of the xenon:
besides the protective layer over the $p$-terphenyl, as described
above, a new $p$-terphenyl-based nanostructured organosilicon
luminophore is also being examined: besides conversion of the xenon
emission to longer wavelengths ($\approx$370~nm instead of
$\approx$330~nm for pure $p$-terphenyl), this would also be less
volatile and could perhaps be used in a clean environment without a
protective layer.

\subsubsection{Back to single phase?}
\label{sec:BackToSinglePhase}

The double-phase technology has already been amply demonstrated for
detection of low energy signals. However, design constraints due to
the need to keep the liquid surface quiet and level, pressure and
temperature stable and uniform, and other technical challenges,
inspire some thinking about single-phase
alternatives~\cite{Majewski06,Giboni11}, which may be cheaper to
operate in the case of very large detectors. The solution must
evidently pass through realization of an amplification process in the
condensed medium. Charge multiplication and secondary scintillation
have been observed in liquid xenon and argon long ago, but the
achievable gains are far from sufficient and the fields required are
extremely high, $\sim$10$^5$~to~10$^6$~V/cm.

The first observation of electron multiplication in condensed noble
gases belongs to Hutchinson in~1948~\cite{Hutchinson48}. He observed
multiplication in solid argon with a gain of $>$12 but it was
transient and disappeared in a short time --- the effect was
attributed to polarization of the solid. No multiplication was found
in the liquid with anode wires as thin as 15~$\mu$m. Later, using much
thinner wires of a few $\mu$m diameter, multiplication in liquid xenon
was observed with gains of~$\sim$10~to~
$\sim$100~\cite{Muller71,Derenzo74,Prunier73,Miyajima76,Miyajima79}. An
attempt to build a multiwire proportional chamber for medical
$\gamma$-ray imaging has been
undertaken~\cite{Zaklad72,Zaklad73}. However, it was found impractical
because of low gain but also due to wire fragility, being unable to
withstand the electrostatic repulsion between individual wires in the
multiwire anode arrangement.

Attempts to obtain stable electron multiplication in liquid argon were
unsuccessful~\cite{Derenzo70}. Although the avalanche did develop in
LAr, it was unstable and not reproducible. In addition, correlation
between the signal amplitude and the deposited energy was not obvious
--- an observation later confirmed by other authors~\cite{Kim02}. This
group used a single sharp needle as an anode to study electron
avalanches in LAr. In pure argon, the avalanche behavior was erratic,
in part due to strong optical feedback owing to the short wavelength
of avalanche photons. This is consistent with the fact that the
avalanche could be stabilized by adding a small amount of xenon; a
gain of~$\sim$100 was reported for this mixture. In tension with these
results, a gain in excess of 100 was reported for pure argon
in~\cite{Bressi91}, using an array of microtips with bending radius
of~$\sim$~0.25~$\mu$m.

Several micro-pattern detectors have also been tested in pursuit of
electron multiplication. A gain of $\approx$15 was reported for
microstrips immersed in liquid
xenon~\cite{Policarpo95,Policarpo97}. Attempts to obtain
multiplication with field emission arrays (Spindt cathodes) were
unsuccessful in argon~\cite{Kim03}, and this was also confirmed in
xenon by one of us (VC).

Secondary scintillation was first observed in high-pressure liquid
xenon in a uniform electric field~\cite{Dolgoshein67}. With increasing
field strength, the scintillation yield induced by $\alpha$-particles
was found to decrease by about 15\% up to $E \sim 70$~kV/cm, as
expected from the suppression of recombination. At higher fields, a
steady increase of luminescence was observed, reaching the zero-field
value at $E \sim 120$~kV/cm. The measurements were conducted at two
temperatures of the liquid, $-5^\circ$C and $+10^\circ$C,
corresponding to liquid densities of about 2.0 and 1.7~g/cm$^3$,
respectively.

Later, proportional scintillation was observed in the non-uniform
field in the vicinity of very thin
wires~\cite{Lansiart76,Miyajima79,Masuda79,Masuda81}. For example,
with a wire of 4~$\mu$m diameter a light yield of
$\sim$10--100~photons per primary electron was
obtained~\cite{Masuda79}. The highest yield was measured at a charge
gain of $\sim$50. It has also been shown that it is possible to obtain
secondary scintillation with thicker wires: a yield of $\sim$5
photons/electron has been estimated for a wire of 20~$\mu$m
diameter~\cite{Doke82} at an anode voltage of 5~kV; for a 50~$\mu$m
wire, $\sim$30~photons/electron was estimated to be achievable at
12~kV~\cite{Policarpo82}.

The ratio of secondary to primary scintillation in single-phase liquid
xenon was originally proposed to discriminate between nuclear and
electron recoils in a dark matter detector some two decades
ago~\cite{Benetti93b}.

Much less information is available on secondary light in liquid
argon. To our knowledge this was observed only recently, in the
channels of a THGEM immersed into this liquid~\cite{Lightfoot09}. At a
THGEM voltage of about 10~kV, signals of $\sim$100 photoelectrons were
recorded from 5.9~keV \mbox{X-rays} with a 1~mm$^2$~SiPM placed behind a
THGEM hole, at a distance of 5~mm. The VUV photons were converted to
the visible range with a wavelength shifter deposited onto the SiPM. A
light yield of $\sim$300 VUV photons per primary electron has been
estimated (this value is similar to that used in double-phase xenon
systems). The amount of detected light increased linearly with applied
voltage indicating the absence of charge multiplication up to at least
10.2~kV. Poor amplitude spectra, much worse than for the double-phase
regime with the THGEM operated in the gas phase, tarnish these
otherwise encouraging results. The reason for that is not quite clear
yet. Optimization of the THGEM geometry can probably bring about some
improvement.

In spite of several rather significant efforts, for now the only
viable alternative to double-phase systems seems to be to explore
scintillation pulse shape discrimination in liquid systems, at zero
electric field, as described in 
Section~\ref{sec:OverviewDetectionPrinciples}.
However, losing the
ionization response has consequences for the achievable energy
threshold for nuclear recoil detection and for event localization in
three dimensions.

We refer to~\cite{LopesChepel05,AprileBook06} and references therein
for a more complete survey on electron multiplication and secondary
scintillation in the liquid phase.

\newpage

\section{Direct dark matter search experiments}
\label{sec:DirectDMExperiments}

In this section we review briefly some dark matter search projects,
aiming to illustrate how noble liquid technologies are implemented in
real experiments. We focus on those which have published WIMP search
results recently or have operated detectors underground, but some
large systems which we consider significant and are close to
deployment are also described. Other technologies, namely cryogenic
bolometers, room temperature scintillators and bubble/droplet chambers
will not be covered here.

With tonne-scale target masses required to probe the whole range of
scattering cross sections favored by theory, a critical consideration
is how these designs can be scaled up whilst maintaining very low
energy threshold for nuclear recoils. As the system size increases,
self-shielding of an inner, fiducial volume from external backgrounds
becomes increasingly effective, and this is one of the key advantages
of the noble liquids. On the other hand, particle discrimination can
be impaired, for example due to increasing difficulty in applying
strong electric fields, which is another important consideration.

We note that noble liquid detectors, some much larger than those
discussed here, have been applied to other physics measurements ---
for example, the MEG experiment features a 2.2~tonne liquid xenon
scintillation calorimeter to search for the lepton-flavor violating
decay $\mu^+ \!\to\! e^+ \gamma$~\cite{Mihara11}; and ICARUS is a
760-tonne liquid argon time projection chamber (TPC) to study cosmic
rays, neutrino oscillations and search for proton
decay~\cite{Rubbia11}. The energy scale of these processes is in the
MeV region or higher. On the other hand, our goals are the detection
of nuclear recoils with energy up to 100~keV at most and hopefully
achieve their discrimination from electron recoil backgrounds. These
are, therefore, very distinct requirements.

\subsection{Liquid xenon detectors}
\label{sec:LXeDetectors}

At the time of writing, five collaborations have operated liquid xenon
detectors built for dark matter searches. We begin by overviewing
these programs, referencing published instrument descriptions and main
WIMP scattering results where appropriate, before describing the
latest experiments in more detail. Other surveys of application of
this technology to WIMP searches have been
published~\cite{AprileDoke09}.

The DAMA team built and operated the first liquid xenon WIMP detector,
DAMA/LXe, at the Gran Sasso Underground Laboratory (LNGS) in Italy,
from the mid 1990s. This was a scintillation chamber with 6.5~kg of
xenon enriched in $^{129}$Xe, read out by three 3.5-inch
photomultipliers~\cite{Bernabei02}. WIMP elastic scattering results
using pulse shape discrimination were published in
1998~\cite{Bernabei98}. The collaboration then opted to pursue room
temperature sodium iodide detectors (DAMA/NaI).

The ZEPLIN program at the Boulby mine (UK) followed from the late
1990s, with ZEPLIN-I publishing final results in 2005~\cite{Alner05};
it featured at its core a PTFE-lined 5-kg chamber with three 8-cm
diameter photomultipliers viewing the liquid xenon
scintillation. ZEPLIN-II and ZEPLIN-III were the first double-phase
systems to be conceived for dark matter searches, and were essentially
developed in parallel and exploiting different technological
solutions. ZEPLIN-II became the first liquid/gas system to operate in
the world, completing in 2007~\cite{Alner07}; it utilized a deep,
high-reflectance PTFE chamber containing 31~kg of liquid xenon with
readout from seven 13-cm diameter PMTs in the gas phase. ZEPLIN-III
concluded the liquid xenon program at Boulby, with science runs in
2008~\cite{Lebedenko09} and, following an upgrade phase, in
2010/11~\cite{Akimov12a}; its design featured 31 2-inch PMTs immersed
in the liquid, viewing a thin disc geometry of 12~kg of liquid xenon
under high electric field~\cite{Akimov07}.

The XENON Collaboration deployed XENON10 at LNGS in 2006. This
double-phase detector contained 14~kg of active xenon in a deep
PTFE-walled chamber, viewed by two arrays of 1-inch PMTs located in
the liquid and gas phases (41 and 48 units,
respectively)~\cite{Aprile11a}. Main WIMP results were published in
2008~\cite{Angle08}. This design was essentially scaled up for
XENON100, in which 62 kg of active mass are viewed by 178
PMTs~\cite{Aprile12c}. A first run of 100 live days of acquisition in
2010 yielded a world leading result on WIMP-nucleon elastic
scattering~\cite{Aprile11b}. This sensitivity lead has been extended
with data from a further 225 live days reported recently
\cite{Aprile12a}. The XENON team is working on the next generation
tonne-scale experiment XENON1T, which will be built at LNGS starting
in 2013.

The LUX350 team has just completed construction of a double-phase
detector with 300~kg active mass, featuring 122 2-inch PMTs equally
divided in bottom and top arrays. The system has been tested in a
surface facility, in preparation for underground deployment in the
Davis complex of the Sanford Underground Laboratory at the Homestake
mine (US) \cite{Akerib12c,Akerib12a}. In collaboration with the
ZEPLIN-III groups, the multi-tonne LZ program is being developed to
occupy the same infrastructure after LUX~\cite{Malling11}.

The latest XMASS detector has the largest liquid xenon WIMP target
running at present: 800~kg. The spherical scintillation chamber is
80~cm in diameter and is instrumented with 642 2-inch PMTs. The
rationale behind this program is to compensate for the poorer nuclear
recoil discrimination achievable in a scintillation-only system with
the significant self-shielding for external backgrounds enabled by the
large target mass; the expected fiducial mass of 100~kg is, however,
comparable to that anticipated in LUX350. XMASS-800 has been fully
operational at the Kamioka Observatory (Japan) since
2011~\cite{Sekiya11}. The Collaboration plans a 26~tonne future
detector aimed also at detecting solar neutrinos and
$\beta\beta$-decay.

The PANDA-X (Particle AND Astrophysical Xenon TPC) is an emerging
effort aiming to deploy a three-stage experiment at the new JinPing
Laboratory (China), the deepest underground facility in the world
(2,513~m depth, 7,500~m water equivalent)~\cite{Giboni11}. Initially,
the experiment will feature a 25~kg fiducial LXe mass in a planar
geometry with 5~kV/cm drift field. The next stage, with 500~kg
fiducial, will reuse the outer infrastructure, but some internal
components will be replaced. A tonne-scale target would require a
larger inner vessel. To reduce background and cost, the team may also
replace the double-phase system by direct readout of proportional
scintillation around thin wires in a liquid-only detector. The light
readout for PANDA-1T might feature gaseous photomultipliers instead of
conventional PMTs.

Finally, we mention briefly DARWIN (DARk matter WImp search with Noble
liquids), a design study for a next-generation, multi-tonne detector
in Europe \cite{Baudis12}. The project focuses on double-phase LAr
and/or LXe targets and is investigating, among other issues,
alternative ionization readouts with LEMs and CMOS pixel detectors
(GridPix \cite{Blanco10}) coupled to electron multipliers, or via
proportional scintillation in the gas phase using gaseous
photomultipliers \cite{Duval11b}.

We now describe in more detail the most competitive systems already
operated or at an advanced stage of construction, highlighting various
design features and how the main technological challenges were
addressed in each detector.

\vspace{5mm}

\begin{figure}
\centerline{\includegraphics[width=0.9\textwidth]{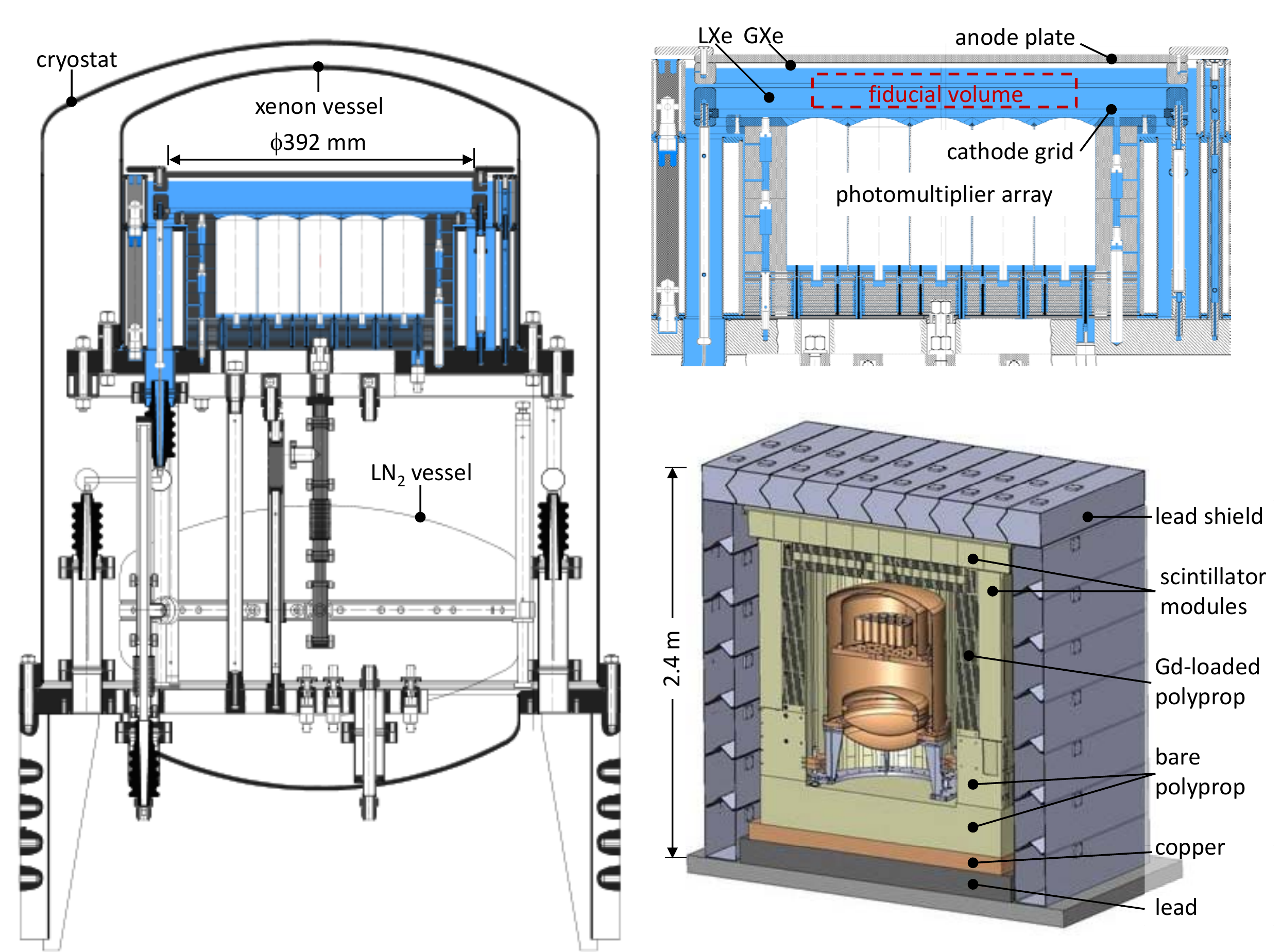}}
\caption{Schematic drawings of the ZEPLIN-III experiment. The WIMP
target is shown left, with liquid xenon in blue; the double-phase
chamber is detailed on the right, with an approximate fiducial volume
indicated by the dashed red rectangle. Also shown on the right is the
fully-shielded configuration at Boulby (including a veto instrument
surrounding the WIMP target). (Adapted from~\cite{Akimov07}
and~\cite{Akimov10b}; courtesy \mbox{ZEPLIN-III} Collaboration; with
permission from Elsevier.)}
\label{fig:ZEPLIN-III}
\end{figure}

{\bf The ZEPLIN Collaboration} operated its third liquid xenon
detector, shown in 
Figure~\ref{fig:ZEPLIN-III},
at the Boulby Underground Laboratory in
the UK (ZEPLIN stands for ZonEd Proportional scintillation in LIquid
Noble gases). The instrument construction is described in detail
in~\cite{Akimov07}. Most elements were built from high purity copper
to minimize background. The outer cryostat vessel enclosed two
chambers; the lower one contained the LN$_2$ coolant, which boiled off
through a heat-exchanger attached to the xenon vessel above it. The
latter housed a 12~kg liquid xenon WIMP target, with immersed
photomultipliers (viewing upward) to maximize detection efficiency for
faint primary scintillation signals. The working volume was formed by
an anode disc 39~cm in diameter and a multi-wire cathode located 4~cm
below it, a few mm above the hexagonal array of 31~PMTs. This volume
was mostly filled with liquid xenon --- leaving a gap of 4~mm above it
where secondary scintillation took place in the cold vapor.

Contrary to ZEPLIN-II, where a wire-grid just below the liquid surface
helped with cross-phase emission, in ZEPLIN-III the planar geometry
allowed application of a strong field to the whole liquid phase with
only two electrodes, thus enhancing the efficiency for charge
extraction from the particle tracks. Typical operation fields were
3--4~kV/cm in the liquid and approximately twice as strong in the
gas~\cite{Lebedenko09,Akimov12a}. A second wire-grid, just above the
photomultiplier windows, shielded the photon detectors from the
external field and also provided a reverse field region to suppress
secondary scintillation signals from low-energy background photons
from the PMTs. Only xenon-friendly, low-outgassing materials were used
within this chamber, in particular avoiding large amounts of PTFE, in
order to maintain sufficient electron lifetime in the liquid without
the need for continuous re-purification. This was achieved, with that
parameter actually improving over one year of operation in the closed
system: from 14~$\mu$s to 45~$\mu$s by the end of the
run~\cite{Majewski12}.
 
In the first science run custom-made photomultipliers D730Q/9829QA
from ETEL were used; these had bialkali photocathodes with metal
fingers deposited on quartz windows under the photocathode for low
temperature operation. The average (cold) quantum efficiency for xenon
light was 30\%~\cite{Araujo04}. For the second dark matter run, those
tubes were replaced with another pin-by-pin compatible model with 40
times lower radioactivity per unit, lowering the overall
electromagnetic background of the experiment to 750~mdru at low energy
(1~dru = 1~kg$^{-1}$day$^{-1}$keV$^{-1}$)~\cite{Araujo12}.
Unfortunately, their optical performance was poorer, with only 26\%
mean quantum efficiency and large gain
dispersion~\cite{Majewski12}. In ZEPLIN-III the PMT array was biased
from a common external voltage ladder, connected to a stack of copper
disks powering each set of PMT dynodes (these can be seen in 
Figure~\ref{fig:ZEPLIN-III}
just below the photomultipliers). Although this reduces the number of
feedthroughs required (lower background), and the need for internal
electronics (better xenon purity and reliability), it does not allow
for gain equalization at the PMT anodes.

Between the two science runs a veto instrument was retrofitted around
the WIMP target, as also shown in 
Figure~\ref{fig:ZEPLIN-III},
replacing some of the
hydrocarbon shielding. This veto included 52 plastic scintillator
modules surrounding a Gd-loaded polypropylene structure tailored for
neutron moderation and efficient radiative
capture~\cite{Akimov10b,Ghag11}. The use of anti-coincidence systems
around WIMP targets is increasingly recognized as important: besides
lowering the effective background of the experiment they provide an
independent measurement of the neutron (and other) background, which
can reduce the systematic error associated with a potential discovery.

As in other double-phase systems, accurate position reconstruction of
particle interactions in three dimensions allows for a very precise
fiducial volume to be defined, well away from any surfaces and
avoiding outer regions with non-uniform electric field and light
collection. The depth coordinate is obtained with precision of a few
tens of $\mu$m from the drift time of the ionization (measured by the
time separation between the primary and the secondary scintillation
signals --- see 
Figure~\ref{fig:Z3Signals}).
The horizontal coordinates are
reconstructed from S2 signals from all PMTs; a spatial resolution of
1.6~mm (FWHM) was achieved for 122~keV
$\gamma$-rays~\cite{Solovov12}. A fiducial volume containing 6.5~kg of
liquid xenon was defined for the 83-day long first run of
\mbox{ZEPLIN-III}~\cite{Lebedenko09}, decreasing to 5.1~kg for the
319-day second run owing to the poorer PMT performance.

The scintillation threshold for WIMP searches was $\approx$7~keV in
both runs~\cite{Akimov12a} --- using recent data for the relative
scintillation efficiency of nuclear recoils~\cite{Horn11}. An
ionization threshold corresponding to the electroluminescence of 5
electrons was defined by the trigger hardware; although ZEPLIN-III
could detect the signature from a single ionization electron --- as
shown in 
Figure~\ref{fig:SEinZ3}
--- in practice these signals could not be used due
to a large single electron background which generated too high a
trigger rate~\cite{Santos11}. An electron recoil rejection efficiency
of 99.99\% was achieved at WIMP-search energies in the first run,
which remains the best reported for double-phase xenon. In both
datasets a handful of events were observed within the signal
acceptance region, consistent with background expectations in both
cases. The combined result excluded a WIMP-nucleon scalar cross
section above 3.9$\cdot$10$^{-8}$~pb (3.9$\cdot$10$^{-44}$~cm$^2$) at
90\% CL for 50~GeV WIMP mass~\cite{Akimov12a,Lebedenko09}.

\vspace{5mm}

\begin{figure}
\centerline{\includegraphics[width=0.9\textwidth]{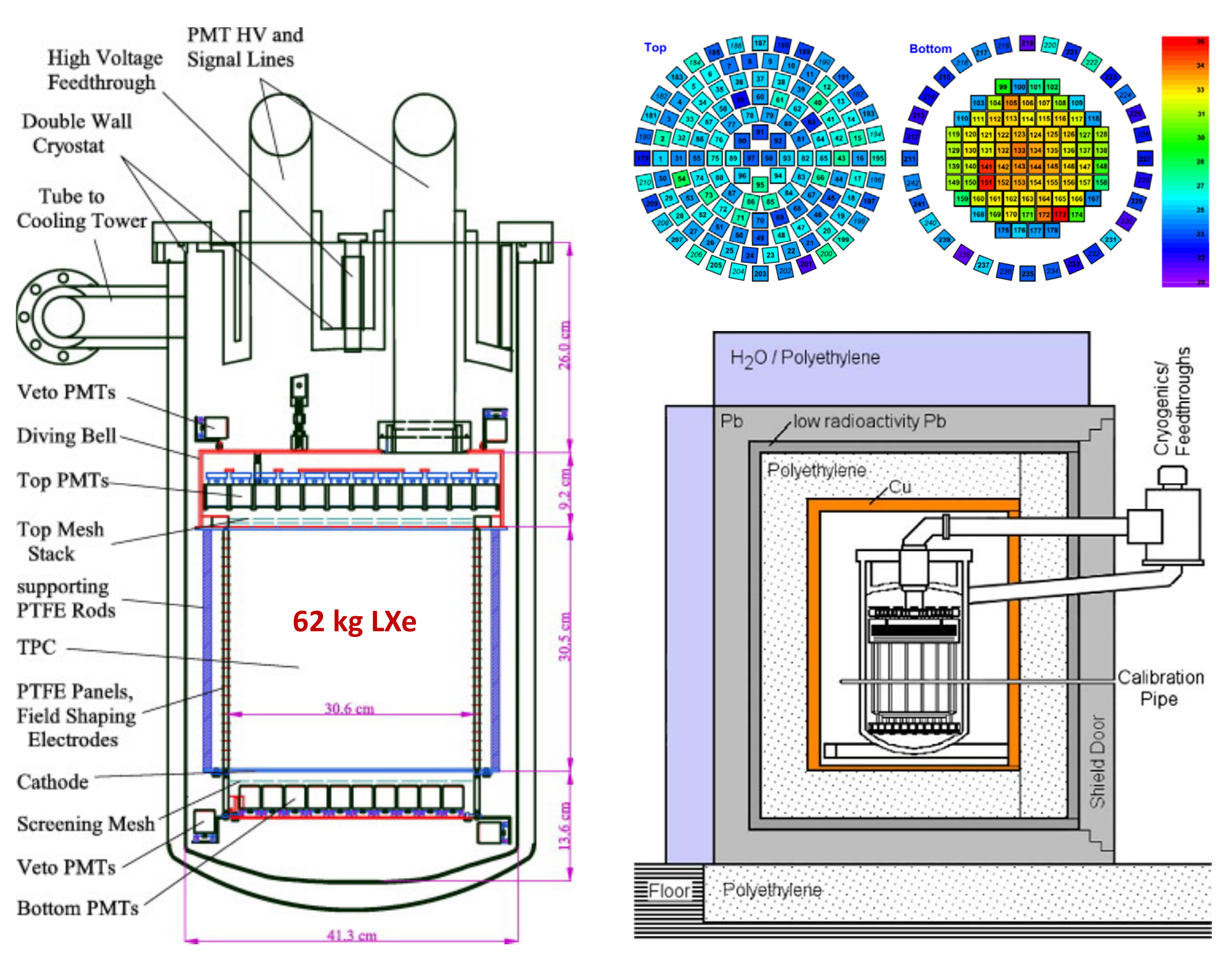}}
\caption{Schematic diagrams of the XENON100 experiment. The main
instrument is shown left; it contains 62~kg of liquid xenon in the
active TPC region and an additional 99~kg in a surrounding `veto'
region. The VUV readout uses two arrays of 1-inch square PMTs with
$\sim$30\% quantum efficiency; the QE distribution and PMT layout are
shown top right. The shielding configuration around the instrument is
also shown on the right. (Adapted from~\cite{Aprile12c}; courtesy
E.~Aprile; with permission from Elsevier.)}
\label{fig:XENON100}
\end{figure}

{\bf The XENON Collaboration} has chosen a bulk detector concept for
XENON10 and its successor XENON100, the latter shown schematically in
Figure~\ref{fig:XENON100}.
The two detectors, very similar in design, are described
in~\cite{Aprile11a} and~\cite{Aprile12c}, respectively. They were
deployed at LNGS (Italy) in sequence, with XENON100 designed to fit
the existing shielding enclosure. Results from a first run of 100 live
days in 2010 have been published~\cite{Aprile11b} and the
Collaboration has just reported from a second run with 225 live-days
of science data \cite{Aprile12a}, following additional purification to
lower further the $^{85}$Kr background.

In XENON100, the 62~kg WIMP target is 31~cm in diameter by 31~cm in
height and is contained within a PTFE cylinder with field shaping
rings enclosed within its walls. Five meshes provide the necessary
electric fields. The cathode at the bottom of the chamber and another
(the `gate') just below the liquid surface define a drift field of
0.53~kV/cm. At this field the ionization drift time for events near
the cathode is 176~$\mu$s, demanding very high purity of the liquid. A
strong extraction field of $\sim$12~kV/cm is formed between the gate
electrode and the anode mesh a few mm above the liquid
surface. Secondary scintillation develops in the gas gap between these
two electrodes. Two additional meshes close to each PMT array are used
to close the electric field lines and thus shield the photomultiplier
tubes. The thickness of the gas phase is defined with the help of a
diving bell, indicated in red in 
Figure~\ref{fig:XENON100},
which is pressurized by
the gas recirculation line. This also allows liquid to be filled to
above the bell for shielding purposes (see below). The active volume
is viewed by two arrays of photomultipliers Hamamatsu R8520-06-Al, one
below the cathode mesh viewing upward and detecting mostly primary
scintillation signals, and the other placed in the gas phase and
viewing downward. The layout of the two arrays is also shown in
Figure~\ref{fig:XENON100}.
The 178 PMTs were selected for low U/Th radioactivity
content; they have 1-inch square window and bialkali photocathodes
with aluminum fingers underneath.

Approximately 4~cm of liquid surrounding the time projection chamber
from all sides (totaling 99~kg of xenon) establishes an active
scintillation `veto' detector for external $\gamma$-rays; this volume
is optically isolated from the main chamber and is viewed by an
additional set of 64~PMTs operated in anti-coincidence with the WIMP
target. The average energy threshold of this volume is $\sim$100~keV
for electron recoils.

Cooling is provided by a pulsed tube refrigerator~\cite{Haruyama05}
and delivered from outside the shield to minimize background. Very
good thermal stability is achieved over several months of
operation. The liquid xenon purity is maintained with active
recirculation through a hot getter.

Very low rates of electron recoil background were achieved at the core
of XENON100 in a preliminary run, with $<$5~mdru reported at low
energy whilst retaining a fiducial mass of 40~kg~\cite{Aprile10}. If
maintained, this rate would lead to only a few background events per
year of operation with a discrimination efficiency at the level of
99.5\%.

Initial WIMP results were derived from 100.9 live days with a 48~kg
fiducial volume, above an 8~keV nuclear recoil threshold. Three
candidate events were observed in the signal region with an expected
background of 1.8$\pm$0.6~events, thus excluding a spin-independent
elastic interaction above 7.0$\cdot$10$^{-9}$~pb
(7.0$\cdot$10$^{-45}$~cm$^2$) for a 50~GeV WIMP with 90\%
confidence~\cite{Aprile11b}.

The sensitivity of this dataset was limited by a higher concentration
of $^{85}$Kr than had been achieved in the preliminary run, which was
attributed to an air leak~\cite{Aprile11b,Aprile11c}. After further
cryogenic distillation a second long run was started in 2011 and
accumulated 225 live days \cite{Aprile12a}. An electron recoil
background rate of $\sim$5~mdru was recovered in a 34~kg fiducial
mass. Two events were observed in the 6.6--30.5~keV recoil energy
range in this exposure, which is consistent with a background
expectation of 1.0$\pm$0.2~events. This excludes at 90\% CL a scalar
interaction above 2.0$\cdot$10$^{-9}$~pb (2.0$\cdot$10$^{-45}$~cm$^2$)
for a 55~GeV WIMP, which constitutes the strongest experimental
constraint on spin-independent WIMP-nucleon scattering to date.

A larger system, XENON1T, is being developed by the Collaboration;
this features a 1-m long TPC containing 2.2~tonnes of xenon. The
original design featured 3-inch QUPID photon detectors
(Section~\ref{sec:HybridDevices}),
but the new 3-inch Hamamatsu R11410 PMTs are also
being considered. The system will be housed in a 10-m diameter water
tank at LNGS, with construction beginning soon.

\begin{figure}
\centerline{\includegraphics[width=0.9\textwidth]{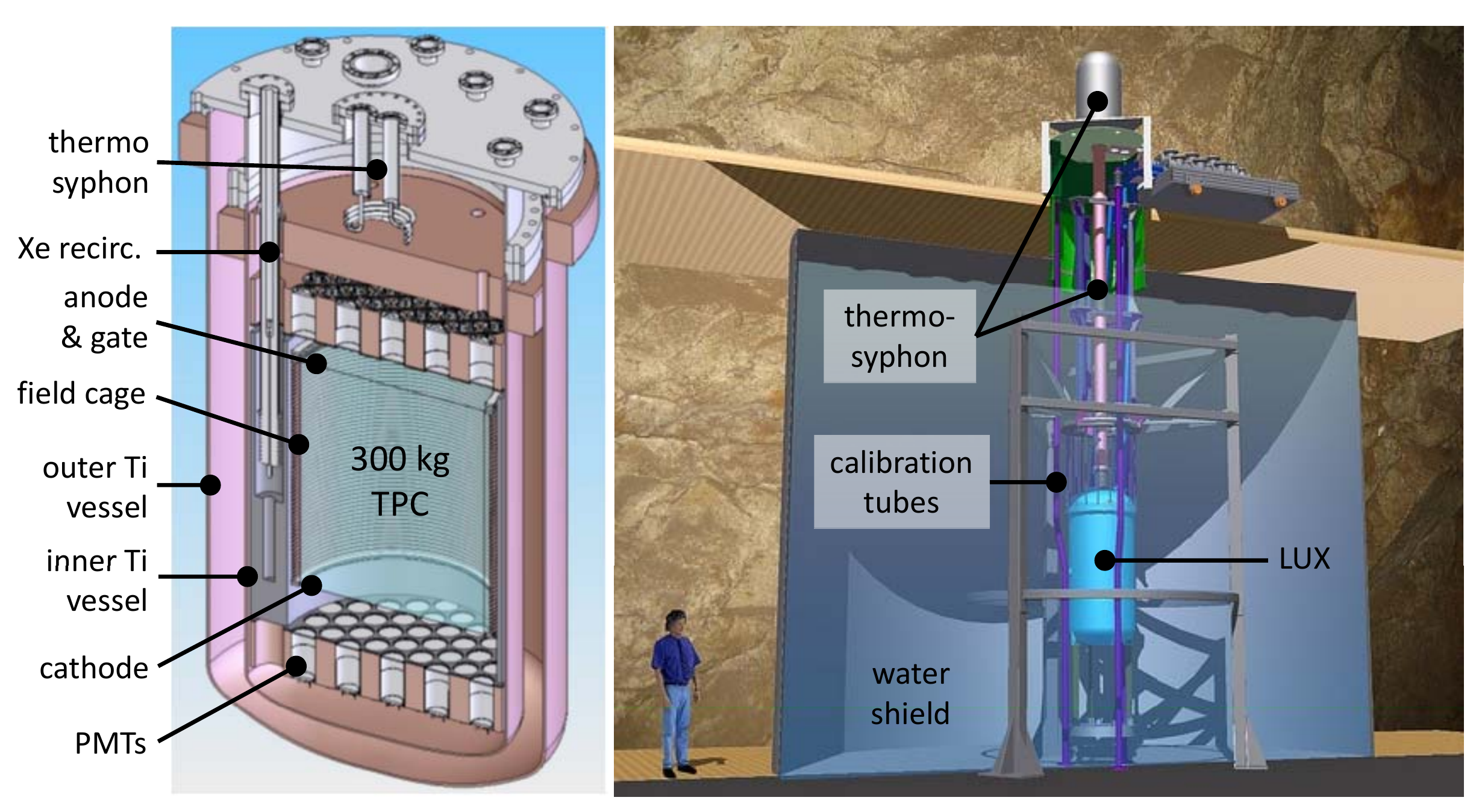}}
\caption{Illustration of the LUX detector and its deployment in the
water tank shield in the Davis complex of the Sanford Underground
Laboratory. The 300-kg time projection chamber, installed within a
double-vessel titanium cryostat, is viewed by two arrays of 61
photomultipliers each. (Adapted from~\cite{Akerib12e}; courtesy LUX
Collaboration.) }
\label{fig:LUX350}
\end{figure}

\vspace{5mm}

{\bf The LUX Collaboration} (LUX stands for Large Underground Xenon
experiment) has built a 350~kg liquid xenon detector to operate at the
Sanford Underground Laboratory (SURF) at the Homestake mine in South
Dakota~\cite{Akerib12c,Akerib12e}. The detector, shown in 
Figure~\ref{fig:LUX350},
exploits the same double-phase principle with measurement of primary
scintillation in the liquid and secondary scintillation in the gas;
the chamber configuration is in general similar to the ZEPLIN-II and
XENON TPCs. The active volume is a cylinder of 49~cm diameter and
59~cm height equipped with 122 Hamamatsu R8778 PMTs (2-inch diameter),
equally distributed in top and bottom arrays.

LUX350 is expected to reach $<0.8$~mdru of electron recoil background
(before discrimination) in an inner 100~kg fiducial mass. Besides
self-shielding, this very low background level is afforded by a
thin-walled titanium cryostat design and operation within an 8-m
diameter water tank. Containing 300~tonnes of ultrapure water
instrumented with 20 ten-inch PMTs (Hamamatsu R7081), this will
provide passive shielding for external $\gamma$-rays and neutrons as
well as an active muon veto.

Cooling and temperature regulation are achieved primarily by
closed-loop N$_2$ thermosyphons connected to a liquid nitrogen
reservoir located above the water tank \cite{Akerib12d}. The
cryogenic system has been designed with the cooling power requirements
of next-generation detectors in mind. The xenon is purified by
continuous circulation and a heat-exchanger system is employed to
allow higher recirculation rates, with the outgoing gas cooling the
incoming xenon.

The detector has been tested in a dedicated surface facility at SURF,
within a 3-m diameter water tank, which provided useful shielding for
detector characterization and allowed full system verification in
realistic underground conditions \cite{Akerib12c}. In these tests the
center of the chamber yielded a very impressive 8 photoelectrons per
keV in scintillation for electron recoils at zero electric field. At
the time of writing the experiment is being commissioned underground
in the redeveloped Davis complex at SURF; initial WIMP results are
expected early in 2013.

For the next phase the LUX and ZEPLIN-III teams are planning the LZ
program at Homestake, where a detector with active mass of several
tonnes can be deployed within the infrastructure developed for
LUX350~\cite{Malling11}. A TPC similar in design to LUX is envisaged,
viewed by Hamamatsu R11410 3-inch photomultipliers, and containing up
to 7~tonnes of active xenon. A liquid scintillator veto is being
designed to fit around a thin-walled titanium cryostat to further
decrease internal backgrounds. Construction is expected to start
around 2014.

\vspace{5mm}

\begin{figure}
\centerline{\includegraphics[width=0.9\textwidth]{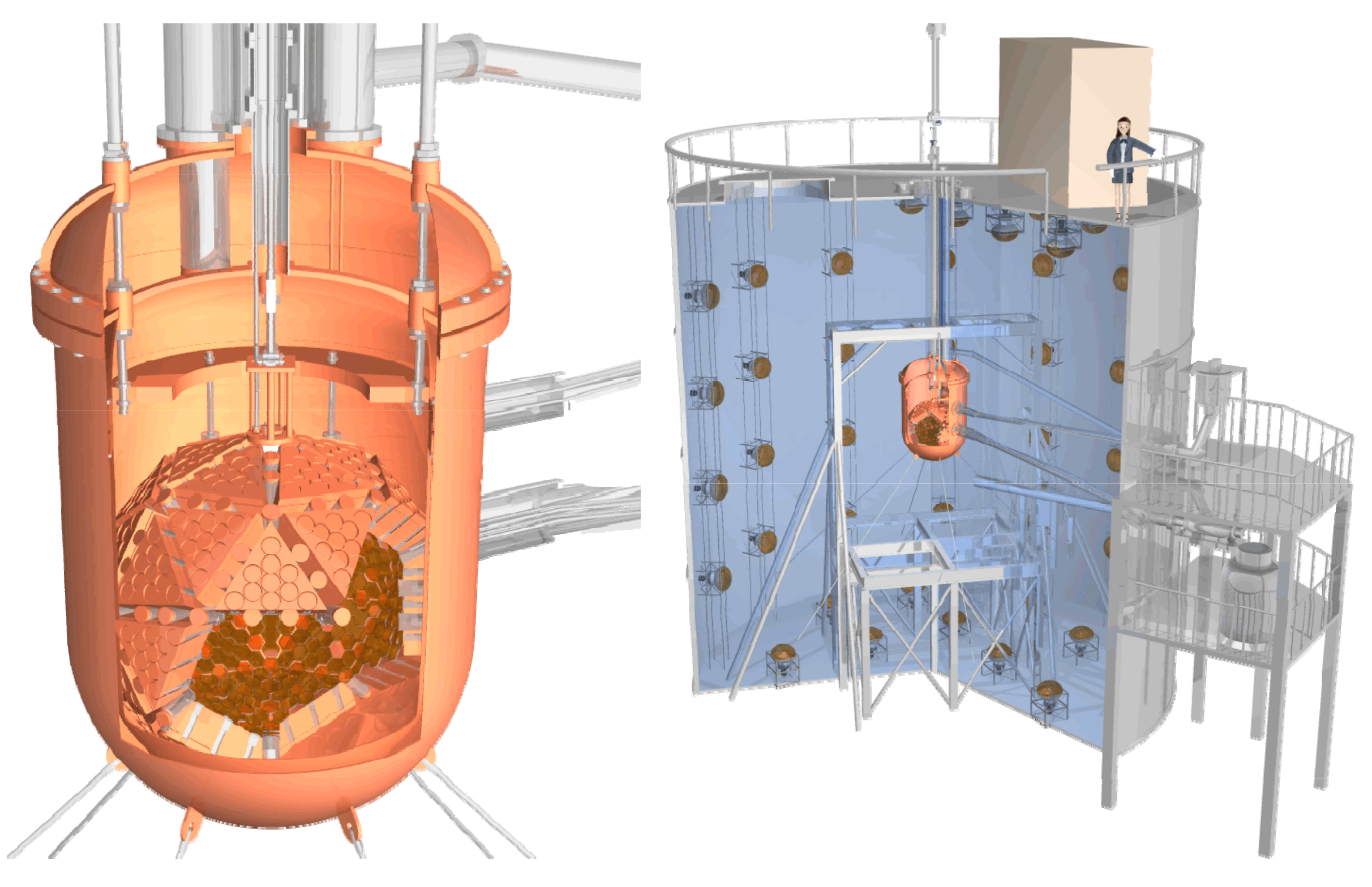}}
\caption{The XMASS scintillation detector is shown left; 800~kg of
liquid xenon are viewed by 642~photomultipliers arranged in a pentakis
dodecahedron geometry. The chamber is located within a 10-m diameter
water shield at Kamioka. (Credits: Kamioka Observatory, ICRR
(Institute for Cosmic Ray Research), The University of Tokyo.)}
\label{fig:XMASS}
\end{figure}

{\bf The XMASS Collaboration} based in Japan has chosen a different
approach, which relies on efficient measurement of primary
scintillation in a single-phase liquid xenon detector 
(Figure~\ref{fig:XMASS}).
Without the need to apply strong electric fields internally, the
design of the target chamber is less constrained and a spherical
geometry, optimal to exploit self-shielding, becomes possible. XMASS
is a multipurpose experimental program which aims to study solar
neutrinos, $\beta$$\beta$-decay and WIMPs~\cite{Suzuki01}. Following
several years of R\&D work with an 80-kg prototype~\cite{Minamino12},
an 800~kg detector has been built and installed at the Kamioka
Observatory; engineering data-taking started in 2011.

The detector is a sphere of 80~cm internal diameter made from OFHC
copper with 642 Hamamatsu R10789 photomultipliers closely packed into
a honeycombed structure. These have hexagonal windows (65~mm diagonal)
and were expressly optimized for low background for this
experiment. They have bialkali photocathodes with metal fingers under
the quartz windows, and quantum efficiencies for xenon light of up to
39\% at liquid xenon temperature. The photocathode coverage fraction
in the chamber is 64\%~\cite{Sekiya11}. A preliminary signal yield of
15.9~photoelectrons per keV has been reported for $^{57}$Co at the
center of the detector~\cite{Liu12} --- 3 times higher than had been
assumed by the XMASS team~\cite{Sekiya11}. This will improve the
previously stated threshold of 5~keV for electron recoils.

The only possibility for electron recoil background rejection is pulse
shape discrimination, as only the scintillation signal is
measured. Therefore, the requirements for both external $\gamma$-ray
shielding and intrinsic contamination are even more stringent than in
double-phase detectors. The planned fiducial volume includes only the
central $\sim$100~kg, with a 20-cm thick outer shell being sacrificed
for shielding from external $\gamma$-rays and neutrons, as well as
from $\beta$- and $\alpha$-particles emitted from the detector
walls. Intrinsic backgrounds, notably $^{85}$Kr and $^{222}$Rn, then
become a critical consideration, and the XMASS Collaboration has
devoted much effort to mitigate these~\cite{Abe09,Sekiya11}. In order
to shield the detector from external radiation, XMASS is immersed into
a \mbox{10-m} diameter tank filled with 800 tonnes of ultrapure water
instrumented with 72 20-inch PMTs. The water shield absorbs
$\gamma$-rays and thermalizes fast neutrons from the rock, besides
providing a Cherenkov muon veto. Overall, an electronic background
rate below 0.1~mdru is expected in the inner 100 kg fiducial volume
and neutron backgrounds are negligible there~\cite{Sekiya11}.

The possibility of using pulse shape discrimination, as in early xenon
dark matter detectors, has been carefully
investigated~\cite{Ueshima11}. The difference in scintillation decay
times for electrons and nuclear recoils is small, as discussed in
Section~\ref{sec:EmissionMechanisms},
but useful nonetheless. That study, conducted with a
small prototype chamber with a high light yield of 20.9 photoelectrons
per keV, reports an electron rejection efficiency of 92\% near 5~keV,
improving to 99.9\% for 15~keV electrons --- both retaining 50\%
nuclear recoil acceptance. A second measurement was performed with
some scintillation light deliberately masked to provide a lower
photoelectron yield, in line with what had been expected for XMASS,
and this confirmed significantly poorer discrimination. However, the
actual performance achieved by XMASS warrants, in our view, a
comparison with the above (unmasked) result. This suggests that only a
few undiscriminated electron recoil events would remain per year of
operation if the claimed nominal background rate can be confirmed.

The 800~kg system is primarily a WIMP detector. As the next step, the
Collaboration plans to scale up the detector to a 2.5~m sphere
containing 20~tonnes of liquid xenon (10~tonnes fiducial). This mass
will allow also detection of solar neutrinos and study
$\beta$$\beta$-decay of $^{136}$Xe as originally
planned~\cite{Suzuki01}.

\subsection{Liquid argon detectors}
\label{sec:LArDetectors}

Dark matter searches with liquid argon have not progressed as quickly
as with liquid xenon, but there has been renewed interest in this
medium in recent years. Experience from large projects, namely ICARUS,
propelled ambitious efforts over a decade ago, but these lost ground
to liquid xenon programs, which adopted a more incremental
approach. Yet, the field is a vibrant area of research once again,
involving double-phase systems as well as scintillation-only detectors
(pulse shape discrimination is very effective in liquid argon, as
discussed in Sections~\ref{sec:OverviewDetectionPrinciples} and 
\ref{sec:EmissionMechanisms}).
Two collaborations have built dark
matter systems, two other detectors are at an advanced stage of
construction, and a new effort is emerging. As in the previous
section, we begin by outlining the main experiments before describing
the most competitive systems in more detail.

The WARP Collaboration started their program at LNGS in the late
1990s, building on R\&D experience from ICARUS, and published in 2008
the only set of dark matter results from liquid argon to
date~\cite{Benetti08}; their 2.3-liter prototype chamber featured
photomultiplier readout and exploited both ionization/scintillation
ratio as well as pulse shape discrimination. A detector with 140~kg
active mass was subsequently built and commissioned at Gran Sasso in
2009~\cite{Acciarri11}, but no published information on the present
status of the project could be found.

The ArDM Collaboration started development of a tonne-scale
double-phase chamber in 2004 \cite{KaufmannRubbia07}; this is
presently being deployed at the Canfranc Underground Laboratory
(Spain)~\cite{EXP-08-2011} following construction and extensive
testing at CERN. The original design features photomultiplier readout of
scintillation in the liquid and detection of emitted ionization using
Large Electron Multipliers in the gas.

Two single-phase detectors are being built by the DEAP/CLEAN
Collaboration to operate at SNOLab (Canada). MiniCLEAN is a 500~kg
scintillation detector using cold PMTs over 4$\pi$ solid
angle~\cite{Hime11}. It has a multiple-target mission: after an
initial 2-year run with natural argon, the Collaboration proposes to
replace the working medium with liquid neon to explore how the rate of
a hypothetical signal varies with atomic mass under similar
backgrounds. The second system, DEAP-3600, features 3.6~tonnes of
liquid argon in a transparent acrylic vessel viewed by warm
photomultipliers~\cite{Boulay12}. It proposes to operate with natural
argon for the first three years followed by use of underground argon
depleted in $^{39}$Ar. MiniCLEAN is due to start data-taking in late
2012, followed by DEAP-3600 one year later.  

Finally, the DarkSide team have returned to double-phase argon but
starting on a smaller scale. A 10~kg prototype has been operated at
LNGS~\cite{Akimov12} and a low-background 50-kg device (33~kg
fiducial) is under construction~\cite{Alton10}.

\vspace{5mm}

\begin{figure}[hb]
\centerline{\includegraphics[width=0.9\textwidth]{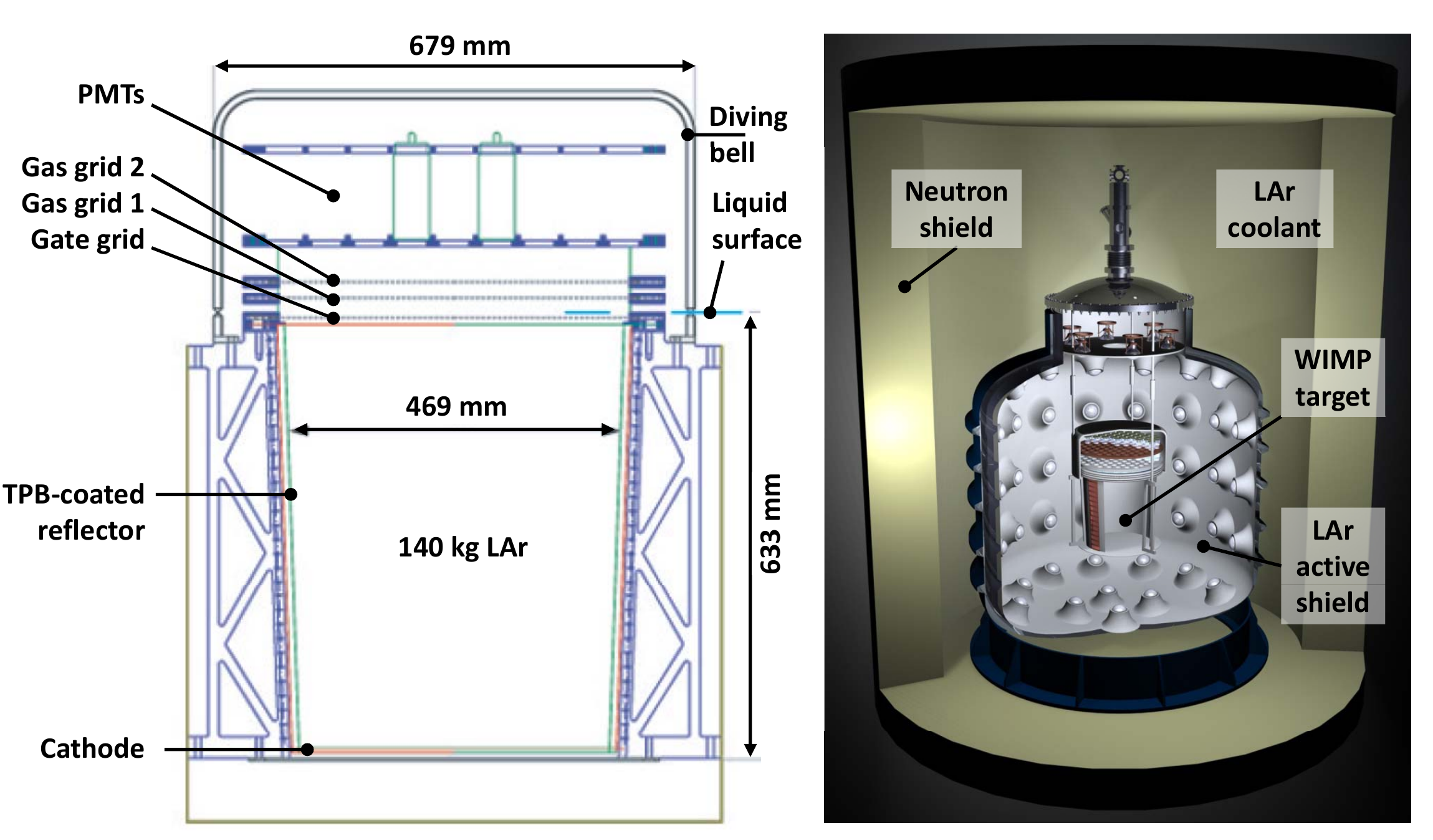}}
\caption{The WARP140 dark matter experiment. The 100-liter WIMP target
is represented left, viewed by 37~PMTs (only two are shown). The
target is surrounded by an active scintillator shield sharing the same
ultra-pure LAr; a stainless steel cryostat (not shown) filled with LAr
coolant contains the nested vessels; this is lined with a polyethylene
layer to shield against cryostat neutrons. (Adapted
from~\cite{Brunetti04}.)}
\label{fig:WARP140}
\end{figure}

{\bf The WIMP Argon Programme (WARP)}~was proposed in 2004, emerging
from R\&D activities with liquid xenon and liquid argon for the ICARUS
experiment. It consisted initially of two stages: an existing
double-phase test chamber with 2.3-liter capacity was modified and
deployed at LNGS; this would prototype a 100-liter detector built from
low background components and using 140~kg of isotopically purified
argon~\cite{Brunetti04}. A future 1.4~tonne detector was envisaged,
occupying some of the infrastructure developed for WARP140, the
100-liter chamber. WIMP search data from the 2.3-liter prototype were
acquired in 2005. With the help of dual parameter discrimination
(S2/S1 ratio and S1 pulse shape, as shown in
Figure~\ref{fig:LArDiscrimination}, right), a WIMP-nucleon
cross-section limit of 1$\cdot$10$^{-6}$~pb
(1$\cdot$10$^{-42}$~cm$^2$) was published in 2008~\cite{Benetti08}.

The WARP140 target, shown schematically in Figure~\ref{fig:WARP140},
has a 60~cm long drift region designed to operate at 0.5--1~kV/cm; the
electric field is conditioned by field-shaping rings embedded in the
chamber walls. The electroluminescence region is defined by two grids
either side of the liquid surface, with the liquid height determined
by a diving bell. The target is viewed by an array of 37 PMTs with 2-
and 3-inch diameter looking down from the gas phase into the active
volume; these models (ETL D750UKFLA and D757UKFLA) have
low-temperature bialkali photocathodes developed by Electron Tubes to
operate at LAr temperature. The voltage divider chains are made from
Kapton printed circuit boards with discrete components selected for
operation at LAr temperature. To wavelength-shift the VUV photons
emitted by the prompt and proportional scintillations of LAr, TPB was
evaporated onto highly reflective plastic foil coating all internal
surfaces. A light yield of 1.6~photoelectrons per keV was reported at
zero electric field with this configuration~\cite{Acciarri11}.

This chamber is immersed in a much larger LAr volume of 8~tonnes,
totally surrounding the inner detector and working as an active veto;
this is viewed by 2- and 3-inch photomultiplier tubes of the above
models, 300 in total. Liquid argon circulates through the WIMP target
and the active shield, driven by an external purification system. The
cryogen is contained within a double-walled stainless steel cryostat,
2.9~m in diameter and 4.45~m in height, lined internally with an
additional 10~cm of polyethylene to shield against neutrons from
stainless steel. External hydrocarbon and lead enclosures complete the
shield.

Construction of the 100-liter detector started in 2005 and by 2009 the
detector had been commissioned at LNGS~\cite{Acciarri11}. However,
problems with high voltage delivery and with the stability of the
wavelength shifting layers halted the first technical run; a second
test was conducted in 2010, but high voltage problems persisted. No
update on WARP140 has been reported recently.

\vspace{5mm}

\begin{figure}
\centerline{\includegraphics[width=0.75\textwidth]{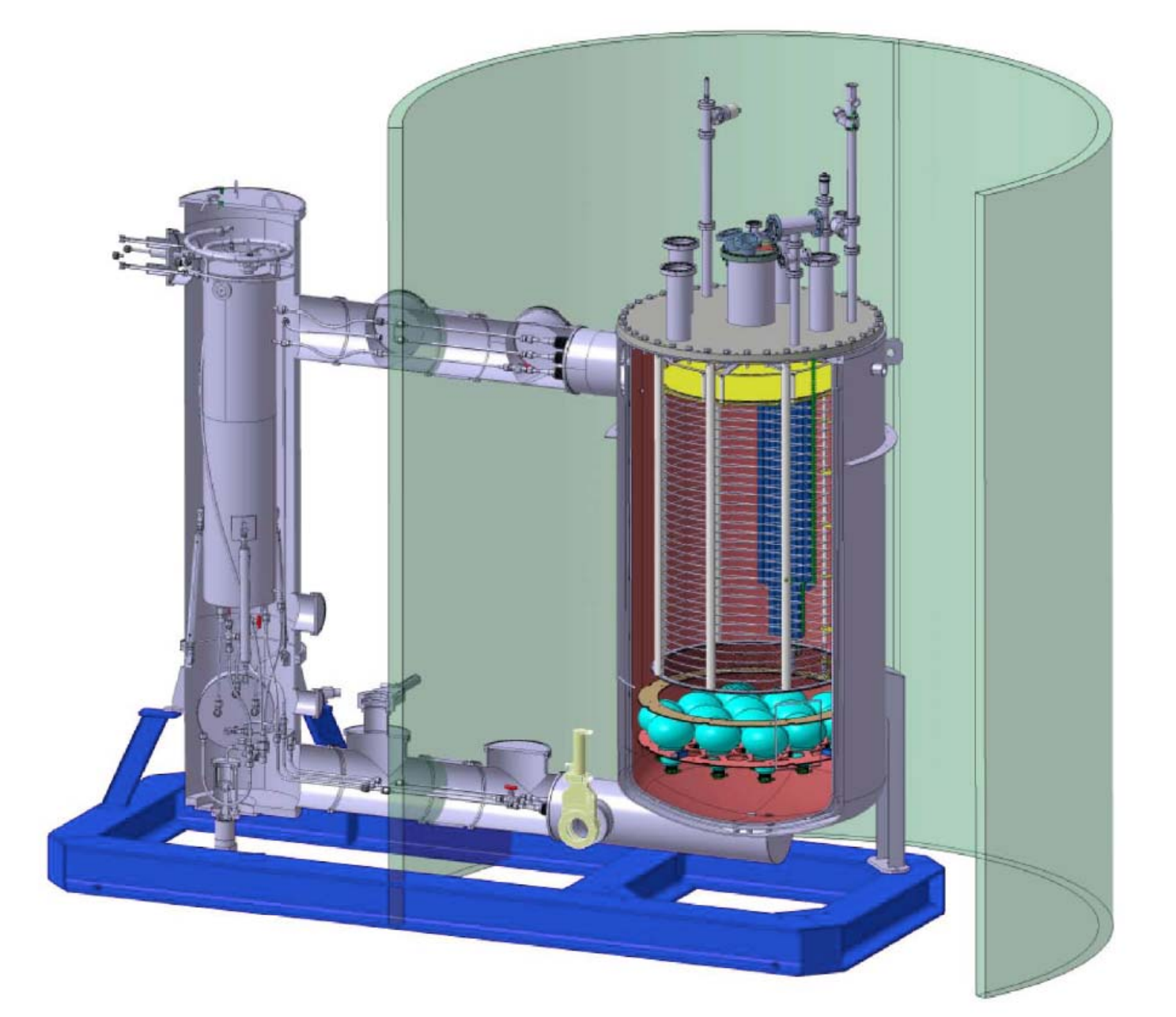}}
\caption{The ArDM inner cryostat containing the 850-kg LAr WIMP
target connected to the liquid argon purification column to the
left. These elements are located within a LAr bath held in an outer
vessel (not shown). 14~PMTs at the bottom of the WIMP chamber view the
wavelength-shifted scintillation light; the ionization charge is
amplified by LEMs in the vapor phase. The HV multiplier can be seen
in dark blue at the back of the chamber. The inner cryostat is
surrounded by a neutron shield. (From~\cite{Marchionni11}; with
permission from IOP Publishing.)}
\label{fig:ArDM}
\end{figure}

{\bf The ArDM (Argon Dark Matter) experiment} aims to search for WIMPs
with a tonne-scale double-phase liquid argon detector deployed at the
Canfranc underground facility (Spain)~\cite{KaufmannRubbia07}. Its
conceptual design is shown in 
Figure~\ref{fig:ArDM}.
The detector contains 850~kg
of active mass of LAr shaped as a cylinder of 80~cm diameter and
120~cm height; the estimated fiducial mass is not known. The active
volume is viewed by 14 hemispherical 8-inch photomultipliers placed at
the bottom (Hamamatsu R5912-02MOD bialkali tubes with platinum
underlay to operate at LAr temperature). TPB is used to
wavelength-shift the argon light; a thin transparent layer is
deposited over the PMT windows, while all other surfaces are covered
with a PTFE fabric (Tetratex) coated with an opaque TPB layer, which
works also as a reflector~\cite{Boccone09}. A light yield of 1
photoelectron per keV at zero electric field was predicted for the
completed chamber based on tests with 7 out of
14~PMTs~\cite{Amsler10}, which would enable a nuclear recoil energy
threshold of 30~keV as originally planned.

The ionization charge is drifted away from the interaction site under
a strong electric field ($>$3~kV/cm) and emitted into the vapor phase
with the help of two electrodes located either side of the liquid
surface. A distinguishing feature of this experiment is the aim to
measure directly the charge extracted from the liquid with Large
Electron Multipliers (LEM) placed in the gas
phase~\cite{KaufmannRubbia07}. A gain of 27 has been achieved in a
dedicated prototype with a 1-mm thick LEM operating in double-phase
argon, with the amplified charge collected by a two-dimensional
projective readout anode providing orthogonal views with 3~mm position
resolution~\cite{Badertscher11}. A cascaded two-stage LEM is expected
to provide readout sensitivity to a few ionization electrons.

Another innovative feature in ArDM is the use of a high voltage (HV)
Greinacher multiplier (also known as Cockcroft-Walton) working in the
LAr to bias the electrode grids as well as the field shaping rings
located along the internal walls~\cite{Horikawa11}. If stable HV
operation is successful this may obviate the need for low-background,
cryogenic HV feedthroughs in future large experiments, which are a
common technical challenge in double-phase detectors. The ArDM
multiplier has been successfully operated at 70~kV, corresponding to a
drift field of 0.5~kV/cm; the stated goal is nearly
400~kV~\cite{Marchionni11}.

The double-phase detector is contained in a 1-m diameter, 2-m tall
cryostat shown in 
Figure~\ref{fig:ArDM}.
This is surrounded by an internal neutron
shield. A cryogenic pump drives a LAr purification column containing
CuO, shown to the left of the cryostat. Electrons have to travel up to
120~cm in the ArDM TPC, which will take $\sim$1~ms even at high
electric field; liquid purity is therefore a critical issue. The
vessel and purification system are fully immersed in a separate LAr
cooling bath, contained within a stainless steel cryostat (not
shown). The temperature of the cooling bath is maintained with the aid
of cryocoolers installed in a recondenser unit sitting on top of the
recirculation/purification section~\cite{Marchionni11}.

The dominant background in ArDM will be the $\beta$ decay of $^{39}$Ar
present in atmospheric argon, which eventually will limit the
sensitivity of the experiment. The rate of these electron recoils
($\sim$1~kBq) demands extremely high discrimination factors, but this
may be compromised by event confusion within the $\sim$1~ms drift
time. Operation with underground argon is envisaged by the
collaboration~\cite{Marchionni11}.

The ArDM detector operated with liquid argon on a surface laboratory
at CERN. A number of tests have been carried out and most sub-systems
have been demonstrated. Commissioning underground at Canfranc is under
way~\cite{EXP-08-2011}.

\vspace{5mm}

{\bf The DEAP/CLEAN Collaboration} is pursuing a single-phase LAr
program at \mbox{SNOLab}, exploiting pulse shape discrimination to
overcome the dominant background of electron recoils from
$^{39}$Ar~\cite{BoulayHime06}. This requires discrimination of at
least a few parts in 10$^9$ for low energy electrons. This program
emerged from the proposal to use LNe and LHe for low energy neutrino
detection~\cite{McKinseyDoyle99,McKinseyCoakley05}. Several R\&D
chambers, including MicroCLEAN and DEAP-1, have been built
specifically to address the discrimination and other key issues, such
as the scintillation efficiency for nuclear recoils in
LAr~\cite{Lippincott08,Boulay09,Gastler11}.

\begin{figure}
\centerline{\includegraphics[width=\textwidth]{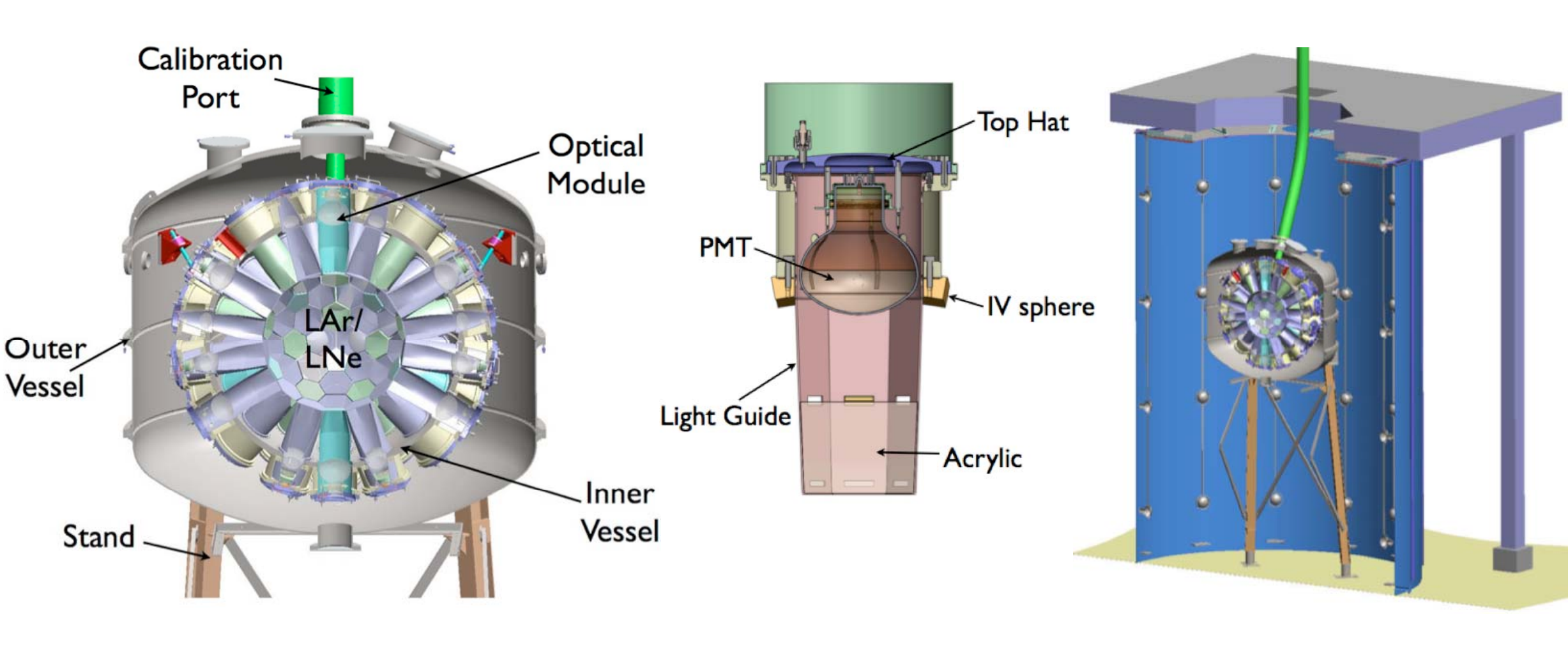}}
\caption{Schematic views of the MiniCLEAN experiment. Left: Cut-out of
outer and inner vessels to reveal the arrangement of optical cassettes
which view the 500~kg LAr (or LNe) WIMP target in 4$\pi$. Center:
detail of optical cassette (here termed `module'), including a
photomultiplier attached to a stainless steel light guide with an
acrylic plug covered in TPB at the front. Right: integrated experiment
immersed in water tank. (From~\cite{Hime11}, courtesy A.~Hime.)  }
\label{fig:MiniCLEAN}
\end{figure}

Two experiments, MiniCLEAN and DEAP-3600, are presently being
installed at SNOLab, with first data expected by the end of
2012~\cite{Rielage12} and 2013~\cite{Boulay12}, respectively. Both
have spherical targets containing natural argon --- in the first
instance --- with the wavelength-shifted LAr scintillation viewed by
photomultipliers over 4$\pi$ solid angle. There are similarities
between the two designs, for which reason we describe the two systems
in parallel, but there are also important differences, with
technological solutions being explored for the design of the much
larger instrument: a future CLEAN experiment (standing for Cryogenic
Low Energy Astrophysics with Noble liquids) may contain 40--120~tonnes
of active target \cite{McKinseyCoakley05,Hime11}.

MiniCLEAN features a 45~cm diameter target containing 500~kg of liquid
argon, with an expected fiducial mass of 100~kg defined by
reconstruction of the scintillation vertex~\cite{Hime11}. The
Collaboration consider replacing the working medium with liquid neon
for a subsequent run in order to exploit the different scaling of
event rates from a possible WIMP signal ($A^2$ dependence) and from
backgrounds. This would also enable sensitivity to solar p-p
neutrinos, which is another stated scientific goal. The optical
readout employs 92 individual `cassettes', each containing an 8-inch
Hamamatsu R5912-02MOD photomultiplier viewing an acrylic plug coated
with TPB wavelength shifter, as shown schematically in 
Figure~\ref{fig:MiniCLEAN}.
These are mounted onto a stainless steel inner vessel. The
cryostat is completed by an outer vessel with 2.6~m diameter for
thermal insulation and containment. The photomultipliers operate cold
within the cryogen~\cite{Nikkel07}. Shielding against PMT neutrons,
one of the dominant backgrounds, is provided by the cryogen and
acrylic layers in front of the PMTs. A second important background
comes from the plating of radon daughters on the wavelength shifter,
with emitters ejecting nuclear recoils into the active volume. This is
mitigated by a strict cleaning procedure performed on the PMT
cassettes {\it ex situ} and transfer to the detector under vacuum.

\begin{figure}[ht]
\centerline{\includegraphics[width=0.6\textwidth]{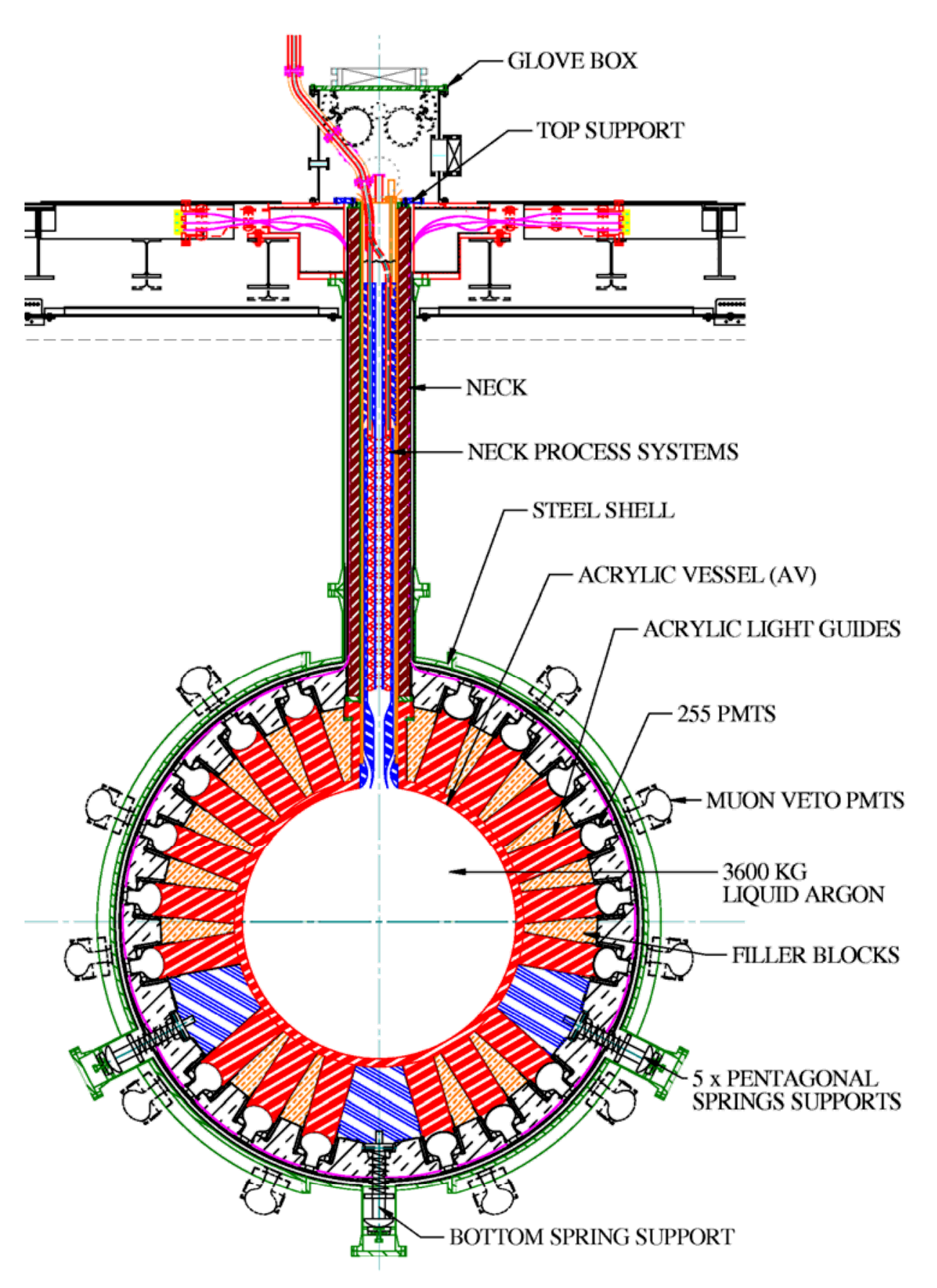}}
\caption{Schematic view of the DEAP-3600 single-phase liquid argon
experiment. The acrylic vessel holds 3,600~kg of LAr, which is viewed
by 255 8-inch photomultipliers through 50-cm light guides. These
provide neutron shielding and thermal insulation between the cryogenic
acrylic vessel and the `warm' PMTs. The inner detector is housed in a
large stainless steel spherical vessel, which itself is immersed in an
8-meter diameter water shielding tank. (From~\cite{Boulay12}, courtesy
M.~Boulay; image credit K.~Dering.) }
\label{fig:DEAP-3600}
\end{figure}

DEAP-3600 (Dark matter experiment with Argon and Pulse shape
discrimination) has been under development since 2005 and is also at
an advanced stage of construction~\cite{Boulay12}. Its 3,600~kg target
(1,000~kg fiducial) of natural LAr is expected to be the first
tonne-scale WIMP detector to operate. The geometry is depicted in
Figure~\ref{fig:DEAP-3600}.
The spherical volume is 85~cm in diameter, defined by a
transparent acrylic vessel coated internally with TPB wavelength
shifter. 255 PMTs (8-inch Hamamatsu R5912) view the target behind
50-cm long acrylic light guides, which keep the PMTs relatively warm
and shield the target against their radioactivity. Plate-out of radon
progeny on the inner surface of the acrylic vessel will be eliminated
before first cool-down by use of a purpose-developed resurfacer device
to remove the top few $\mu$m of acrylic prior to deposition of the TPB
{\it in situ}. An outer containment vessel made from stainless steel
completes the cryostat.

The two instruments will be installed within individual water tanks
side by side at SNOLab's Cube Hall. MiniCLEAN is cooled with a
pulse-tube refrigerator with heat exchanger, whereas DEAP-3600 employs
a liquid nitrogen thermosyphon to liquefy the argon. Both water tanks
are instrumented with photomultipliers to provide a muon veto.  

The success of these experiments is predicated on the achievement of
sufficient discrimination power for the efficient rejection of the
very high rate of electron recoils expected from the decay of
$^{39}$Ar --- a total activity of 3.6~kBq is expected in DEAP-3600,
for example. Work with the MicroCLEAN and DEAP-1 prototypes has
established useful levels of discrimination ($\sim$10$^{-6}$) but the
baseline levels required for the full scale WIMP experiments have so
far only been inferred through model-based extrapolations of
statistically-limited datasets~\cite{Lippincott10,Boulay09}. Work with
DEAP-1 underground is still ongoing, exploring more sophisticated
pulse shape analyses than the fraction of prompt light used initially.

Pulse shape discrimination depends critically on the number of
detected photoelectrons and so the light yield of these chambers is
extremely important. From the performance achieved in the prototype
chambers, along with Monte Carlo simulations of their experimental
set-up, MiniCLEAN expect $\sim$6 photoelectrons per keV for electron
recoils~\cite{Rielage12} and aim for a 50~keV WIMP-search threshold
for 10$^{-9}$ discrimination. For DEAP-3600 the baseline is 8
photoelectrons per keV, along with a 60~keV threshold for nuclear
recoils required to achieve 10$^{-10}$ discrimination~\cite{Boulay12}.

After a 3-year run of DEAP-3600, expected to be free of background,
the project may operate the detector with underground-sourced argon,
affording some gain in WIMP-search threshold due to the lower
requirements on discrimination, and so confirm any signal hints
observed in the first run. MiniCLEAN have adopted a different
strategy: a first 2-year long run with natural argon might be followed
by operation with liquid neon, from which a lower WIMP signal rate is
expected in the canonical elastic scattering model. This will provide
a distinct signature from dominant background sources, which are
expected to increase in rate (e.g.~neutrons) or to remain constant
(surface contamination).

\newpage

\section{Detecting coherent neutrino scattering with the noble liquids}
\label{sec:NeutrinoDetection}

Some progress towards a first measurement of coherent neutrino-nucleus
elastic scattering using liquefied noble gases has also taken place in
recent years. Efforts have focused on two types of very intense source
of low and intermediate energy neutrinos: nuclear reactors and
accelerator-driven neutron spallation sources.

Nuclear power reactors produce an average of 6 antineutrinos per
$^{235}$U fission, with a total of 5$\cdot$10$^{20}$ $\bar{\nu}$/s
emitted from a facility with 3~GW thermal output --- of which several
dozen operate presently around the world~\cite{IAEA}. At standoff
distances of 30~m down to 10~m the antineutrino flux is
$\sim$5--50$\cdot$10$^{12}$~$\bar{\nu}$/cm$^2$/s. Antineutrinos offer
interesting prospects for nuclear non-proliferation activities since
no amount of line-of-sight shielding can mask the count rates observed
by a nearby detector; direct monitoring of the rate of production of
fissile material or the degree of fuel burn-up becomes
possible~\cite{Bernstein02}.

A few suitable spallation sources exist around the world and, to our
knowledge, CNS has been considered only at two, namely the new
Spallation Neutron Source (SNS) at Oak Ridge National Laboratory (US)
and the ISIS facility at the Rutherford Appleton Laboratory (UK). The
potential for neutrino physics at such facilities --- including CNS
--- has long been recognized
(e.g.~\cite{Armbruster98,Avignone03,Scholberg06,Bolozdynya12a}). 
The SNS yields $\sim$10$^{15}$~$\nu$/s (a factor of $\sim$10 higher than at
ISIS). These neutrinos (equal fluxes of
$\nu_e$,$\nu_\mu$,$\bar{\nu}_\mu$) come from the decay-at-rest of
pions and muons produced by the interaction of high-energy protons on
a spallation target. Although reactors provide much higher fluxes, the
pulsed nature of the proton beam (typically $\sim$0.1--1~$\mu$s pulse
width at 40--60~Hz repetition) and higher neutrino energies
(e.g.~30~MeV for $\nu_{\mu}$ versus a few MeV for reactor $\bar{\nu}$)
are clear advantages in this instance. Similar distances to the
spallation target are viable, but shielding against the very
penetrating spallation neutrons must be a prime consideration,
influencing the distance as well as angle to the incident proton beam.

In nature, core-collapse supernovae radiate the overwhelming majority
of their energy as neutrinos and, were one to occur in our galaxy, a
multi-second burst containing multiple events would be observed by the
next generation of DM detectors operating
underground~\cite{Horowitz03}. In general, $^8$B solar neutrinos will
pose a nuclear recoil background for low-threshold DM searches
($\leq$3~keV) already at WIMP-nucleon cross sections of
$\sim$10$^{-10}$~pb (10$^{-46}$~cm$^2$)~\cite{Munroe07}. For higher
thresholds, atmospheric neutrinos and diffuse supernova neutrino
background (DSNB) --- the past history of all supernova explosions ---
will dominate below $\sim$10$^{-12}$~pb
(10$^{-48}$~cm$^2$)~\cite{Strigari09}.

The experimental challenge is manifest in 
Figure~\ref{fig:WIMPrates},
which depicts expected CNS event rates at ISIS and at a nuclear reactor. 
Extremely low thresholds are required to achieve sensible event rates at either
location. The trade-off mentioned in 
Section~\ref{sec:OverviewDetectionPrinciples}
between scattering rate
($\propto E_\nu^2\,N^2/A$) and the recoil spectrum ($E_r^{max} \propto
E_\nu^2/A$), with $N$ and $A$ the neutron and atomic numbers of the
target, respectively, is patent in the figure. Xenon works best with
the harder neutrino spectrum of the spallation source ($E_\nu <
53$~MeV), whereas argon (or, better still, neon) would be better
suited for reactor antineutrinos ($E_\nu \lesssim 10$~MeV).

Naturally, both types of facility are extremely powerful neutron
sources too. Extensive shielding is required, and an anti-coincidence
neutron detector is probably essential. Difficulties associated with a
surface deployment compound the problem. Atmospheric muons may be
disruptive, especially for double-phase detectors: a relatively small
detector with 60~cm diameter would experience $\sim$50~$\mu$/s
depositing tens of MeV per event, whilst conducting a search for rare
events at keV energies. In addition, cosmic-ray neutrons as well as
those arising in muon-initiated cascades are serious backgrounds too.

The prospect of deploying noble liquid detectors specifically for
measuring CNS has been studied by a few groups around the world. We
mention briefly the most advanced below --- whilst noting that other
measurement technologies, notably cryogenic germanium, are also
contenders for a first observation.

\vspace{5mm}

\begin{figure}[t]
\centerline{\includegraphics[width=0.9\textwidth]{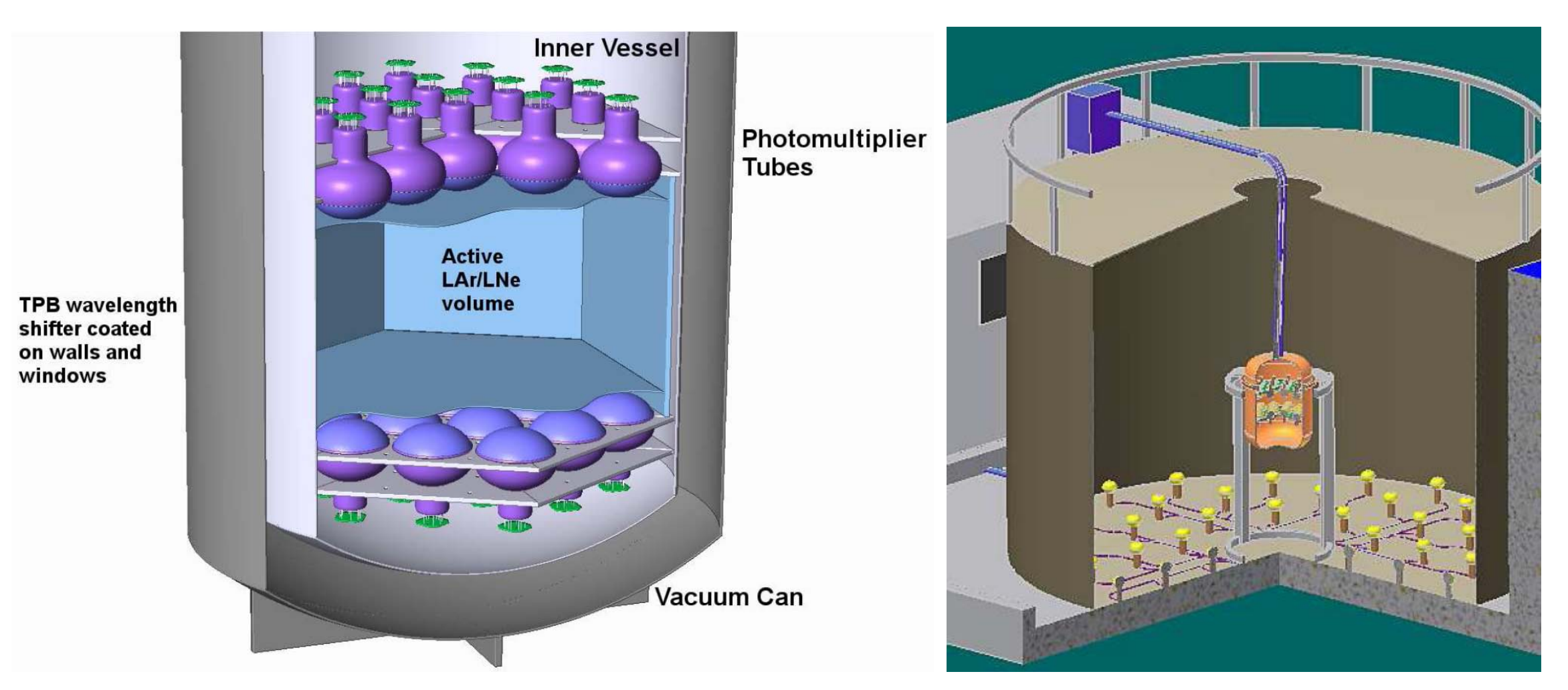}}
\caption{Diagram of the CLEAR experiment proposed for SNS at Oak
Ridge. Left: the LAr/LNe scintillation target is defined by a
TPB-coated chamber viewed by 38~PMTs immersed in the liquid, contained
in a vacuum cryostat (image credit J.~Nikkel). Right: the detector
would be located in an instrumented water tank for shielding and muon
veto (image credit J.~Fowler). (From~\cite{Scholberg09}, courtesy
K.~Scholberg.)}
\label{fig:CLEAR}
\end{figure}

{\bf The CLEAR Experiment} (Coherent Low Energy A (Nuclear) Recoils)
was conceived for deployment at SNS, at a distance of 46~m to the
spallation target and in the backward direction with respect to the
proton beam. A single-phase, scintillation-only design was adopted,
allowing interchangeable liquid argon and liquid neon targets (456~kg
and 391~kg, respectively), within an active volume 60~cm in diameter
and 44~cm in height~\cite{Scholberg09}. Significant recoil energies
can result in these targets, especially in LNe where the recoil
spectrum extends to $>$100~keV, allowing use of pulse shape
discrimination from the scintillation signal. The VUV light is
wavelength-shifted by TPB evaporated onto the PTFE chamber walls as
well as two fused silica or acrylic plates, and detected by
38~Hamamatsu R5912-02MOD photomultipliers 
(Figure~\ref{fig:CLEAR}).
The detector
would be located inside an instrumented water tank for shielding and
cosmic-ray veto.

A significant event rate is expected in this configuration. Assuming
SNS operation at a power of 1.4~MW, a live running time of
2.4$\cdot$10$^7$~s/yr for each target material, and a beam timing cut
of 6$\cdot$10$^{-4}$ efficiency, CLEAR should register
$\sim$600~events/year with LAr above a 20~keV nuclear recoil
threshold, and $\sim$250~events/year for a LNe run with 30~keV
threshold (including 50\% recoil acceptance after pulse shape
discrimination). The dominant background is from $^{39}$Ar, PMT
$\gamma$-rays, radon daughters and neutrons produced from beam loss in
the transport line (which were found to be higher than those from the
spallation target itself). Although there has been no progress
recently due to lack of funding, the Collaboration still deem the
project to be technically viable.

\vspace{5mm}

\begin{figure}[ht]
\centerline{\includegraphics[width=0.43\textwidth]{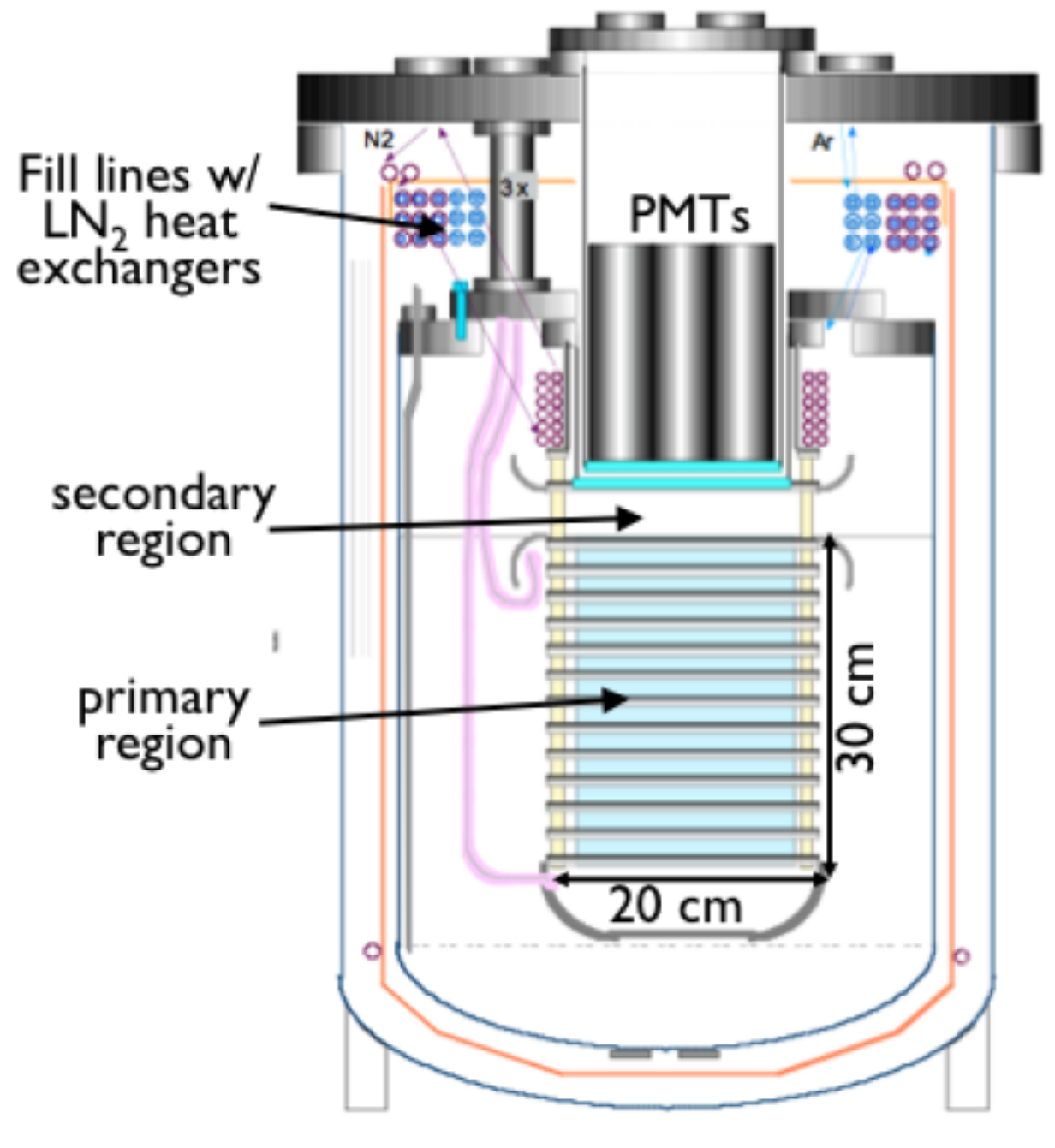}}
\caption{Possible design of a 10-kg double-phase Ar detector for
deployment at a nuclear power reactor. Measurement of CNS will be
attempted in the 1--10 ionization electron regime. (Courtesy
A.~Bernstein.)  }
\label{fig:Bernstein}
\end{figure}

{\bf A very different and innovative approach was proposed in
2004~\cite{Hagmann04}}: if measurement of single ionization electrons
could be achieved through proportional scintillation in double-phase
argon, this would open the way to utilizing these detectors in
electron counting mode (i.e.~sub-threshold in S1). This started a
LAr-based programme at Lawrence Livermore National Laboratory (US)
aimed specifically at nuclear reactor monitoring. Deployment in the
`tendon gallery' of a 3.4~GW$_\mathrm{th}$ reactor, 25~m away from the
core, is being considered. A schematic of a possible prototype
detector is shown in 
Figure~\ref{fig:Bernstein}.
This might achieve $\sim$80 counts/day
above a threshold of 2~ionization electrons measured by proportional
scintillation. $^{39}$Ar is again a problematic background also in
this regime, which would require use of underground argon. Preliminary
work with a smaller chamber is underway, in order to characterize the
response to single electrons and to measure the ionization yield in
LAr at very low energies~\cite{Bernstein11,Sangiorgio10}.

\vspace{5mm}

\begin{figure}[ht]
\centerline{\includegraphics[width=0.6\textwidth]{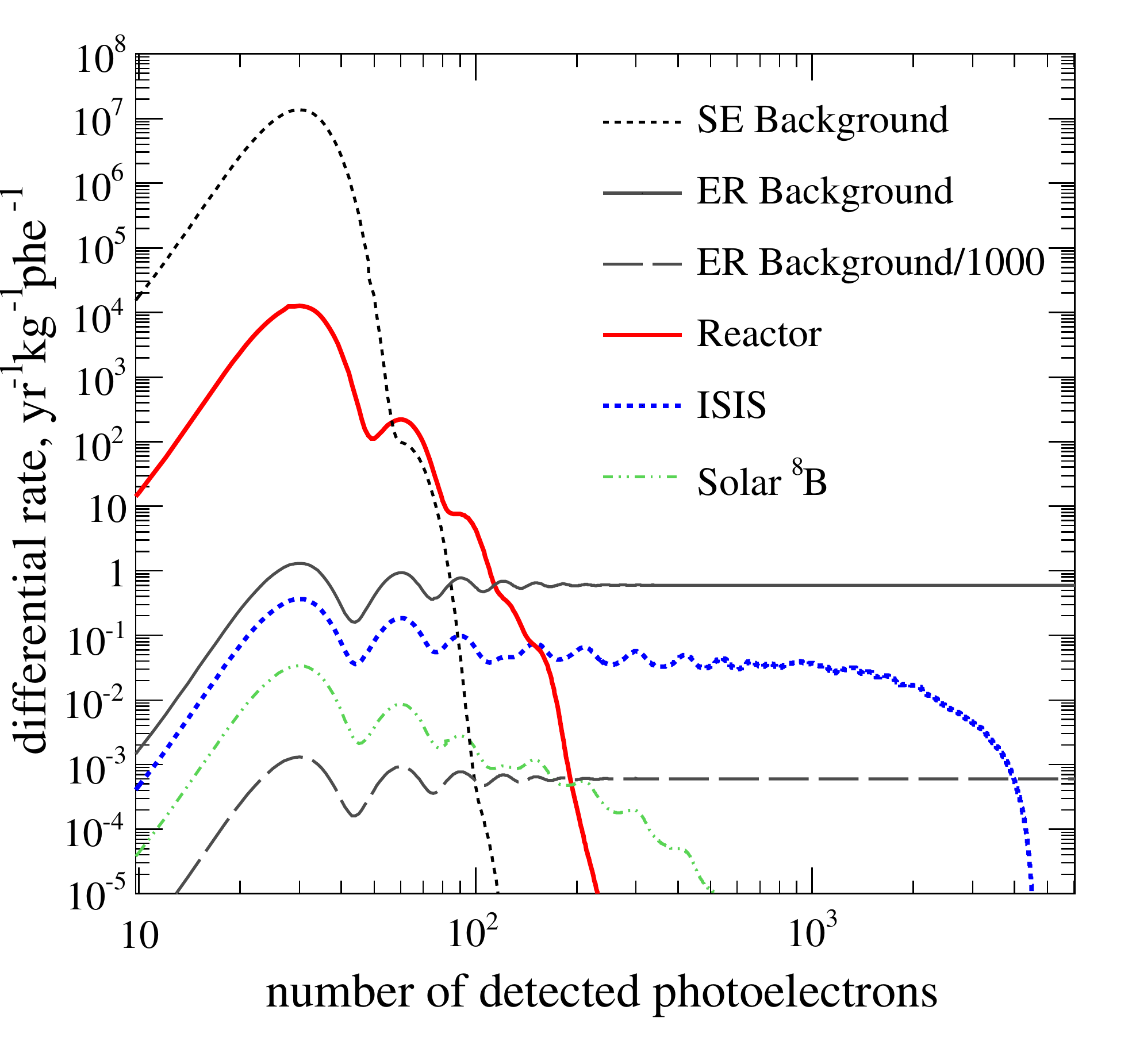}}
\caption{Neutrino-induced ionization spectra expected in the liquid
target of ZEPLIN-III at a nuclear reactor and at the ISIS stopped pion
source. CNS from solar $^8$B is also shown for reference. The peak
structure reflects discrete numbers of ionization electrons measured
by electroluminescence. Single electron background and internal
electron recoil backgrounds (1~dru) are also shown; the latter is
reduced by $\sim$1000 times at ISIS due to the beam-coincident
measurement~\cite{Santos11}. (Courtesy ZEPLIN-III Collaboration; with
permission from Springer Science and Business Media.)  }
\label{fig:CNSRatesZ3}
\end{figure}

{\bf With this type of measurement in mind, the ZEPLIN program}
characterized the single electron signature and its production
mechanisms in double--phase xenon: first using ZEPLIN-II data~\cite{Edwards08}, then with
ZEPLIN-III~\cite{Santos11} (see also 
Figure~\ref{fig:SEinZ3}).
Similar work with a surface prototype was conducted at ITEP
(Russia)~\cite{Burenkov09}. The excellent signal-to-noise ratio
achieved on the single electron signature and the low background
measured at two electrons and above led the ZEPLIN-III Collaboration to
explore further a deployment of this detector at a nuclear reactor and
at the ISIS facility (UK)~\cite{Santos11}. The predicted CNS signal
and dominant internal backgrounds, as measured through proportional
scintillation with a yield of 30~photoelectrons per electron, are
reproduced in 
Figure~\ref{fig:CNSRatesZ3}.
These assume ionization yields as measured
down to 4~keV recoil energy~\cite{Horn11} and extrapolated to zero
in an {\it ad hoc} way below that.

It was found that the reactor signal was salient over background above
a 3-electron threshold, with $\gtrsim$2,000~events expected in
10~kg$\cdot$yr live exposure (at 10~m from a 3~GW$_\mathrm{th}$
reactor core). These signals are below the scintillation threshold, so
no discrimination can be relied upon.

The prospects for detecting this signature with a beam-coincident
measurement at the stopped pion source seemed also encouraging (at
10~m from the spallation target, with a neutrino flux of
$\sim$3.6$\cdot$10$^6$~cm$^{-2}$s$^{-1}$ per flavor). Owing to the
much harder energy spectrum, approximately half of the detected events
should produce measurable S1 and S2 signals, thus allowing
three-dimensional position reconstruction and electron/nuclear recoil
discrimination; this would provide an important cross-check for the
part of the spectrum detected through ionization only. At ISIS,
non-beam related backgrounds could be significantly reduced with a
beam coincidence cut with 10$^{-3}$ efficiency --- corresponding to a
measurement window of 20~$\mu$s every 20~ms: $\sim$700 signal events
were expected above 3~emitted electrons in a 10~kg$\cdot$yr exposure.


\begin{figure}[b]
\centerline{\includegraphics[width=0.8\textwidth]{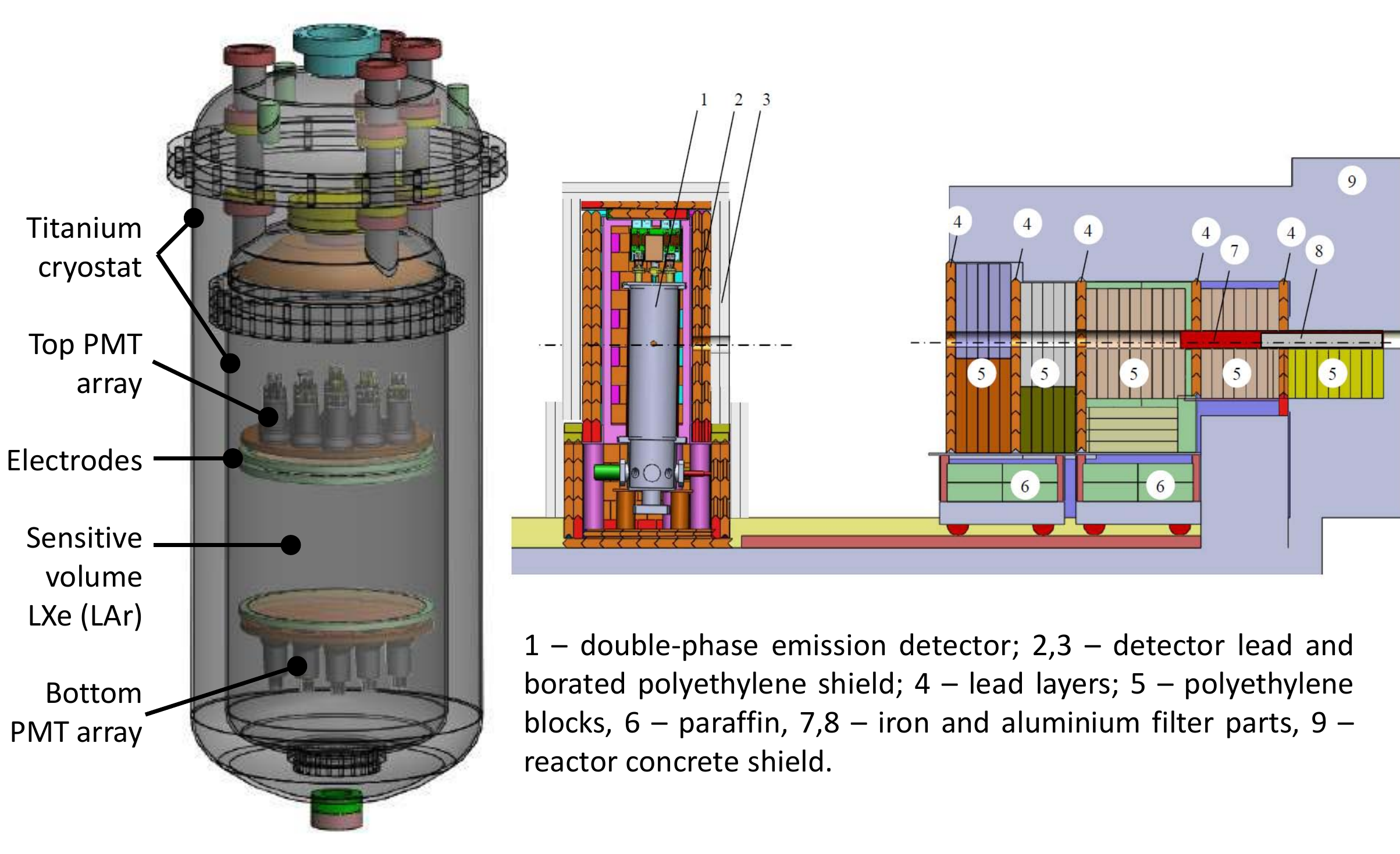}}
\caption{The RED-100 double-phase detector is shown left; it will
contain 200(100)~kg of LXe(LAr). The right panel shows the smaller
prototype chamber exposed to 24~keV neutrons at the MEPhI nuclear
reactor, to study the response to very low energy nuclear
recoils. (Courtesy RED Collaboration.)}
\label{fig:RED}
\end{figure}

The possibility of an ISIS deployment was subsequently studied in more
detail by one of us (HA), after a possible location for the experiment
was identified near the spallation target, at 90$^\mathrm{o}$ to the
proton beam. The target-target distance would be 8.66~m, with the
line-of-sight shielding including an existing $\approx$5~m of iron and
$\approx$1~m of concrete; additional neutron shielding would be added
to the full shield used underground at Boulby (including the neutron
veto). With realistic parameters considered for the accelerator, the
spallation target and the experiment, the signal rate expected in a
6~kg fiducial LXe volume was only 52~events/year. Cosmogenic and
internal neutrons were studied by Monte Carlo simulation; they were
found to contribute $\sim$30~unvetoed events/year in the recoil energy
range 0.1--10~keV within the beam coincidence window, but these could
be further subtracted with an off-beam measurement. Electron recoils
backgrounds would add a further $\sim$10~events/year before
discrimination. However, beam-coincident spallation neutrons above
20~MeV were found to overwhelm a measurement at that location,
irrespective of any additional shielding. A 1-$\mu$s delayed
measurement targeting follow-on muon-decay neutrinos only (pion decay
neutrinos are essentially prompt) was proposed to solve this problem
for SNS neutrino studies~\cite{nuSNS05}. This would eliminate
high-energy beam-induced backgrounds, as delayed neutrons can be
shielded effectively. However, this is only possible for events
containing both S1 and S2 pulses (since S1 carries the event timing
information). The overall timing cut would retain $\sim$25\% signal
acceptance, leaving 13~events per year. This rate was deemed too low
to justify pursuing the experiment further at that particular location
--- at least with that target mass, which was the main limiting factor
in this instance.

\vspace{5mm}

{\bf Finally, the RED Collaboration} (Russia) is pursuing a promising
effort for a first detection of CNS. They propose a LXe/LAr experiment
named RED-100 (Russian Emission Detector). The baseline design
features 200~kg of liquid xenon, with approximately 100~kg of fiducial
mass; liquid argon is also a possible target, with 100~kg of total
mass for a 50~kg fiducial.

The detector design is illustrated in 
Figure~\ref{fig:RED}.
The sensitive volume
is 45~cm in diameter and 45~cm in height, defined by optically
transparent mesh electrodes and lateral field-shaping rings. The
envisaged drift field is 0.5--1~kV/cm in the liquid and 7--10~kV/cm in
the 1-cm thick gas phase. The team plan to use two arrays of
low-background Hamamatsu R11410-10 photomultipliers (19~phototubes
each). The mean expected number of photoelectrons per electron in the
secondary scintillation region is 80~\cite{Akimov12e}.

To study the response of LXe and LAr to sub-keV nuclear recoils, the
teams are preparing a setup at the MEPhI research reactor. This will
include a neutron filter to prepare a quasi-monochromatic 24~keV
neutron beam, irradiating a smaller double-phase chamber (this was one
of the prototypes for ZEPLIN-III, with 7~photomultipliers). This
experiment will allow the measurement of the ionization and
scintillation yields with maximum recoil energies of $\sim$0.7~keV for
Xe and $\sim$2.3~keV for Ar.

Two deployments of RED-100 are envisaged. Firstly, at 19~m from the
core of the Kalininskaya nuclear power station, where the antineutrino
flux is 1.35$\cdot$10$^{13}$~cm$^{-2}$s$^{-1}$; $\sim$20 events/day
are expected in 100~kg of LXe above a 2~ionization electron threshold,
and $\sim$200~events/day in 100~kg of LAr. An experiment at SNS is
also being considered, 40~m from the spallation target and 10~m
underground; the neutrino flux there would be
5$\cdot$10$^6$~cm$^{-2}$s$^{-1}$ (per flavor). The expected count
rates are $\sim$1,400 and $\sim$400~events/100~kg/year for LXe and
LAr, respectively.

\section{Conclusion}
\label{sec:Conclusion}

Liquefied noble gas technology has been developed into well
established radiation and particle detection techniques, having found
a wide range of application. Technical challenges have been overcome
over the last decade to demonstrate that these media can provide low
energy threshold, low background, target mass scalability and stable
operation over long periods. In particular, double-phase liquid/gas
detectors are now at the forefront of low energy particle
physics. They are top contenders for a first direct detection of WIMP
dark matter as well as of coherent neutrino-nucleus elastic
scattering.

\begin{figure}[ht]
\vspace{-5mm}
\centerline{\includegraphics[width=0.7\textwidth]{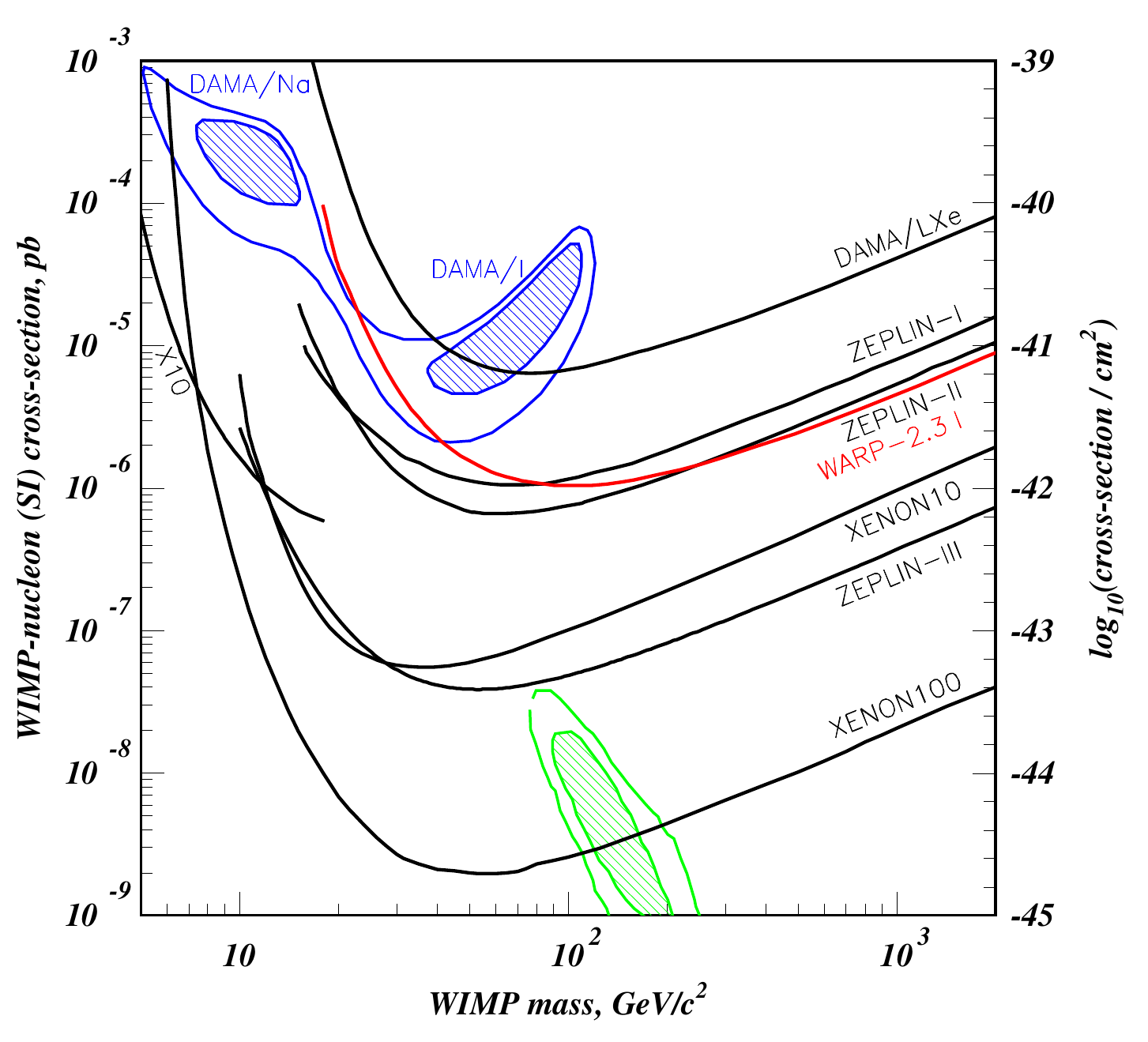}}
\vspace{-3mm}
\caption{Spin-independent WIMP-nucleon scattering cross-section limits
(90\% CL) published from liquid xenon (black) and liquid argon
experiments (red); in order of publication:
DAMA/LXe~\cite{Bernabei98}, \mbox{ZEPLIN-I}~\cite{Alner05},
\mbox{ZEPLIN-II}~\cite{Alner07}, \mbox{WARP~2.3-l}~\cite{Benetti08},
XENON10~\cite{Angle08,Aprile09} (low energy analysis~\cite{Angle11}),
\mbox{ZEPLIN-III}~\cite{Akimov12a} and XENON100~\cite{Aprile12a}. The
parameter space favored by Constrained MSSM after 1~fb$^{-1}$ of LHC
data is shown in green~\cite{Buchmueller11}. The DAMA/NaI annual
modulation result~\cite{Bernabei08} interpreted as a nuclear recoil
WIMP signal in~\cite{Gondolo09} is also shown (in blue) for
reference.}
\label{fig:DMResultsSI}
\end{figure}

In dark matter searches, xenon detectors have made the most progress
so far, but liquid argon experiments are also close to data-taking
underground; liquid neon setups are being considered. We present in
Figure~\ref{fig:DMResultsSI}
WIMP-nucleon exclusion limits reported by liquid noble gas
experiments. These have covered over three orders of magnitude in
scalar cross-section sensitivity in just over a decade, and there are
signs that this rate is accelerating as self-shielding begins to pay
off with the larger target masses now being deployed. XENON100 offers
presently the tightest experimental constraint from any technology.

Prospects for a first detection of CNS have also improved due to the
excellent sensitivity afforded by the ionization response in
double-phase detectors. This quest goes hand in hand with the search
for light WIMPs, both involving sub-keV detection thresholds in
scalable target technologies. A first measurement of CNS may be close,
although some challenges remain in controlling neutron backgrounds at
the neutrino sources.

Practically all current dark matter experiments using the double-phase
technique rely on measurement of the ionization signal via secondary
scintillation in a uniform electric field detected by a large number
of photomultiplier tubes. The quest for sensitivity stimulated
development of new PMTs capable of operating at cryogenic temperatures
and having high quantum efficiency for xenon VUV light. In the last
two decades, the quantum efficiency at these wavelengths has
doubled. Progress in reducing the radioactivity background has been
even more remarkable --- a factor of $\sim$1,000 has been achieved over
this period.

Alternative readout techniques, such as solid state photon detectors
and micro-pattern detectors for charge amplification, have continued
to be developed. The case for these approaches over the traditional
PMT readout has not been made so far --- at least at low energies.
However, at higher energies and especially for very large detectors
new solutions may be advantageous or even necessary (e.g.~LAr TPCs for
neutrino detection).

The success of experiments that rely on the double-phase technique
depends on a good understanding of the physics involved in the
detection process. In our view most processes are now well understood;
nevertheless, some knowledge gaps remain in the keV and sub-keV energy
regions, in particular regarding the response to nuclear
recoils. Exploration of this energy regime will be essential to enable
a first measurement of CNS, and these experiments are now contributing
to this effort.

In the year 2012 we celebrated the silver jubilee of direct dark
matter searches~\cite{Ahlen87}. The noble liquids have now taken a
clear lead in trying to answer this most important of scientific
questions. The authors conclude by reaffirming their belief that the
double-phase technique will maintain a strong position in low energy
particle physics in the future.


\acknowledgments

The authors are grateful to E.~Aprile, D.~McKinsey,  A.~Mangiarotti 
and A.~Hitachi for fruitful discussions.
We also wish to thank R.~Ferreira~Marques, P.~Mendes, A.~Currie 
and C.~Ghag for reading the manuscript and for useful suggestions. 
A special thank you is due to A.~Breskin, who
inspired this work and patiently supported it through the whole
writing process.

\bibliographystyle{MJ1}
\bibliography{RevBib.v4}

\end{document}